\documentclass[journal=jcisd8,manuscript=article]{achemso}
\setkeys{acs}{articletitle = true, maxauthors = 0}

\usepackage[version=3]{mhchem} %
\usepackage{amsfonts}
\usepackage{makecell}
\usepackage{graphicx}
\usepackage{caption}
\usepackage{subcaption}
\usepackage[hidelinks=true]{hyperref}
\usepackage{placeins}
\usepackage{longtable}
\usepackage{mathrsfs}
\usepackage{url}
\usepackage{todonotes}
\usepackage{mathtools}
\usepackage{multirow}
\usepackage{gensymb}

\usepackage{xr}

\makeatletter
\newcommand*{\addFileDependency}[1]{%
  \typeout{(#1)}
  \@addtofilelist{#1}
  \IfFileExists{#1}{}{\typeout{No file #1.}}
}
\makeatother

\newcommand*{\myexternaldocument}[1]{%
    \externaldocument{#1}%
    \addFileDependency{#1.tex}%
    \addFileDependency{#1.aux}%
}

\myexternaldocument{supp_standalone}

\usepackage{color}

\author{Katherine S. Lim}
\affiliation[MIT]{Department of Electrical Engineering and Computer Science, Massachusetts Institute of Technology, 77 Massachusetts Avenue, Cambridge, Massachusetts 02139, United States}
\alsoaffiliation[MIT3]{Department of Biology, Massachusetts Institute of Technology, 77 Massachusetts Avenue, Cambridge, Massachusetts 02139, United States}

\author{Andrew G. Reidenbach}
\affiliation[Broad]{Chemical Biology and Therapeutics Science Program, Broad Institute, 415 Main Street, Cambridge, Massachusetts 02142, United States}

\author{Bruce K. Hua}
\affiliation[Harvard]{Department of Chemistry and Chemical Biology, Harvard University, 12 Oxford Street, Cambridge, Massachusetts 02138, United
States}
\alsoaffiliation[Broad]{Chemical Biology and Therapeutics Science Program, Broad Institute, 415 Main Street, Cambridge, Massachusetts 02142, United States}

\author{Jeremy W. Mason}
\affiliation[Broad]{Chemical Biology and Therapeutics Science Program, Broad Institute, 415 Main Street, Cambridge, Massachusetts 02142, United States}
\alsoaffiliation[Novartis]{Novartis Institutes for BioMedical Research, Cambridge, Massachusetts 02139, United States}

\author{Christopher J. Gerry}
\affiliation[Harvard]{Department of Chemistry and Chemical Biology, Harvard University, 12 Oxford Street, Cambridge, Massachusetts 02138, United
States}
\alsoaffiliation[Broad]{Chemical Biology and Therapeutics Science Program, Broad Institute, 415 Main Street, Cambridge, Massachusetts 02142, United States}

\author{Paul A. Clemons}
\email{pclemons@broadinstitute.org}
\affiliation[Broad]{Chemical Biology and Therapeutics Science Program, Broad Institute, 415 Main Street, Cambridge, Massachusetts 02142, United States}

\author{Connor W. Coley}
\email{ccoley@mit.edu}
\affiliation[MIT]{Department of Electrical Engineering and Computer Science, Massachusetts Institute of Technology, 77 Massachusetts Avenue, Cambridge, Massachusetts 02139, United States}
\alsoaffiliation[MIT2]{Department of Chemical Engineering, Massachusetts Institute of Technology, 77 Massachusetts Avenue, Cambridge, Massachusetts 02139, United States}
\alsoaffiliation[Broad]{Chemical Biology and Therapeutics Science Program, Broad Institute, 415 Main Street, Cambridge, Massachusetts 02142, United States}
\title{Machine learning on  DNA-encoded library count data using an uncertainty-aware probabilistic loss function} %

\begin{document}

\begin{abstract}

DNA-encoded library (DEL) screening and quantitative structure-activity relationship (QSAR) modeling are two techniques used in drug discovery to find novel small molecules that bind a protein target. Applying QSAR modeling to DEL selection data can facilitate the selection of compounds for off-DNA synthesis and evaluation. Such a combined approach has been done very recently by training binary classifiers to learn DEL enrichments of aggregated ``disynthons'' in order to accommodate the sparse and noisy nature of DEL data. However, a binary classification model cannot distinguish between different levels of enrichment, and information is potentially lost during disynthon aggregation. 
Here we demonstrate a regression approach to learning DEL enrichments of individual molecules, using a custom negative-log-likelihood loss function, that effectively denoises DEL data and introduces opportunities for visualization of learned structure-activity relationships. Our approach explicitly models the Poisson statistics of the sequencing process used in the DEL experimental workflow under a frequentist view. %
We illustrate this approach on a DEL dataset of 108,528 compounds screened against carbonic anhydrase (CAIX), and a dataset of 5,655,000 compounds screened against soluble epoxide hydrolase (sEH) and SIRT2. Due to the treatment of uncertainty in the data through the negative-log-likelihood loss used during training, the models can ignore low-confidence outliers. While our approach does not demonstrate a benefit for extrapolation to novel structures, %
we expect our denoising and visualization pipeline to be useful in identifying structure-activity trends and highly enriched pharmacophores in DEL data. Further, this approach to uncertainty-aware regression modeling is applicable to other sparse or noisy datasets where the nature of stochasticity is known or can be modeled; in particular, the Poisson enrichment ratio metric we use can apply to other settings that compare sequencing count data between two experimental conditions.

\end{abstract}

\section{Introduction}

The discovery of new small-molecule therapeutics or chemical probes often starts with finding compounds with affinity to a protein of interest. %
Most existing medicines depend on such interactions between small molecules and therapeutically relevant proteins \cite{schreiber_chemical_2019,imming_drugs_2006}. To find  molecules that selectively bind a protein of interest, it can be valuable to synthesize and screen diverse libraries of small molecules. DNA-encoded libraries (DELs) are one technology for achieving this goal \cite{clark_design_2009, kleiner_small-molecule_2011, goodnow_dna-encoded_2017, flood_dna_2020}. 

\begin{figure}
    \centering
        \includegraphics[scale=1]{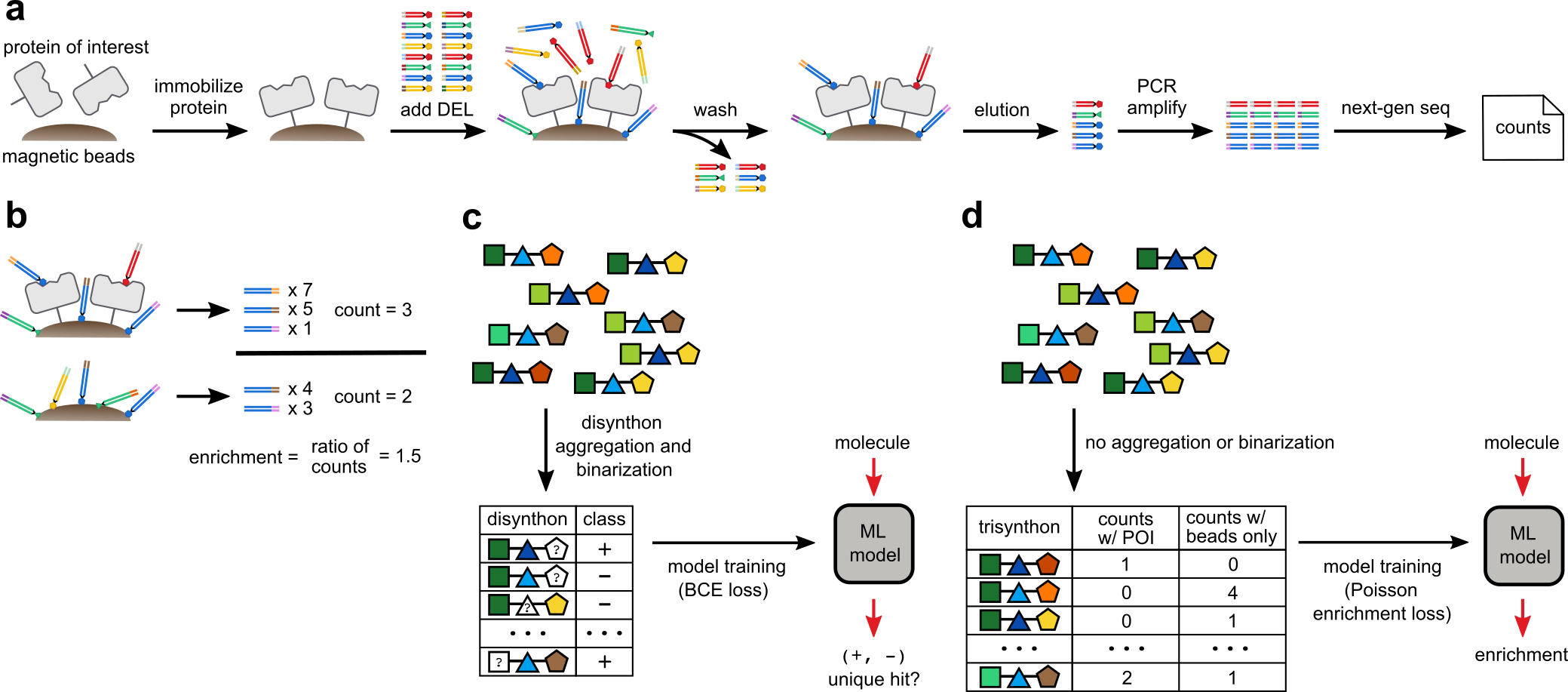}
    \caption{Approaches to analyzing DNA-encoded library (DEL) selection data to infer structure-activity relationships. \textbf{(a)} DEL experimental workflow. \textbf{(b)} Classic way of getting hits: raw normalized counts. \textbf{(c)} \citeauthor{mccloskey_machine_2020}'s approach: binary classification of ``disynthons''\cite{mccloskey_machine_2020}. \textbf{(d)} Our approach: regression task on ``trisynthons,'' taking uncertainty into account. The trisynthon drawing style was adapted from \citeauthor{mccloskey_machine_2020}).}
    \label{fig:approach_formulation}
\end{figure}

DELs are constructed by split-pool synthesis, which involves generating a combinatorially large number (up to trillions) of small molecules attached to single- or double-stranded DNA “barcodes” used to identify
the molecules. These DNA-tagged molecules are then screened for binding against a protein of interest (POI). To do this, the POI is immobilized, %
and the DEL molecules are %
incubated with the POI. Molecules that  have weaker affinity for the POI (non-binders) are removed by washing, and the remaining stronger binders are eluted. The binders are then identified via next-generation sequencing of their DNA tags after amplification by PCR, and the sequencing reads are processed into \emph{counts} (number of observations) for each barcode (Figure~\ref{fig:approach_formulation}a). These \emph{counts} are used to calculate an \emph{enrichment} value for each small molecule after normalization (e.g., to a beads-only control condition), with a higher enrichment value ostensibly indicating higher binding affinity to the POI (Figure~\ref{fig:approach_formulation}b), although a direct quantitative link has not been established \cite{franzini_identification_2015, denton_robustness_2018-1}, potentially due to confounding factors such as chemical yields during library synthesis, baseline abundances of library members, stringency of washing, and PCR sensitivity \cite{satz_dna_2015-1, satz_simulated_2016, satz_analysis_2017, sannino_quantitative_2019, hall_simple_2020, foley_selecting_2021}.
A subset of compounds are selected for follow-up investigation via on- or off-DNA validation binding assays. These compounds may be determined based on various factors, such as high enrichments, raw counts, selective enrichment across related protein targets, enrichment across multiple protein concentrations, chemical yield, structural similarity to other predicted hits, and physicochemical properties \cite{clark_design_2009, franzini_interrogating_2015, kung_characterization_2020, hall_simple_2020, petersen_novel_2016, foley_selecting_2021}. %
However, count values in DEL screens are generally noisy and sparse, meaning that exclusively following up on the molecules with highest enrichment values (potentially with additional filtering based on other properties, as mentioned above) may leave out many promising molecules and overlook potentially useful structure-activity relationship (SAR) information. Aggregation of enrichment data, e.g., at the level of ``monosynthons'' (all molecules with one building block in common) or ``disynthons'' (all molecules with two building blocks in common), may help denoise counts; however, information about individual molecules (i.e., ``trisynthons'' for libraries constructed from three families of building blocks) is lost in the process of aggregation. In addition, the top hits may be challenging to synthesize off DNA, have poor predicted solubility, or be the product of prospective side reactions, limiting the ability of researchers to validate their perceived affinity.

To address these issues, quantitative structure-activity relationship (QSAR) modeling can
enable a more automated and  holistic way of codifying relationships between
molecule structure and protein-binding activity. QSAR modeling typically involves learning a mathematical relationship based on labeled (structure, activity) data \cite{cherkasov_qsar_2014, muratov_qsar_2020-1}. In the case of DEL data, QSAR modeling might  provide a smooth, continuous mapping from molecular structure to enrichment as a proxy measure of protein-binding affinity, with the caveat that the intended structure of a DEL member may not have been what was actually synthesized. Such a model could be used in a virtual screening setting to evaluate libraries of candidate molecules, including ones not present in the original DEL. %

While QSAR modeling is ubiquitous in small-molecule drug discovery, it had not been applied to DELs
until very recently \cite{mccloskey_machine_2020}. %
McCloskey et al. applied QSAR modeling to DEL data %
by grouping molecules and using their aggregated enrichment data to label each group as a competitive binder to the POI or not---a binary classification problem (Figure~\ref{fig:approach_formulation}c)\cite{mccloskey_machine_2020}. Aggregating molecules at the level of disynthons partially mitigates the sparsity of DEL data and increases the certainty of assigned labels. QSAR models were trained on good/bad disynthons and used to screen compounds in a virtual make-on-demand library, ultimately leading to the successful identification of %
hits with micromolar activity. However, two potential downsides to this modeling approach are that (a) models cannot distinguish between different levels of enrichment (e.g., values of 1.5 and 10.0 are both ``enriched'' in a binary sense but not equally so) and (b) information about the enrichment of individual molecules is lost during aggregation. %

\begin{figure}
    \centering
        \includegraphics[scale=1]{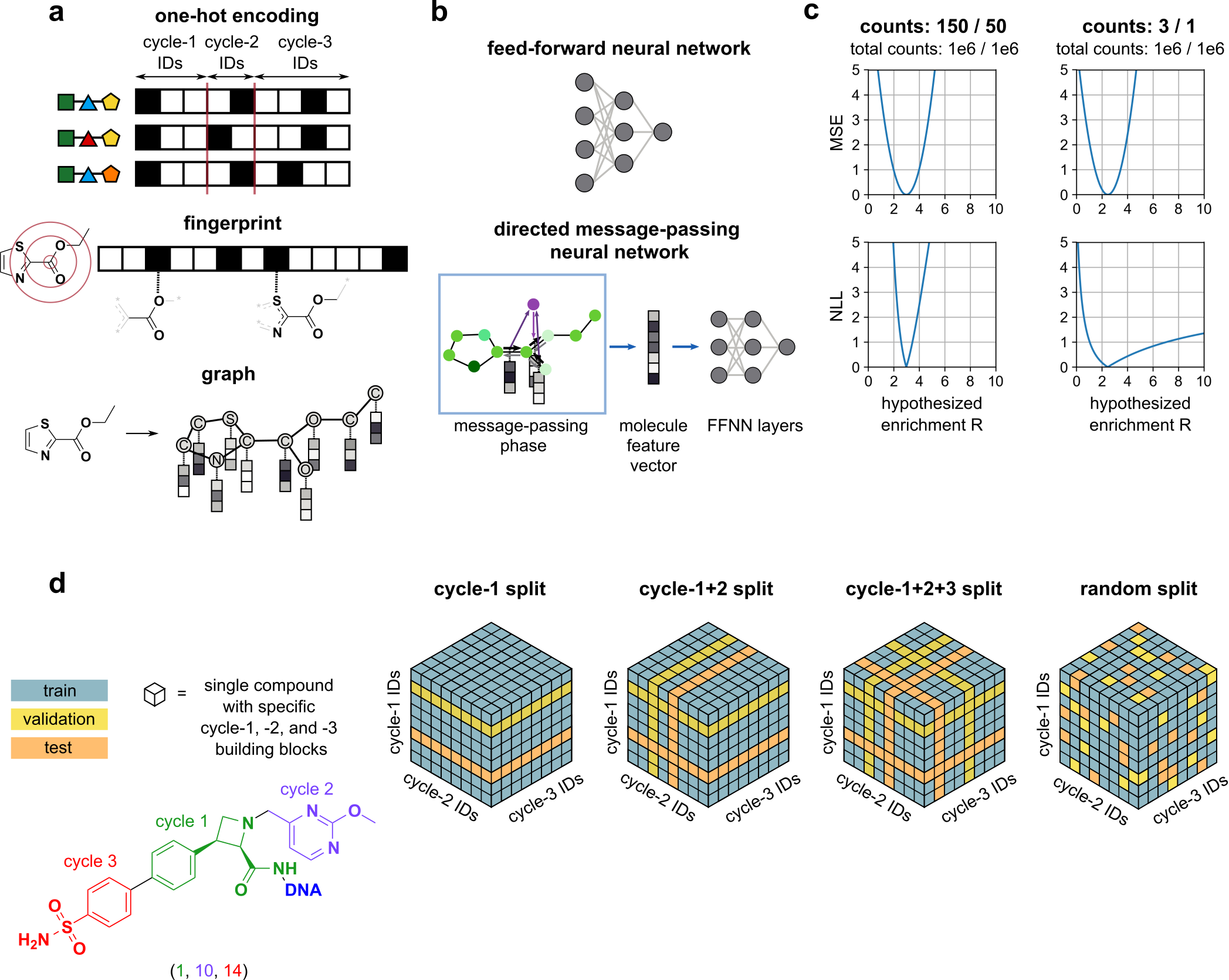}
    \caption{\textbf{(a)} Molecule representations. \textbf{(b)} Model architectures (message passing network drawing style adapted from \citeauthor{wu_moleculenet_2018}). \textbf{(c)} Training objectives / loss functions for theoretical count values. \textbf{(d)} How molecules in a dataset are divided for model training and evaluation (data splits). NLL: negative-log likelihood; MSE: mean-squared error.}
    \label{fig:reps_models_loss_fns_splits}
\end{figure}

In this work, we explore an alternate approach to combining QSAR modeling and DEL data in a regression formulation that avoids both the binarization and aggregation of enrichment data (Figure~\ref{fig:approach_formulation}d). While we do not demonstrate ability to generalize to new chemical structures as \citeauthor{mccloskey_machine_2020}\cite{mccloskey_machine_2020} achieved, we do show that the models can effectively denoise DEL data due to the novel uncertainty-aware loss function used during training. Such smoothing of DEL data may allow for the elucidation of patterns in SAR and weaker binders that would otherwise be obscured by noise. We also present a pipeline for visualization of the learned SAR of our fingerprint-based models, which in general can be used to improve the utility of such models' predictions for medicinal chemists. Specifically, understanding the learned SAR of such models increases confidence in top predicted compounds and allows for rational modification of structures to improve synthetic accessibility or solubility as needed to facilitate experimental validation off-DNA. In our approach, we use a probabilistic loss function that we argue is more suitable to modeling DEL data than standard regression loss functions like mean-squared error (MSE) due to the stochastic nature of how the data are acquired. Because the next-generation sequencing step of DEL synthesis corresponds well with a Poisson distribution \cite{bentley_accurate_2008,simons_convergence_1971, kuai_randomness_2018}, we can define a loss function that accounts for the uncertainty 
that results from Poisson sampling. Mathematically, the goal is for the model to predict enrichment values that minimize the negative log-likelihood (NLL) of failing to reject a null hypothesis that the true enrichment ratio equals the predicted enrichment ratio  under a two-sided test given  the observed barcode counts that we measure in a DEL experiment; the true enrichment value for each molecule corresponds to the (unknown) ratio of the molecule's relative abundance after affinity selection to the molecule's relative abundance in a beads-only control experiment. We provide an example of this loss function's treatment of uncertainty (Figure~\ref{fig:reps_models_loss_fns_splits}c; Methods).

We evaluated and compared model performance using three different molecule representations (Figure~\ref{fig:reps_models_loss_fns_splits}a; one-hot encoding of building blocks, Morgan circular fingerprints \cite{rogers_extended-connectivity_2010}, and molecular graphs)%
, two different model types (Figure~\ref{fig:reps_models_loss_fns_splits}b; feed-forward neural networks and directed message-passing networks \cite{yang_analyzing_2019}), and eight data splits---ways of dividing molecules in a dataset for model training and evaluation (Figure~\ref{fig:reps_models_loss_fns_splits}d; Methods; random split and various cycle splits). 
We trained and evaluated these QSAR models on two DELs containing 108,528 and 5,655,000 compounds corresponding to DOS-DEL-1 from \citeauthor{gerry_dna_2019}\cite{gerry_dna_2019} (abbreviated here as DD1S; in this paper, the names DD1S and DOS-DEL-1 are used interchangeably) and a derivative of DEL-A from \citeauthor{clark_design_2009}\cite{clark_design_2009} (referred to here as the triazine library) (Table~\ref{tbl:lib_sizes_counts}). 
DOS-DEL-1 was tested against carbonic anhydrase IX (CAIX) as a model protein target, for which we expected to confirm that benzenesulfonamides are active pharmacophores; the triazine library was tested against soluble epoxide hydrolase (sEH) and SIRT2. We show that models trained with a NLL loss function outperform baseline models trained using a MSE loss  without treatment of data uncertainty, in addition to random and weak (k-nearest-neighbors) baselines. The NLL-trained models also reveal an ability to act as structural regularizers, smoothing noisy enrichment values for individual compounds in a DEL, and enable visualizations of learned SAR to inform follow-up medicinal chemistry. Due to the general applicability of NLL loss functions to training regression models in maximum likelihood settings, we expect our approach to apply to regression modeling on other sparse or noisy datasets where the stochasticity is known or can be modeled. %

\section{Results}

\subsection{Single-task model performance is improved when trained on the NLL}

\begin{figure}
    \centering
        \includegraphics[scale=1]{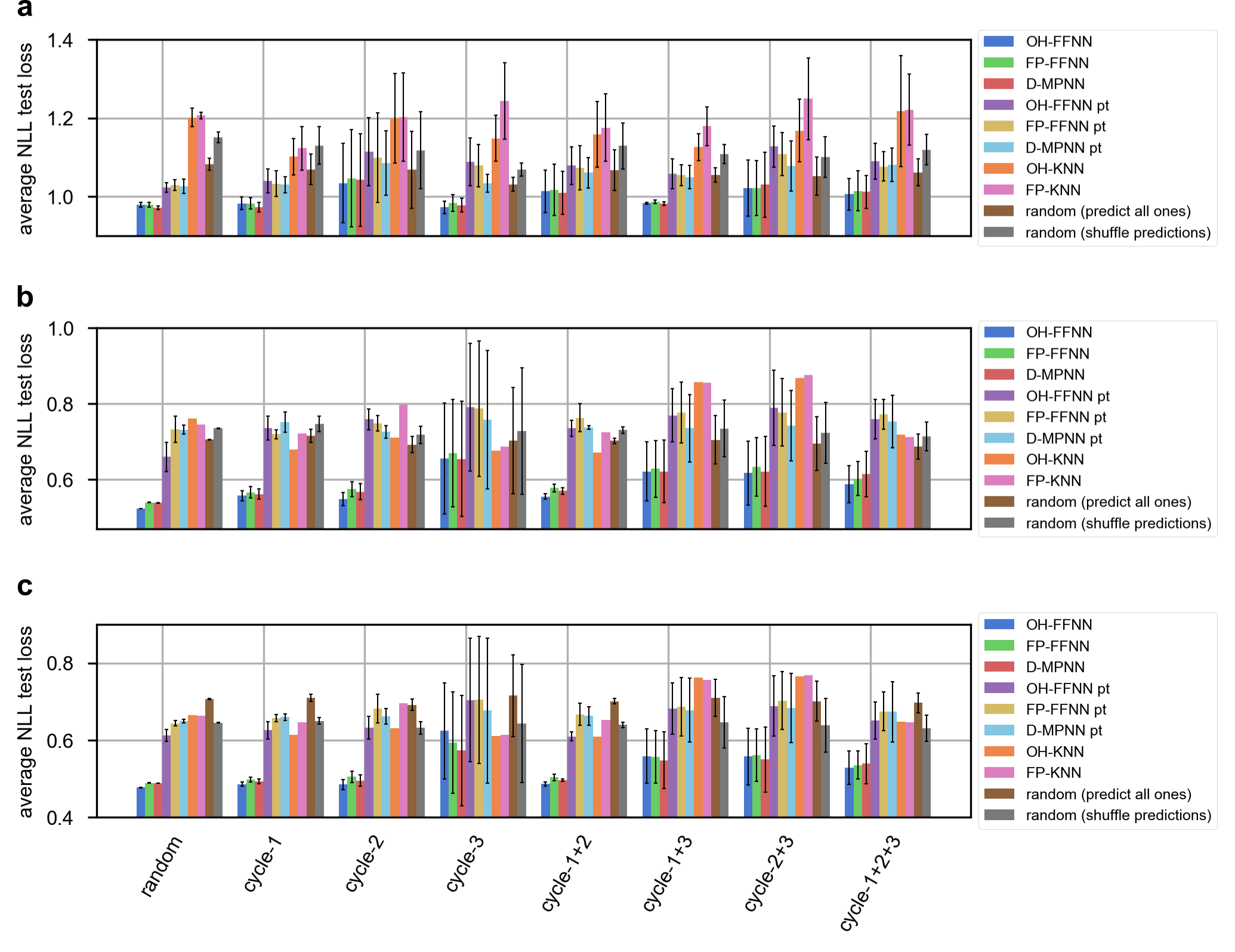}
    \caption{Comparison of model performance, as measured by negative log-likelihood (NLL) loss. OH: one-hot; FP: fingerprint; FFNN: feed-forward neural network; D-MPNN: directed message-passing neural network; KNN: k-nearest neighbors. The NLL test losses of the NLL-trained models (OH-FFNN, FP-FFNN, D-MPNN) are compared to those of the baseline point-prediction-trained models (OH-FFNN pt, FP-FFNN pt, D-MPNN pt), k-nearest-neighbors models (OH-KNN, FP-KNN), and random models (predict all ones, shuffle predictions), for various data splits (\emph{cf.} Figure~\ref{fig:reps_models_loss_fns_splits}d) on the \textbf{(a)} DD1S CAIX, \textbf{(b)} triazine sEH, \textbf{(c)} triazine SIRT2 datasets. Error bars represent $\pm$ one standard deviation. OH-FFNN, FP-FFNN, OH-FFNN pt, FP-FFNN pt, random (predict all ones), and random (shuffle predictions) results are averaged over five trials for each dataset; D-MPNN and D-MPNN pt results are averaged over five trials for the DD1S CAIX dataset and over three trials for the triazine sEH and triazine SIRT2 datasets; OH-KNN and FP-KNN results are averaged over five trials for the DD1S CAIX dataset and are single trials on a random 10\% of the test set for the triazine sEH and triazine SIRT2 datasets. The result of each trial is shown separately in the SI (Figure~\ref{fig:NLL_test_losses_scatter_plot}).}
    \label{fig:single_task_test_losses}
\end{figure}

The first evaluation was intended to directly compare the models trained using the NLL loss function to models trained using the MSE loss and a point estimate of the calculated enrichment (``pt'') without regard for uncertainty, in addition to k-nearest-neighbors (KNN) and random baselines that also use these point estimates. For each dataset, model type, and data split type, the average test NLL was calculated over several trials, where each trial used a different random seed for data splitting (Figures~\ref{fig:single_task_test_losses},~\ref{fig:NLL_test_losses_scatter_plot}; Tables~\ref{tbl:CAIX_NLL_test_losses},~\ref{tbl:sEH_NLL_test_losses},~\ref{tbl:SIRT2_NLL_test_losses}). Overall, the models explicitly trained using the uncertainty-aware Poisson enrichment loss function show higher performance than the baseline point prediction, KNN, and random models when evaluated in terms of that same metric. 
For completeness, we also calculated the test MSE loss and rank correlation coefficient for the same models when treating the calculated enrichments as precise measurements, which we know not to be the case (Figures~\ref{fig:MSE_test_losses_bar_graph},~\ref{fig:MSE_test_losses_scatter_plot},~\ref{fig:rank_corr_coeffs_bar_graph},~\ref{fig:rank_corr_coeffs_scatter_plot}; Tables~\ref{tbl:CAIX_MSE_test_losses},~\ref{tbl:sEH_MSE_test_losses},~\ref{tbl:SIRT2_MSE_test_losses},~\ref{tbl:CAIX_rank_corr_coeffs},~\ref{tbl:sEH_rank_corr_coeffs},~\ref{tbl:SIRT2_rank_corr_coeffs}). 
 We note that the KNN baseline models tend to show poorer performance than the random baselines; this is reasonable given that the KNN model is likely trying to fit to noisy data, whereas the predict-all-ones random baseline benefits from a smoothing effect since the average enrichment across all compounds in the test set is likely close to 1. The shuffle-predictions random baseline (randomly shuffling the test-set predictions of a FP-FFNN model) similarly benefits from the FP-FFNN's predicted enrichments likely being close to the average. %

The three molecular representations used---a one-hot encoded feed-forward network, a fingerprint-based feed-forward network, and the graph-based directed message-passing neural network---show similar performance within each data split. Considering that one-hot models are incapable of generalizing to new chemical structures, it is surprising to see them perform comparably to the fingerprint-based feed-forward networks and message-passing networks on the cycle splits (\emph{vide infra}). Whereas the random split only requires a model to interpolate to new combinations of chemical building blocks seen during training, cycle splits require the model to generalize to chemical building blocks not included in either the training or validation sets. 

In addition, the models trained using a cycle split generally have larger variability in performance than the models trained using a random split over the 5 repeated trials. This variability is attributable to models getting ``lucky'' or ``unlucky'' in terms of which building blocks happen to appear in their test sets and whether the distribution of enrichments differs between the training and test sets. In particular, prediction on the test set may be difficult for certain cycle splits if the test-set building blocks are structurally dissimilar to the building blocks used for training, since the model will not generalize well from the structures it has seen during training to make predictions for the test-set compounds. %
Prediction on the test set may also be difficult if the test- and training-set building blocks are similar in structure but have very different effects on enrichment (e.g., as in activity cliffs, where it would be unreasonable to expect a model to detect the sharp changes in activity, having never seen the structures corresponding to them). For instance, for the DD1S CAIX dataset, there is an activity cliff for compounds with a benzenesulfonamide; prediction on the test set for this dataset may be difficult if the test set includes building blocks with a benzenesulfonamide while the training set does not. 

As such, among the various cycle splits, certain cycle splits have considerably higher variability in test loss, depending on the dataset. For example, the DD1S CAIX dataset shows the highest variability for test loss using the cycle-2 split, mostly due to high test losses for one of the trials (the trial using random seed 4; Figure~\ref{fig:NLL_test_losses_scatter_plot}). The higher test loss for this trial may be explained by a distributional shift in calculated enrichments for the compounds in the trial's test set (Figure~\ref{fig:CAIX_cycle2_distrib_shift}). 
For the triazine sEH and triazine SIRT2 datasets, the cycle-3 split (and to a lesser extent the cycle-1+3 and cycle-2+3 splits) shows the greatest variability. Certain cycle-3 building blocks (building blocks 58, 157, and 179) are uniquely in the test set for the cycle-3, cycle-1+3, and cycle-2+3 splits for one of the five random seeds; the models' abilities to extrapolate to these particular structures seem to be poor. %

\subsubsection{Performance on DOS-DEL-1 for CAIX shows denoising ability and illustrates connection between counts and uncertainty levels}

\begin{figure}
    \newgeometry{left=0.1cm, top=0.1cm} %
    \centering
        \includegraphics[scale=1]{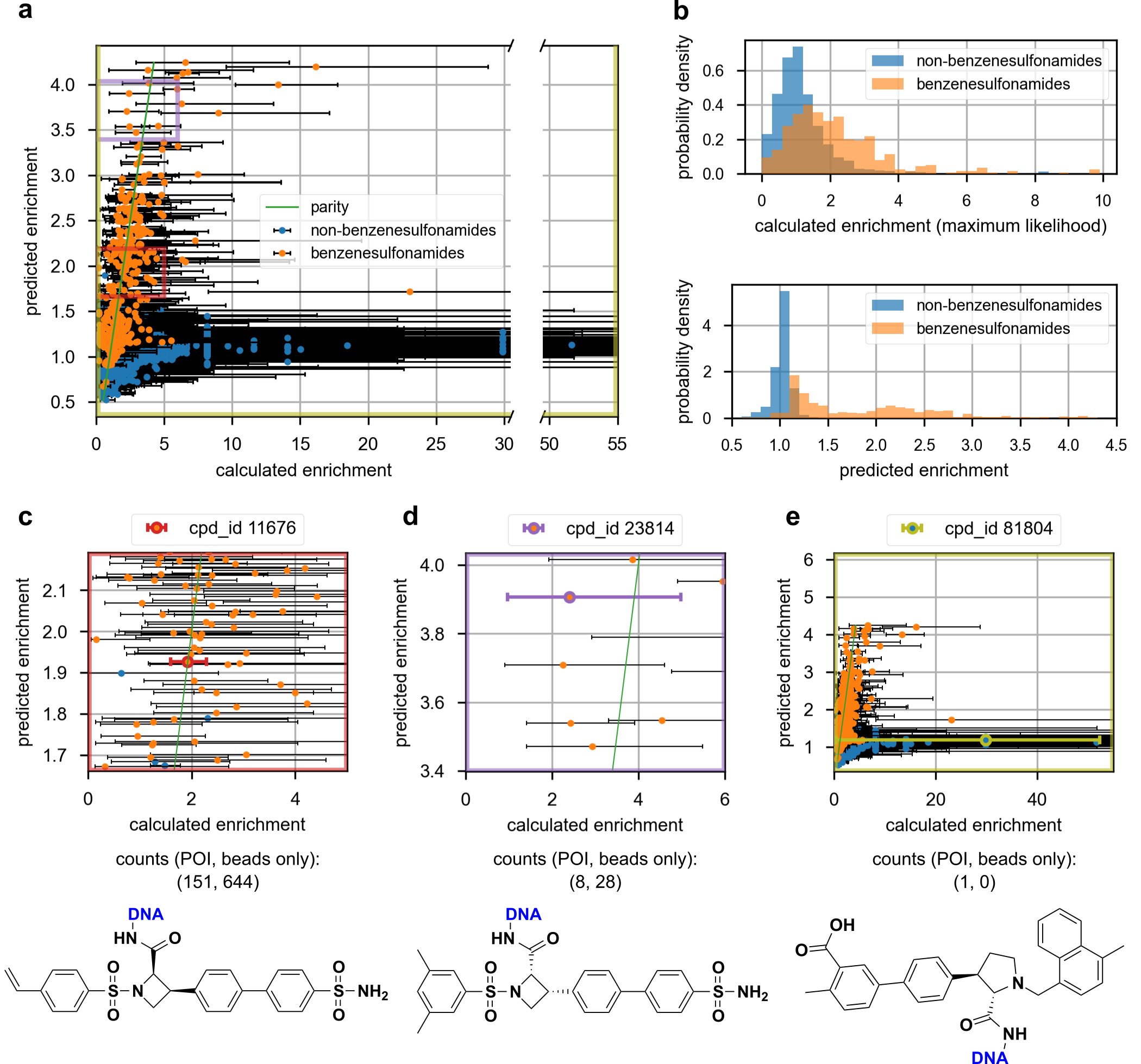}
    \caption{\textbf{(a)} Scatter plot of predicted and calculated enrichments for the test-set compounds of a FP-FFNN on a random split (\emph{cf.} Figure~\ref{fig:reps_models_loss_fns_splits}d) of the DD1S CAIX dataset. The green parity line is the identity function, for reference. \textbf{(b)} Histograms of calculated and predicted enrichments for the test-set compounds of a FP-FFNN on a random split (\emph{cf.} Figure~\ref{fig:reps_models_loss_fns_splits}d) of the DD1S CAIX dataset. The horizontal axis cutoff of 10 in the histogram of calculated enrichments is arbitrary, for the sake of legibility. \textbf{(c)} Close-up of a compound (ID 11676) with high counts (POI, beads only: 151, 644) and low uncertainty. The predicted enrichment of 1.93 approximates the calculated enrichment of 1.92. \textbf{(d)} Close-up of a compound (ID 23814) with low counts (POI, beads only: 8, 28) and high uncertainty. The predicted enrichment of 3.91 is high relative to the calculated enrichment of 2.41. \textbf{(e)} Close-up of a compound (ID 81804) with low counts (POI, beads only: 1, 0) and high uncertainty. The predicted enrichment of 1.19 is low relative to the calculated enrichment of 29.89. The total barcode counts in this dataset are 638,831 and 5,208,230 for the POI and beads-only conditions, respectively. Error bars represent 95\% confidence intervals for calculated enrichments; the horizontal axis values of the scatter plot datapoints are maximum-likelihood calculated enrichments (calculated using $z=0$; Methods). Compound IDs (``cpd\_id'') are sequential based on building block cycle numbers.}
    \label{fig:DD1S_CAIX_parity_plot_1D_histograms}
    \restoregeometry
\end{figure}

In addition to the quantitative metrics (Figure~\ref{fig:single_task_test_losses}) that summarize overall performance, we sought a more qualitative evaluation of model fit. The  correlation between the predicted and calculated enrichments of the test-set compounds for a random split using a fingerprint-based model indicates good agreement (Figures~\ref{fig:DD1S_CAIX_parity_plot_1D_histograms}a,~\ref{fig:CAIX_parity_plot_full}). Further, there is visible separation in predicted enrichments between the compounds with a benzenesulfonamide substructure and the compounds without a benzenesulfonamide substructure (Figure~\ref{fig:DD1S_CAIX_parity_plot_1D_histograms}b), which is known to be an important motif for binding CAIX's zinc atom in the active site \cite{gerry_dna_2019, li_versatile_2018, buller_selection_2011}. %
We provide analogous visualizations of predicted enrichments for benzenesulfonamides versus non-benzenesulfonamides for an MSE-loss-trained fingerprint-based model on the same split of the data (Figure~\ref{fig:CAIX_MSE_parity_plot}).
Disynthon aggregation results in a lesser extent of separation in the enrichments of compounds with or without a benzenesulfonamide, further suggesting that our trisynthon-based approach is particularly useful for denoising counts without obscuring the underlying structure-activity trends (Figure~\ref{fig:CAIX_disynthon_parity_plot_histograms}).

Examples of individual compounds illustrate the connection between counts and  levels of uncertainty in their calculated enrichments (Figure~\ref{fig:DD1S_CAIX_parity_plot_1D_histograms}cde). For compounds with higher counts and lower uncertainty, reflected by smaller error bars, the predicted enrichment tends to be closer to the calculated enrichment. For compounds with lower counts and higher uncertainty, the predicted enrichment may be further from the calculated enrichment, but not significantly so. For each of the three example compounds, model predictions fall within the 95\% confidence interval estimated from the count data (Figure~\ref{fig:DD1S_CAIX_parity_plot_1D_histograms}cde); averaging over all random splits, 92.94\% of model predicted enrichments fall within the 95\% confidence interval (Figure~\ref{fig:DD1S_CAIX_parity_plot_1D_histograms}cde; Table~\ref{tbl:DD1S_CAIX_FP-FFNN_coverage_probs}).

The low-count high-uncertainty datapoints illustrate a key motivation for using an uncertainty-aware loss function during training. By taking into account the high uncertainty in the calculated enrichments of compounds with low counts during training, the model avoids giving too much weight to such outliers. While these compounds could potentially correspond to potent binders, there is not as much evidence about their enrichment as there is about the enrichment of high-count, low-uncertainty compounds. Further, comparison of example outliers with their nearest neighbors in the DD1S CAIX dataset suggests that these outliers are unlikely to be potent binders, since structurally similar compounds in the dataset have relatively low calculated enrichments (Table~\ref{tbl:DD1S_CAIX_outliers}).%

\subsubsection{Performance on the triazine library for sEH and SIRT2 shows a coarse correlation between predicted and calculated enrichments}

\begin{figure} %
    \centering
            \includegraphics[scale=1]{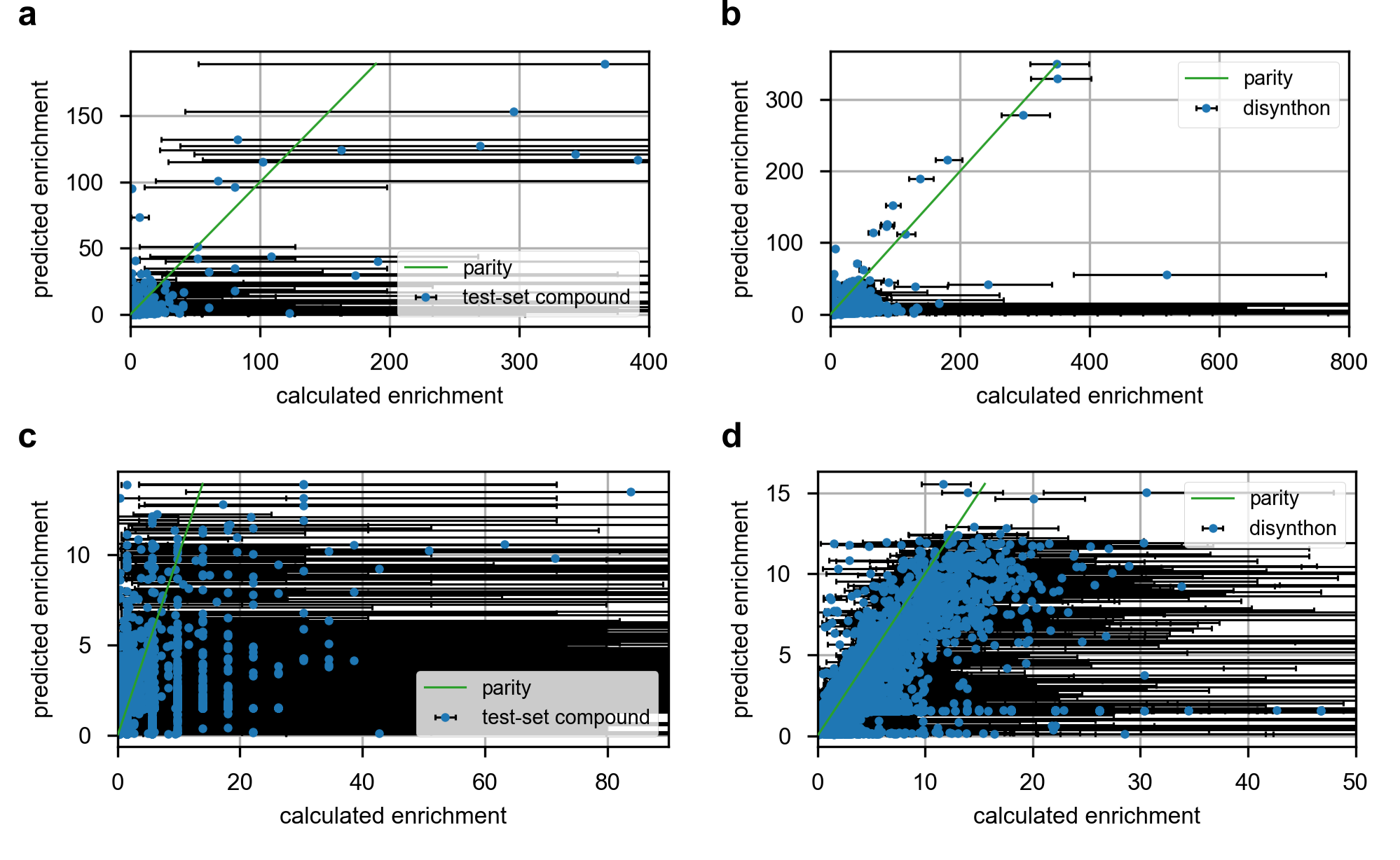}
    \caption{Scatter plot of predicted and calculated enrichments for a random subset (20,000 compounds) of the test set of a FP-FFNN on a random split (\emph{cf.} Figure~\ref{fig:reps_models_loss_fns_splits}d) of the \textbf{(a)} triazine sEH, \textbf{(c)} triazine SIRT2 dataset, and for all disynthons in the \textbf{(b)} triazine sEH, \textbf{(d)} triazine SIRT2 dataset. The green parity line is the identity function, for reference. Error bars represent 95\% confidence intervals for calculated enrichments; the horizontal axis values of the datapoints are maximum-likelihood calculated enrichments (calculated using $z=0$; Methods).}
    \label{fig:sEH_SIRT2_parity_scatter_plots}
\end{figure}

A similar visual inspection of the correlation between predicted and calculated trisynthon enrichments is less informative for the sEH and SIRT2 datasets due to the much larger uncertainties of calculated enrichments. The increase in uncertainties stems from the much larger size of the triazine library (5.7 million compounds compared to 109k for the DD1S CAIX dataset). Since the sequencing depth is not proportionately scaled up, the increase in number of compounds results in fewer average counts per compound (Table~\ref{tbl:lib_sizes_counts}). 

Parity plots for a fingerprint-based feed-forward network on a random split of the triazine sEH and triazine SIRT2 datasets (Figures~\ref{fig:sEH_SIRT2_parity_scatter_plots}ac,~\ref{fig:triazine_parity_plots_full}ac)  show a coarse linear correlation between predicted and calculated enrichments; the same data can be visualized as 2D histograms %
(Figures~\ref{fig:CAIX_2D_hists},~\ref{fig:sEH_2D_hists},~\ref{fig:SIRT2_2D_hists}). %
We also compared predicted and calculated enrichments of aggregated disynthons on the same datasets (Figures~\ref{fig:sEH_SIRT2_parity_scatter_plots}bd,~\ref{fig:triazine_parity_plots_full}bd; Methods) and found a clearer linear correlation, suggesting that the coarse correlation for individual compounds may be attributed to the noisiness of the data.

\subsubsection{Binary classification baseline comparisons show no evidence of a benefit to regression modeling}

As a baseline comparison motivated by \citeauthor{mccloskey_machine_2020}\cite{mccloskey_machine_2020}'s binary classification approach, we trained binary classifiers on each of the datasets, and also evaluated the trained regression models as classifiers. The results (Figures~\ref{fig:bin_plots_fixed_threshold},~\ref{fig:bin_plots_multiple_thresholds_PR_AUC},~\ref{fig:bin_plots_multiple_thresholds_ROC_AUC}; Tables~\ref{tbl:CAIX_AUCs},~\ref{tbl:sEH_AUCs},~\ref{tbl:SIRT2_AUCs}) are inconclusive; based on the PR and ROC AUC metrics for classification, there is no evidence of a benefit to regression modeling of these data using a NLL loss function.

\subsection{Visualization of learned SAR reveals substructures driving enrichment}
\begin{figure}
    \centering
        \includegraphics[scale=1]{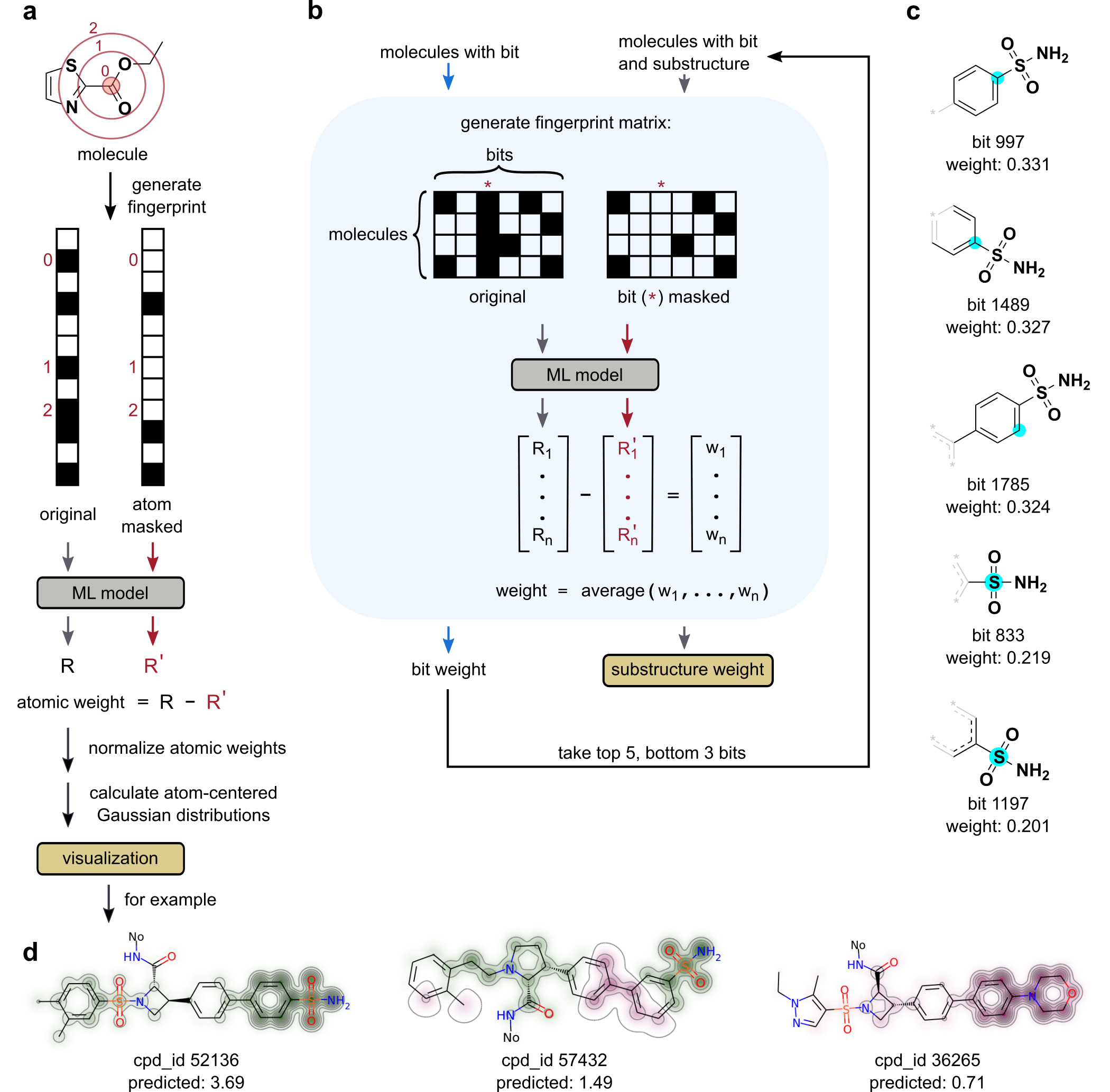}
    \caption{\textbf{(a)} Workflow for the generation of atom-centered Gaussian visualizations. \textbf{(b)} Workflow for calculating fingerprint bit and substructure importance. \textbf{(c)} Top 5 substructures and \textbf{(d)} example visualizations for compounds in the DD1S CAIX dataset based on the predictions of a FP-FFNN trained on a random split (\emph{cf.} Figure~\ref{fig:reps_models_loss_fns_splits}d) of the DD1S CAIX dataset. In the example visualizations, atoms contributing positively to enrichment are highlighted in green, and atoms contributing negatively to enrichment are highlighted in pink, with color intensity corresponding to the level of contribution to enrichment. ``No'' represents the DNA linker attachment point. Compound IDs (``cpd\_id'') are sequential based on building block cycle numbers.
}
    \label{fig:DD1S_CAIX_learned_SAR_vis}
\end{figure}

The ability to interpret what has been learned by a SAR model can improve the utility of its predictions for medicinal chemists \cite{deokar_qsar_2018,mahipal_3d_2010,gao_3d_2011,alam_3d-qsar_2017}%
. In particular, SAR model interpretation allows researchers to understand the rationale behind top predicted compounds and may facilitate the rational design of more potent binders that can be experimentally validated. As fingerprint-based models are especially conducive to substructure-based interpretation, we explored two ways of visualizing the learned SAR of our fingerprint-based feed-forward networks: atom-centered Gaussian visualizations \cite{riniker_similarity_2013} (Figure~\ref{fig:DD1S_CAIX_learned_SAR_vis}a) and fingerprint bit and substructure importance (Figure~\ref{fig:DD1S_CAIX_learned_SAR_vis}b). Interpretation of the message-passing neural networks through atom masking, integrated gradients, or similar substructure importance estimation techniques is left to future work.

\subsubsection{Atom-centered Gaussian visualizations provide a qualitative evaluation of model learning and generalization performance}

\begin{figure} %
    \centering
        \includegraphics[scale=1]{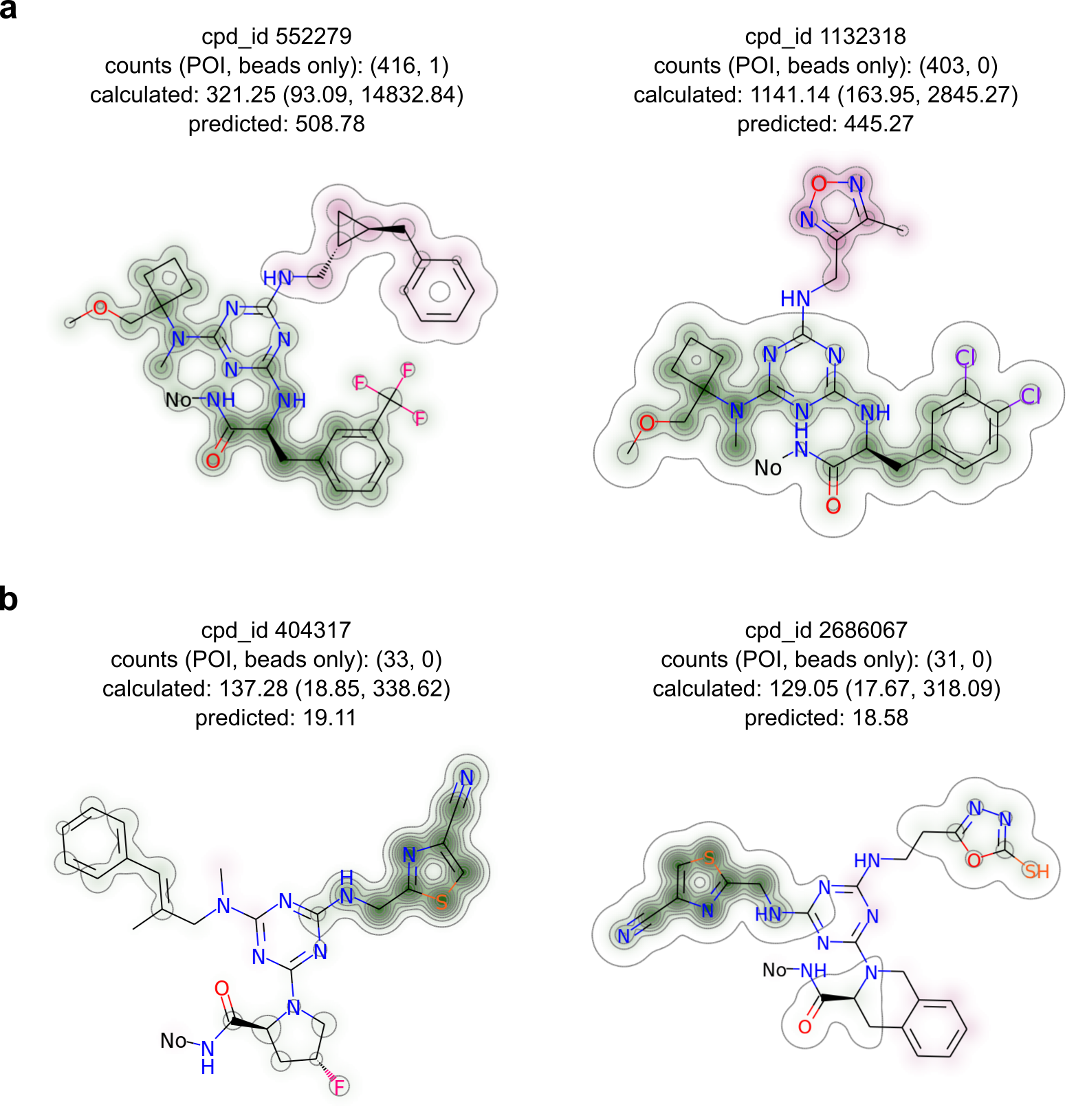}
    \caption{Atom-centered Gaussian visualizations for example compounds in the test set of a FP-FFNN trained on a random split (\emph{cf.} Figure~\ref{fig:reps_models_loss_fns_splits}d) of the \textbf{(a)} triazine sEH, \textbf{(b)} triazine SIRT2 dataset. Atoms contributing positively to enrichment are highlighted in green, and atoms contributing negatively to enrichment are highlighted in pink, with color intensity corresponding to the level of contribution to enrichment. ``No'' represents the DNA linker attachment point. Compound IDs (``cpd\_id'') are sequential based on building block cycle numbers.
    }
    \label{fig:random_split_learned_SAR_vis}
\end{figure}

Atom-centered Gaussian visualizations allow for visualization of atomic contributions to model predictions for a given molecule. Each visualization shows the structure of the molecule, with atoms contributing positively to enrichment highlighted in green, and atoms contributing negatively to enrichment highlighted in pink (Figure~\ref{fig:DD1S_CAIX_learned_SAR_vis}a; Methods). %
Visualizations of several compounds' enrichments for fingerprint-based feed-forward networks trained on a random split of each dataset are shown as examples (Figures~\ref{fig:DD1S_CAIX_learned_SAR_vis}d,~\ref{fig:CAIX_top_predicted_cpds},~\ref{fig:random_split_learned_SAR_vis}). For the DD1S CAIX dataset, the visualization algorithm highlights benzenesulfonamides as strongly enriched pharmacophores, as expected (Figure~\ref{fig:DD1S_CAIX_learned_SAR_vis}d).

The atom-centered Gaussian visualizations for the top test-set compounds predicted by a random split model on the triazine sEH dataset provide qualitative evidence that the models are learning reasonable SAR (Figure~\ref{fig:sEH_top_predicted_cpds}). We consistently see a trifluoromethylphenyl or dichlorophenyl group as a strongly enriched motif in these top predicted compounds, which is reasonable given that many potent sEH binders reported in the literature contain these substructures, among other fluorinated or chlorinated phenyl substituents \cite{anandan_unsymmetrical_2009,mcelroy_qsar_2003,kramer_discovery_2019, lee_optimized_2014}. %
A similar visual evaluation of top predicted compounds for the triazine SIRT2 dataset reveals a cyanothiazole as a strongly enriched motif (Figure~\ref{fig:SIRT2_top_predicted_cpds}), although this substructure lacks literature support.

We also used these visualizations to qualitatively evaluate our fingerprint-based models’ empirical abilities to generalize on the triazine sEH and triazine SIRT2 datasets (\emph{cf.} Discussion). More specifically, we generated atom-centered Gaussian visualizations for the top predicted compounds for a model extrapolating on a cycle-1+2+3 split, as well as visualizations of the same compounds for a corresponding model interpolating on a random split of the data (Figures~\ref{fig:sEH_new_BB_cpds_full_vis},~\ref{fig:SIRT2_new_BB_cpds_full_vis}). For the top predicted compounds, we did not observe much similarity between the structures that the cycle-1+2+3 split model and the random split model predict as contributing positively or negatively to enrichment. This dissimilarity in learned SAR suggests that the cycle-1+2+3 split models, having never seen any of the three building blocks, lack the random split models' understanding of the aspects of the building blocks' structures that will positively or negatively contribute to overall enrichment. %

\subsubsection{Bit and substructure importance confirms previously found SAR for CAIX and identifies substructures driving enrichment for sEH and SIRT2}

The second approach to SAR visualization involved identifying the fingerprint bits and substructures that contribute most heavily to the predicted enrichment (Figure~\ref{fig:DD1S_CAIX_learned_SAR_vis}b; \emph{cf.} Methods). This process involved iterating through bit-masked predictions to find the most important fingerprint bits and from those bits identifying the most important corresponding substructures without hand-selecting individual molecules to look at. As Morgan circular fingerprints were used in this study, the result of this analysis is a set of atom-centered neighborhoods of up to radius 3 and their importance weights.

For the DD1S CAIX dataset, the top 5 substructures contributing to enrichment (Figures~\ref{fig:DD1S_CAIX_learned_SAR_vis}c,~\ref{fig:CAIX_onebit_seed_0_bars},~\ref{fig:CAIX_onebit_seed_0_hists}; Table~\ref{tbl:SMARTS_DD1S}) all correspond to an arylsulfonamide, with the first and third corresponding specifically to a \emph{para}-substituted benzenesulfonamide, the second to a \emph{para}-substituted arylsulfonamide, the fourth to any arylsulfonamide, and the fifth to \emph{para}- and \emph{meta}-substituted arylsulfonamides; note, however, that the only two arylsulfonamide motifs in DD1S are \emph{para}- and \emph{meta}-substituted benzenesulfonamides. These results are consistent with prior observations that the benzenesulfonamide is the most important motif for binding affinity to carbonic anhydrase, and more specifically that compounds with a \emph{para}-substituted benzenesulfonamide are more highly enriched than compounds with a \emph{meta}-substituted benzenesulfonamide \cite{gerry_dna_2019, li_versatile_2018, buller_selection_2011}. %
The top 5 bits (and corresponding substructures) from this analysis are consistent across two other random splits of the DD1S CAIX dataset (Figures~\ref{fig:CAIX_onebit_seed_1_bars},~\ref{fig:CAIX_onebit_seed_1_hists},~\ref{fig:CAIX_onebit_seed_2_bars},~\ref{fig:CAIX_onebit_seed_2_hists}), although the bottom 3 bits---corresponding to substructures most strongly contributing to \emph{lower} enrichments---vary. This variance is unsurprising given that low enrichments are associated with low counts and higher uncertainty. %

For the triazine sEH dataset, the top substructure (Figures~\ref{fig:sEH_onebit_bars},~\ref{fig:sEH_onebit_hists}; Table~\ref{tbl:SMARTS_sEH}) consists mostly of the triazine core (which is shared by all molecules in the dataset) with a methylated tertiary nitrogen. The tertiary nitrogen is also present in several of the other substructures with the highest weights. Lastly for the triazine SIRT2 dataset (Figures~\ref{fig:SIRT2_onebit_bars},~\ref{fig:SIRT2_onebit_hists}; Table~\ref{tbl:SMARTS_SIRT2}), the top five substructures correspond to the cyanothiazole in cycle-3 building block 67 (and equivalently cycle-2 building block 66). %

In addition to analyzing the most important individual substructures, we also investigated contributions of pairs of substructures to model predictions (Figures~\ref{fig:CAIX_twobits_seed_0_bars},~\ref{fig:CAIX_twobits_seed_1_bars},~\ref{fig:CAIX_twobits_seed_2_bars},~\ref{fig:sEH_twobits_bars},~\ref{fig:SIRT2_twobits_bars}, ~\ref{fig:CAIX_twobits_seed_0_hists},~\ref{fig:CAIX_twobits_seed_1_hists},~\ref{fig:CAIX_twobits_seed_2_hists},~\ref{fig:sEH_twobits_hists},~\ref{fig:SIRT2_twobits_hists}; Tables~\ref{tbl:SMARTS_twobits_DD1S},~\ref{tbl:SMARTS_twobits_DD1S_seed_1},~\ref{tbl:SMARTS_twobits_DD1S_seed_2},~\ref{tbl:SMARTS_twobits_sEH},~\ref{tbl:SMARTS_twobits_SIRT2}). Whereas enrichment of a single substructure indicates that one specific motif may be sufficient for high affinity, enrichment of a pair of substructures would indicate that a specific combination or arrangement of motifs may account for affinity. SAR for such combinations of structural motifs have previously been explored
\cite{wawer_automated_2014, eberle_improved_2011, wassermann_directed_2012}. %

\subsection{Multitask models show no significant difference in performance} 

We also investigated whether multitask modeling would result in improved model performance for the sEH and SIRT2 protein targets on the triazine dataset. Models' output layers were extended to a dimension of two to simultaneously predict enrichment for both protein targets. One would expect improved model performance on the multitask setting \emph{if} the factors that contribute to enrichment or lack thereof are similar across the tasks. Since sEH and SIRT2 both have hits from the triazine library, we explored training a model simultaneously on the two targets, for both a random split and a cycle-1+2+3 split (Table~\ref{tab:multi-task_random}; Figure~\ref{fig:multi_task_test_loss_bar_graphs}). There is no significant difference in model performance for the single task versus multitask models on either data split, suggesting that the chemical features responsible for affinity to sEH and SIRT2 are distinct. Note that one could still consider extending this approach to a greater number of protein targets to assist in the prediction of \emph{selective} enrichment.

\begin{table}[H]
\caption{Test-set losses (mean $\pm$ standard deviation) for single-task and multi-task NLL models on the triazine sEH and SIRT2 datasets. OH-FFNN and FP-FFNN results are averaged over five trials; D-MPNN results are averaged over three trials.}
\label{tab:multi-task_random}
\begin{tabular}{c|cccc|}
\cline{2-5}
\multicolumn{1}{l|}{}                & \multicolumn{4}{c|}{\textbf{sEH}}                                                                                                                    \\ \cline{2-5} 
\multicolumn{1}{l|}{}                & \multicolumn{2}{c|}{\textit{random split}}                                          & \multicolumn{2}{c|}{\textit{cycle-1+2+3 split}}                \\ \hline
\multicolumn{1}{|c|}{\textbf{model}} & \multicolumn{1}{c|}{single-task loss}    & \multicolumn{1}{c|}{multi-task loss}     & \multicolumn{1}{c|}{single-task loss}    & multi-task loss     \\ \hline
\multicolumn{1}{|c|}{OH-FFNN}        & \multicolumn{1}{c|}{$0.5240 \pm 0.0002$} & \multicolumn{1}{c|}{$0.5236 \pm 0.0005$} & \multicolumn{1}{c|}{$0.5882 \pm 0.0486$} & $0.5884 \pm 0.0491$ \\ \hline
\multicolumn{1}{|c|}{FP-FFNN}        & \multicolumn{1}{c|}{$0.5400 \pm 0.0004$} & \multicolumn{1}{c|}{$0.5398 \pm 0.0004$} & \multicolumn{1}{c|}{$0.6031 \pm 0.0445$} & $0.6045 \pm 0.0435$ \\ \hline
\multicolumn{1}{|c|}{D-MPNN}         & \multicolumn{1}{c|}{$0.5390 \pm 0.0006$} & \multicolumn{1}{c|}{$0.5393 \pm 0.0006$} & \multicolumn{1}{c|}{$0.6155 \pm 0.0598$} & $0.6162 \pm 0.0568$ \\ \hline
\multicolumn{1}{l|}{}                & \multicolumn{4}{c|}{\textbf{SIRT2}}                                                                                                                  \\ \cline{2-5} 
\multicolumn{1}{l|}{}                & \multicolumn{2}{c|}{\textit{random split}}                                          & \multicolumn{2}{c|}{\textit{cycle-1+2+3 split}}                \\ \hline
\multicolumn{1}{|c|}{\textbf{model}} & \multicolumn{1}{c|}{single-task loss}    & \multicolumn{1}{c|}{multi-task loss}     & \multicolumn{1}{c|}{single-task loss}    & multi-task loss     \\ \hline
\multicolumn{1}{|c|}{OH-FFNN}        & \multicolumn{1}{c|}{$0.4770 \pm 0.0005$} & \multicolumn{1}{c|}{$0.4770 \pm 0.0007$} & \multicolumn{1}{c|}{$0.5282 \pm 0.0431$} & $0.5270 \pm 0.0441$ \\ \hline
\multicolumn{1}{|c|}{FP-FFNN}        & \multicolumn{1}{c|}{$0.4884 \pm 0.0008$} & \multicolumn{1}{c|}{$0.4886 \pm 0.0007$} & \multicolumn{1}{c|}{$0.5353 \pm 0.0363$} & $0.5368 \pm 0.0351$ \\ \hline
\multicolumn{1}{|c|}{D-MPNN}         & \multicolumn{1}{c|}{$0.4882 \pm 0.0007$} & \multicolumn{1}{c|}{$0.4881 \pm 0.0007$} & \multicolumn{1}{c|}{$0.5388 \pm 0.0521$} & $0.5407 \pm 0.0494$ \\ \hline
\end{tabular}
\end{table}

\section{Discussion}

\subsection{Ability to smooth the response surface}

To evaluate how the models handle the noise in the DEL data, we compared the distributions for the calculated enrichments and model-predicted enrichments on the DD1S CAIX dataset. It is well established that the main structural motif determining binding affinity to carbonic anhydrase is the presence of a benzenesulfonamide \cite{gerry_dna_2019, li_versatile_2018, buller_selection_2011}; compounds in DD1S with a benzenesulfonamide have very high affinity for the protein, whereas other compounds have comparatively low affinity. Compared to the calculated enrichments for the compounds in the dataset, we noticed better separation between the model-predicted enrichments for compounds with a benzenesulfonamide and  compounds without a benzenesulfonamide (Figure~\ref{fig:DD1S_CAIX_parity_plot_1D_histograms}b). This greater separation in enrichments shows that the models are able to act as structural regularizers and smooth noisy enrichment values for individual compounds. Such denoising of enrichments may improve the selection of compounds for follow-up investigation by enabling the detection of SAR patterns and reducing the number of false-positive compounds chosen solely based on count data.

\subsection{Generalization performance}

Quantitative performance (as measured by the NLL loss function) on the cycle splits is similar between the one-hot feed-forward networks, which by design cannot generalize, and the models using structure-based representations (\emph{cf.} Figure~\ref{fig:single_task_test_losses}). Therefore, the NLL metric alone does not reveal evidence of the models generalizing to new chemical structures. It has been noted \cite{wawer_automated_2014} that the gain in generality with structure-based representations is accompanied by a loss of synthetic accessibility if the input representation is directly optimized to propose new molecular candidates; synthon-based descriptors (e.g., one-hot encoding of building blocks) denote a synthesis recipe, while generic fingerprints and other structure-based representations rely on abstract features that may warrant custom synthesis routes for corresponding products.

To confirm our models' generalization performance, we analyzed the predictions of models trained on a cycle-1+2+3 split of the triazine sEH and triazine SIRT2 datasets, focusing on the test-set compounds that do not share a building block with \emph{any} of the compounds in the training or validation sets. %
Due to structural symmetry in the triazine library, where a subset of cycle-2 building blocks are structurally identical to a cycle-3 building block and \emph{vice versa}, in our analysis we excluded compounds with a test-set building block that is structurally identical to a training-set or validation-set building block. We did not observe high correlation between the predicted and calculated enrichments for these compounds with all building blocks uniquely in the test set, making it unclear whether the models are generalizing (Figures~\ref{fig:sEH_c123_parity_plots},~\ref{fig:SIRT2_c123_parity_plots}). The one-hot feed-forward networks, which again cannot generalize, predict relatively similar values for all the molecules, as expected; the spread in the predicted enrichments is due to random weight initializations. %

We also investigated the top predicted compounds for the fingerprint-based feed-forward networks by comparing their atom-centered Gaussian visualizations for both the model generalizing on a cycle-1+2+3 split, and a corresponding model interpolating on a random split (Figures~\ref{fig:sEH_new_BB_cpds_full_vis},~\ref{fig:SIRT2_new_BB_cpds_full_vis}). We did not observe much similarity in learned SAR between the cycle-1+2+3 split and random split models---this qualitative evaluation did not reveal evidence of the models extrapolating. %
We also noted that for a given model, the coloring of the same building block in the atom-centered Gaussian visualizations of different molecules may differ. This disparity may be explained by the fact that the atom-level weights used to generate the atom-centered Gaussian visualizations are normalized on a per-molecule basis, as opposed to being normalized across the entire dataset.

\subsection{Domain of applicability and chemical diversity}

\begin{figure}[H]
    \centering
        \includegraphics[width=\linewidth]{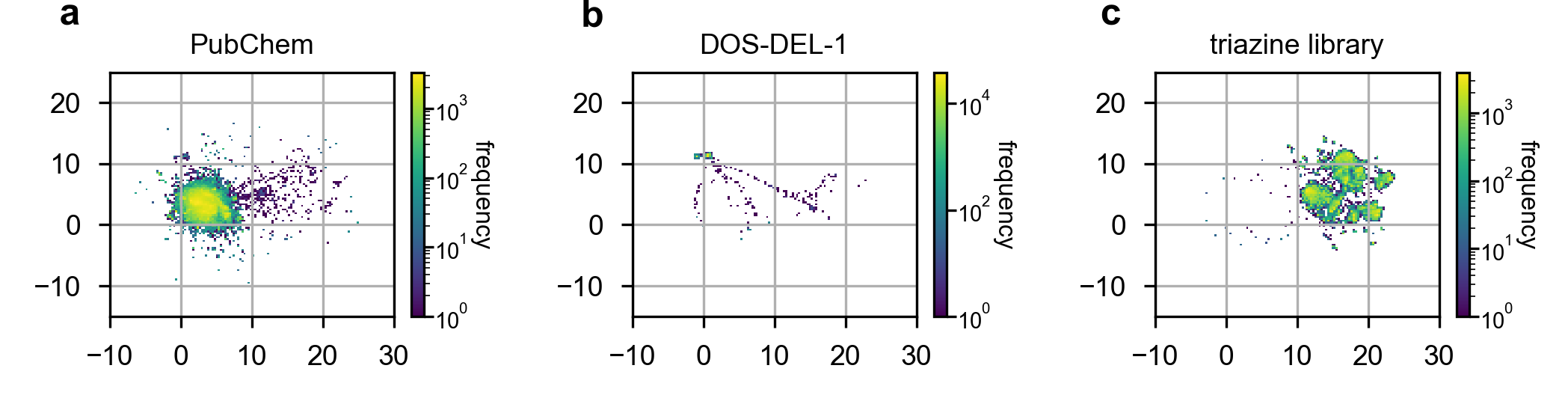}
    \caption{UMAP projection for \textbf{(a)} a random sample of 600k compounds from PubChem, \textbf{(b)} DOS-DEL-1, and \textbf{(c)} a random sample of 10\% of the compounds in the triazine library. The UMAP embedding was fit to all three sets of compounds (using a random 10\% of the compounds in DOS-DEL-1) simultaneously (\emph{cf.} Methods). The coordinates of each plot represent the two dimensions to which the molecular fingerprints were projected.}
    \label{fig:chem_space_plots}
\end{figure}

Any QSAR model’s domain of applicability will depend on the diversity of the compounds in the training data. The triazine library we used contains all combinations of 78 building blocks for cycle 1, 290 building blocks for cycle 2, and 250 building blocks for cycle 3, attached to a common triazine core. These compounds may not cover enough chemical space for the models to generalize to new scaffolds, particularly those not containing the triazine core. In particular, the models’ domain of applicability is likely limited compared to that of the models built by \citeauthor{mccloskey_machine_2020}, which were trained on 31 to 42 pooled libraries with more extensive coverage of chemical space and successfully generalized to new structures as confirmed experimentally \cite{mccloskey_machine_2020}. UMAP projections (Figure \ref{fig:chem_space_plots}; Methods) of both of the libraries we used (DOS-DEL-1 and the triazine library) and of a random sample of 600,000 compounds from PubChem show that our datasets cover relatively localized areas of chemical space, with relatively sparse coverage compared to compounds in PubChem. This sparse chemical coverage suggests that our models would likely have trouble generalizing to compounds dissimilar to those in these DELs. Further, little chemical similarity was found between compounds in the Enamine compound collections (REAL database and screening collection) and the top five compounds (by lower bound of calculated enrichment) in the DD1S CAIX dataset, likely because commercially available compounds lack the scaffolds used in DOS-DEL-1 (Table~\ref{tbl:enamine_comparison}).

\subsection{Generality of the loss function}

Using a NLL loss function is a general way to train a regression model in a maximum likelihood setting. As such, we expect our approach to apply to modeling on other sparse or noisy datasets where the stochasticity is known or can be modeled; in our case, we model the stochasticity of the DEL data using Poisson distributions. %
The Poisson enrichment ratio metric is relevant to other settings that compare sequencing count data between two experimental conditions (e.g., a treatment condition versus a control condition), and it admits extension across multiple off-target controls using comparison metrics based on multiple Poisson ratio test values and confidence intervals. Moreover, it can be used at multiple levels of resolution (e.g., mono-, di-, and tri-synthons) which suggests opportunities for combining synthetic history-based analyses with structure-based cheminformatics. We acknowledge that this is an active area of investigation \cite{mccloskey_machine_2020, ma_regression_2021}, including a graph neural network (GNN)-based regression model to predict enrichment scores in noisy data accounting for confounding factors \cite{ma_regression_2021}. In the context of DELs, we expect our modeling approach to be especially useful when learning from selection data for larger libraries with higher levels of noise due to low sequencing coverage relative to the size of the library.

\section{Conclusion}

Our use of a Poisson enrichment ratio loss function to train enrichment-predicting regression models on DEL data for individual compounds introduces a novel approach to QSAR modeling on DEL data. Training to optimize the uncertainty-aware NLL directly leads to improved performance over a MSE loss as measured by the consistency between the model and data. Analysis of model predictions on the DD1S CAIX dataset, in the context of known SAR for carbonic anhydrase, shows that these models can act as structural regularizers and smooth noisy enrichment values for individual compounds in a DEL, which may reduce the risk of selecting a false-positive trisynthon based solely on its count data. However, based on a comparison of quantitative performance across the NLL model types, the calculated and predicted enrichments for compounds with chemical building blocks not used to train the models, and visualizations of learned SAR for the fingerprint-based feed-forward networks, it is unclear whether the models are generalizing. Nevertheless, the visualization procedure does reveal the substructures the model relies on for making its predictions and allows for a retrospective analysis of SAR trends without requiring monosynthon or disynthon aggregation. Additionally, it is possible that the lack of apparent generalization ability is a limitation of the data (i.e., the sparse chemical coverage of the libraries used to train the models) as opposed to the methods.

Further work can be done to explore more model types, more datasets, and other ways of improving model training (including pretraining). One of the more interesting extensions would be to connect to off-DNA affinity data by determining dissociation constants of the top predicted molecules from our models. Although this study does not include any off-DNA affinity data or experimental validation, on-DNA activity is already presumed to have some correlation (albeit noisy) with off-DNA activity based on previous successes in hit finding \cite{eidam_analysis_2016,arico-muendel_haystack_2016,favalli_del_2018,kunig_dna-encoded_2018}. Another extension might include treating model predictions themselves as uncertain \cite{hirschfeld_uncertainty_2020}. This modification would require a different loss function, since log likelihood calculation assumes that the enrichment is a fixed but unknown parameter, and that the data are what are stochastic (i.e., a frequentist view).

In summary, this work builds upon that of \citeauthor{mccloskey_machine_2020}, which demonstrates that binary classification of aggregated disynthons can identify diverse, drug-like hits with micromolar activity \cite{mccloskey_machine_2020}. We provide a proof of concept for an alternate approach applying regression modeling with a NLL loss function to individual compounds of a DEL, and anticipate that such an approach may prove useful for future work in this emerging field of machine learning on DEL data to improve the selection of compounds for follow-up investigation.

\section{Methods} %

\subsection{Problem formulation: regression} 

In our approach to QSAR modeling on DEL data, we treated learning DEL enrichments as a regression task of predicting enrichments of individual compounds. More specifically, the enrichment we predicted for each molecule is the ratio of the relative abundance of the DNA barcode for that molecule in the condition with the POI to the relative abundance of the barcode in the beads-only control. This enrichment served as a proxy measure of the molecule’s affinity for the protein of interest. Such a regression approach to modeling with DEL data had not been done before, likely due to the fact that DEL data is generally noisy and sparse, with high uncertainty in enrichments calculated for individual compounds. Usually when training regression models, the MSE is used, but this loss function does not consider uncertainty in the training data. As such, the MSE loss function is not appropriate for modeling with DEL data.

\subsection{Problem formulation: loss function} 

Our regression modeling approach uses an uncertainty-aware NLL loss function. Uncertainty in DEL data arises in part from the stochastic nature of the sequencing step in the DEL experimental workflow, which we accounted for by modeling sequencing as random sampling (of DNA barcodes to be ligated onto the oligos bound to the chip) without replacement. Since there is an excess of DNA post-amplification relative to the number of measured sequencing reads (i.e., the number of sampled barcodes), we approximated sampling without replacement as sampling with replacement, yielding a binomial distribution that in the limit of low probabilities can be closely approximated by a Poisson distribution \cite{simons_convergence_1971}. In the regime of DEL data, where the total number of sequencing reads is large and the relative abundances of individual DNA barcodes are small, the probability mass functions of the Poisson and binomial distributions are nearly identical.

Modeling barcode count distributions as Poisson distributions allows us to use a Poisson ratio test to evaluate the consistency of  the  barcode counts observed in a DEL experiment with a hypothesized enrichment ratio, $R$. %
To perform the ratio test, we compared two known counts ($k_1$, $k_2$) from two experiments (one with the POI, one with beads only) with different total counts ($n_1$, $n_2$) drawn from two unknown Poisson distributions (described by $\lambda_1$, $\lambda_2$) using the following hypothesis test:
\[\text{null hypothesis: }H_0=\frac{\lambda_1}{\lambda_2}\leq R_0\]
\[\text{alternate hypothesis: }H_1=\frac{\lambda_1}{\lambda_2} > R_0\]
where $R_0$ is an enrichment threshold that we can freely specify for the purposes of hypothesis testing. 
Based on a variance stabilizing square root transformation \cite{gu_testing_2008}
, we can calculate a $z$-score as:
\[z = 2\frac{\sqrt{k_1+\frac{3}{8}}-\sqrt{\left(\frac{n_1}{n_2}R_0\right)\left(k_2+\frac{3}{8}\right)}}{\sqrt{1+\left(\frac{n_1}{n_2}R_0\right)}} \sim N(0,1)\]

In practice, we convert this calculation to a probability score for a two-sided alternate hypothesis, following \citeauthor{gu_testing_2008}:
\[p(k_1,k_2|n_1,n_2,R_0=R) \propto 1-\phi\left( |z|  \right)\]

where $\phi$ is the standard normal cumulative distribution function. 
Finally, we defined the loss function as the negative log likelihood of the probability that we would fail to reject the null hypothesis given the observed data:
\[\text{Loss}(R)=-\log p(k_1,k_2|n_1,n_2,R_0=R)\]

This loss function's treatment of uncertainty in the training data is illustrated in plots of the NLL as a function of predicted enrichment for various theoretical count values (Figure~\ref{fig:NLL_plots}). Note that with this loss function, the model will learn to predict a value of $R$ given a molecular structure and will not separately predict $\lambda_1$ and $\lambda_2$. The total counts ($n_1$, $n_2$) are calculated as the sum of all observed counts \emph{after} any deduplication or thresholding based on UMI abundances.

\subsection{Problem formulation: calculation of enrichment}

We can invert the problem and solve for the $R_0$ for various values of $z$, roughly corresponding to different levels of confidence. Here we used $z=-2$ for the lower bound of calculated enrichment, $z=0$ for the maximum likelihood calculated enrichment, and $z=2$ for the upper bound of calculated enrichment.

\subsection{Baseline formulation}

Five different baselines were implemented: (a) one-hot encoded feed-forward networks, (b) point prediction (``pt’’) models, (c) k-nearest-neighbors (KNN) regression models, (d) random models, and (e) binary classifiers.

The one-hot encoded feed-forward networks are incapable of generalization, and were thus used as one way of evaluating the extrapolative abilities of the models using structure-based representations.

The point prediction (``pt’’) models were trained using the MSE loss function and a point estimate of the calculated enrichment, without treatment of uncertainty. These models were used to evaluate whether there is a benefit (as measured by NLL) to using an uncertainty-aware loss function during training.

Two types of random models were implemented: one with all predicted enrichments of test-set compounds set to 1 (``predict all ones''), and one where the predictions of a FP-FFNN on the same data split were randomly shuffled (``shuffle predictions'').

The binary classifiers were trained on random splits of the data, using binary cross entropy (BCE) loss and a fixed threshold for enrichment; we defined the top 0.5\% and 0.01\% of compounds in the training set to be enriched for the DD1S CAIX and triazine library datasets, respectively. This baseline was included to evaluate whether regression modeling yields any benefit to model performance (as measured by PR / ROC AUC) over a binary classification approach.

\subsection{Models and representations tested} 

We used three different molecule representations (one-hot encoding, Morgan circular fingerprints \cite{rogers_extended-connectivity_2010}, and molecular graphs) and two model types (feed-forward neural networks and directed message-passing networks \cite{yang_analyzing_2019}). The one-hot and fingerprint-based feed-forward neural networks are abbreviated as OH-FFNN and FP-FFNN respectively, and the graph-based directed message-passing neural networks are abbreviated as D-MPNN.  RDKit \cite{rdkit} was used to generate the Morgan circular fingerprints, using 2048 bits and a radius of 3. To generate a fingerprint for each molecule, atomic neighborhoods of radius 0 to 3 centered at each atom in the molecule were hashed to integer identifiers, which were then hashed to 2048 bits to produce the molecular fingerprint as a bit vector. PyTorch was used to build the models, and the Chemprop \cite{yang_analyzing_2019} package was specifically used to build the directed message-passing networks. 
We note that this study is not meant to be an exhaustive exploration of representations and model architectures. %

\subsection{Hyperparameter optimization}
To evaluate and compare model performance for different representations, model types, and loss functions, we performed hyperparameter optimization via the Optuna Python package, using loss on the validation set as the objective function. From each run of optimization, we took the performance (as measured by NLL, MSE, rank correlation coefficient, PR AUC, or ROC AUC) of the model with the lowest validation loss. %
In particular for the FFNNs, we optimized each model’s dropout rate, hidden layer sizes, and initial learning rate (Table~\ref{tbl:DD1S_hyperparam_vals}). For the D-MPNNs, we optimized the dropout rate, hidden layer size, number of message-passing steps, number of feed-forward layers, and initial learning rate (Table~\ref{tbl:triazine_hyperparam_vals}; Figures~\ref{fig:CAIX_D-MPNN_hyperparams},~\ref{fig:CAIX_D-MPNN_bin_hyperparams}; \emph{cf.} Supporting Information). For the KNN baseline models, we optimized each model's number of nearest neighbors, k (Table~\ref{tbl:DD1S_hyperparam_vals}; Figure~\ref{fig:CAIX_KNN_k_opt_histograms}).

\subsection{Data} 

Two DEL datasets were used for model training and evaluation. The first is from a screen of DOS-DEL-1 (DD1S), which contains 108,528 compounds and is a resynthesis of the library from \citeauthor{gerry_dna_2019}\cite{gerry_dna_2019}, with carbonic anhydrase IX (CAIX) as the protein target. The primary structural motif driving enrichment in this dataset is well established in the literature: compounds with a benzenesulfonamide have very high affinity for carbonic anhydrase, whereas other compounds have comparatively low affinity \cite{gerry_dna_2019, li_versatile_2018, buller_selection_2011}. As such, this dataset was mainly used as a tool to understand what the models were learning and to validate our analyses. %
While modeled after the experiment in \citeauthor{gerry_dna_2019}, this dataset has not been previously published.

The second dataset is from a screen of a triazine library of 5,655,000 compounds, modeled after \citeauthor{clark_design_2009}'s  DEL-A \cite{clark_design_2009}, with the protein targets sEH and SIRT2, both of which have enriched compounds belonging to the triazine library. This experiment and dataset have not been previously published.

Each combination of building blocks within a library was assigned a unique integer compound ID (``cpd\_id'') for tracking. Further details on the synthesis and screening of the DELs are provided below.

\subsection{DEL synthesis}

DOS-DEL-1 \cite{gerry_dna_2019} and the triazine library \cite{clark_design_2009} were synthesized following the previously reported conditions. The eight cycle-1 building blocks (i.e., scaffolds) used in DOS-DEL-1 were prepared according to the reported methods \cite{gerry_dna_2019, verho_stereospecic_2017}, and the remainder of the building blocks for DOS-DEL-1 and the triazine library were sourced from commercial vendors and used without further purification. The DNA “headpiece” (Figure~\ref{fig:headpiece_structure}) and DNA cycle tags (Figure~\ref{fig:cycle_tags}) were purchased as custom orders from LGC Biosearch Technologies as lyophilized powders with all 5’ ends phosphorylated, and the double-stranded oligonucleotides containing primer-binding sites (Figure~\ref{fig:forward_primer_binding_site_duplex}) were purchased as custom orders from IDT as lyophilized powders with all 5’ ends phosphorylated. For both libraries, the DNA headpiece was additionally extended using Fmoc-15-amino-4,7,10,13-tetraoxapentadecanoic acid, the “AOP linker.”

DOS-DEL-1 was synthesized according to protocols adapted from previously reported conditions \cite{gerry_dna_2019} (modifications detailed below). Following each synthesis step, the material was purified by ethanol precipitation and reverse-phase ISCO (RediSep Rf Gold C18 columns) rather than semi-preparative reverse-phase HPLC. The cycle-4 tag (i.e., library identifier) and closing primer (Figures~\ref{fig:cycle_4_tag_general},~\ref{fig:cycle_4_tag_seqs}) were installed in a single ligation step rather than two separate steps. Following ligation of the cycle-4 tag and closing primer, the final screening construct was additionally purified using the Model 491 prep cell (Bio-Rad).

The triazine library was synthesized according to protocols adapted from previously reported conditions for “DEL-A” \cite{clark_design_2009} (modifications detailed below). The library was constructed using a set of 78 Fmoc-amino acids as the cycle-1 building blocks, 290 amines as the cycle-2 building blocks, and 250 amines as the cycle-3 building blocks; these building blocks differed from those used by \citeauthor{clark_design_2009} \cite{clark_design_2009}. Following ligation of the cycle-4 tag and closing primer, the final screening construct was purified using the Model 491 prep cell (Bio-Rad).

\subsection{DEL affinity selection and sequencing}

Preparation of protein and library were performed on ice and all other steps were performed at room temperature unless otherwise indicated. Affinity selection was carried out with a KingFisher Duo Prime (ThermoFisher Scientific). CAIX (R\&D Systems 2188-CA-010), sEH (Cayman Chemical 10011669-50), and SIRT2 (Cayman Chemical 20011191-100) were purchased from commercial vendors. Each protein was screened in duplicate. Buffers used are as follows: B Buffer [25 mM HEPES pH 7.4, 150 mM NaCl, 0.05\% Tween-20 (w/v)] and S Buffer [25 mM HEPES pH 7.4, 150 mM NaCl, 0.05\% Tween-20 (w/v), 0.3 mg/mL Ultrapure Salmon Sperm DNA (ThermoFisher Scientific 15632011)]. Dynabeads™ MyOne™ Streptavidin C1 (ThermoFisher Scientific 65001) were washed three times with B buffer before protein immobilization. 40 $\mu$L of beads were used for each sample. A Tris-NTA biotin adapter (Sigma 75543) was used (231 ng per sample) to immobilize the His-tagged proteins. Six molar equivalents of NiCl$_2$ were added to the adapter and incubated for 5 min. This solution was added to the strep beads agitated for 15 min; then the beads were washed three times with B buffer. Each protein (100 $\mu$L, 111 pmol, 1.11 $\mu$M) was immobilized (1 h, medium mix) and then washed twice with 200 $\mu$L B Buffer before DEL addition. 1 million copies of each library member were diluted in S Buffer and were added to the immobilized protein and incubated (60 min, medium mix). The DEL-protein mixture was washed once with S Buffer (200 $\mu$L, 3 min, medium mix) and twice with B Buffer (200 $\mu$L, 3 min, medium mix) before heat elution in B Buffer (100 $\mu$L, 10 min, 90 \degree C). 20 $\mu$L of each elution was restriction digested with StuI (NEB R0187) (2 h, 37 \degree C) and cleaned up using ChargeSwitch PCR Clean-Up Kit (ThermoFisher Scientific CS12000). $\sim4\times10^8$ molecules of DNA (19 $\mu$L) were used for PCR reactions that contained 25 $\mu$L of Invitrogen Platinum™ Hot Start PCR Master Mix (2X) (Invitrogen 13000012), 3 $\mu$L i5 index primer (10 $\mu$M stock) and 3 $\mu$L i7 index primer (10 $\mu$M stock concentration). The PCR method is as follows: 95 \degree C for 2 min; [20 cycles of 95 \degree C (15 s), 55 \degree C (15 s), 72 \degree C (30 s)]; 72 \degree C for 7 min; hold at 4 \degree C. PCR products were cleaned up with ChargeSwitch PCR Clean-Up Kit, pooled in equimolar amounts, and the 187 bp amplicon was gel purified using a 2\% E-Gel™ EX Agarose Gels (ThermoFisher Scientific G401002) and QIAquick Gel Extraction Kit (Qiagen 28704). DNA concentration was obtained using the Qubit dsDNA BR assay kit and sequenced using a HiSeq SBS v4 50 cycle kit (Illumina FC-401-4002) and HiSeq SR Cluster Kit v4 (Illumina GD-401-4001) on a HiSeq 2500 instrument (Illumina) in a single 50-base read with custom primer CTTAGCTCCCAGCGACCTGCTTCAATGTCGGATAGTG and 8-base index read using custom primer CTGATGGAGGTAGAAGCCGCAGTGAGCATGGT.

\subsection{Data processing}
Reads observed during sequencing were processed into counts for each barcoded small molecule through a custom Python script. Inexact matches  were allowed for each of the cycle-1, cycle-2, cycle-3, and library tags if observed sequences could be matched unambiguously to the most likely expected sequence. For DOS-DEL-1, up to 1 error was allowed in the cycle-1 tag and up to 2 errors in the library tag. For the triazine library, up to 1 error was allowed in the library tag. Unique molecular identifiers (UMIs) were used to de-duplicate reads using the directed graph approach implemented in the \verb+umi_tools+ package \cite{smith_umi-tools:_2017}. Due to the high sequencing depth of DOS-DEL-1 relative to the size of the library, a minimum read threshold (minimum number of reads per UMI) of 5 was imposed to prevent counts from being inflated by erroneous reads or PCR errors in the UMI region that were not identified through the directed graph approach (Figure~\ref{fig:UMI_effects}).

\subsection{Data splits}

\begin{table}[H]
\caption{Sizes of the training, validation, and test sets for each data split on the DD1S CAIX and triazine library datasets.}
\begin{tabular}{|c|l|c|c|c|}
\hline
dataset                    & \multicolumn{1}{c|}{data split} & training set size & validation set size & test set size \\ \hline
\multirow{8}{*}{DD1S} & random                          & 75969             & 10853               & 21706         \\ \cline{2-5} 
                           & cycle 1                         & 67830             & 13566               & 27132         \\ \cline{2-5} 
                           & cycle 2                         & 75208             & 11424               & 21896         \\ \cline{2-5} 
                           & cycle 3                         & 75696             & 10944               & 21888         \\ \cline{2-5} 
                           & cycle 1+2                       & 47005             & 17969               & 43554         \\ \cline{2-5} 
                           & cycle 1+3                       & 47310             & 17670               & 43548         \\ \cline{2-5} 
                           & cycle 2+3                       & 52456             & 16704               & 39368         \\ \cline{2-5} 
                           & cycle 1+2+3                     & 32785             & 19085               & 56658         \\ \hline
\multirow{8}{*}{triazine}  & random                          & 3958499           & 565501              & 1131000       \\ \cline{2-5} 
                           & cycle 1                         & 3915000           & 580000              & 1160000       \\ \cline{2-5} 
                           & cycle 2                         & 3958500           & 565500              & 1131000       \\ \cline{2-5} 
                           & cycle 3                         & 3958500           & 565500              & 1131000       \\ \cline{2-5} 
                           & cycle 1+2                       & 2740500           & 855500              & 2059000       \\ \cline{2-5} 
                           & cycle 1+3                       & 2740500           & 855500              & 2059000       \\ \cline{2-5} 
                           & cycle 2+3                       & 2770950           & 848250              & 2035800       \\ \cline{2-5} 
                           & cycle 1+2+3                     & 1918350           & 958450              & 2778200       \\ \hline
\end{tabular}
\label{tbl:data_split_sizes}
\end{table}

To split the data into training, validation, and test sets, we used both random splits and various cycle splits (Figure~\ref{fig:reps_models_loss_fns_splits}d; Table~\ref{tbl:data_split_sizes}). Random splits involved a 70\%/10\%/20\% partitioning of the data  into train/validation/test. For the cycle splits, we randomly split the building-block IDs for different cycles into training, validation, and test sets (again using a 70\%/10\%/20\% split for each set of cycle building blocks). For each cycle split along one cycle $A$, the compounds in the dataset were randomly split by their cycle-$A$ building-block ID into training, validation, and test sets. For each cycle split along two cycles $A$ and $B$, the cycle-$A$ and cycle-$B$ building-block IDs were each split into training, validation, and test sets. Compounds with a building-block ID in at least one of the two test building-block ID sets made up the final test set; compounds that did not have a test-set building-block ID but did have a validation-set building-block ID made up the final validation set; all remaining compounds made up the final training set. Similarly for cycle split along all three cycles, the compounds were first split separately by cycle-1, cycle-2, and cycle-3 building-block IDs into training, validation, and test sets. The union of the three test sets was the final test set; the union of the three validation sets minus the test set was the final validation set; all remaining compounds went to the final training set. All possible (unordered) combinations of cycles were used for the cycle splits, for a total of seven different cycle splits. We note that for these splits, building blocks were distinguished by ID rather than structure.

\subsection{Quantitative evaluation metrics}

We evaluated the performance of the regression models using the NLL loss of the test set, to match the loss function of the experimental NLL models. For the baseline point prediction (pt) models, the point value of enrichment was used to calculate loss during training. For completeness, we also evaluated the models using the MSE loss and rank correlation coefficient when treating the calculated enrichment as a precise measurement, although we believe the NLL loss is still the more relevant metric due to the stochastic nature of how sequencing data are obtained.

For the baseline binary classifiers and regression models evaluated as classifiers (using a fixed threshold for enrichment), we evaluated model performance using the PR AUC and ROC AUC of the test set. Since with DEL data we are typically interested only in the top predicted compounds in highly imbalanced datasets, PR AUC is the more relevant metric.

\subsection{Disynthon analysis}

To investigate whether the coarse correlation between predicted and calculated enrichments for the triazine sEH and triazine SIRT2 datasets (Figure~\ref{fig:sEH_SIRT2_parity_scatter_plots}ac) may be attributed to the noisiness of the data, we compared the predicted and calculated enrichments of aggregated disynthons on the same datasets (Figure~\ref{fig:sEH_SIRT2_parity_scatter_plots}bd). For each disynthon, the calculated enrichment was obtained using the summed counts for all constituent trisynthons. The predicted disynthon enrichment was a simple average of all constituent trisynthons' predicted enrichments. This simple average provides a qualitative understanding of enrichment; for future work, we note that for a more accurate measure of enrichment, the average should be weighted to correct for the non-uniform baseline abundances of the compounds, which affect the disynthon enrichment calculated by pooling counts. For example, consider the theoretical case of a disynthon that has two constituent trisynthons with counts (POI, beads only) of (4, 1) and (16, 2), which give maximum likelihood enrichment ratios of 4 and 8, respectively. Pooling the trisynthon counts gives a disynthon enrichment of 20/3 $\approx$ 6.67. Taking a simple average of the individual estimates of trisynthon enrichments, however, results in a different disynthon enrichment of 6.

\subsection{Visualization}

The first approach we used to visualize learned SAR was through atom-centered Gaussian visualizations (Figure~\ref{fig:DD1S_CAIX_learned_SAR_vis}a). These visualizations were generated by first calculating a weight for each atom in the molecule. The bits of the molecular fingerprint corresponding to the atom and the atomic neighborhoods (with radius up to 3) centered at that atom were masked, and the model’s prediction on the masked fingerprint was subtracted from the model’s prediction on the original fingerprint. The weights were then normalized by dividing by the absolute value of the weight with the greatest magnitude. Finally, these normalized weights were used to calculate atom-centered Gaussian distributions and generate a map of the molecule indicating which parts of the molecule contribute positively or negatively to enrichment.

The second approach we explored for visualization was calculating bit and substructure importance (Figure~\ref{fig:DD1S_CAIX_learned_SAR_vis}b). To do this, we first calculated a weight for each of the 2048 fingerprint bits, corresponding to the bit’s contribution to model predictions. We note that there were fingerprint bit collisions where multiple substructures were mapped to the same bit; to quantify the contribution of each specific substructure, even among substructures mapped to the same bit, we calculated substructure weights in addition to bit weights. To calculate each bit weight, we first computed molecule-level weights---for each molecular fingerprint with the bit, we took the model’s prediction on the original fingerprint, and subtracted the model’s prediction on the same fingerprint but with the bit masked. We averaged these molecule-level weights to get the final bit weight. Afterwards, we analyzed the substructures mapped to the 5 bits with the highest weights (to investigate positive SAR) and the 3 bits with the lowest weights (to investigate negative SAR). More specifically, we calculated a weight for each substructure (corresponding to the substructure’s contribution to model predictions) by averaging the molecule-level weights corresponding to molecules with the substructure.

\subsection{UMAP details}

We trained a UMAP embedding on 4096-bit radius-3 Morgan circular fingerprints of a combination of 600k compounds randomly sampled from PubChem, a random 10\% of DOS-DEL-1, and a random 10\% of the triazine library. We increased the number of fingerprint bits from 2048 (as used for the QSAR models) to 4096 in order to reduce bit collisions and thus encode more structural information in the fingerprints. The trained embedding was applied separately to each of the three libraries, i.e., the 600k PubChem compounds used for training, all compounds in DOS-DEL-1, and the subset of the triazine library used for training (Figure~\ref{fig:chem_space_plots}). As for UMAP parameters, the \texttt{metric} parameter was set to \texttt{`jaccard’} since we were using bit-based fingerprints. Otherwise, default UMAP parameters (\texttt{n\_neighbors=15, min\_dist=0.1, n\_components=2}) were used without tuning.

\begin{acknowledgement}

This work has been funded in part through the National Institute of General Medical Sciences (R35GM127045 awarded to S.L.S.) and the NIBR Scholar's Program (co-led by Karin Briner, Fred Zicri, Cindy Hon, and Stuart Schreiber) and we are grateful to Stuart Schreiber for his support of this work. KSL thanks the MIT UROP Office for additional funding support. We thank Alicia Lindeman (Novartis) for performing the next-generation sequencing. We thank the GCP research credits program for supporting the computational costs of this work run on Google Cloud Platform. The authors acknowledge the MIT SuperCloud and Lincoln Laboratory Supercomputing Center for providing HPC resources that have contributed to the research results reported within this paper. 
\end{acknowledgement}

\section{Data and Software Availability}
All code, data, and tabulated results used in this study can be found at \url{https://github.com/coleygroup/del_qsar}. 

\begin{suppinfo}

The Supporting Information contains additional methods relating to construction of the NLL loss function, calculation of bit and substructure importance for SAR visualization with substructure pairs, and further details of the DELs, hyperparameter optimization, and data processing for the DD1S CAIX dataset. We also include results relating to further comparisons of model performance on both scaffold-based and random splits of the data, and we provide tables with all raw performance numbers which appear in the charts in this paper. In addition, we analyze the class balance of classification datasets, analyze outliers and compare DD1S compounds to on-demand libraries, and provide further visualizations and a list of the RDKit calculated features used by our models.

\end{suppinfo}

\bibliography{main}

\begin{tocentry}

\centering
\includegraphics[height=1.5in]{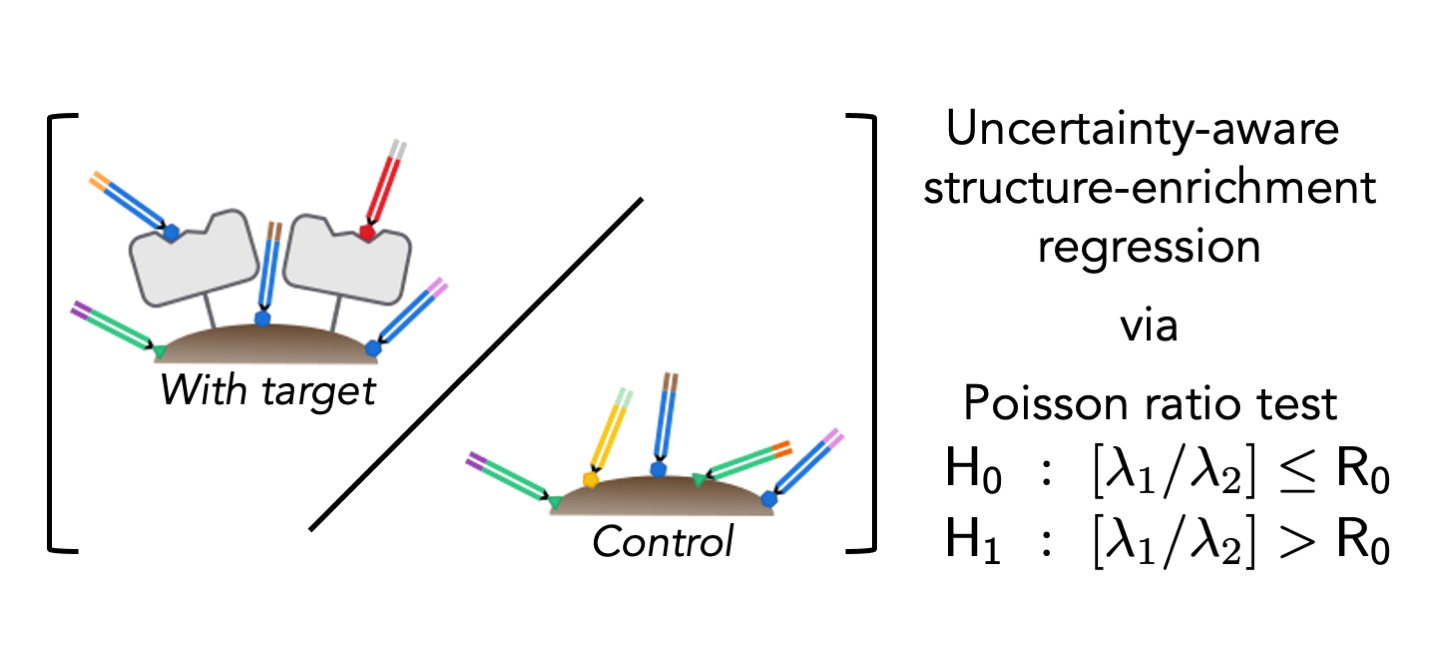}

\end{tocentry}

\end{document}


\renewcommand{\thesection}{S\arabic{section}}
\renewcommand{\thefigure}{S\arabic{figure}}
\renewcommand{\thetable}{S\arabic{table}}
\renewcommand{\thepage}{S-\arabic{page}}

\setcounter{figure}{0} 
\setcounter{table}{0} 
\setcounter{page}{1}

\captionsetup{font=footnotesize}
\newgeometry{left=0.5in, right=0.5 in, top=0.75in, bottom=0.75in}

\section{Additional Methods}

\subsection{NLL loss function}

We assume that the counts for each barcode are distributed according to a Poisson distribution \cite{kuai_randomness_2018}. Poisson distributions are characterized by a single parameter $\lambda$ that defines a rate of seeing a certain event under a continuous observation interval. Because the total number of counts is large ($n > 100,000$) and the number of counts for any individual barcode will be relatively small ($p < 0.01$), we approximate the counts/reads as a pseudo-continuous observation interval. The parameters describing the POI and beads-only distributions, $\lambda_1$ and $\lambda_2$, are unknown. There is also some known ratio between these distributions, $R = \lambda_1 / \lambda_2$. \textbf{This $R$ is what we refer to as the enrichment ratio. }

With our $k_1$ counts from $n_1$ observations for the POI, we can estimate $\hat \lambda_1$ as $k_1/n_1$---this is the most likely value of $\lambda_1$ given what we have observed, but it is unlikely to be exact. For example, if our true $\lambda_1 = 0.01$ and we have $n_1 =1000$ observations, we would see $k_1 = 10$ \emph{on average}; sometimes we'd see fewer counts, sometimes we'd see greater counts. The maximum likelihood estimates for $\lambda_2$ is similarly $\hat \lambda_2 = k_2 / n_2$.

A naive enrichment metric could therefore be $\hat R = \hat \lambda_1 / \hat \lambda_2$---what is the ratio of the most likely values for these two Poisson distributions? This works okay most of the time, but does not take into account uncertainty well. Particularly in the low-count regime, things fall apart. If we observed $k_1 = 1$ in the experiment and $k_2 = 0$ in the beads-only, then we would estimate $\hat R = \infty$.

Instead of looking at the ratio of the most likely $\lambda$ values for each population, we look at the most likely ratio of the $\lambda$ values. The distinction is a little subtle, but the calculation and confidence intervals look quite different. Formally, we define a hypothesis test:

\begin{equation} H_0:~~\frac{\lambda_1}{\lambda_2} \le R_0 ~~~~~~~~~~ H_1:~~\frac{\lambda_1}{\lambda_2} > R_0 \end{equation}

We would like to make an inference about the ratio between the true rates for the POI and beads-only case. Several metrics have been proposed for this test, but here we use one based on a variance stabilizing square root transformation \cite{gu_testing_2008}. 
We calculate a $z$-score as:

\begin{equation}
    z = 2 \frac{ \left( k_1 + \frac{3}{8} \right)^{1/2} - \left( k_2 + \frac{3}{8} \right)^{1/2} \left( \frac{ n_1}{n_2}R_0 \right)^{1/2}}{\left( 1 + \frac{n_1}{n_2}R_0 \right)^{1/2}}
\end{equation}
This $z$ should be normally distributed with a mean of 0 and variance of 1. Now, for specific values of $R_0$, we can see what this ratio test will do.

Consider the simplifying case where $n_1 = n_2$ and we're testing $R_0 = 0$. The hypothesis test is therefore asking whether we believe the ratio between the true Poisson rates to be greater than $R_0 = 1$. Our $z$ score simplifies to
\begin{equation}
    z = \sqrt{2} \left( \left( k_1 + \frac{3}{8} \right)^{1/2} - \left( k_2 + \frac{3}{8} \right)^{1/2} \right)~~~~~~~n_1 = n_2, ~~R_0 = 1
\end{equation}
Clearly, if $k_1 = k_2$, we have no reason to believe that the true enrichment ratio is greater (or less) than one, so we will get $z = 0$. If we observe $k_1 > k_2$, we will start to believe there has been some enrichment ($z>0$). The greater the discrepancy between $k_1$ and $k_2$, the more significant this seems to be, and the more likely we will reject the null hypothesis. 

We can finally invert the problem and solve for the $R_0$ for various values of $z$, roughly corresponding to different levels of confidence. Solving for $z=0$ gives us the most likely enrichment ratio; we can calculate a confidence interval by solving for $z=\pm 2$. %

\begin{figure}[H] 
    \centering
        \includegraphics[scale=1]{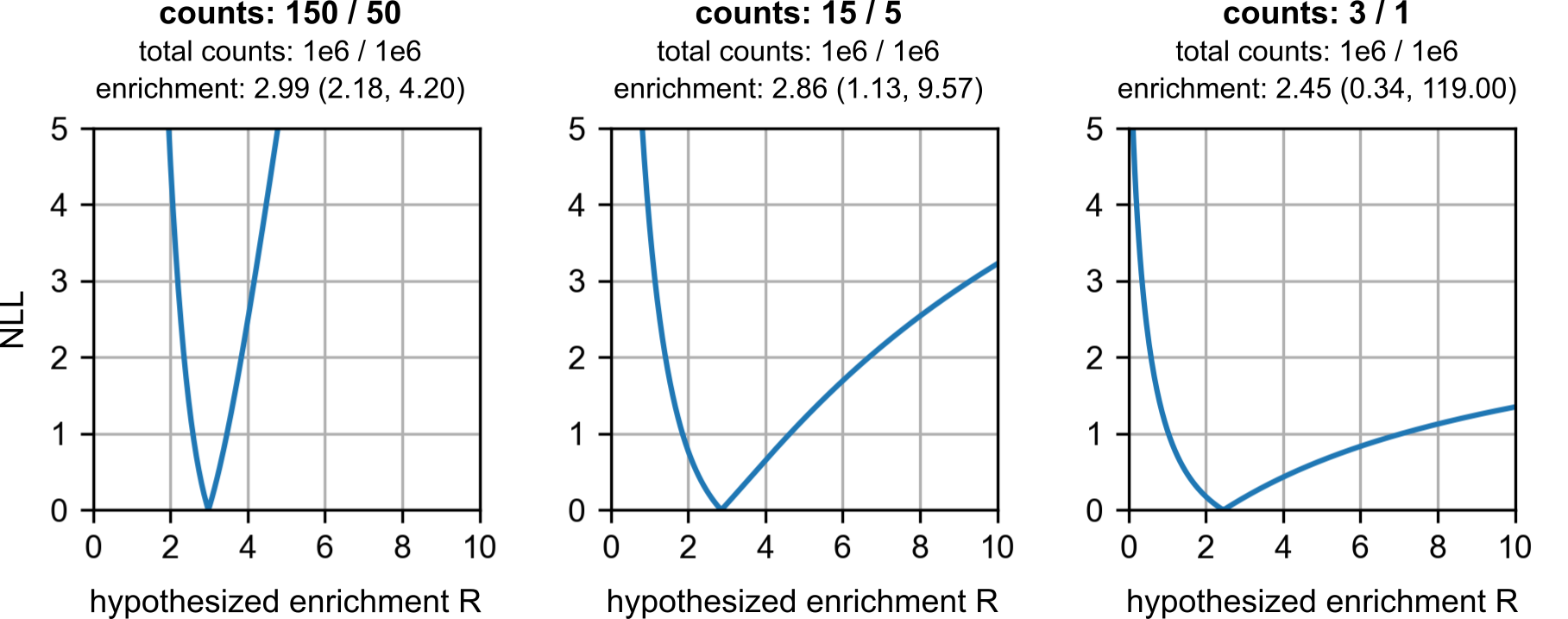}
    \caption{Plots of NLL as a function of predicted enrichment for various theoretical counts.}
    \label{fig:NLL_plots}
\end{figure}

As a toy example, consider a few cases where there seems to have been some enrichment. Take $n_1 = n_2 = 1,000,000$, and the following ($k_1$, $k_2$) values: (150, 50), (15, 5), and (3, 1). That is, we have a million total counts for the POI and beads-only experiments, and for a particular barcode, we have (POI, beads-only) counts of (150, 50), (15, 5), or (3, 1). The calculated $R_0$ for each of these cases is 2.99, 2.86, and 2.45, respectively, with confidence intervals of [2.18, 4.20], [1.13, 9.57], and [0.34, 119.00]. %
We can look at the negative-log likelihood (NLL) of failing to reject the null hypothesis that the true enrichment ratio is $R$, given $k_1$, $n_1$, $k_2$, and $n_2$, as a function of $R$ (Figure~\ref{fig:NLL_plots}).

These plots demonstrate the NLL loss function's treatment of uncertainty---with higher barcode counts, there is less uncertainty about the enrichment, and thus the loss function sharply penalizes model predictions that stray from the most likely enrichment. On the other hand, with lower barcode counts, there is greater uncertainty, so the loss function is more lenient and tolerates predictions within a wider plausible range around the most likely enrichment.

We note that the enrichment calculation itself (using the Poisson ratio test) can aggregate over different sets of synthon combinations (e.g., A, B, C, AB, BC, AC, ABC, where A/B/C are monosynthons, AB/BC/AC are disynthons, and ABC is a trisynthon). With the higher counts after aggregation, one would expect and observe reduced uncertainties. Whether this is a `better' treatment or how one might use that information in the surrogate structure-enrichment model is not totally clear. Training on monosynthons alone would not achieve enough chemical diversity (for the DELs in this study, at least) to generalize beyond the ca. hundreds of structures. We could assemble disynthon structures as done by \citeauthor{mccloskey_machine_2020} and train the regression model on all AB, BC, AC disynthons, but it may not make sense to combine trisynthons, disynthons, and monosynthons into a single surrogate model, particularly because the number of trisynthons far exceeds the number of disynthons or monosynthons in a typical library.

\subsection{Library sizes and total sequencing counts}

\begin{table}[H]
\caption{Library sizes and total counts.}
\begin{tabular}{|m{3cm}|m{3.5cm}|m{5cm}|m{5cm}|}
\hline
dataset & number of\newline compounds & total number of counts in\newline experimental condition\newline (with POI) & total number of counts in\newline beads-only control \\ \hline
DD1S CAIX & 108,528 & 638,831 & 5,208,230 \\ \hline
triazine sEH & \multirow{2}{*}{5,655,000} & 5,085,569 & \multirow{2}{*}{5,395,114} \\ \cline{1-1} \cline{3-3}
triazine SIRT2 &  & 3,497,768 &  \\ \hline
\end{tabular}
\label{tbl:lib_sizes_counts}
\end{table}

\subsection{Hyperparameter optimization}

On the DD1S CAIX dataset, we performed 5 differently seeded runs of 100 iterations of Bayesian hyperparameter optimization (using loss on the validation set as the objective function) for each model type and data split type (random split and the various cycle splits for the regression models; just random split for the binary classifiers), using the Optuna Python package. 

On the larger triazine sEH and triazine SIRT2 datasets, we reduced the hyperparameter search space and performed 5 differently seeded runs of 25 iterations of optimization for each FFNN and 4 iterations of optimization for representative D-MPNN(s), due to computational cost. For the binary classifiers, we used one representative D-MPNN for optimization; for the regression models, we used one experimental D-MPNN (trained using NLL) and one baseline D-MPNN pt model (trained using MSE), both on a random split of the data. Specifically for the FFNNs, we increased the smallest possible hidden layer size to ensure adequate model capacity, and decreased the range of values for the initial learning rate based on hyperparameter optimization results for the FFNNs on the DD1S CAIX dataset. For the representative D-MPNNs, we similarly decreased the range of values for the initial learning rate and fixed other hyperparameters based on results for the D-MPNNs on the DD1S CAIX dataset. We used the optimization results to fix the initial learning rate for the other D-MPNNs (separately for the NLL and pt regression models), and performed a total of 3 differently seeded runs of training and evaluation for the D-MPNNs on each data split type used (including the optimization runs for random split). The regression D-MPNNs trained using NLL on the triazine SIRT2 dataset were notably sensitive to the learning rate, and so we re-ran training and evaluation for these models at a lower fixed learning rate.

For the baseline KNN models, the number of neighbors (k) was optimized on the DD1S CAIX dataset.

\begin{table}[H]
\caption{Ranges and value types for hyperparameter optimization on the DD1S CAIX dataset.}
\begin{tabular}{|c|l|c|c|l}
\cline{1-1} \cline{3-4}
DD1S & \multicolumn{1}{c|}{} & \multicolumn{2}{c|}{range of values} & \multicolumn{1}{c}{} \\ \hline
model & \multicolumn{1}{c|}{hyperparameter} & lower bound & upper bound & \multicolumn{1}{c|}{value type} \\ \hline
\multirow{5}{*}{FFNN} & initial learning rate & $1\mathrm{e}{-5}$ & $1\mathrm{e}{-1}$ & \multicolumn{1}{l|}{\begin{tabular}[c]{@{}l@{}}continuous \\ (logarithmic scale)\end{tabular}} \\ \cline{2-5} 
 & number of hidden layers & 1 & 3 & \multicolumn{1}{l|}{\begin{tabular}[c]{@{}l@{}}discrete \\ (step size = 1)\end{tabular}} \\ \cline{2-5} 
 & layer size architecture & N/A & N/A & \multicolumn{1}{l|}{\begin{tabular}[c]{@{}l@{}}categorical \\ (flat, pyramid with\\ factor of 2 or 4)\end{tabular}} \\ \cline{2-5} 
 & largest hidden layer size & 16 & 1024 & \multicolumn{1}{l|}{\begin{tabular}[c]{@{}l@{}}discrete \\ (base 2 logarithmic\\ scale, step size = 1)\end{tabular}} \\ \cline{2-5} 
 & dropout rate & 0 & 0.5 & \multicolumn{1}{l|}{\begin{tabular}[c]{@{}l@{}}discrete\\ (step size = 0.05)\end{tabular}} \\ \hline
\multirow{5}{*}{D-MPNN} & initial learning rate & $1\mathrm{e}{-5}$ & $2\mathrm{e}{-3}$ & \multicolumn{1}{l|}{\begin{tabular}[c]{@{}l@{}}continuous\\ (logarithmic scale)\end{tabular}} \\ \cline{2-5} 
 & number of message-passing steps & 2 & 6 & \multicolumn{1}{l|}{\begin{tabular}[c]{@{}l@{}}discrete \\ (step size = 1)\end{tabular}} \\ \cline{2-5} 
 & number of FFN hidden layers & 1 & 3 & \multicolumn{1}{l|}{\begin{tabular}[c]{@{}l@{}}discrete \\ (step size = 1)\end{tabular}} \\ \cline{2-5} 
 & size of FFN hidden layers & 300 & 2400 & \multicolumn{1}{l|}{\begin{tabular}[c]{@{}l@{}}discrete \\ (step size = 100)\end{tabular}} \\ \cline{2-5} 
 & dropout rate & 0 & 0.5 & \multicolumn{1}{l|}{\begin{tabular}[c]{@{}l@{}}discrete \\ (step size = 0.05)\end{tabular}} \\ \hline
 \multirow{1}{*}{KNN} & number of nearest neighbors (k) & $1$ & $9$ & \multicolumn{1}{l|}{\begin{tabular}[c]{@{}l@{}}discrete\\ (step size = 2)\end{tabular}} \\ \hline
\end{tabular}
\label{tbl:DD1S_hyperparam_vals}
\end{table}

\begin{table}[H]
\caption{Ranges and value types for hyperparameter optimization on the triazine datasets.}
\begin{tabular}{|c|l|c|c|l}
\cline{1-1} \cline{3-4}
triazine & \multicolumn{1}{c|}{} & \multicolumn{2}{c|}{range of values} & \multicolumn{1}{c}{} \\ \hline
model & \multicolumn{1}{c|}{hyperparameter} & lower bound & upper bound & \multicolumn{1}{c|}{value type} \\ \hline
\multirow{5}{*}{FFNN} & initial learning rate & $1\mathrm{e}{-5}$ & $3\mathrm{e}{-2}$ & \multicolumn{1}{l|}{\begin{tabular}[c]{@{}l@{}}continuous \\ (logarithmic scale)\end{tabular}} \\ \cline{2-5} 
 & number of hidden layers & 1 & 3 & \multicolumn{1}{l|}{\begin{tabular}[c]{@{}l@{}}discrete \\ (step size = 1)\end{tabular}} \\ \cline{2-5} 
 & layer size architecture & N/A & N/A & \multicolumn{1}{l|}{\begin{tabular}[c]{@{}l@{}}categorical \\ (flat, pyramid with\\ factor of 2 or 4)\end{tabular}} \\ \cline{2-5} 
 & \begin{tabular}[c]{@{}l@{}}smallest hidden layer size\\ (largest possible is 1024)\end{tabular} & 64 & 1024 & \multicolumn{1}{l|}{\begin{tabular}[c]{@{}l@{}}discrete \\ (base 2 logarithmic\\ scale, step size = 1)\end{tabular}} \\ \cline{2-5} 
 & dropout rate & 0 & 0.5 & \multicolumn{1}{l|}{\begin{tabular}[c]{@{}l@{}}discrete\\ (step size = 0.05)\end{tabular}} \\ \hline
\begin{tabular}[c]{@{}c@{}}D-MPNN\\ (random split)\end{tabular} & initial learning rate & $1\mathrm{e}{-5}$ & $3\mathrm{e}{-4}$ & \multicolumn{1}{l|}{\begin{tabular}[c]{@{}l@{}}categorical\\ ($1\mathrm{e}{-5}$, $3\mathrm{e}{-5}$,\\ $1\mathrm{e}{-4}$, $3\mathrm{e}{-4}$)\end{tabular}} \\ \hline
\end{tabular}
\label{tbl:triazine_hyperparam_vals}
\end{table}

\subsection{Bit and substructure importance}

To ensure accurate counting of substructures, we used custom SMARTS patterns for the substructures. These are shown in tables in the Additional Results section.

\subsubsection{Substructure-pair analysis}

In this analysis we investigated contributions of pairs of substructures to model predictions. To do this, we took the top substructure $a$ (mapped to bit $A$) from the single-substructure analysis, and looked just at the molecules with that substructure. For each fingerprint bit $B$ not corresponding to the top substructure, we calculated a bit weight by first calculating molecule-level weights (the model’s prediction on the original molecular fingerprint minus the model’s prediction on same fingerprint but with bits $A$ and $B$ masked). We then averaged over these molecule-level weights to get the final weight for the bit. Afterwards, we analyzed the substructures mapped to the 5 bits with the highest weights (to investigate positive SAR) and the 3 bits with the lowest weights (to investigate negative SAR). To do this, we calculated a weight for each substructure $b$ of interest by averaging the molecule-level weights corresponding to the molecules that have both substructures $a$ and $b$.

\begin{figure} [H]
    \centering
        \includegraphics[scale=1]{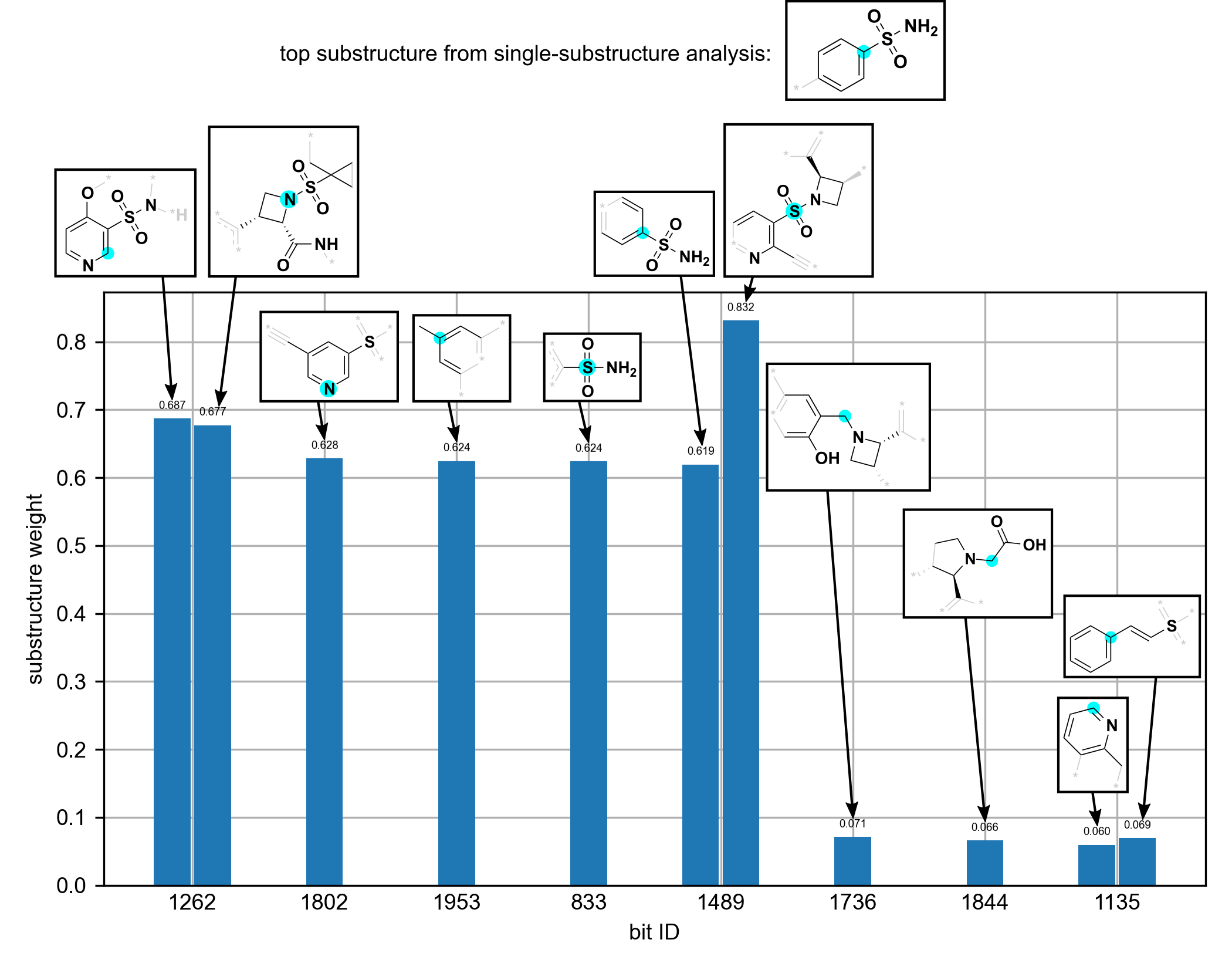}
    \caption{Substructure-pair analysis on the DD1S CAIX dataset (random split, seed 0; \emph{cf.} Figure~\ref{fig:reps_models_loss_fns_splits}d), based on the top substructure from the single-substructure analysis.}
    \label{fig:CAIX_twobits_seed_0_bars}
\end{figure}

\begin{figure} [H]
    \centering
        \includegraphics[scale=1]{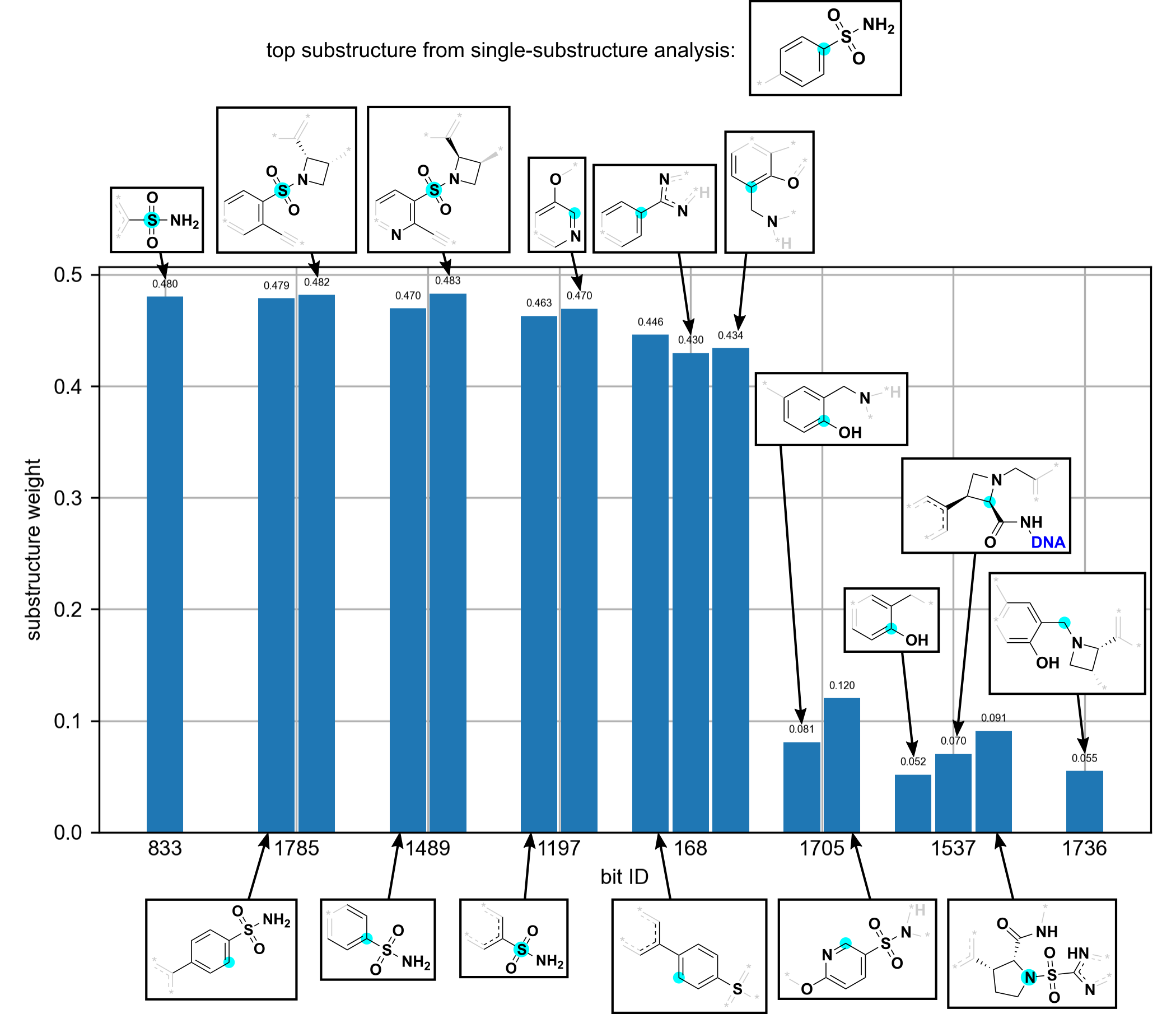}
    \caption{Substructure-pair analysis on the DD1S CAIX dataset (random split, seed 1; \emph{cf.} Figure~\ref{fig:reps_models_loss_fns_splits}d), based on the top substructure from the single-substructure analysis.}
    \label{fig:CAIX_twobits_seed_1_bars}
\end{figure}

\begin{figure} [H]
    \centering
        \includegraphics[scale=1]{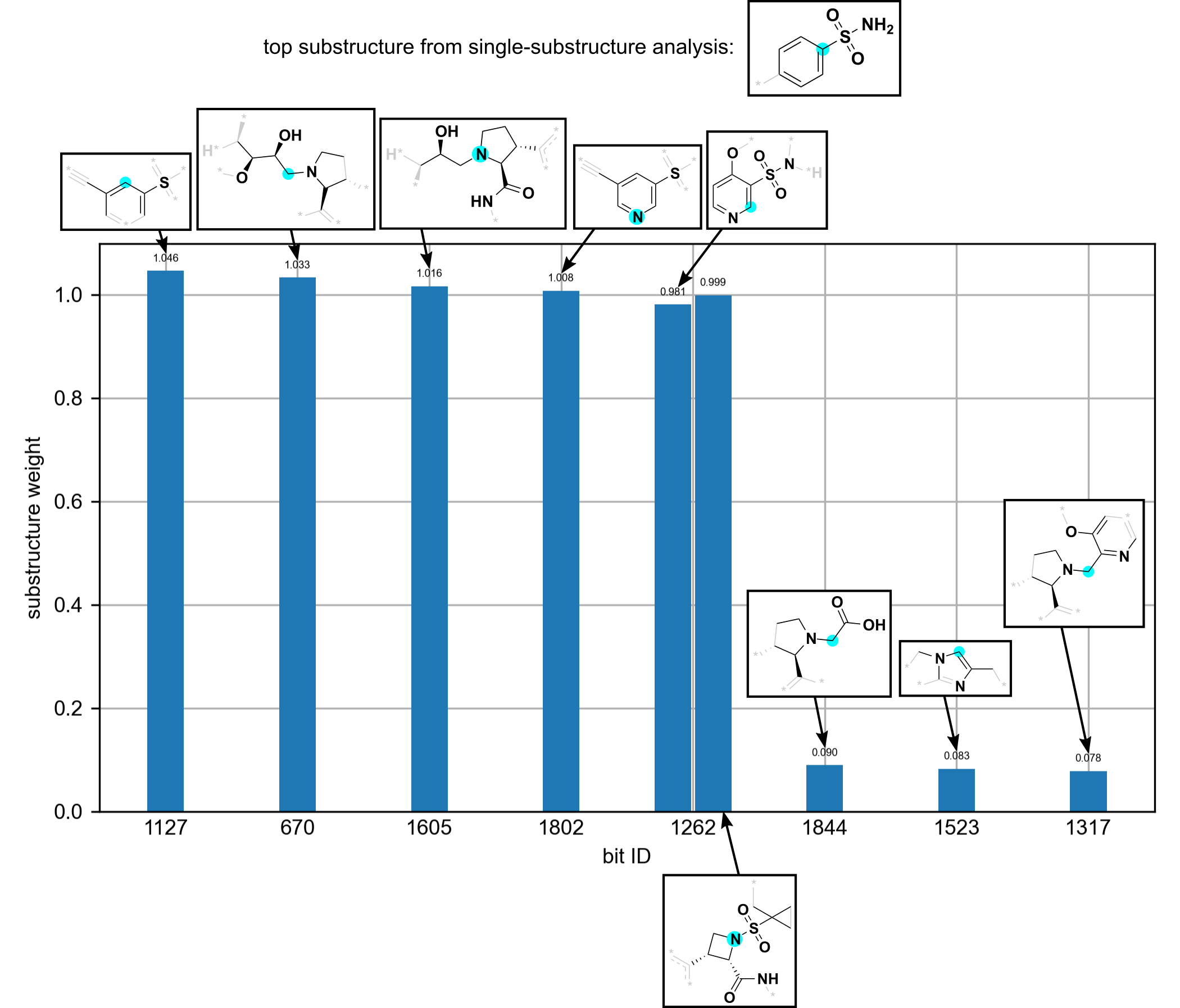}
    \caption{Substructure-pair analysis on the DD1S CAIX dataset (random split, seed 2; \emph{cf.} Figure~\ref{fig:reps_models_loss_fns_splits}d), based on the top substructure from the single-substructure analysis.}
    \label{fig:CAIX_twobits_seed_2_bars}
\end{figure}

\begin{figure} [H]
    \centering
        \includegraphics[scale=1]{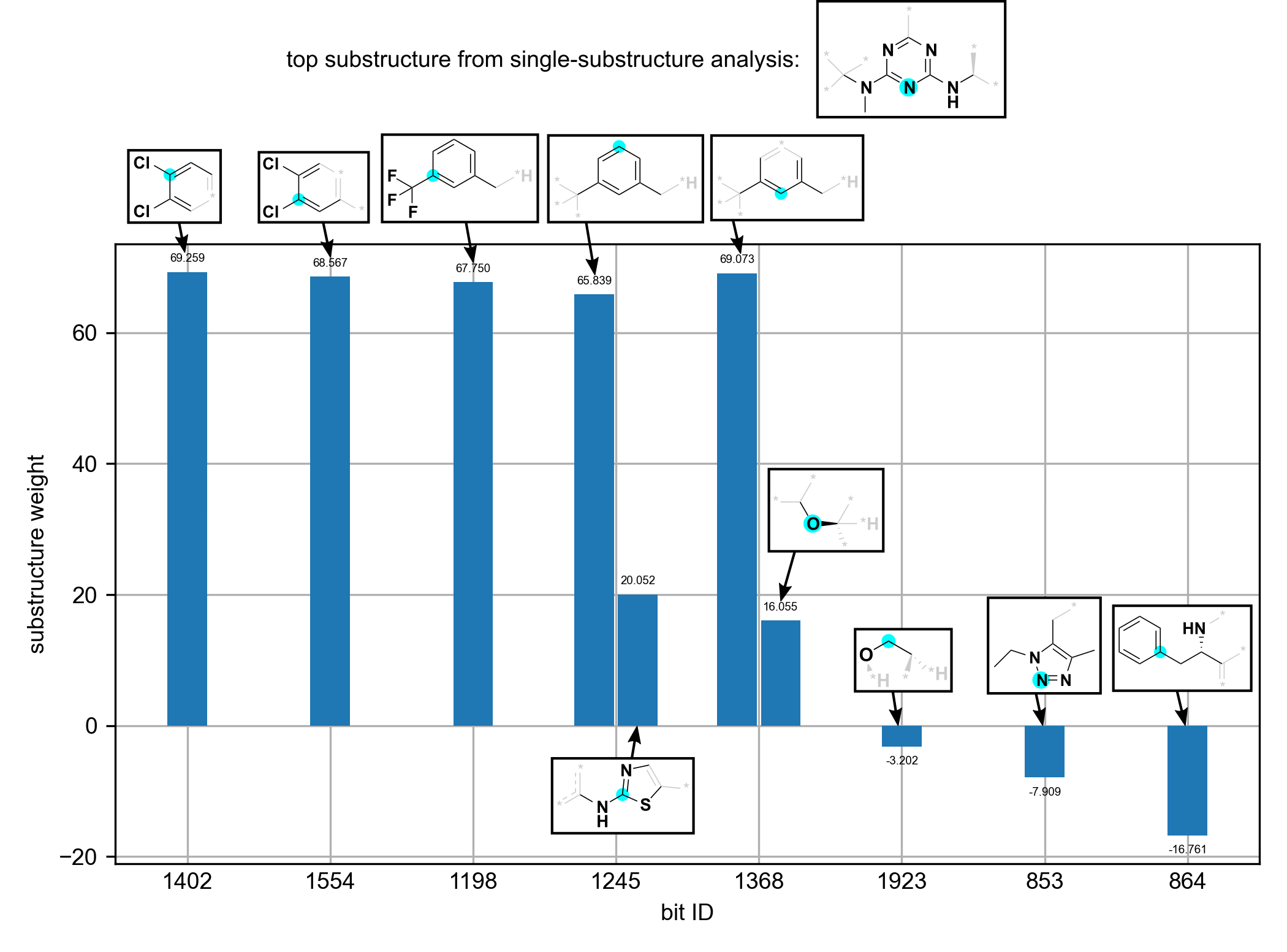}
    \caption{Substructure-pair analysis on the triazine sEH dataset (random split, seed 0; \emph{cf.} Figure~\ref{fig:reps_models_loss_fns_splits}d), based on the top substructure from the single-substructure analysis.}
    \label{fig:sEH_twobits_bars}
\end{figure}

\begin{figure} [H]
    \centering
        \includegraphics[scale=1]{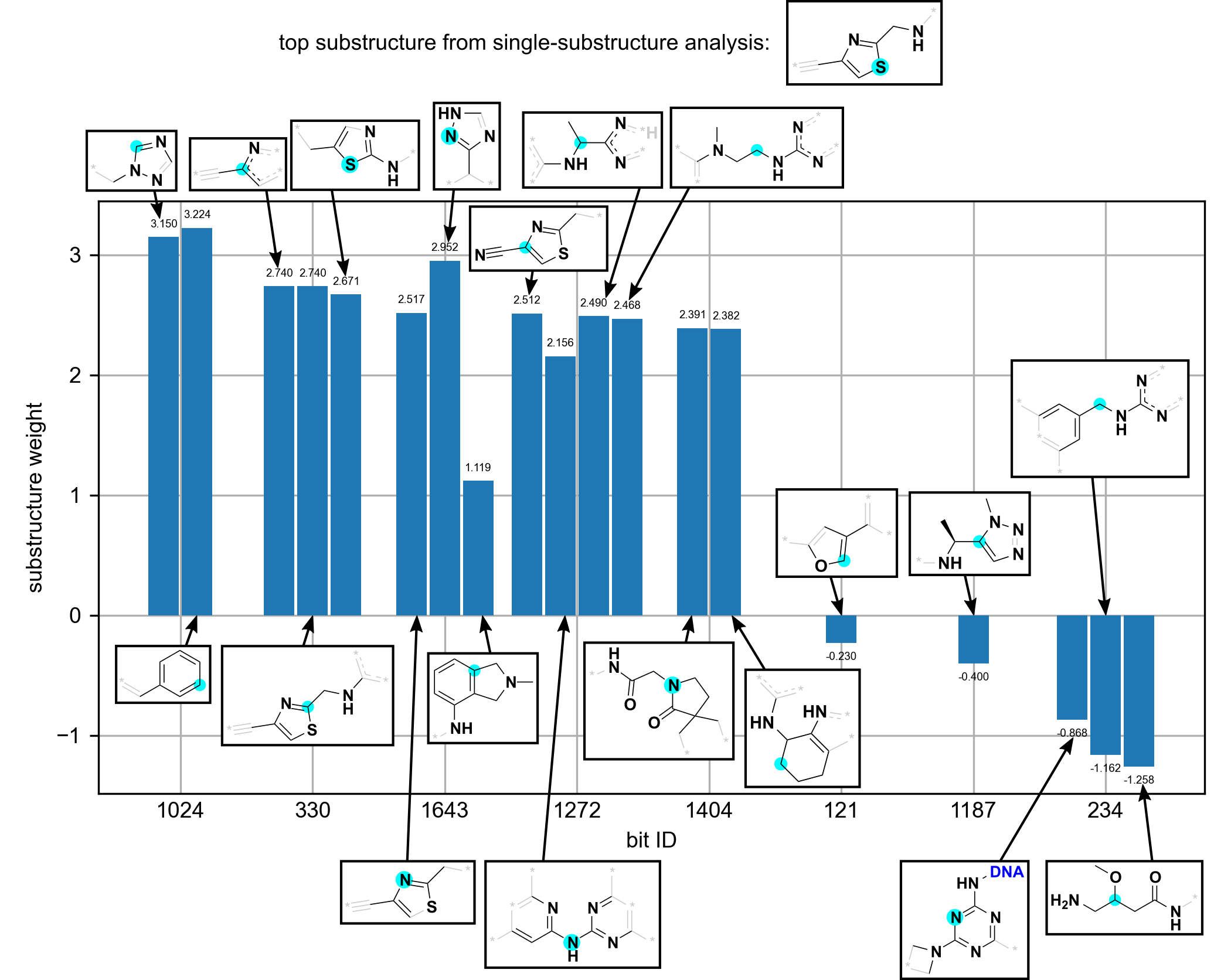}
    \caption{Substructure-pair analysis on the triazine SIRT2 dataset (random split, seed 0; \emph{cf.} Figure~\ref{fig:reps_models_loss_fns_splits}d), based on the top substructure from the single-substructure analysis.}
    \label{fig:SIRT2_twobits_bars}
\end{figure}

The results for each dataset are shown in the Additional Results section (Figures~\ref{fig:CAIX_twobits_seed_0_bars},~\ref{fig:CAIX_twobits_seed_1_bars},~\ref{fig:CAIX_twobits_seed_2_bars},~\ref{fig:sEH_twobits_bars}, ~\ref{fig:SIRT2_twobits_bars}).

\begin{figure} [H]
    \centering
        \includegraphics[scale=1]{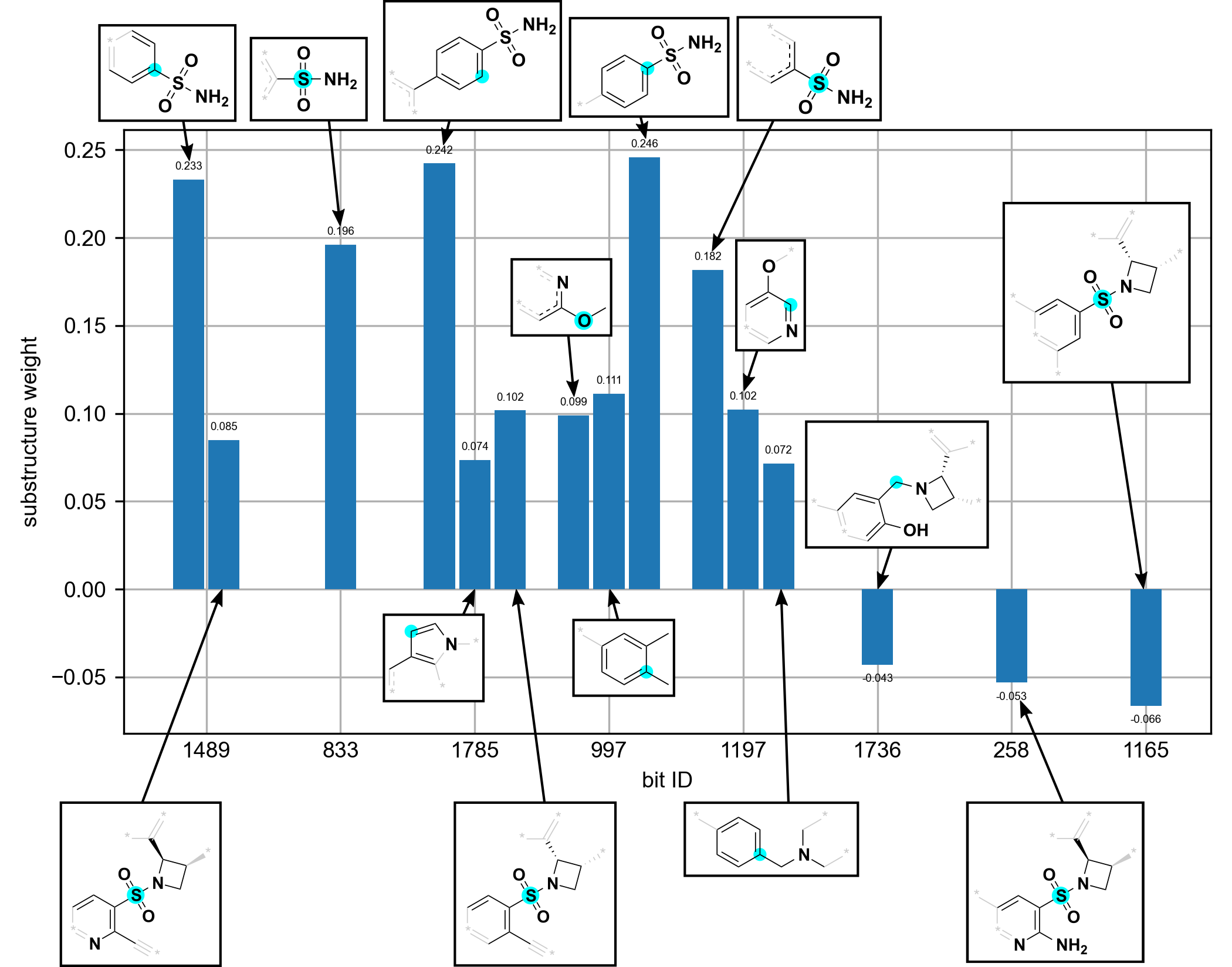}
    \caption{Single-substructure analysis on the DD1S CAIX dataset (random split, seed 1; \emph{cf.} Figure~\ref{fig:reps_models_loss_fns_splits}d), including substructures mapped to the top 5 and bottom 3 bits.}
    \label{fig:CAIX_onebit_seed_1_bars}
\end{figure}

We note that since we only considered molecules with the top substructure $a$ from the single-substructure analysis, not all substructures considered for a certain bit in the single-substructure analysis may be included for the same bit in this substructure-pair analysis. For instance, the middle substructure shown for bit 1785 in the DD1S CAIX (seed 1) single-substructure analysis (Figure~\ref{fig:CAIX_onebit_seed_1_bars}) is not included for the same bit in the substructure-pair analysis on the same dataset (Figure~\ref{fig:CAIX_twobits_seed_1_bars}), since that substructure corresponds to a cycle-3 building block, and the top substructure from the single-substructure analysis also corresponds to a cycle-3 building block.

We also note that because a given building block or moiety has many atomic neighborhoods of radius 0 to 3, some of the top bits for this analysis may correspond to other atomic neighborhoods of the same moiety that the top substructure corresponds to (e.g., a benzenesulfonamide for the DD1S CAIX dataset).

\subsection{DEL synthesis}

\begin{figure}[H] 
    \centering
        \includegraphics[scale=1]{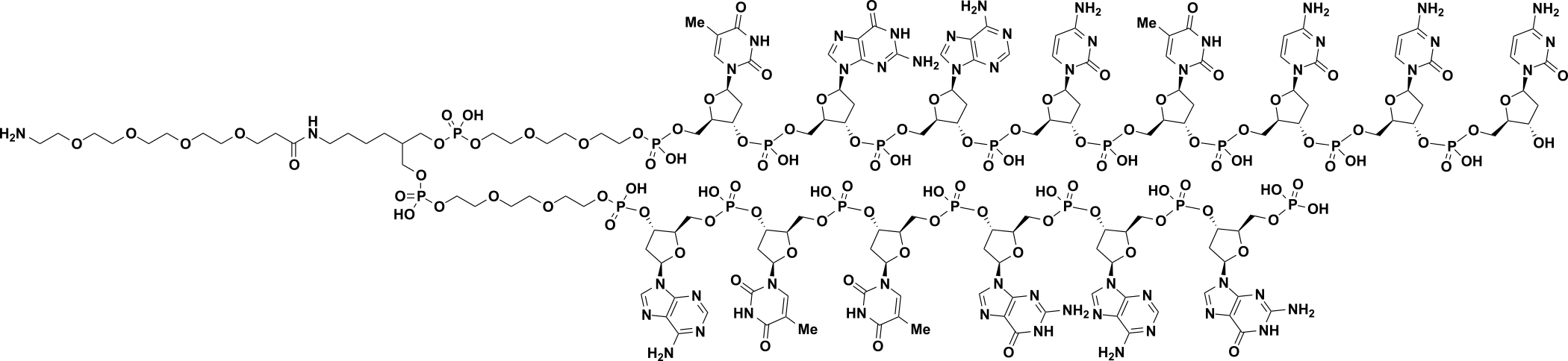}
    \caption{Chemical structure of the ``headpiece'' with the ``AOP-linker.''}
    \label{fig:headpiece_structure}
\end{figure}

\begin{figure}[H] 
    \centering
        \includegraphics[scale=1]{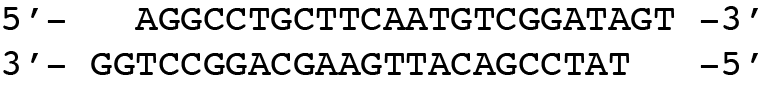}
    \caption{Forward primer binding site duplex.}
    \label{fig:forward_primer_binding_site_duplex}
\end{figure}

\begin{figure}[H] 
    \centering
        \includegraphics[scale=1]{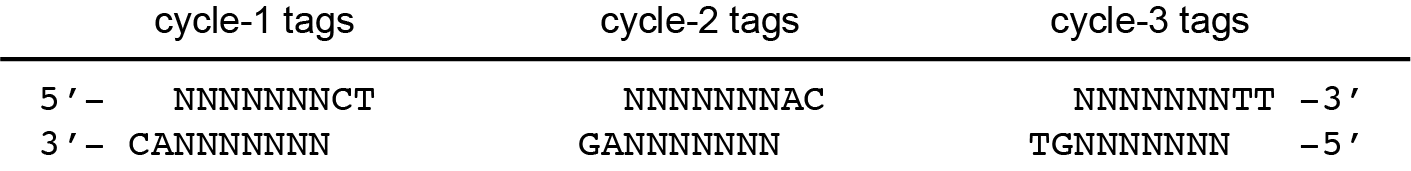}
    \caption{Structures of DNA oligonucleotides used as cycle tags.}
    \label{fig:cycle_tags}
\end{figure}

\begin{figure}[H] 
    \centering
        \includegraphics[scale=1]{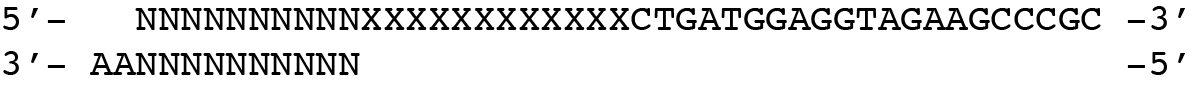}
    \caption{DNA duplex containing the cycle-4 tag (library identifier, designated with N’s), unique molecular identifier region (designated with X’s), and closing primer binding site.}
    \label{fig:cycle_4_tag_general}
\end{figure}

\begin{figure}[H] 
    \centering
        \includegraphics[scale=1]{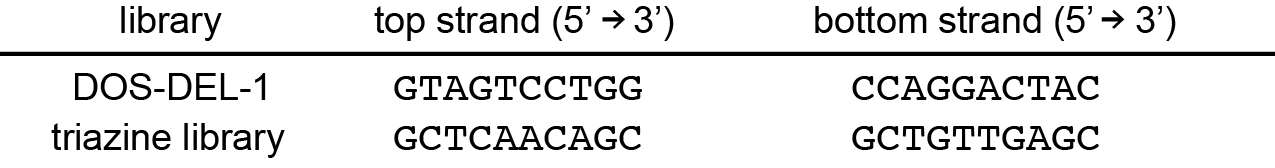}
    \caption{Cycle-4 tag sequences for DOS-DEL-1 and the triazine library.}
    \label{fig:cycle_4_tag_seqs}
\end{figure}

\subsection{Data processing}

\begin{figure}[H] 
    \centering
        \includegraphics[scale=1]{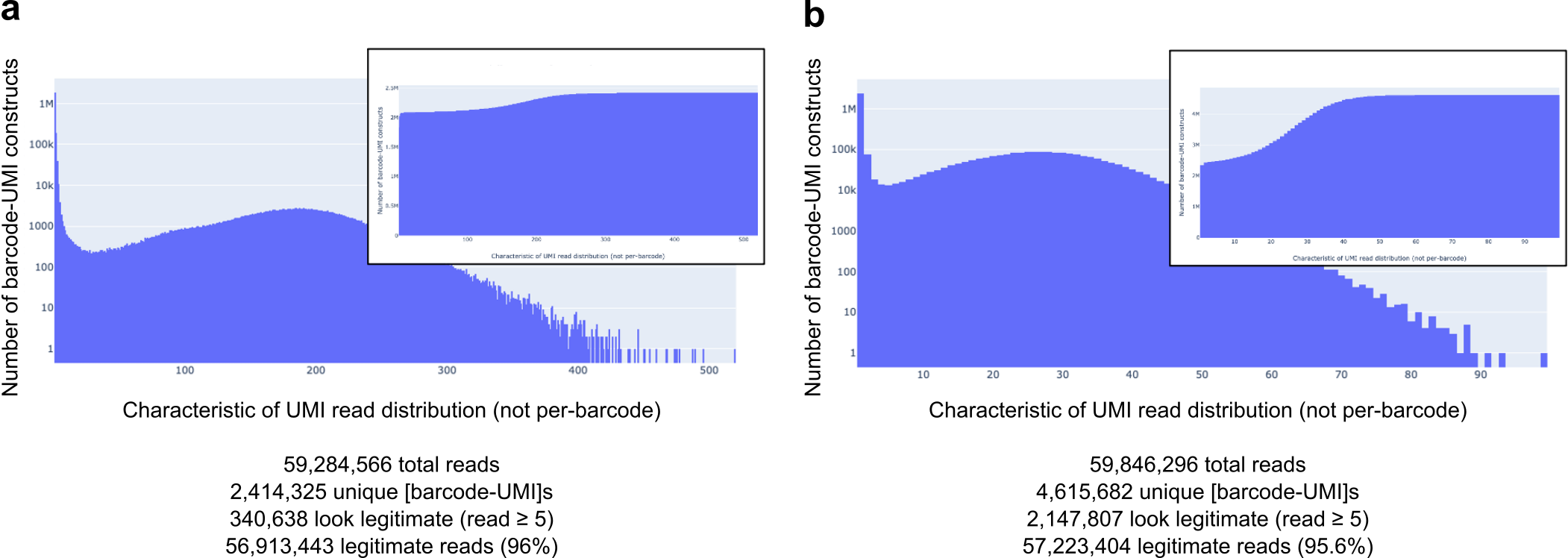}
    \caption{Abundances of UMIs for any ([umi-barcode] construct) for the DD1S CAIX \textbf{(a)} experimental (with POI) and \textbf{(b)} beads (without POI) data. Insets are cumulative histograms.}
    \label{fig:UMI_effects}
\end{figure}

\section{Additional Results}

\subsection{Hyperparameter optimization}

\begin{figure}[H] 
    \centering
        \includegraphics[scale=1]{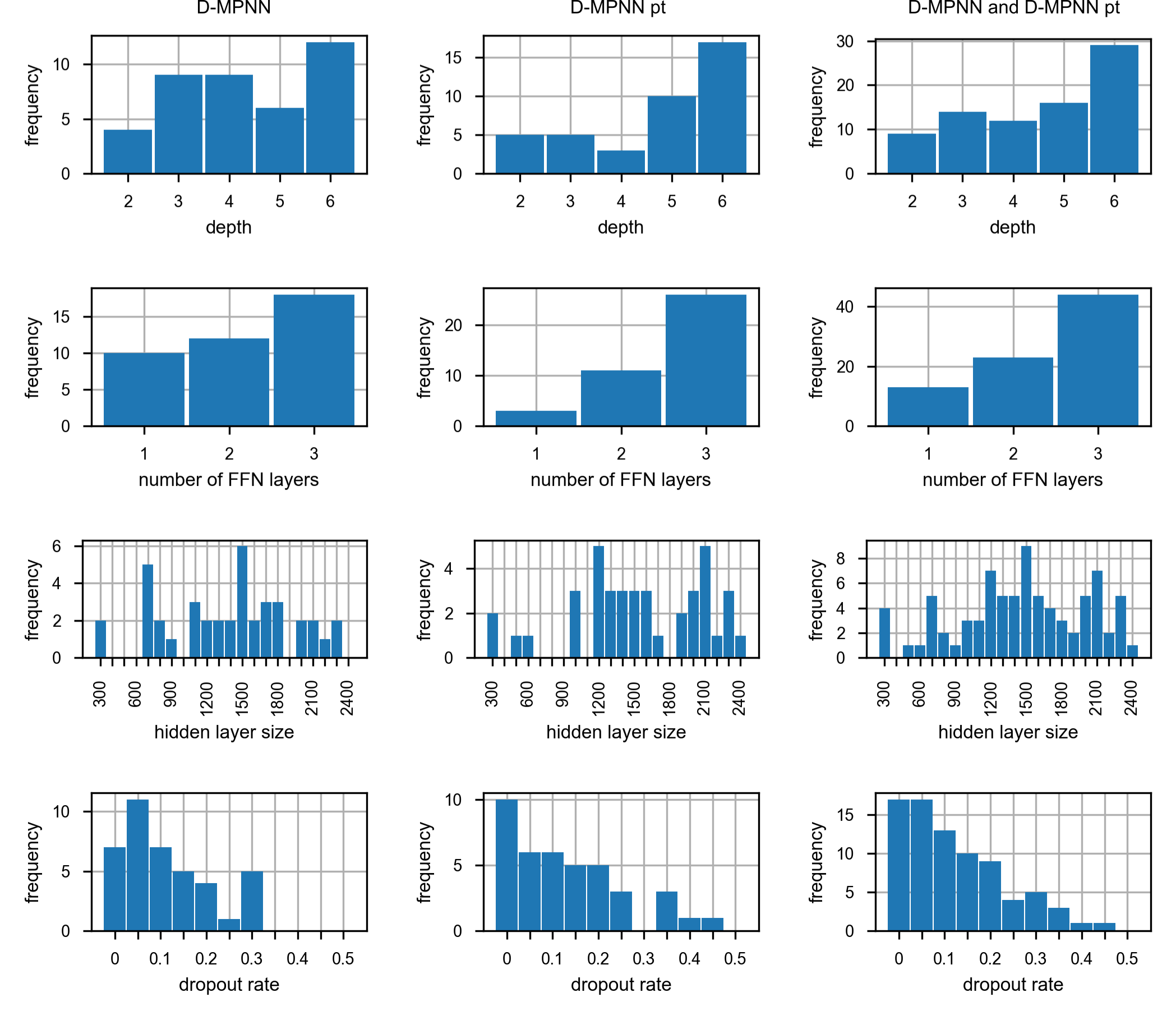}
    \caption{Histograms of optimized hyperparameter values for the D-MPNN and D-MPNN pt regression models (separately and aggregated) on the DD1S CAIX dataset.}
    \label{fig:CAIX_D-MPNN_hyperparams}
\end{figure}

\begin{figure}[H] 
    \centering
        \includegraphics[scale=1]{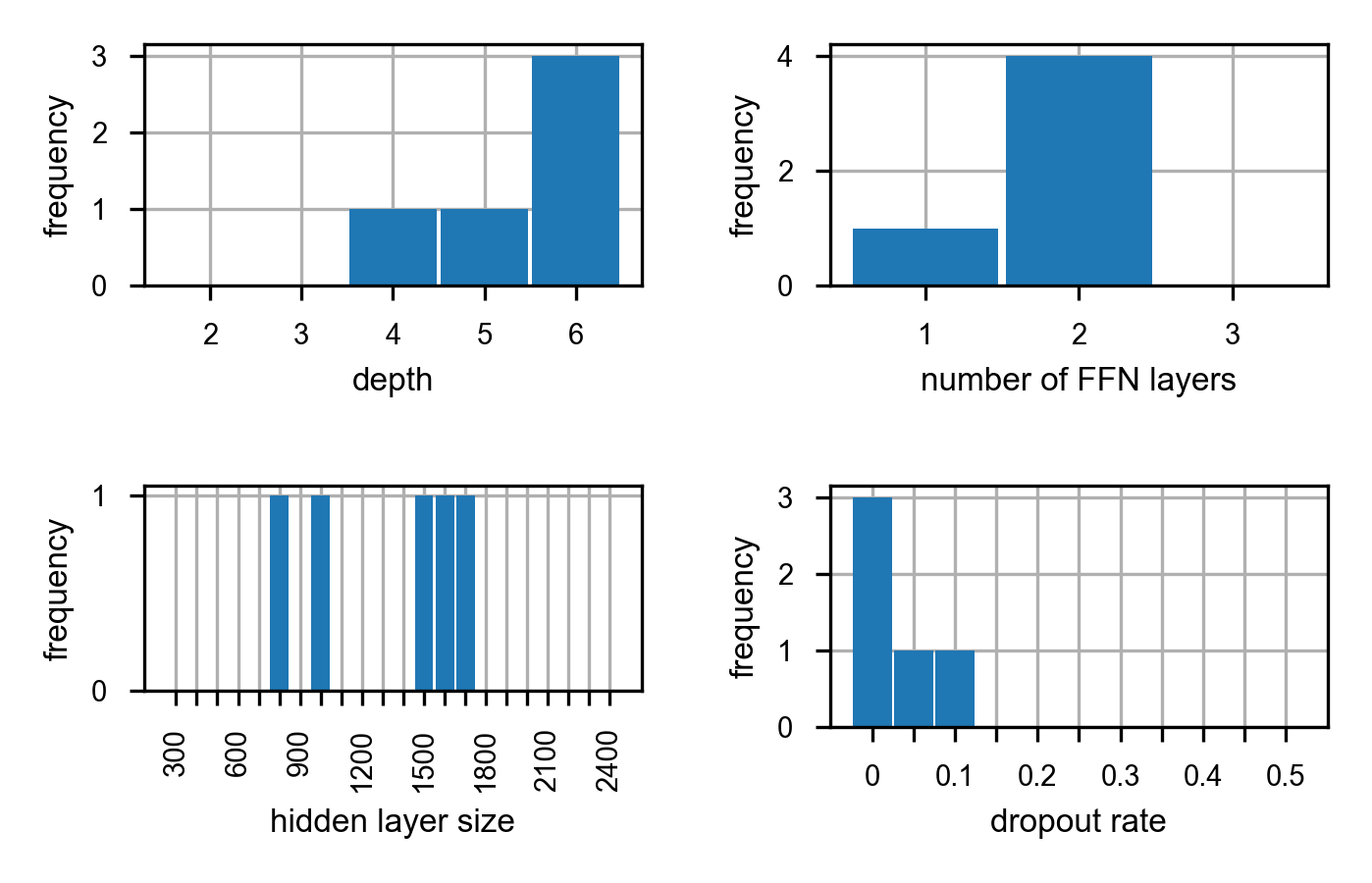}
    \caption{Histograms of optimized hyperparameter values for the D-MPNN binary classifiers on the DD1S CAIX dataset.}
    \label{fig:CAIX_D-MPNN_bin_hyperparams}
\end{figure}

The depth (number of message-passing steps), number of FFN layers, hidden layer size, and dropout rate were fixed for the directed message-passing networks on the triazine datasets, based on the hyperparameter optimization results for the message-passing networks on the DD1S CAIX dataset. For the D-MPNN and D-MPNN pt regression models, we took the mode for the depth (6), number of FFN layers (3), and hidden layer size (1500) based on the aggregated results for the D-MPNN and D-MPNN pt models, and out of the two modes for the dropout rate, we chose 0.05 (Figure~\ref{fig:CAIX_D-MPNN_hyperparams}). For the D-MPNN binary classifiers, we similarly took the mode for the depth (6), number of FFN layers (2), and dropout rate (0), and took the median for the hidden layer size (1500) (Figure~\ref{fig:CAIX_D-MPNN_bin_hyperparams}).

Based on the results from optimization of the number of nearest neighbors (k) for the baseline KNN models on the DD1S CAIX dataset, the k-value was fixed as 9 for all KNN models trained on the triazine datasets.

\begin{figure}[H] 
    \centering
        \includegraphics[scale=1]{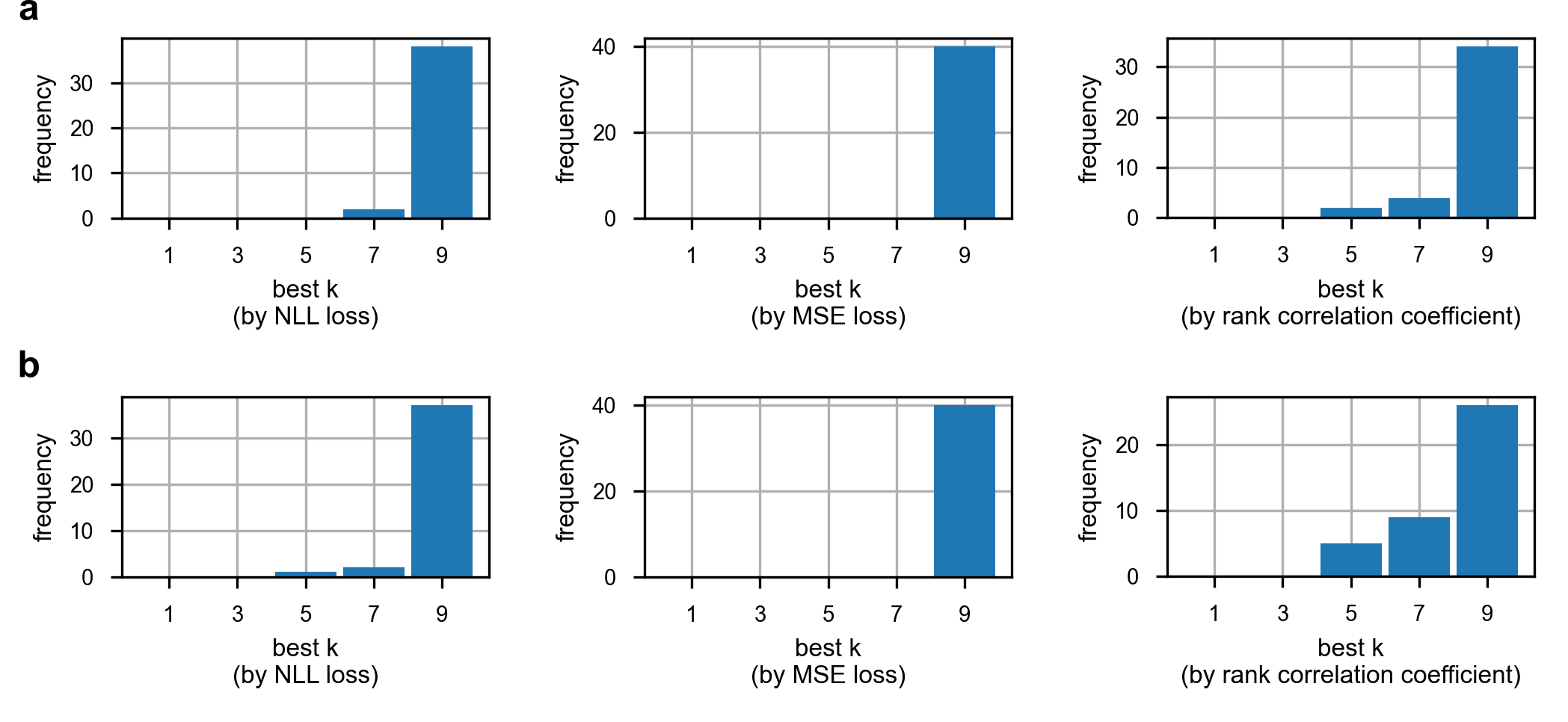}
    \caption{Histograms of optimized k-values for the \textbf{(a)} OH-KNN, \textbf{(b)} FP-KNN baseline models on the DD1S CAIX dataset.}
    \label{fig:CAIX_KNN_k_opt_histograms}
\end{figure}

\subsection{Model performance}

\subsubsection{NLL test loss}

\begin{figure}[H] 
    \centering
        \includegraphics[scale=1]{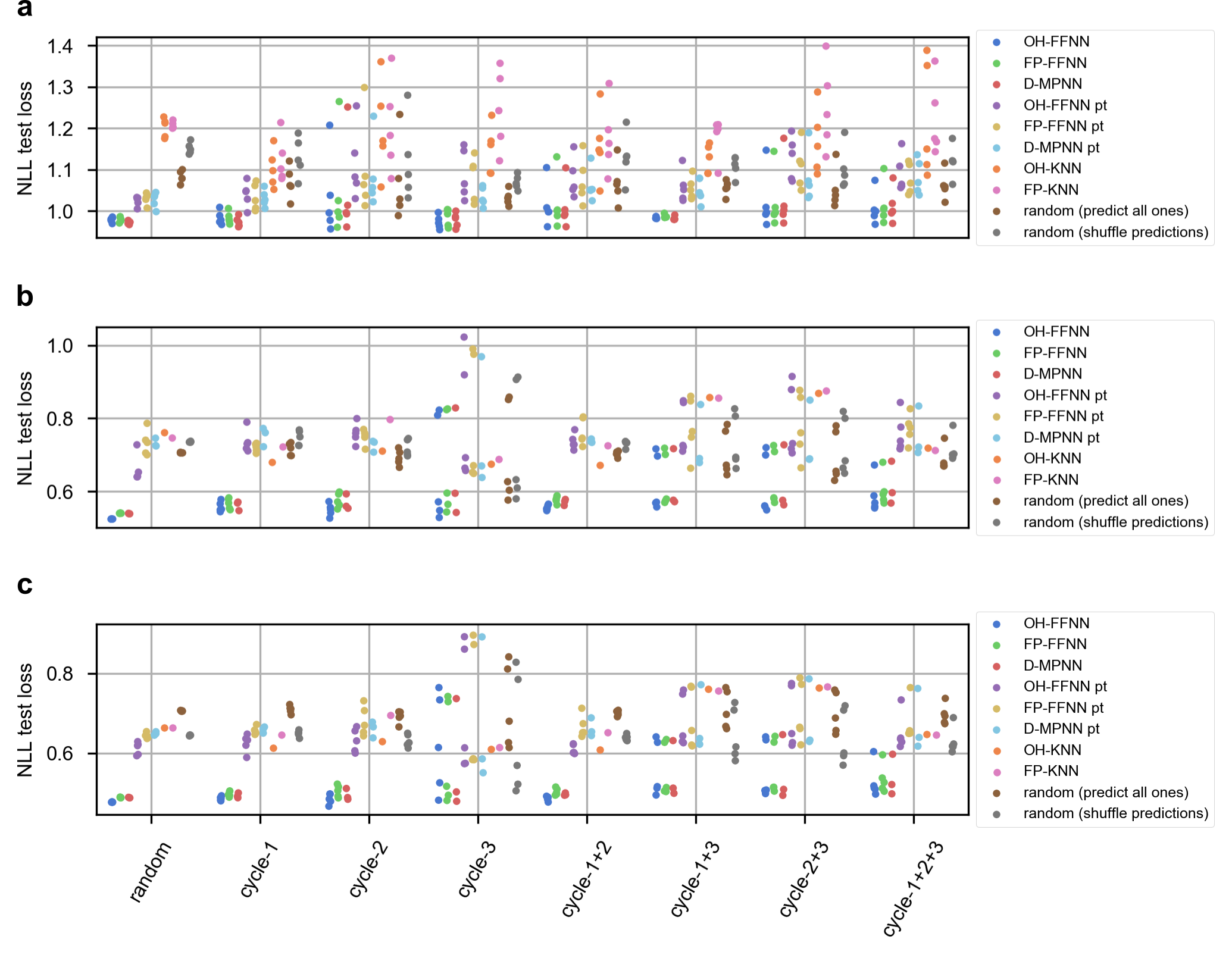}
    \caption{Comparison of model performance, as measured by negative-log likelihood (NLL), for the NLL-trained models (OH-FFNN, FP-FFNN, D-MPNN) versus the baseline point-prediction-trained models (OH-FFNN pt, FP-FFNN pt, D-MPNN pt) for various data splits (\emph{cf.} Figure~\ref{fig:reps_models_loss_fns_splits}d) on the \textbf{(a)} DD1S CAIX, \textbf{(b)} triazine sEH, \textbf{(c)} triazine SIRT2 datasets. For each dataset, data split type, and model type, the individual test loss for each trial is shown. Five trials were performed for OH-FFNN, FP-FFNN, OH-FFNN pt, FP-FFNN pt, and the random baselines on each dataset; for D-MPNN and D-MPNN pt, five trials were performed on the DD1S CAIX dataset and three trials were performed on the triazine sEH and triazine SIRT2 datasets; for OH-KNN and FP-KNN, five trials were performed on the DD1S CAIX dataset and a single trial (evaluated on a random 10\% of the test set) was performed on the triazine sEH and triazine SIRT2 datasets.}
    \label{fig:NLL_test_losses_scatter_plot}
\end{figure}

\begin{table}[H]
\caption{NLL test losses (mean $\pm$ standard deviation) for regression models and baseline models (KNN and random) on the DD1S CAIX dataset. Results are averaged over five trials.}
\begin{tabular}{|c|c|c|c|c|}
\cline{1-1} \cline{3-5}
DD1S CAIX &  & \multicolumn{3}{c|}{cycle split} \\ \hline
model & random split & 1 & 2 & 3 \\ \hline
OH-FFNN & $0.9791\pm0.0063$ & $0.9831\pm 0.0163$ & $1.0348\pm 0.1008$ & $0.9730\pm 0.0161$ \\ \hline
FP-FFNN & $0.9795\pm 0.0061$ & $0.9832\pm 0.0145$ & $1.0465\pm 0.1240$ & $0.9845\pm 0.0211$ \\ \hline
D-MPNN & $0.9719\pm 0.0039$ & $0.9735\pm 0.0120$ & $1.0429\pm 0.1179$ & $0.9780\pm 0.0174$ \\ \hline
OH-FFNN pt & $1.0237\pm 0.0116$ & $1.0396\pm 0.0304$ & $1.1146\pm 0.0873$ & $1.0883\pm 0.0607$ \\ \hline
FP-FFNN pt & $1.0295\pm 0.0138$ & $1.0332\pm 0.0322$ & $1.0992\pm 0.1144$ & $1.0793\pm 0.0543$ \\ \hline
D-MPNN pt & $1.0264\pm 0.0186$ & $1.0305\pm 0.0202$ & $1.0856\pm 0.0825$ & $1.0342\pm 0.0232$ \\ \hline
OH-KNN & $1.2018\pm 0.0232$ & $1.1024\pm 0.0463$ & $1.1995\pm 0.1135$ & $1.1486\pm 0.0589$\\ \hline
FP-KNN & $1.2067\pm 0.0085$ & $1.1233\pm 0.0559$ & $1.2032\pm 0.1126$ & $1.2440\pm 0.0969$\\ \hline
random (predict all ones) & $1.0828\pm 0.0147$ & $1.0694\pm 0.0385$ & $1.0684\pm 0.0977$ & $1.0318\pm 0.0179$\\ \hline
random (shuffle predictions) & $1.1511\pm 0.0136$ & $1.1302\pm 0.0477$ & $1.1184\pm 0.0981$ & $1.0694\pm 0.0171$ \\ \hline
\end{tabular}

\begin{tabular}{c|c|c|c|c|}
\cline{2-5}
 & \multicolumn{4}{c|}{cycle split} \\ \hline
\multicolumn{1}{|c|}{model} & 1+2 & 1+3 & 2+3 & 1+2+3 \\ \hline
\multicolumn{1}{|c|}{OH-FFNN} & $1.0140\pm 0.0537$ & $0.9837\pm 0.0025$ & $1.0221\pm 0.0713$ & $1.0063\pm 0.0402$ \\ \hline
\multicolumn{1}{|c|}{FP-FFNN} & $1.0171\pm 0.0654$ & $0.9867\pm 0.0044$ & $1.0222\pm 0.0694$ & $1.0150\pm 0.0508$ \\ \hline
\multicolumn{1}{|c|}{D-MPNN} & $1.0099\pm 0.0548$ & $0.9822\pm 0.0046$ & $1.0304\pm 0.0826$ & $1.0124\pm 0.0415$ \\ \hline
\multicolumn{1}{|c|}{OH-FFNN pt} & $1.0791\pm 0.0482$ & $1.0584\pm 0.0385$ & $1.1282\pm 0.0524$ & $1.0902\pm 0.0455$ \\ \hline
\multicolumn{1}{|c|}{FP-FFNN pt} & $1.0739\pm 0.0561$ & $1.0553\pm 0.0266$ & $1.1082\pm 0.0545$ & $1.0772\pm 0.0372$ \\ \hline
\multicolumn{1}{|c|}{D-MPNN pt} & $1.0610\pm 0.0389$ & $1.0497\pm 0.0295$ & $1.0781\pm 0.0642$ & $1.0812\pm 0.0421$ \\ \hline
\multicolumn{1}{|c|}{OH-KNN} & $1.1589\pm 0.0841$ & $1.1262\pm 0.0344$ & $1.1681\pm 0.0800$ & $1.2174\pm 0.1414$ \\ \hline
\multicolumn{1}{|c|}{FP-KNN} & $1.1762\pm 0.0856$ & $1.1794\pm 0.0499$ & $1.2496\pm 0.1044$ & $1.2216\pm 0.0903$ \\ \hline
\multicolumn{1}{|c|}{random (predict all ones)} & $1.0676\pm 0.0510$ & $1.0556\pm 0.0178$ & $1.0528\pm 0.0488$ & $1.0620\pm 0.0341$ \\ \hline
\multicolumn{1}{|c|}{random (shuffle predictions)} & $1.1293\pm 0.0582$ & $1.1082\pm 0.0246$ & $1.1010\pm 0.0521$ & $1.1199\pm 0.0393$ \\ \hline
\end{tabular}
\label{tbl:CAIX_NLL_test_losses}
\end{table}

\begin{table}[H]
\caption{NLL test losses (mean $\pm$ standard deviation) for regression models and baseline models on the triazine sEH dataset. OH-FFNN, FP-FFNN, OH-FFNN pt, FP-FFNN pt, random (predict all ones), and random (shuffle predictions) results are averaged over five trials; D-MPNN and D-MPNN pt results are averaged over three trials; OH-KNN and FP-KNN results are single trials evaluated on a random 10\% of the test set.}
\begin{tabular}{|c|c|c|c|c|}
\cline{1-1} \cline{3-5}
triazine sEH &  & \multicolumn{3}{c|}{cycle split} \\ \hline
model & random split & 1 & 2 & 3 \\ \hline
OH-FFNN & $0.5240\pm0.0002$ & $0.5576\pm0.0139$ & $0.5490\pm0.0170$ & $0.6556\pm0.1469$ \\ \hline
FP-FFNN & $0.5400\pm0.0004$ & $0.5665\pm0.0146$ & $0.5752\pm0.0198$ & $0.6705\pm0.1420$ \\ \hline
D-MPNN & $0.5390\pm0.0006$ & $0.5619\pm0.0129$ & $0.5685\pm0.0213$ & $0.6551\pm0.1526$ \\ \hline
OH-FFNN pt & $0.6601\pm0.0381$ & $0.7363\pm0.0312$ & $0.7594\pm0.0279$ & $0.7909\pm0.1688$ \\ \hline
FP-FFNN pt & $0.7331\pm0.0345$ & $0.7197\pm0.0117$ & $0.7494\pm0.0206$ & $0.7874\pm0.1787$ \\ \hline
D-MPNN pt & $0.7319\pm0.0120$ & $0.7518\pm0.0262$ & $0.7263\pm0.0165$ & $0.7588\pm0.1827$ \\ \hline
OH-KNN & $0.7622$ & $0.6801$ & $0.7109$ & $0.6759$\\ \hline
FP-KNN & $0.7463$ & $0.7220$ & $0.7980$ & $0.6879$\\ \hline
random (predict all ones) & $0.7058\pm 0.0008$ & $0.7156\pm 0.0171$ & $0.6924\pm 0.0211$ & $0.7030\pm 0.1398$\\ \hline
random (shuffle predictions) & $0.7357\pm 0.0012$ & $0.7466\pm 0.0204$ & $0.7185\pm 0.0228$ & $0.7281\pm 0.1669$ \\ \hline
\end{tabular}

\begin{tabular}{c|c|c|c|c|}
\cline{2-5}
 & \multicolumn{4}{c|}{cycle split} \\ \hline
\multicolumn{1}{|c|}{model} & 1+2 & 1+3 & 2+3 & 1+2+3 \\ \hline
\multicolumn{1}{|c|}{OH-FFNN} & $0.5554\pm0.0074$ & $0.6218\pm0.0778$ & $0.6174\pm0.0846$ & $0.5882\pm0.0486$ \\ \hline
\multicolumn{1}{|c|}{FP-FFNN} & $0.5780\pm0.0099$ & $0.6284\pm0.0745$ & $0.6333\pm0.0770$ & $0.6031\pm0.0445$ \\ \hline
\multicolumn{1}{|c|}{D-MPNN} & $0.5706\pm0.0085$ & $0.6214\pm0.0827$ & $0.6219\pm0.0916$ & $0.6155\pm0.0598$ \\ \hline
\multicolumn{1}{|c|}{OH-FFNN pt} & $0.7361\pm0.0211$ & $0.7698\pm0.0695$ & $0.7893\pm0.0993$ & $0.7595\pm0.0523$ \\ \hline
\multicolumn{1}{|c|}{FP-FFNN pt} & $0.7635\pm0.0370$ & $0.7768\pm0.0805$ & $0.7777\pm0.0890$ & $0.7723\pm0.0394$ \\ \hline
\multicolumn{1}{|c|}{D-MPNN pt} & $0.7380\pm0.0050$ & $0.7355\pm0.0888$ & $0.7423\pm0.0932$ & $0.7538\pm0.0697$ \\ \hline
\multicolumn{1}{|c|}{OH-KNN} & $0.6719$ & $0.8572$ & $0.8686$ & $0.7189$ \\ \hline
\multicolumn{1}{|c|}{FP-KNN} & $0.7253$ & $0.8559$ & $0.8756$ & $0.7126$ \\ \hline
\multicolumn{1}{|c|}{random (predict all ones)} & $0.7024\pm 0.0077$ & $0.7054\pm 0.0638$ & $0.6951\pm 0.0704$ & $0.6871\pm 0.0329$ \\ \hline
\multicolumn{1}{|c|}{random (shuffle predictions)} & $0.7309\pm 0.0090$ & $0.7353\pm 0.0750$ & $0.7233\pm 0.0800$ & $0.7138\pm 0.0378$ \\ \hline
\end{tabular}
\label{tbl:sEH_NLL_test_losses}
\end{table}

\begin{table}[H]
\caption{NLL test losses (mean $\pm$ standard deviation) for regression models and baseline models on the triazine SIRT2 dataset. OH-FFNN, FP-FFNN, OH-FFNN pt, FP-FFNN pt, and random baseline results are averaged over five trials; D-MPNN and D-MPNN pt results are averaged over three trials; OH-KNN and FP-KNN results are single trials evaluated on a random 10\% of the test set.}
\begin{tabular}{|c|c|c|c|c|}
\cline{1-1} \cline{3-5}
triazine SIRT2 &  & \multicolumn{3}{c|}{cycle split} \\ \hline
model & random split & 1 & 2 & 3 \\ \hline
OH-FFNN & $0.4770\pm0.0005$ & $0.4858\pm0.0052$ & $0.4850\pm0.0131$ & $0.6243\pm0.1246$ \\ \hline
FP-FFNN & $0.4884\pm0.0008$ & $0.4975\pm0.0067$ & $0.5046\pm0.0149$ & $0.5930\pm0.1316$ \\ \hline
D-MPNN & $0.4882\pm0.0007$ & $0.4931\pm0.0063$ & $0.4946\pm0.0146$ & $0.5730\pm0.1429$ \\ \hline
OH-FFNN pt & $0.6122\pm0.0159$ & $0.6254\pm0.0225$ & $0.6321\pm0.0295$ & $0.7034\pm0.1602$ \\ \hline
FP-FFNN pt & $0.6433\pm0.0070$ & $0.6567\pm0.0097$ & $0.6812\pm0.0374$ & $0.7049\pm0.1649$ \\ \hline
D-MPNN pt & $0.6495\pm0.0044$ & $0.6594\pm0.0079$ & $0.6611\pm0.0204$ & $0.6765\pm0.1882$ \\ \hline
OH-KNN & $0.6648$ & $0.6141$ & $0.6304$ & $0.6103$\\ \hline
FP-KNN & $0.6638$ & $0.6460$ & $0.6956$ & $0.6142$\\ \hline
random (predict all ones) & $0.7065\pm 0.0009$ & $0.7101\pm 0.0095$ & $0.6918\pm 0.0154$ & $0.7152\pm 0.1058$\\ \hline
random (shuffle predictions) & $0.6449\pm 0.0009$ & $0.6498\pm 0.0080$ & $0.6315\pm 0.0162$ & $0.6423\pm 0.1533$ \\ \hline
\end{tabular}

\begin{tabular}{c|c|c|c|c|}
\cline{2-5}
 & \multicolumn{4}{c|}{cycle split} \\ \hline
\multicolumn{1}{|c|}{model} & 1+2 & 1+3 & 2+3 & 1+2+3 \\ \hline
\multicolumn{1}{|c|}{OH-FFNN} & $0.4862\pm0.0059$ & $0.5583\pm0.0700$ & $0.5574\pm0.0733$ & $0.5282\pm0.0431$ \\ \hline
\multicolumn{1}{|c|}{FP-FFNN} & $0.5033\pm0.0087$ & $0.5569\pm0.0679$ & $0.5607\pm0.0683$ & $0.5353\pm0.0363$ \\ \hline
\multicolumn{1}{|c|}{D-MPNN} & $0.4965\pm0.0031$ & $0.5477\pm0.0731$ & $0.5499\pm0.0842$ & $0.5388\pm0.0521$ \\ \hline
\multicolumn{1}{|c|}{OH-FFNN pt} & $0.6096\pm0.0122$ & $0.6814\pm0.0667$ & $0.6884\pm0.0778$ & $0.6507\pm0.0474$ \\ \hline
\multicolumn{1}{|c|}{FP-FFNN pt} & $0.6668\pm0.0285$ & $0.6863\pm0.0757$ & $0.7023\pm0.0746$ & $0.6746\pm0.0507$ \\ \hline
\multicolumn{1}{|c|}{D-MPNN pt} & $0.6625\pm0.0236$ & $0.6774\pm0.0826$ & $0.6833\pm0.0901$ & $0.6735\pm0.0783$ \\ \hline
\multicolumn{1}{|c|}{OH-KNN} & $0.6090$ & $0.7621$ & $0.7646$ & $0.6477$ \\ \hline
\multicolumn{1}{|c|}{FP-KNN} & $0.6523$ & $0.7566$ & $0.7682$ & $0.6459$ \\ \hline
\multicolumn{1}{|c|}{random (predict all ones)} & $0.7008\pm 0.0065$ & $0.7094\pm 0.0483$ & $0.7011\pm 0.0523$ & $0.6967\pm 0.0255$ \\ \hline
\multicolumn{1}{|c|}{random (shuffle predictions)} & $0.6392\pm 0.0071$ & $0.6464\pm 0.0669$ & $0.6385\pm 0.0700$ & $0.6301\pm 0.0340$ \\ \hline
\end{tabular}
\label{tbl:SIRT2_NLL_test_losses}
\end{table}

\begin{figure}[H] 
    \centering
        \includegraphics[scale=1]{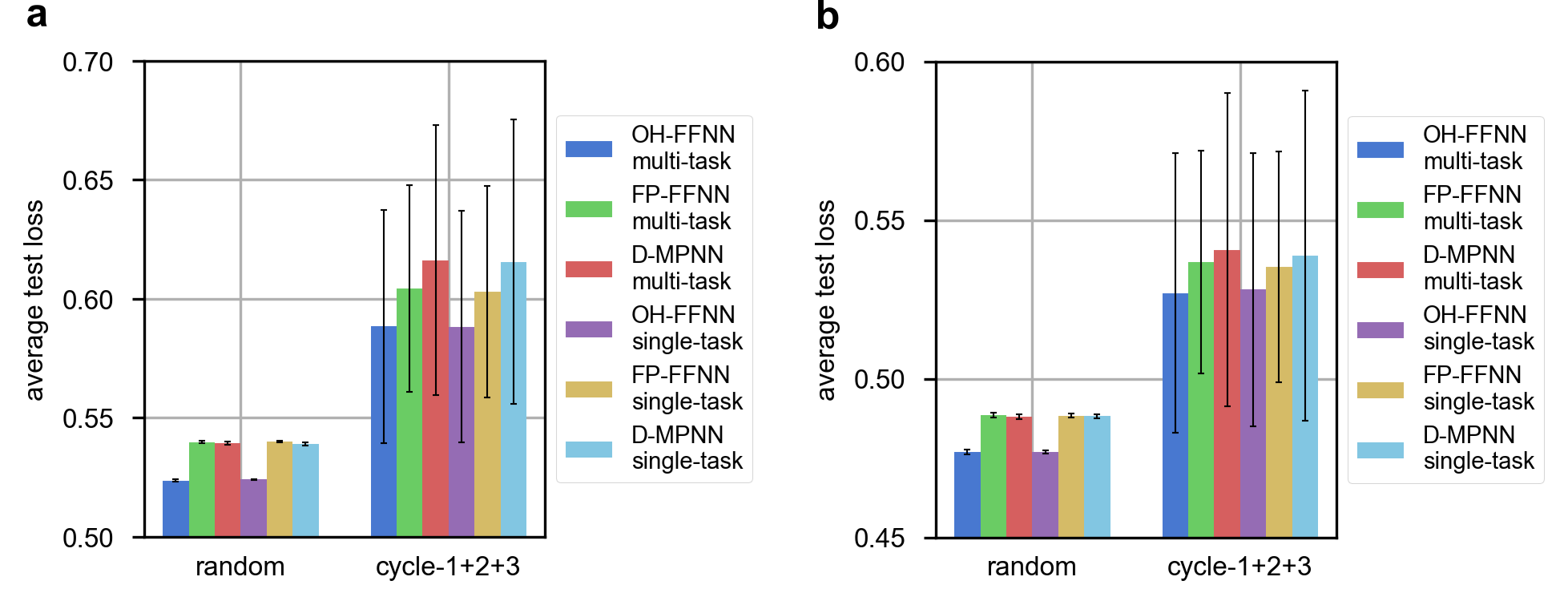}
    \caption{Comparison of model performance, as measured by negative-log likelihood, for the multi-task versus single-task models on the \textbf{(a)} triazine sEH and \textbf{(b)} triazine SIRT2 datasets. Error bars represent $\pm$ one standard deviation. OH-FFNN and FP-FFNN results are averaged over five trials; D-MPNN results are averaged over three trials.}
    \label{fig:multi_task_test_loss_bar_graphs}
\end{figure}

\subsubsection{MSE test loss}

\begin{figure}[H] 
    \centering
        \includegraphics[scale=1]{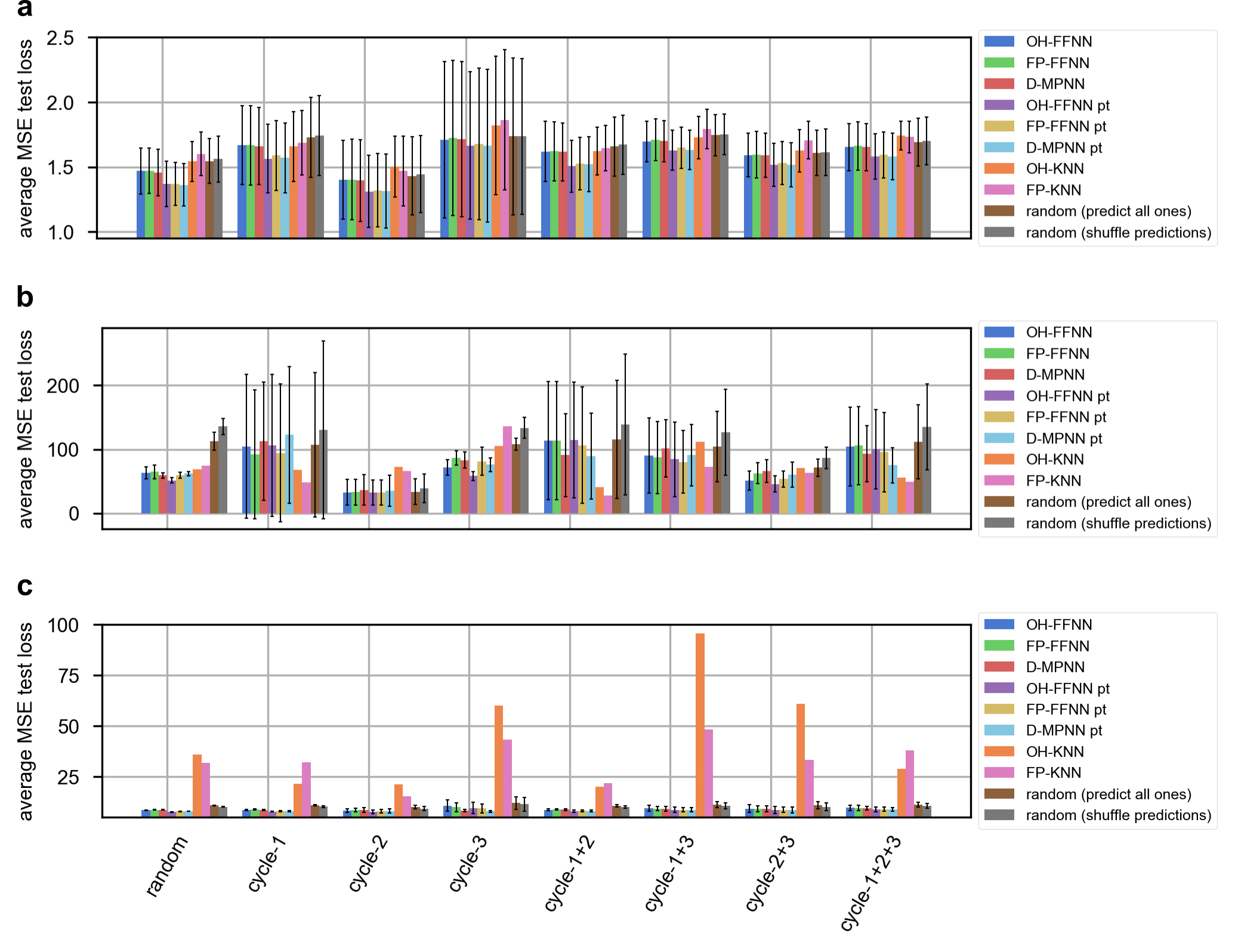}
    \caption{Comparison of model performance, as measured by mean-squared error (MSE). OH: one-hot; FP: fingerprint; FFNN: feed-forward neural network; D-MPNN: directed message-passing neural network; KNN: k-nearest neighbors. The MSE test losses of the negative-log-likelihood-trained models (OH-FFNN, FP-FFNN, D-MPNN) are compared to those of the baseline point-prediction-trained models (OH-FFNN pt, FP-FFNN pt, D-MPNN pt), k-nearest-neighbors models (OH-KNN, FP-KNN), and random models (predict all ones, shuffle predictions), for various data splits (\emph{cf.} Figure~\ref{fig:reps_models_loss_fns_splits}d) on the \textbf{(a)} DD1S CAIX, \textbf{(b)} triazine sEH, \textbf{(c)} triazine SIRT2 datasets. Error bars represent $\pm$ one standard deviation. OH-FFNN, FP-FFNN, OH-FFNN pt, FP-FFNN pt, random (predict all ones), and random (shuffle predictions) results are averaged over five trials for each dataset; D-MPNN and D-MPNN pt results are averaged over five trials for the DD1S CAIX dataset and over three trials for the triazine sEH and triazine SIRT2 datasets; OH-KNN and FP-KNN results are averaged over five trials for the DD1S CAIX dataset and are single trials (evaluated on a random 10\% of the test set) for the triazine sEH and triazine SIRT2 datasets. The result of each trial is shown separately below (Figure~\ref{fig:MSE_test_losses_scatter_plot}).}
    \label{fig:MSE_test_losses_bar_graph}
\end{figure}

\begin{figure}[H] 
    \centering
        \includegraphics[scale=1]{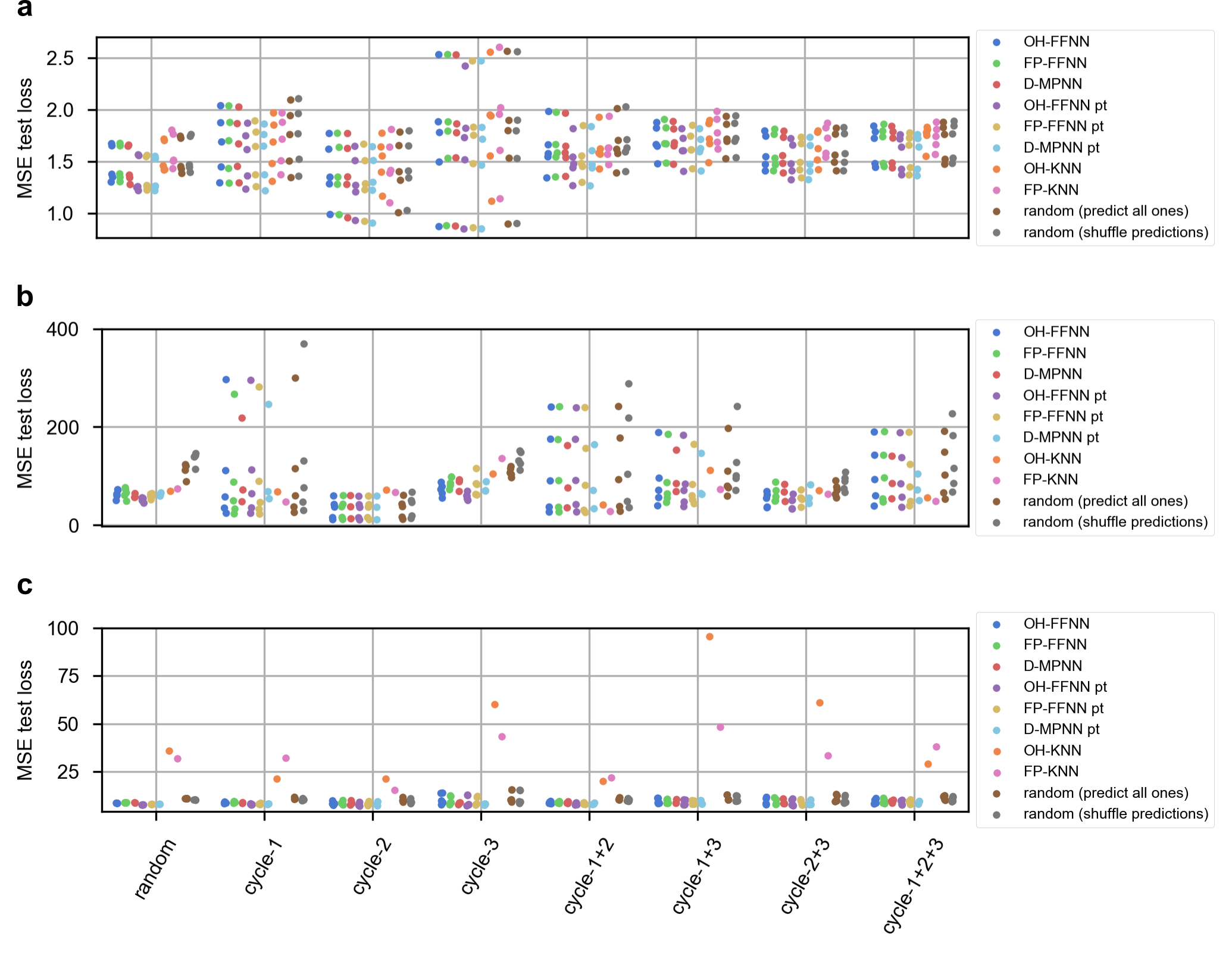}
    \caption{Comparison of model performance, as measured by mean-squared error (MSE), for the negative-log-likelihood-trained models (OH-FFNN, FP-FFNN, D-MPNN) versus the baseline point-prediction-trained models (OH-FFNN pt, FP-FFNN pt, D-MPNN pt) for various data splits (\emph{cf.} Figure~\ref{fig:reps_models_loss_fns_splits}d) on the \textbf{(a)} DD1S CAIX, \textbf{(b)} triazine sEH, \textbf{(c)} triazine SIRT2 datasets. For each dataset, data split type, and model type, the individual test loss for each trial is shown. Five trials were performed for OH-FFNN, FP-FFNN, OH-FFNN pt, FP-FFNN pt, and the random baselines on each dataset; for D-MPNN and D-MPNN pt, five trials were performed on the DD1S CAIX dataset and three trials were performed on the triazine sEH and triazine SIRT2 datasets; for OH-KNN and FP-KNN, five trials were performed on the DD1S CAIX dataset and a single trial (evaluated on a random 10\% of the test set) was performed on the triazine sEH and triazine SIRT2 datasets.}
    \label{fig:MSE_test_losses_scatter_plot}
\end{figure}

\begin{table}[H]
\caption{MSE test losses (mean $\pm$ standard deviation) for regression models and baseline models on the DD1S CAIX dataset. Results are averaged over five trials.}
\begin{tabular}{|c|c|c|c|c|}
\cline{1-1} \cline{3-5}
DD1S CAIX                    &                    & \multicolumn{3}{c|}{cycle split}                             \\ \hline
model                        & random split       & 1                  & 2                  & 3                  \\ \hline
OH-FFNN                      & $1.4690\pm 0.1772$ & $1.6675\pm 0.3039$ & $1.4007\pm 0.3047$ & $1.7111\pm 0.6044$ \\ \hline
FP-FFNN                      & $1.4707\pm 0.1771$ & $1.6665\pm 0.3078$ & $1.4037\pm 0.3096$ & $1.7228\pm 0.5988$ \\ \hline
D-MPNN                       & $1.4578\pm 0.1803$ & $1.6602\pm 0.2975$ & $1.3951\pm 0.3172$ & $1.7153\pm 0.5970$ \\ \hline
OH-FFNN pt                   & $1.3682\pm 0.1753$ & $1.5639\pm 0.2651$ & $1.3107\pm 0.2782$ & $1.6662\pm 0.5680$ \\ \hline
FP-FFNN pt                   & $1.3693\pm 0.1674$ & $1.5886\pm 0.2696$ & $1.3207\pm 0.2848$ & $1.6775\pm 0.5846$ \\ \hline
D-MPNN pt                    & $1.3610\pm 0.1626$ & $1.5704\pm 0.2699$ & $1.3143\pm 0.2839$ & $1.6645\pm 0.5886$ \\ \hline
OH-KNN                       & $1.5420\pm 0.1523$ & $1.6580\pm 0.2689$ & $1.5044\pm 0.2347$ & $1.8214\pm 0.5335$ \\ \hline
FP-KNN                       & $1.6015\pm 0.1684$ & $1.6870\pm 0.2493$ & $1.4687\pm 0.2694$ & $1.8638\pm 0.5405$ \\ \hline
random (predict all ones)    & $1.5452\pm 0.1732$ & $1.7276\pm 0.3077$ & $1.4313\pm 0.3036$ & $1.7359\pm 0.6049$ \\ \hline
random (shuffle predictions) & $1.5606\pm 0.1756$ & $1.7418\pm 0.3075$ & $1.4432\pm 0.2969$ & $1.7356\pm 0.6020$ \\ \hline
\end{tabular}

\begin{tabular}{c|c|c|c|c|}
\cline{2-5}
                                                   & \multicolumn{4}{c|}{cycle split}                                                  \\ \hline
\multicolumn{1}{|c|}{model}                        & 1+2                & 1+3                & 2+3                & 1+2+3              \\ \hline
\multicolumn{1}{|c|}{OH-FFNN}                      & $1.6202\pm 0.2337$ & $1.6965\pm 0.1580$ & $1.5918\pm 0.1693$ & $1.6539\pm 0.1822$ \\ \hline
\multicolumn{1}{|c|}{FP-FFNN}                      & $1.6221\pm 0.2279$ & $1.7102\pm 0.1625$ & $1.5955\pm 0.1798$ & $1.6635\pm 0.1866$ \\ \hline
\multicolumn{1}{|c|}{D-MPNN}                       & $1.6168\pm 0.2230$ & $1.6996\pm 0.1598$ & $1.5893\pm 0.1697$ & $1.6541\pm 0.1821$ \\ \hline
\multicolumn{1}{|c|}{OH-FFNN pt}                   & $1.5067\pm 0.2005$ & $1.6290\pm 0.1548$ & $1.5166\pm 0.1676$ & $1.5807\pm 0.1744$ \\ \hline
\multicolumn{1}{|c|}{FP-FFNN pt}                   & $1.5249\pm 0.2019$ & $1.6482\pm 0.1582$ & $1.5302\pm 0.1666$ & $1.5926\pm 0.1791$ \\ \hline
\multicolumn{1}{|c|}{D-MPNN pt}                    & $1.5215\pm 0.2131$ & $1.6301\pm 0.1520$ & $1.5165\pm 0.1712$ & $1.5804\pm 0.1785$ \\ \hline
\multicolumn{1}{|c|}{OH-KNN}                       & $1.6243\pm 0.1839$ & $1.7265\pm 0.1643$ & $1.6260\pm 0.1639$ & $1.7424\pm 0.1125$ \\ \hline
\multicolumn{1}{|c|}{FP-KNN}                       & $1.6456\pm 0.1746$ & $1.7935\pm 0.1512$ & $1.7069\pm 0.1454$ & $1.7312\pm 0.1222$ \\ \hline
\multicolumn{1}{|c|}{random (predict all ones)}    & $1.6585\pm 0.2275$ & $1.7451\pm 0.1585$ & $1.6086\pm 0.1767$ & $1.6910\pm 0.1836$ \\ \hline
\multicolumn{1}{|c|}{random (shuffle predictions)} & $1.6714\pm 0.2285$ & $1.7514\pm 0.1580$ & $1.6133\pm 0.1784$ & $1.7011\pm 0.1846$ \\ \hline
\end{tabular}
\label{tbl:CAIX_MSE_test_losses}
\end{table}

\begin{table}[H]
\caption{MSE test losses (mean $\pm$ standard deviation) for regression models and baseline models on the triazine sEH dataset. OH-FFNN, FP-FFNN, OH-FFNN pt, FP-FFNN pt, random (predict all ones), and random (shuffle predictions) results are averaged over five trials; D-MPNN and D-MPNN pt results are averaged over three trials; OH-KNN and FP-KNN results are single trials evaluated on a random 10\% of the test set.}
\begin{tabular}{|c|c|c|c|c|}
\cline{1-1} \cline{3-5}
triazine sEH                 &                       & \multicolumn{3}{c|}{cycle split}                                       \\ \hline
model                        & random split          & 1                      & 2                     & 3                     \\ \hline
OH-FFNN                      & $63.1165\pm 9.1153$   & $104.6138\pm 112.4744$ & $32.6997\pm 19.9417$  & $71.4771\pm 12.4136$  \\ \hline
FP-FFNN                      & $64.6382\pm 10.9077$  & $91.7278\pm 100.8977$  & $33.1806\pm 20.0352$  & $86.6032\pm 11.1850$  \\ \hline
D-MPNN                       & $59.0148\pm 4.4403$   & $112.3067\pm 92.4066$  & $36.5230\pm 23.5934$  & $82.9586\pm 12.7764$  \\ \hline
OH-FFNN pt                   & $51.3925\pm 4.4271$   & $105.9992\pm 111.1797$ & $ 32.1739\pm 19.7865$ & $ 58.0676\pm 7.1649$  \\ \hline
FP-FFNN pt                   & $59.3360\pm 4.9986$   & $94.1664\pm 107.8041$  & $32.5065\pm  19.7674$ & $81.0961\pm  21.9021$ \\ \hline
D-MPNN pt                    & $62.1430\pm 3.3269$   & $122.5217\pm 107.2855$ & $35.2012\pm 24.1319$  & $75.8201\pm 10.7562$  \\ \hline
OH-KNN                       & $68.8859$             & $67.8826$              & $72.0830$             & $105.1172$            \\ \hline
FP-KNN                       & $73.9428$             & $47.8342$              & $66.3675$             & $136.0269$            \\ \hline
\begin{tabular}[c]{@{}c@{}}random\\ (predict all ones)\end{tabular}    & $112.7630\pm 14.3542$ & $107.2573\pm 112.9778$ & $33.5890\pm 20.0749$  & $108.0191\pm 9.1461$  \\ \hline
\begin{tabular}[c]{@{}c@{}}random\\ (shuffle predictions)\end{tabular} & $135.7934\pm 12.7459$ & $130.3497\pm 139.0098$ & $38.8306\pm 22.4695$  & $133.3429\pm 16.3413$ \\ \hline
\end{tabular}

\begin{tabular}{c|c|c|c|c|}
\cline{2-5}
                                                   & \multicolumn{4}{c|}{cycle split}                                                                \\ \hline
\multicolumn{1}{|c|}{model}                        & 1+2                    & 1+3                   & 2+3                    & 1+2+3                 \\ \hline
\multicolumn{1}{|c|}{OH-FFNN}                      & $113.7291\pm 92.2577$  & $90.0097\pm 58.8666$  & $51.1786\pm 14.8026$   & $104.5738\pm 61.6597$ \\ \hline
\multicolumn{1}{|c|}{FP-FFNN}                      & $113.7116\pm 92.3884$  & $87.1879\pm 56.6259$  & $ 62.6018\pm  16.3096$ & $105.7660\pm 60.8721$ \\ \hline
\multicolumn{1}{|c|}{D-MPNN}                       & $90.8070\pm 64.8965$   & $101.6983\pm 44.9100$ & $65.8889\pm 17.5284$   & $92.7588\pm  44.0416$ \\ \hline
\multicolumn{1}{|c|}{OH-FFNN pt}                   & $114.5469\pm 90.5483$  & $84.1884\pm 58.1733$  & $45.7241\pm 12.8903$   & $100.4466\pm 62.0703$ \\ \hline
\multicolumn{1}{|c|}{FP-FFNN pt}                   & $106.4167\pm 91.1926$  & $80.3312\pm 49.2571$  & $53.4952\pm 12.6520$   & $95.4048\pm 61.8960$  \\ \hline
\multicolumn{1}{|c|}{D-MPNN pt}                    & $89.0670\pm 67.3121$   & $90.8949\pm 47.8299$  & $60.0897\pm 19.7125$   & $74.9849\pm 27.3220$  \\ \hline
\multicolumn{1}{|c|}{OH-KNN}                       & $40.8193$              & $112.1089$            & $70.5626$              & $55.3985$             \\ \hline
\multicolumn{1}{|c|}{FP-KNN}                       & $27.5076$              & $72.3482$             & $63.1585$              & $48.6848$             \\ \hline
\multicolumn{1}{|c|}{\begin{tabular}[c]{@{}c@{}}random\\ (predict all ones)\end{tabular}}    & $115.4427\pm 92.2479$  & $104.1369\pm 55.0694$ & $71.2054\pm 13.5378$   & $111.8880\pm 57.8742$ \\ \hline
\multicolumn{1}{|c|}{\begin{tabular}[c]{@{}c@{}}random\\ (shuffle predictions)\end{tabular}} & $138.5909\pm 110.4645$ & $126.9835\pm 67.2980$ & $86.4051\pm 16.5700$   & $135.2677\pm 67.4508$ \\ \hline
\end{tabular}
\label{tbl:sEH_MSE_test_losses}
\end{table}

\begin{table}[H]
\caption{MSE test losses (mean $\pm$ standard deviation) for regression models and baseline models on the triazine SIRT2 dataset. OH-FFNN, FP-FFNN, OH-FFNN pt, FP-FFNN pt, random (predict all ones), and random (shuffle predictions) results are averaged over five trials; D-MPNN and D-MPNN pt results are averaged over three trials; OH-KNN and FP-KNN results are single trials evaluated on a random 10\% of the test set.}
\begin{tabular}{|c|c|c|c|c|}
\cline{1-1} \cline{3-5}
triazine SIRT2               &                    & \multicolumn{3}{c|}{cycle split}                             \\ \hline
model                        & random split       & 1                  & 2                  & 3                  \\ \hline
OH-FFNN                      & $0.4770\pm 0.0005$ & $0.4858\pm 0.0052$ & $0.4850\pm 0.0131$ & $0.6243\pm 0.1246$ \\ \hline
FP-FFNN                      & $0.4884\pm 0.0008$ & $0.4975\pm 0.0067$ & $0.5046\pm 0.0149$ & $0.5930\pm 0.1316$ \\ \hline
D-MPNN                       & $0.4882\pm 0.0007$ & $0.4931\pm 0.0063$ & $0.4946\pm 0.0146$ & $0.5730\pm 0.1429$ \\ \hline
OH-FFNN pt                   & $0.6122\pm 0.0159$ & $0.6254\pm 0.0225$ & $0.6321\pm 0.0295$ & $0.7034\pm 0.1602$ \\ \hline
FP-FFNN pt                   & $0.6433\pm 0.0070$ & $0.6567\pm 0.0097$ & $0.6812\pm 0.0374$ & $0.7049\pm 0.1649$ \\ \hline
D-MPNN pt                    & $0.6495\pm 0.0044$ & $0.6594\pm 0.0079$ & $0.6611\pm 0.0204$ & $0.6765\pm 0.1882$ \\ \hline
OH-KNN                       & $0.6648$           & $0.6141$           & $0.6304$           & $0.6103$           \\ \hline
FP-KNN                       & $0.6638$           & $0.6460$           & $0.6956$           & $0.6142$           \\ \hline
random (predict all ones)    & $0.7065\pm 0.0009$ & $0.7101\pm 0.0095$ & $0.6918\pm 0.0154$ & $0.7152\pm 0.1058$ \\ \hline
random (shuffle predictions) & $0.6449\pm 0.0009$ & $0.6498\pm 0.0080$ & $0.6315\pm 0.0162$ & $0.6423\pm 0.1533$ \\ \hline
\end{tabular}

\begin{tabular}{c|c|c|c|c|}
\cline{2-5}
                                                   & \multicolumn{4}{c|}{cycle split}                                                  \\ \hline
\multicolumn{1}{|c|}{model}                        & 1+2                & 1+3                & 2+3                & 1+2+3              \\ \hline
\multicolumn{1}{|c|}{OH-FFNN}                      & $0.4862\pm 0.0059$ & $0.5583\pm 0.0700$ & $0.5574\pm 0.0733$ & $0.5282\pm 0.0431$ \\ \hline
\multicolumn{1}{|c|}{FP-FFNN}                      & $0.5033\pm 0.0087$ & $0.5569\pm 0.0679$ & $0.5607\pm 0.0683$ & $0.5353\pm 0.0363$ \\ \hline
\multicolumn{1}{|c|}{D-MPNN}                       & $0.4965\pm 0.0031$ & $0.5477\pm 0.0731$ & $0.5499\pm 0.0842$ & $0.5388\pm 0.0521$ \\ \hline
\multicolumn{1}{|c|}{OH-FFNN pt}                   & $0.6096\pm 0.0122$ & $0.6814\pm 0.0667$ & $0.6884\pm 0.0778$ & $0.6507\pm 0.0474$ \\ \hline
\multicolumn{1}{|c|}{FP-FFNN pt}                   & $0.6668\pm 0.0285$ & $0.6863\pm 0.0757$ & $0.7023\pm 0.0746$ & $0.6746\pm 0.0507$ \\ \hline
\multicolumn{1}{|c|}{D-MPNN pt}                    & $0.6625\pm 0.0236$ & $0.6774\pm 0.0826$ & $0.6833\pm 0.0901$ & $0.6735\pm 0.0783$ \\ \hline
\multicolumn{1}{|c|}{OH-KNN}                       & $0.6090$           & $0.7621$           & $0.7646$           & $0.6477$           \\ \hline
\multicolumn{1}{|c|}{FP-KNN}                       & $0.6523$           & $0.7566$           & $0.7682$           & $0.6459$           \\ \hline
\multicolumn{1}{|c|}{random (predict all ones)}    & $0.7008\pm 0.0065$ & $0.7094\pm 0.0483$ & $0.7011\pm 0.0523$ & $0.6967\pm 0.0255$ \\ \hline
\multicolumn{1}{|c|}{random (shuffle predictions)} & $0.6392\pm 0.0071$ & $0.6464\pm 0.0669$ & $0.6385\pm 0.0700$ & $0.6301\pm 0.0340$ \\ \hline
\end{tabular}
\label{tbl:SIRT2_MSE_test_losses}
\end{table}

\subsubsection{Rank correlation coefficient}

\begin{figure}[H] 
    \centering
        \includegraphics[scale=1]{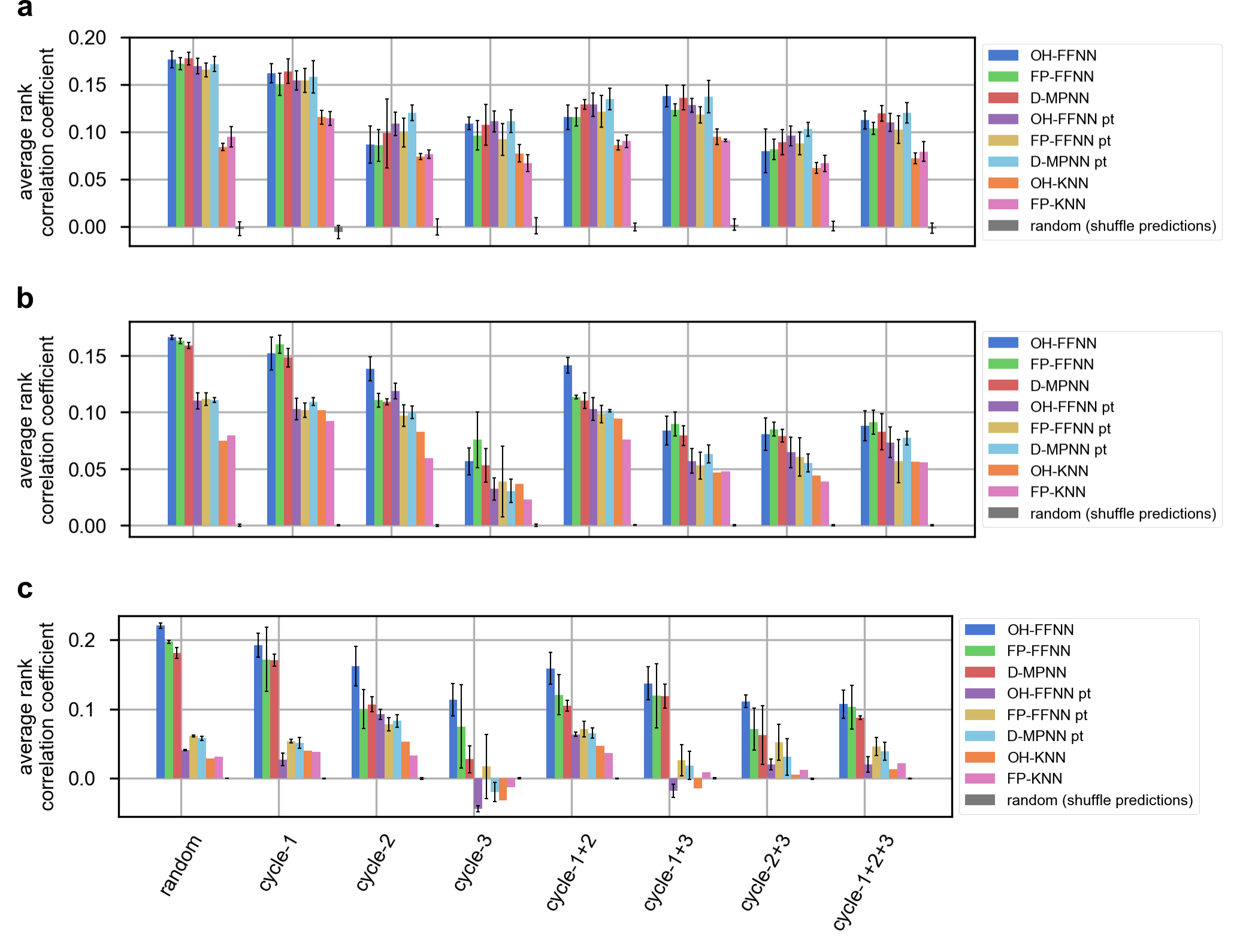}
    \caption{Comparison of model performance, as measured by rank correlation coefficient. OH: one-hot; FP: fingerprint; FFNN: feed-forward neural network; D-MPNN: directed message-passing neural network; KNN: k-nearest neighbors. The rank correlation coefficients of the negative-log-likelihood-trained models (OH-FFNN, FP-FFNN, D-MPNN) are compared to the baseline point-prediction-trained models (OH-FFNN pt, FP-FFNN pt, D-MPNN pt), k-nearest-neighbors models (OH-KNN, FP-KNN), and random models (predict all ones, shuffle predictions), for various data splits (\emph{cf.} Figure~\ref{fig:reps_models_loss_fns_splits}d) on the \textbf{(a)} DD1S CAIX, \textbf{(b)} triazine sEH, \textbf{(c)} triazine SIRT2 datasets. Error bars represent $\pm$ one standard deviation. OH-FFNN, FP-FFNN, OH-FFNN pt, FP-FFNN pt, random (predict all ones), and random (shuffle predictions) results are averaged over five trials for each dataset; D-MPNN and D-MPNN pt results are averaged over five trials for the DD1S CAIX dataset and over three trials for the triazine sEH and triazine SIRT2 datasets; OH-KNN and FP-KNN results are averaged over five trials for the DD1S CAIX dataset and are single trials (evaluated on a random 10\% of the test set) for the triazine sEH and triazine SIRT2 datasets. The result of each trial is shown separately below (Figure~\ref{fig:rank_corr_coeffs_scatter_plot}).}
    \label{fig:rank_corr_coeffs_bar_graph}
\end{figure}

\begin{figure}[H] 
    \centering
        \includegraphics[scale=1]{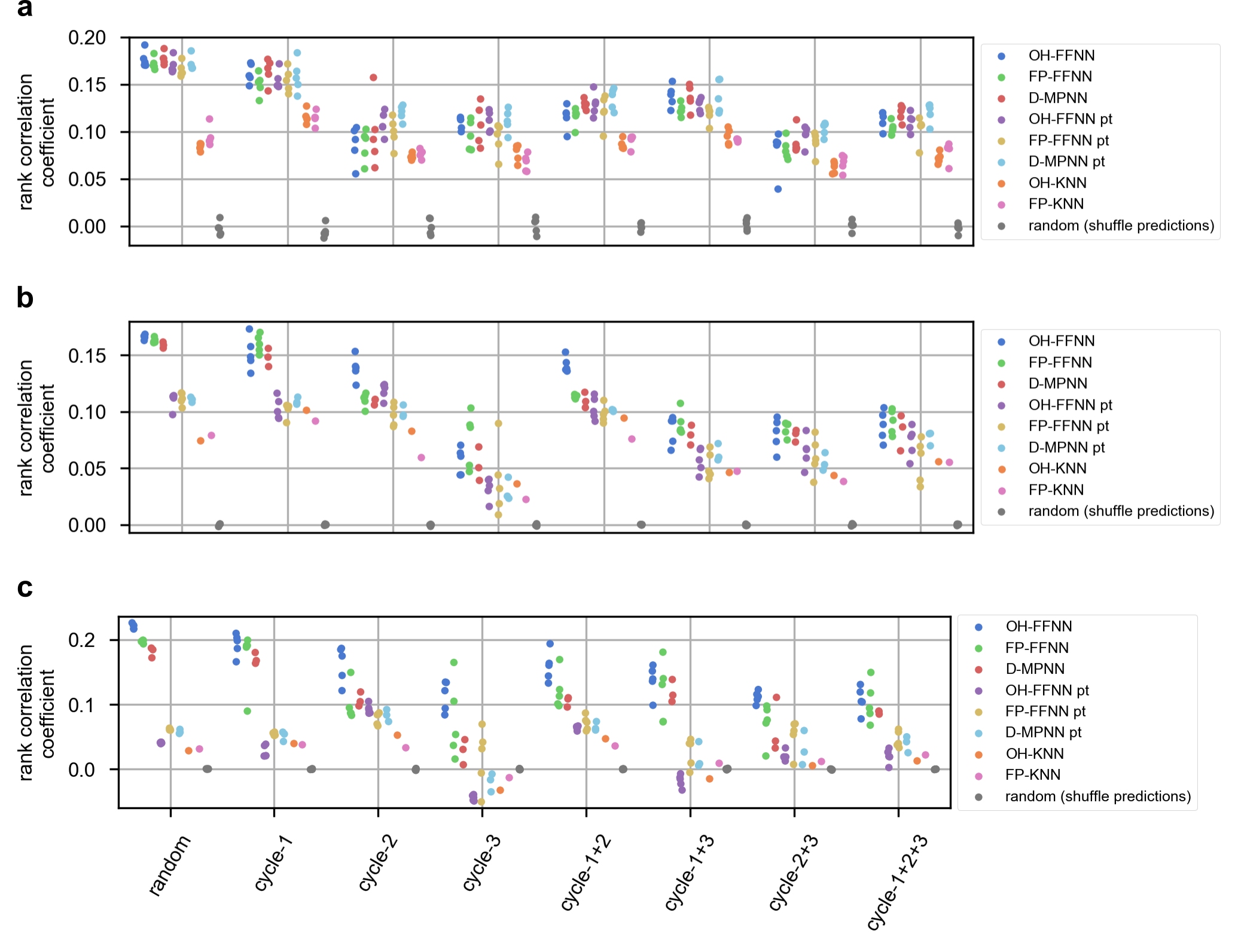}
    \caption{Comparison of model performance, as measured by rank correlation coefficient, for the negative-log-likelihood-trained models (OH-FFNN, FP-FFNN, D-MPNN) versus the baseline point-prediction-trained models (OH-FFNN pt, FP-FFNN pt, D-MPNN pt), k-nearest-neighbors models (OH-KNN, FP-KNN), and random models (predict all ones, shuffle predictions), for various data splits (\emph{cf.} Figure~\ref{fig:reps_models_loss_fns_splits}d) on the \textbf{(a)} DD1S CAIX, \textbf{(b)} triazine sEH, \textbf{(c)} triazine SIRT2 datasets. For each dataset, data split type, and model type, the individual rank correlation coefficient for each trial is shown. Five trials were performed for OH-FFNN, FP-FFNN, OH-FFNN pt, FP-FFNN pt, and the random baselines (predict all ones, shuffle predictions) on each dataset; for D-MPNN and D-MPNN pt, five trials were performed on the DD1S CAIX dataset and three trials were performed on the triazine sEH and triazine SIRT2 datasets; for OH-KNN and FP-KNN, five trials were performed on the DD1S CAIX dataset and one trial (evaluated on a random 10\% of the test set) was performed on the triazine sEH and triazine SIRT2 datasets.}
    \label{fig:rank_corr_coeffs_scatter_plot}
\end{figure}

\begin{table}[H]
\caption{Rank correlation coefficients (mean $\pm$ standard deviation) for regression models and baseline models on the DD1S CAIX dataset. Results are averaged over five trials.}
\begin{tabular}{|c|c|c|c|c|}
\cline{1-1} \cline{3-5}
DD1S CAIX                    &                     & \multicolumn{3}{c|}{cycle split}                               \\ \hline
model                        & random split        & 1                   & 2                   & 3                  \\ \hline
OH-FFNN                      & $0.1767\pm 0.0089$  & $0.1619\pm 0.0102$  & $0.0867\pm 0.0197$  & $0.1093\pm 0.0067$ \\ \hline
FP-FFNN                      & $0.1720\pm 0.0064$  & $0.1503\pm 0.0115$  & $0.0859\pm 0.0167$  & $0.0965\pm 0.0155$ \\ \hline
D-MPNN                       & $0.1776\pm 0.0065$  & $0.1641\pm 0.0131$  & $0.0986\pm 0.0362$  & $0.1076\pm 0.0216$ \\ \hline
OH-FFNN pt                   & $0.1699\pm 0.0082$  & $0.1547\pm 0.0102$  & $0.1088\pm 0.0123$  & $0.1112\pm 0.0108$ \\ \hline
FP-FFNN pt                   & $0.1655\pm 0.0074$  & $0.1546\pm 0.0124$  & $0.0996\pm 0.0153$  & $0.0923\pm 0.0167$ \\ \hline
D-MPNN pt                    & $0.1718\pm 0.0078$  & $0.1584\pm 0.0171$  & $0.1203\pm 0.0080$  & $0.1116\pm 0.0122$ \\ \hline
OH-KNN                       & $0.0844\pm 0.0037$  & $0.1157\pm 0.0071$  & $0.0739\pm 0.0032$  & $0.0777\pm 0.0091$ \\ \hline
FP-KNN                       & $0.0949\pm 0.0108$  & $0.1144\pm 0.0071$  & $0.0766\pm 0.0045$  & $0.0673\pm 0.0088$ \\ \hline
random (shuffle predictions) & $-0.0021\pm 0.0072$ & $-0.0054\pm 0.0069$ & $-0.0001\pm 0.0085$ & $0.0011\pm 0.0084$ \\ \hline
\end{tabular}

\begin{tabular}{c|c|c|c|c|}
\cline{2-5}
                                                   & \multicolumn{4}{c|}{cycle split}                                                    \\ \hline
\multicolumn{1}{|c|}{model}                        & 1+2                 & 1+3                & 2+3                & 1+2+3               \\ \hline
\multicolumn{1}{|c|}{OH-FFNN}                      & $0.1157\pm 0.0128$  & $0.1380\pm 0.0115$ & $0.0801\pm 0.0231$ & $0.1126\pm 0.0093$  \\ \hline
\multicolumn{1}{|c|}{FP-FFNN}                      & $0.1159\pm 0.0097$  & $0.1233\pm 0.0064$ & $0.0817\pm 0.0108$ & $0.1041\pm 0.0064$  \\ \hline
\multicolumn{1}{|c|}{D-MPNN}                       & $0.1290\pm 0.0051$  & $0.1364\pm 0.0128$ & $0.0893\pm 0.0133$ & $0.1197\pm 0.0083$  \\ \hline
\multicolumn{1}{|c|}{OH-FFNN pt}                   & $0.1290\pm 0.0123$  & $0.1284\pm 0.0072$ & $0.0962\pm 0.0103$ & $0.1103\pm 0.0097$  \\ \hline
\multicolumn{1}{|c|}{FP-FFNN pt}                   & $0.1219\pm 0.0166$  & $0.1182\pm 0.0087$ & $0.0882\pm 0.0118$ & $0.1027\pm 0.0143$  \\ \hline
\multicolumn{1}{|c|}{D-MPNN pt}                    & $0.1350\pm 0.0113$  & $0.1376\pm 0.0172$ & $0.1030\pm 0.0073$ & $0.1205\pm 0.0106$  \\ \hline
\multicolumn{1}{|c|}{OH-KNN}                       & $0.0863\pm 0.0051$  & $0.0951\pm 0.0083$ & $0.0622\pm 0.0060$ & $0.0723\pm 0.0056$  \\ \hline
\multicolumn{1}{|c|}{FP-KNN}                       & $0.0905\pm 0.0066$  & $0.0912\pm 0.0014$ & $0.0671\pm 0.0084$ & $0.0794\pm 0.0104$  \\ \hline
\multicolumn{1}{|c|}{random (shuffle predictions)} & $-0.0001\pm 0.0040$ & $0.0023\pm 0.0059$ & $0.0009\pm 0.0053$ & $-0.0014\pm 0.0052$ \\ \hline
\end{tabular}
\label{tbl:CAIX_rank_corr_coeffs}
\end{table}

\begin{table}[H]
\caption{Rank correlation coefficients (mean $\pm$ standard deviation) for regression models and baseline models on the triazine sEH dataset. OH-FFNN, FP-FFNN, OH-FFNN pt, FP-FFNN pt, random (predict all ones), and random (shuffle predictions) results are averaged over five trials; D-MPNN and D-MPNN pt results are averaged over three trials; OH-KNN and FP-KNN results are single trials evaluated on a random 10\% of the test set.}
\begin{tabular}{|c|c|c|c|c|}
\cline{1-1} \cline{3-5}
triazine sEH                 &                    & \multicolumn{3}{c|}{cycle split}                              \\ \hline
model                        & random split       & 1                  & 2                   & 3                  \\ \hline
OH-FFNN                      & $0.1660\pm 0.0020$ & $0.1517\pm 0.0146$ & $0.1384\pm 0.0106$  & $0.0566\pm 0.0119$ \\ \hline
FP-FFNN                      & $0.1629\pm 0.0022$ & $0.1600\pm 0.0080$ & $0.1104\pm 0.0062$  & $0.0755\pm 0.0244$ \\ \hline
D-MPNN                       & $0.1589\pm 0.0027$ & $0.1480\pm 0.0080$ & $0.1091\pm 0.0028$  & $0.0529\pm 0.0149$ \\ \hline
OH-FFNN pt                   & $0.1101\pm 0.0071$ & $0.1028\pm 0.0096$ & $0.1185\pm 0.0069$  & $0.0322\pm 0.0098$ \\ \hline
FP-FFNN pt                   & $0.1115\pm 0.0054$ & $0.1016\pm 0.0064$ & $0.0969\pm 0.0094$  & $0.0387\pm 0.0314$ \\ \hline
D-MPNN pt                    & $0.1106\pm 0.0023$ & $0.1092\pm 0.0033$ & $0.0996\pm 0.0055$  & $0.0303\pm 0.0103$ \\ \hline
OH-KNN                       & $0.0747$           & $0.1017$           & $0.0828$            & $0.0365$           \\ \hline
FP-KNN                       & $0.0795$           & $0.0919$           & $0.0595$            & $0.0229$           \\ \hline
random (shuffle predictions) & $0.0001\pm 0.0010$ & $6\mathrm{e}{-5}\pm 0.0005$   & $-0.0002\pm 0.0007$ & $-6\mathrm{e}{-5}\pm 0.0009$  \\ \hline
\end{tabular}

\begin{tabular}{c|c|c|c|c|}
\cline{2-5}
                                                   & \multicolumn{4}{c|}{cycle split}                                                   \\ \hline
\multicolumn{1}{|c|}{model}                        & 1+2                & 1+3                 & 2+3                & 1+2+3              \\ \hline
\multicolumn{1}{|c|}{OH-FFNN}                      & $0.1414\pm 0.0069$ & $0.0837\pm 0.0129$  & $0.0805\pm 0.0141$ & $0.0878\pm 0.0133$ \\ \hline
\multicolumn{1}{|c|}{FP-FFNN}                      & $0.1134\pm 0.0015$ & $0.0895\pm 0.0106$  & $0.0850\pm 0.0063$ & $0.0911\pm 0.0107$ \\ \hline
\multicolumn{1}{|c|}{D-MPNN}                       & $0.1100\pm 0.0068$ & $0.0793\pm 0.0087$  & $0.0791\pm 0.0053$ & $0.0828\pm 0.0158$ \\ \hline
\multicolumn{1}{|c|}{OH-FFNN pt}                   & $0.1029\pm 0.0101$ & $0.0569\pm 0.0107$  & $0.0645\pm 0.0135$ & $0.0733\pm 0.0135$ \\ \hline
\multicolumn{1}{|c|}{FP-FFNN pt}                   & $0.0983\pm 0.0077$ & $0.0527\pm 0.0120$  & $0.0605\pm 0.0168$ & $0.0567\pm 0.0192$ \\ \hline
\multicolumn{1}{|c|}{D-MPNN pt}                    & $0.1012\pm 0.0009$ & $0.0630\pm 0.0077$  & $0.0551\pm 0.0079$ & $0.0771\pm 0.0062$ \\ \hline
\multicolumn{1}{|c|}{OH-KNN}                       & $0.0945$           & $0.0467$            & $0.0437$           & $0.0559$           \\ \hline
\multicolumn{1}{|c|}{FP-KNN}                       & $0.0759$           & $0.0477$            & $0.0386$           & $0.0554$           \\ \hline
\multicolumn{1}{|c|}{random (shuffle predictions)} & $0.0001\pm 0.0002$ & $-0.0002\pm 0.0006$ & $5\mathrm{e}{-5}\pm 0.0007$   & $0.0002\pm 0.0005$ \\ \hline
\end{tabular}
\label{tbl:sEH_rank_corr_coeffs}
\end{table}

\begin{table}[H]
\caption{Rank correlation coefficients (mean $\pm$ standard deviation) for regression models and baseline models on the triazine SIRT2 dataset. OH-FFNN, FP-FFNN, OH-FFNN pt, FP-FFNN pt, random (predict all ones), and random (shuffle predictions) results are averaged over five trials; D-MPNN and D-MPNN pt results are averaged over three trials; OH-KNN and FP-KNN results are single trials evaluated on a random 10\% of the test set.}
\begin{tabular}{|c|c|c|c|c|}
\cline{1-1} \cline{3-5}
triazine SIRT2               &                    & \multicolumn{3}{c|}{cycle split}                                \\ \hline
model                        & random split       & 1                  & 2                   & 3                    \\ \hline
OH-FFNN                      & $0.2207\pm 0.0040$ & $0.1928\pm 0.0172$ & $0.1625\pm 0.0284$  & $0.1136\pm 0.0233$   \\ \hline
FP-FFNN                      & $0.1974\pm 0.0023$ & $0.1720\pm 0.0461$ & $0.1002\pm 0.0279$  & $0.0752\pm 0.0600$   \\ \hline
D-MPNN                       & $0.1813\pm 0.0081$ & $0.1706\pm 0.0087$ & $0.1072\pm 0.0110$  & $0.0278\pm 0.0195$   \\ \hline
OH-FFNN pt                   & $0.0409\pm 0.0010$ & $0.0275\pm 0.0093$ & $0.0929\pm 0.0074$  & $ -0.0440\pm 0.0044$ \\ \hline
FP-FFNN pt                   & $0.0616\pm 0.0012$ & $0.0541\pm 0.0024$ & $0.0785\pm 0.0094$  & $0.0174\pm 0.0465$   \\ \hline
D-MPNN pt                    & $0.0580\pm 0.0033$ & $0.0515\pm 0.0076$ & $0.0832\pm 0.0093$  & $-0.0195\pm 0.0140$  \\ \hline
OH-KNN                       & $0.0292$           & $0.0399$           & $0.0534$            & $-0.0315$            \\ \hline
FP-KNN                       & $0.0318$           & $0.0384$           & $0.0332$            & $ -0.0129$           \\ \hline
random (shuffle predictions) & $0.0004\pm 0.0004$ & $1\mathrm{e}{-5}\pm 0.0005$   & $-0.0004\pm 0.0013$ & $8\mathrm{e}{-5}\pm 0.0009$     \\ \hline
\end{tabular}

\begin{tabular}{c|c|c|c|c|}
\cline{2-5}
                                                   & \multicolumn{4}{c|}{cycle split}                                                    \\ \hline
\multicolumn{1}{|c|}{model}                        & 1+2                & 1+3                 & 2+3                 & 1+2+3              \\ \hline
\multicolumn{1}{|c|}{OH-FFNN}                      & $0.1590\pm 0.0231$ & $0.1373\pm 0.0237$  & $0.1115\pm 0.0092$  & $0.1073\pm 0.0199$ \\ \hline
\multicolumn{1}{|c|}{FP-FFNN}                      & $0.1209\pm 0.0289$ & $0.1197\pm 0.0462$  & $0.0715\pm 0.0305$  & $0.1032\pm 0.0315$ \\ \hline
\multicolumn{1}{|c|}{D-MPNN}                       & $0.1051\pm 0.0078$ & $0.1192\pm 0.0175$  & $0.0624\pm 0.0424$  & $0.0880\pm 0.0025$ \\ \hline
\multicolumn{1}{|c|}{OH-FFNN pt}                   & $0.0639\pm 0.0030$ & $-0.0179\pm 0.0097$ & $0.0204\pm 0.0075$  & $0.0201\pm 0.0112$ \\ \hline
\multicolumn{1}{|c|}{FP-FFNN pt}                   & $0.0713\pm 0.0111$ & $0.0261\pm 0.0224$  & $0.0521\pm 0.0259$  & $0.0460\pm 0.0128$ \\ \hline
\multicolumn{1}{|c|}{D-MPNN pt}                    & $0.0657\pm 0.0070$ & $0.0190\pm 0.0205$  & $0.0312\pm 0.0266$  & $0.0394\pm 0.0126$ \\ \hline
\multicolumn{1}{|c|}{OH-KNN}                       & $0.0472$           & $-0.0142$           & $0.0057$            & $0.0132$           \\ \hline
\multicolumn{1}{|c|}{FP-KNN}                       & $0.0367$           & $0.0094$            & $0.0123$            & $0.0224$           \\ \hline
\multicolumn{1}{|c|}{random (shuffle predictions)} & $0.0002\pm 0.0006$ & $0.0003\pm 0.0009$  & $-0.0005\pm 0.0008$ & $-7\mathrm{e}{-5}\pm 0.0004$  \\ \hline
\end{tabular}
\label{tbl:SIRT2_rank_corr_coeffs}
\end{table}

\subsection{Comparison of predicted enrichments to 95\% confidence intervals for the FP-FFNNs on all random splits of the DD1S CAIX dataset}

\begin{table}[H]
\caption{Comparison of test-set predicted enrichments to 95\% confidence intervals (estimated from the count data) for the FP-FFNNs on all random splits of the DD1S CAIX dataset.}
\begin{tabular}{|c|c}
\cline{1-1}
DD1S CAIX (random split) & \multicolumn{1}{l}{}                      \\ \hline
random seed              & \multicolumn{1}{c|}{\begin{tabular}[c]{@{}c@{}}percentage of test-set predicted\\ enrichments within 95\% confidence interval\end{tabular}} \\ \hline
0                        & \multicolumn{1}{c|}{93.02\%}              \\ \hline
1                        & \multicolumn{1}{c|}{92.82\%}              \\ \hline
2                        & \multicolumn{1}{c|}{92.91\%}              \\ \hline
3                        & \multicolumn{1}{c|}{92.93\%}              \\ \hline
4                        & \multicolumn{1}{c|}{93.03\%}              \\ \hline
\end{tabular}
\label{tbl:DD1S_CAIX_FP-FFNN_coverage_probs}
\end{table}

\subsection{Parity plots to evaluate correlation between predicted and calculated enrichments}

\begin{figure}[H] 
    \centering
        \includegraphics[scale=1]{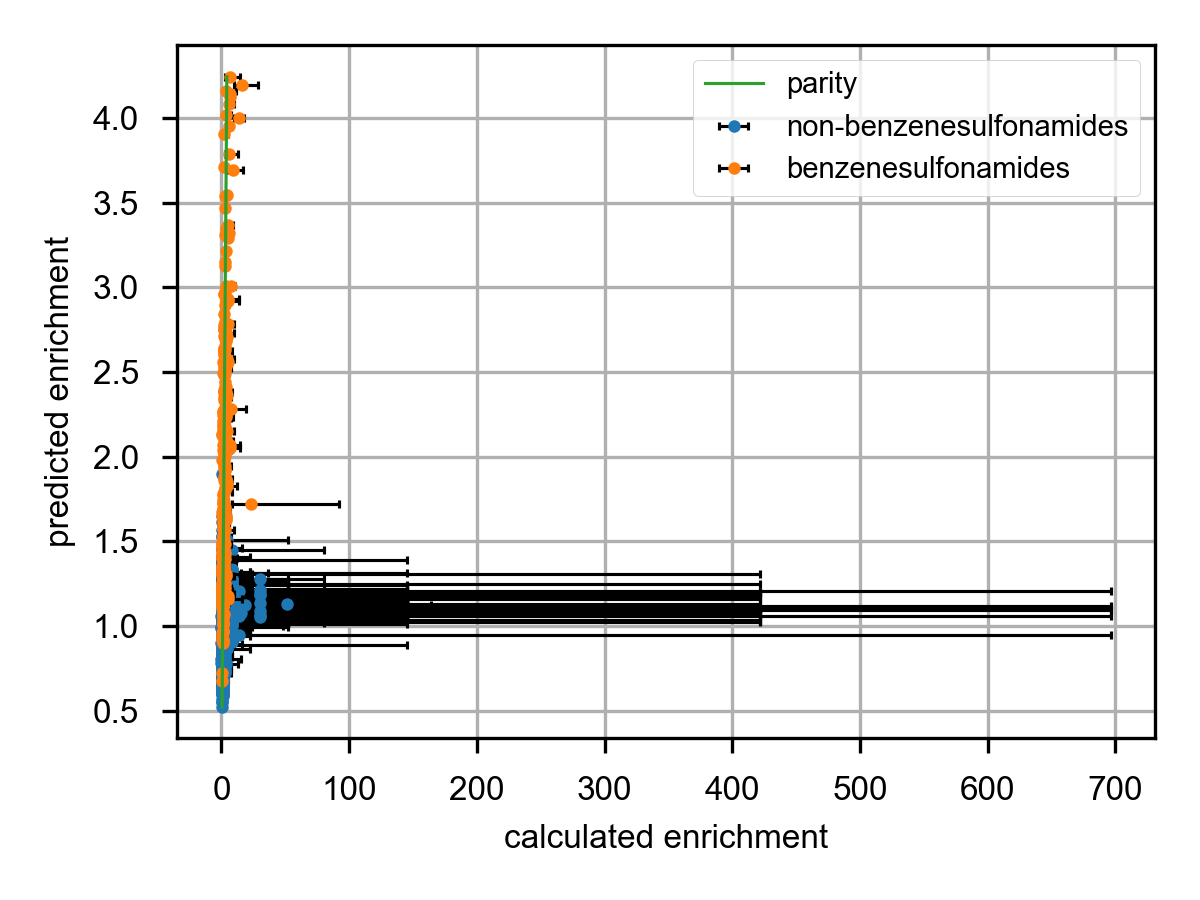}
    \caption{Full scatter plot of predicted and calculated enrichments for the test-set compounds of a FP-FFNN on a random split (\emph{cf.} Figure~\ref{fig:reps_models_loss_fns_splits}d) of the DD1S CAIX dataset (as shown zoomed-in in Figure~\ref{fig:DD1S_CAIX_parity_plot_1D_histograms}a). The green parity line is the identity function, for reference. Error bars represent 95\% confidence intervals for calculated enrichments.}
    \label{fig:CAIX_parity_plot_full}
\end{figure}

\begin{figure}[H] 
    \centering
        \includegraphics[scale=1]{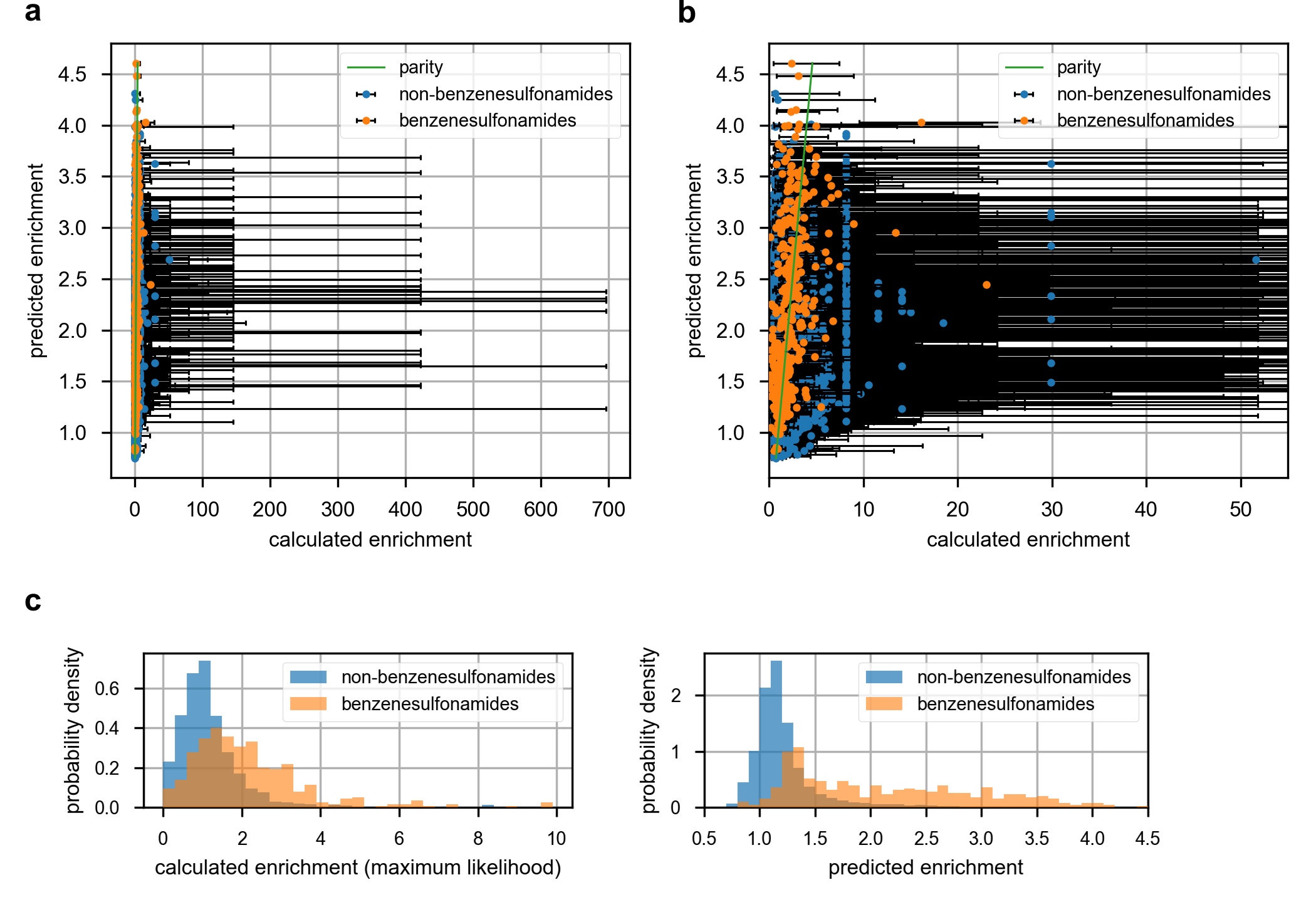}
    \caption{\textbf{(a)} Full and \textbf{(b)} zoomed-in scatter plot of predicted and calculated enrichments for the test-set compounds of a FP-FFNN pt model (trained using MSE loss) on a random split (\emph{cf.} Figure~\ref{fig:reps_models_loss_fns_splits}d) of the DD1S CAIX dataset. The green parity line is the identity function, for reference. Error bars represent 95\% confidence intervals for calculated enrichments. \textbf{(c)} Histograms of calculated and predicted enrichments for the test-set compounds of a FP-FFNN pt model on a random split (\emph{cf.} Figure~\ref{fig:reps_models_loss_fns_splits}d) of the DD1S CAIX dataset. The horizontal axis cutoff of 10 in the histogram of calculated enrichments is arbitrary, for the sake of legibility.}
    \label{fig:CAIX_MSE_parity_plot}
\end{figure}

\begin{figure}[H]
    \centering
        \includegraphics[scale=1]{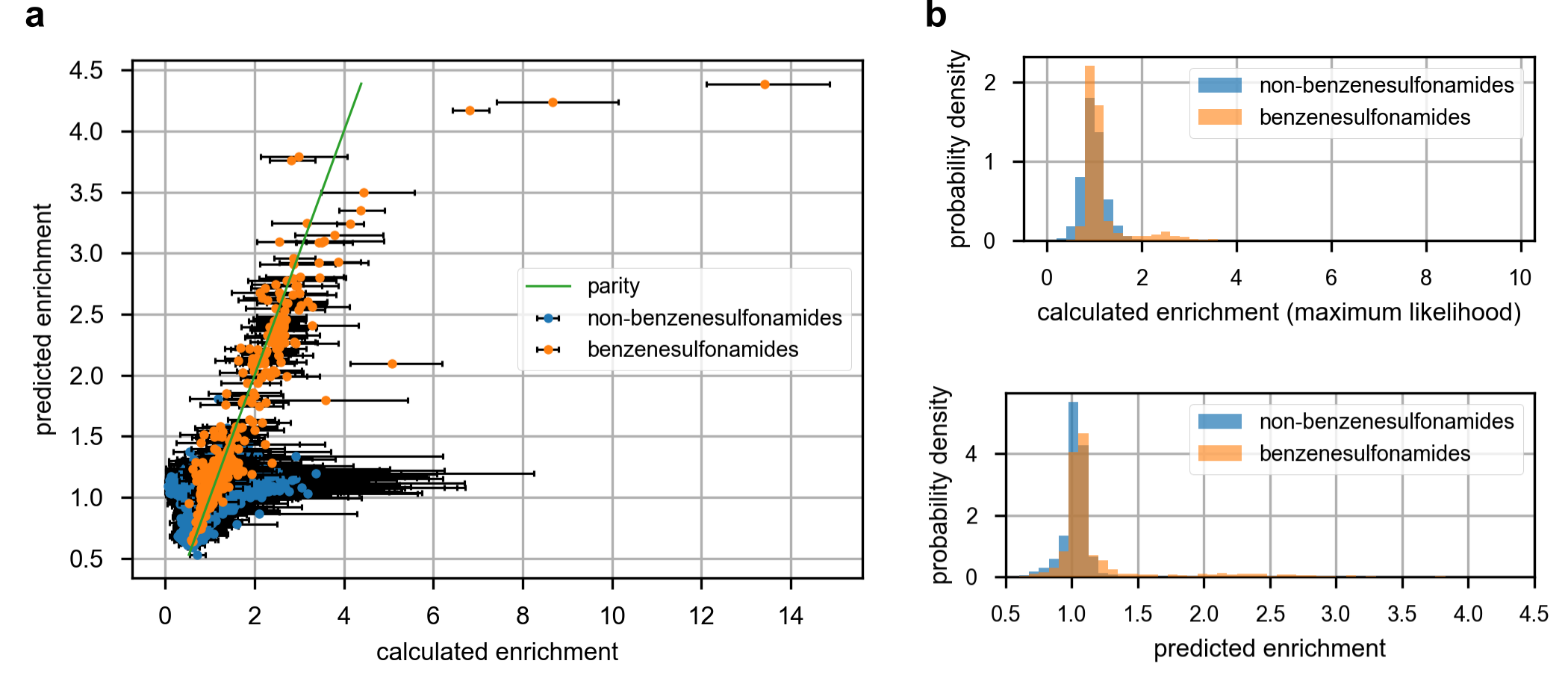}
    \caption{\textbf{(a)} Scatter plot of predicted and calculated enrichments for all disynthons in the DD1S CAIX dataset, using the predictions of a FP-FFNN on a random split (\emph{cf.} Figure~\ref{fig:reps_models_loss_fns_splits}d) of the DD1S CAIX dataset. \textbf{(b)} Histograms of calculated and predicted enrichments for all disynthons in the DD1S CAIX dataset, using the predictions of a FP-FFNN on a random split (\emph{cf.} Figure~\ref{fig:reps_models_loss_fns_splits}d) of the DD1S CAIX dataset.}
    \label{fig:CAIX_disynthon_parity_plot_histograms}
\end{figure}

\begin{figure}[H] 
    \centering
        \includegraphics[scale=1]{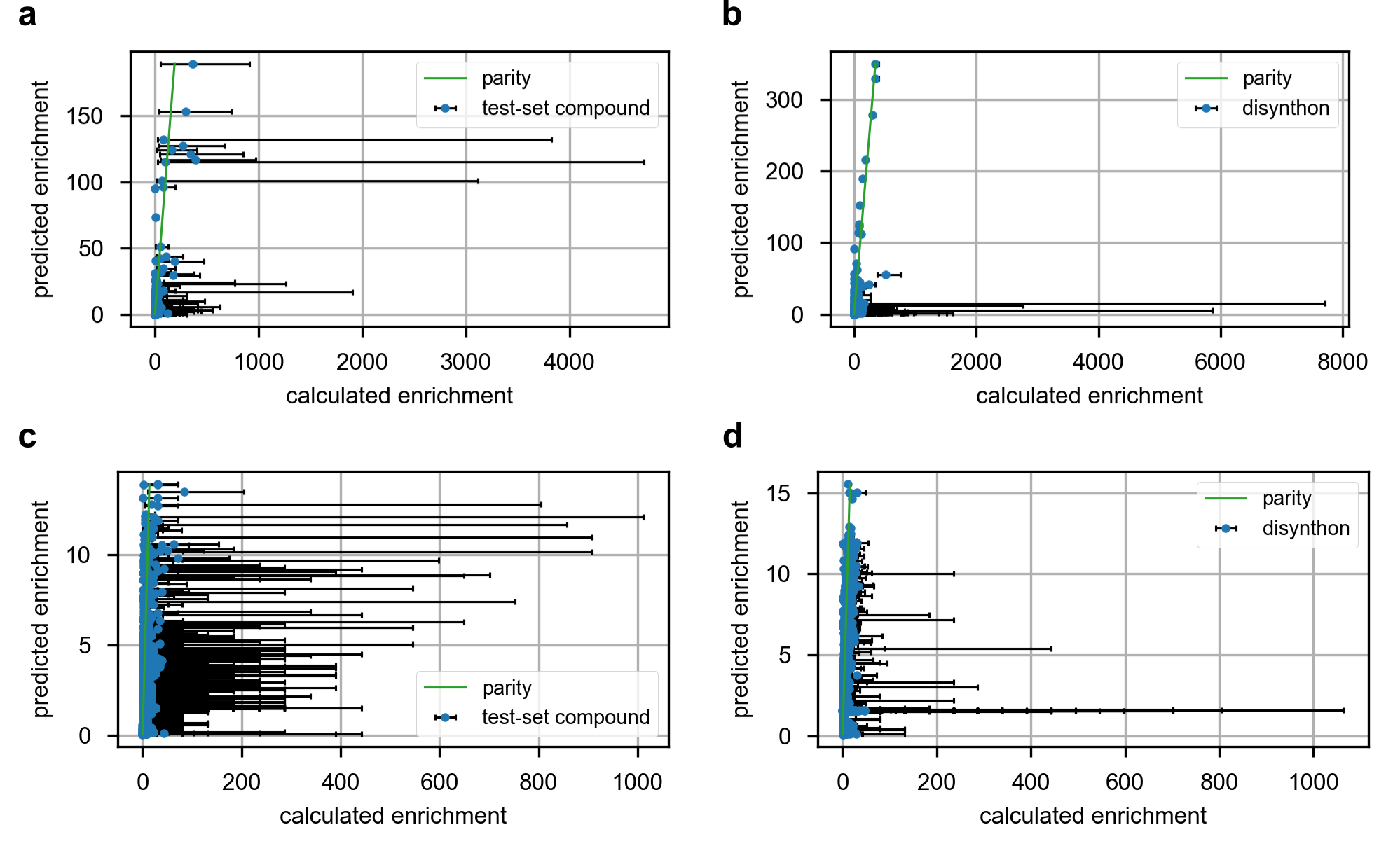}
    \caption{Full scatter plot of predicted and calculated enrichments for a subset (20,000 compounds) of the test set of a FP-FFNN on a random split (\emph{cf.} Figure~\ref{fig:reps_models_loss_fns_splits}d) of the \textbf{(a)} triazine sEH, \textbf{(c)} triazine SIRT2 dataset, and for all disynthons in the \textbf{(b)} triazine sEH, \textbf{(d)} triazine SIRT2 dataset (as shown zoomed-in in Figure~\ref{fig:sEH_SIRT2_parity_scatter_plots}). The green parity line is the identity function, for reference. Error bars represent 95\% confidence intervals for calculated enrichments.}
    \label{fig:triazine_parity_plots_full}
\end{figure}

\subsection{Distributional shift in calculated enrichment for the DD1S CAIX dataset (cycle-2 split, seed 4)}

\begin{figure}[H] 
    \centering
        \includegraphics[scale=1]{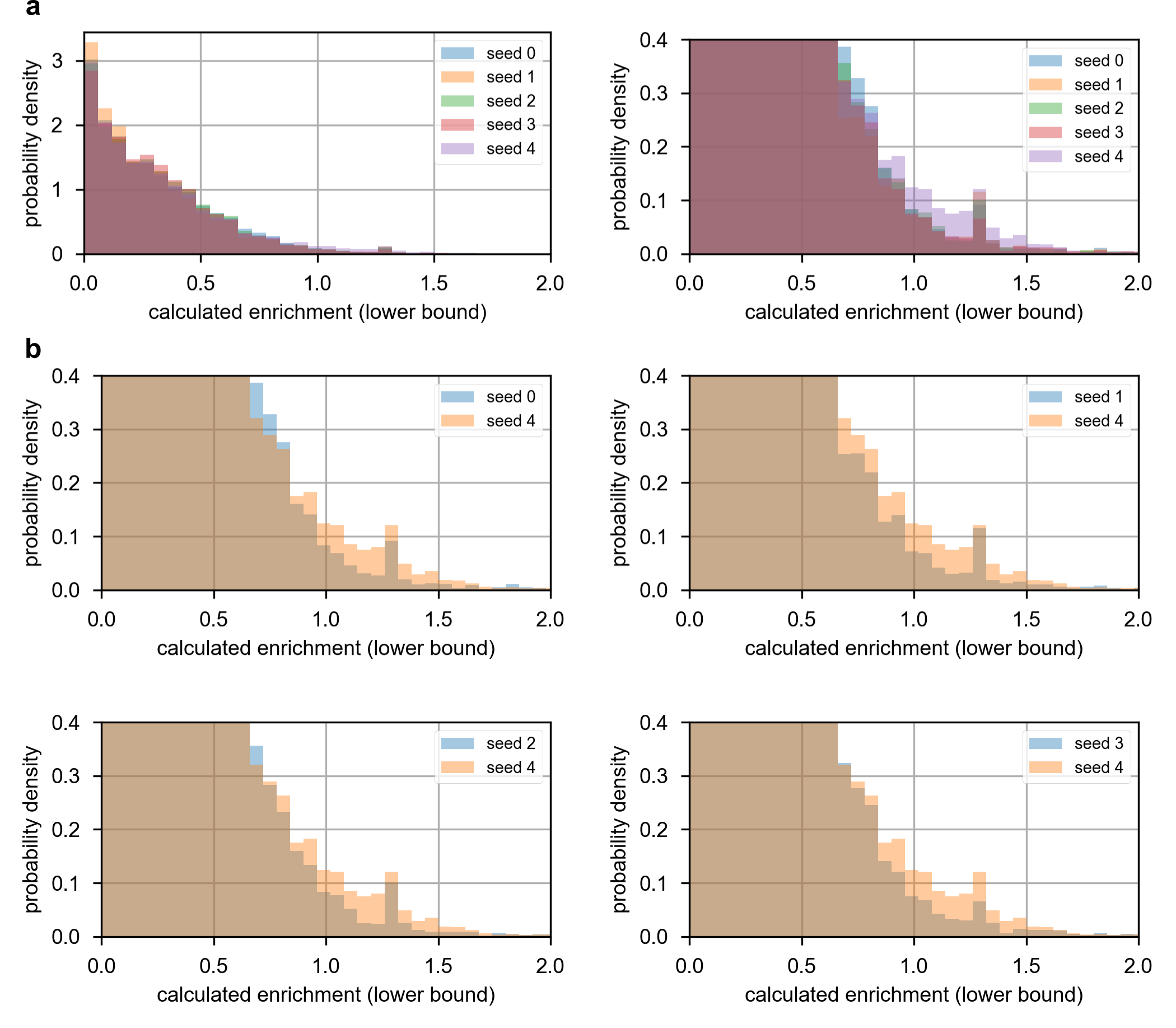}
    \caption{Histograms (full and zoomed-in) of the lower bound of calculated enrichment for the test set of the DD1S CAIX dataset split along cycle 2 (\emph{cf.} Figure~\ref{fig:reps_models_loss_fns_splits}d). Each seed represents a cycle-2 split using a different random seed. \textbf{(a)} shows the overlaid distributions for all five splits; \textbf{(b)} shows individual comparisons of the distribution for the seed-4 split with the distribution for each of the other splits.}
    \label{fig:CAIX_cycle2_distrib_shift}
\end{figure}

\subsection{Binary classification baseline comparisons (fixed threshold)}

\begin{figure}[H] 
    \centering
        \includegraphics[scale=1]{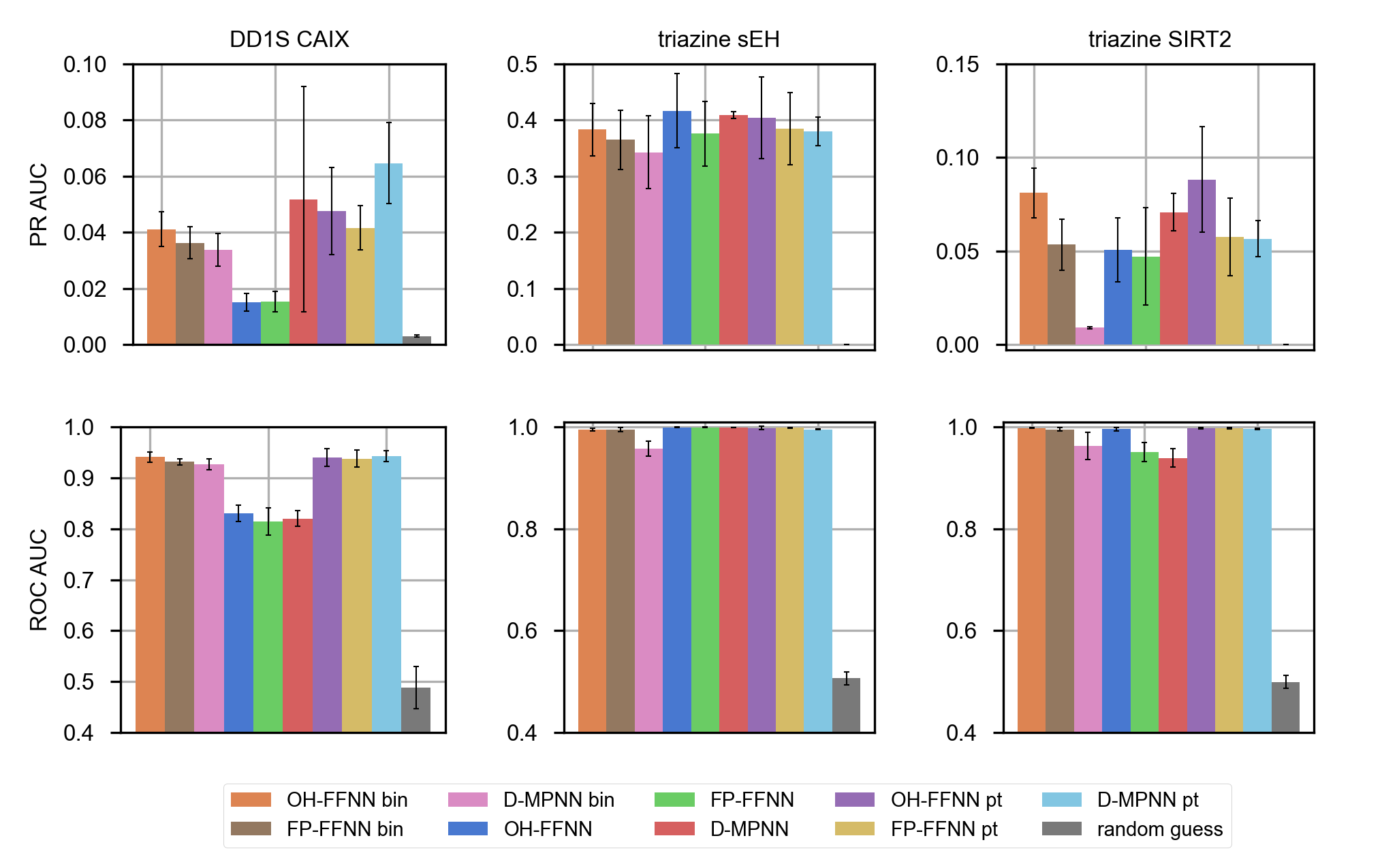}
    \caption{PR and ROC AUC scores (mean $\pm$ standard deviation) for the binary classifiers (``bin'') and regression models evaluated as classifiers, at fixed thresholds. The top 0.5\% and 0.01\% of compounds in the training set were defined as enriched for the DD1S CAIX and triazine datasets, respectively. OH-FFNN bin, FP-FFNN bin, OH-FFNN, FP-FFNN, OH-FFNN pt, FP-FFNN pt, and random guess results are averaged over five trials for each dataset; D-MPNN bin, D-MPNN, and D-MPNN pt results are averaged over five trials for the DD1S CAIX dataset and over three trials for the triazine sEH and triazine SIRT2 datasets. For each trial, the random-guess baseline was generated by randomly shuffling the predictions of the FP-FFNN for that trial.}
    \label{fig:bin_plots_fixed_threshold}
\end{figure}

\begin{table}[H]
\caption{PR and ROC AUCs (mean $\pm$ standard deviation) for the regression models, binary classifiers (``bin''), and baseline random-guess (random shuffling of the predictions of the FP-FFNNs) on the DD1S CAIX dataset. The top 0.5\% of compounds in the training set were defined as enriched. Results are averaged over five trials.}
\begin{tabular}{|c|cc}
\cline{1-1}
DD1S CAIX (random split) &  &  \\ \hline
model & \multicolumn{1}{c|}{PR AUC} & \multicolumn{1}{c|}{ROC AUC} \\ \hline
OH-FFNN & \multicolumn{1}{c|}{$0.0150\pm0.0032$} & \multicolumn{1}{c|}{$0.8304\pm0.0160$} \\ \hline
FP-FFNN & \multicolumn{1}{c|}{$0.0153\pm0.0036$} & \multicolumn{1}{c|}{$0.8137\pm0.0268$} \\ \hline
D-MPNN & \multicolumn{1}{c|}{$0.0517\pm0.0402$} & \multicolumn{1}{c|}{$0.8202\pm0.0155$} \\ \hline
OH-FFNN pt & \multicolumn{1}{c|}{$0.0475\pm0.0155$} & \multicolumn{1}{c|}{$0.9401\pm0.0170$} \\ \hline
FP-FFNN pt & \multicolumn{1}{c|}{$0.0416\pm0.0079$} & \multicolumn{1}{c|}{$0.9375\pm0.0170$} \\ \hline
D-MPNN pt & \multicolumn{1}{c|}{$0.0646\pm0.0144$} & \multicolumn{1}{c|}{$0.9428\pm0.0106$} \\ \hline
OH-FFNN bin & \multicolumn{1}{c|}{$0.0411\pm0.0062$} & \multicolumn{1}{c|}{$0.9409\pm0.0101$} \\ \hline
FP-FFNN bin & \multicolumn{1}{c|}{$0.0362\pm0.0058$} & \multicolumn{1}{c|}{$0.9312\pm0.0064$} \\ \hline
D-MPNN bin & \multicolumn{1}{c|}{$0.0337\pm0.0058$} & \multicolumn{1}{c|}{$0.9264\pm0.0105$} \\ \hline
random guess & \multicolumn{1}{c|}{$0.0030\pm0.0003$} & \multicolumn{1}{c|}{$0.4878\pm0.0413$} \\ \hline
\end{tabular}
\label{tbl:CAIX_AUCs}
\end{table}

\begin{table}[H]
\caption{PR and ROC AUCs (mean $\pm$ standard deviation) for the regression models, binary classifiers (``bin''), and baseline random-guess (random shuffling of the predictions of the FP-FFNNs) on the triazine sEH dataset. The top 0.01\% of compounds in the training set were defined as enriched. OH-FFNN bin, FP-FFNN bin, OH-FFNN, FP-FFNN, OH-FFNN pt, FP-FFNN pt, and random guess results are averaged over five trials; D-MPNN bin, D-MPNN, and D-MPNN pt results are averaged over three trials.}
\begin{tabular}{|c|cc}
\cline{1-1}
triazine sEH (random split) &  &  \\ \hline
model & \multicolumn{1}{c|}{PR AUC} & \multicolumn{1}{c|}{ROC AUC} \\ \hline
OH-FFNN & \multicolumn{1}{c|}{$0.4168\pm0.0660$} & \multicolumn{1}{c|}{$0.9997\pm0.0003$} \\ \hline
FP-FFNN & \multicolumn{1}{c|}{$0.3757\pm0.0581$} & \multicolumn{1}{c|}{$0.9997\pm0.0003$} \\ \hline
D-MPNN & \multicolumn{1}{c|}{$0.4087\pm0.0058$} & \multicolumn{1}{c|}{$0.9998\pm4\mathrm{e}{-5}$} \\ \hline
OH-FFNN pt & \multicolumn{1}{c|}{$0.4042\pm0.0729$} & \multicolumn{1}{c|}{$0.9981\pm0.0031$} \\ \hline
FP-FFNN pt & \multicolumn{1}{c|}{$0.3845\pm0.0641$} & \multicolumn{1}{c|}{$0.9990\pm0.0007$} \\ \hline
D-MPNN pt & \multicolumn{1}{c|}{$0.3800\pm0.0258$} & \multicolumn{1}{c|}{$0.9956\pm0.0007$} \\ \hline
OH-FFNN bin & \multicolumn{1}{c|}{$0.3832\pm0.0471$} & \multicolumn{1}{c|}{$0.9952\pm0.0030$} \\ \hline
FP-FFNN bin & \multicolumn{1}{c|}{$0.3649\pm0.0531$} & \multicolumn{1}{c|}{$0.9953\pm0.0036$} \\ \hline
D-MPNN bin & \multicolumn{1}{c|}{$0.3426\pm0.0649$} & \multicolumn{1}{c|}{$0.9578\pm0.0144$} \\ \hline
random guess & \multicolumn{1}{c|}{$0.0001\pm2\mathrm{e}{-5}$} & \multicolumn{1}{c|}{$0.5065\pm0.0125$} \\ \hline
\end{tabular}
\label{tbl:sEH_AUCs}
\end{table}

\begin{table}[H]
\caption{PR and ROC AUCs (mean $\pm$ standard deviation) for the regression models, binary classifiers (``bin''), and baseline random-guess (random shuffling of the predictions of the FP-FFNNs) on the triazine SIRT2 dataset. The top 0.01\% of compounds in the training set were defined as enriched. OH-FFNN bin, FP-FFNN bin, OH-FFNN, FP-FFNN, OH-FFNN pt, FP-FFNN pt, and random guess results are averaged over five trials; D-MPNN bin, D-MPNN, and D-MPNN pt results are averaged over three trials.}
\begin{tabular}{|c|cc}
\cline{1-1}
triazine SIRT2 (random split) & \multicolumn{2}{c}{} \\ \hline
model & \multicolumn{1}{c|}{PR AUC} & \multicolumn{1}{c|}{ROC AUC} \\ \hline
OH-FFNN & \multicolumn{1}{c|}{$0.0506\pm0.0172$} & \multicolumn{1}{c|}{$0.9965\pm0.0033$} \\ \hline
FP-FFNN & \multicolumn{1}{c|}{$0.0471\pm0.0260$} & \multicolumn{1}{c|}{$0.9509\pm0.0187$} \\ \hline
D-MPNN & \multicolumn{1}{c|}{$0.0708\pm0.0099$} & \multicolumn{1}{c|}{$0.9393\pm0.0180$} \\ \hline
OH-FFNN pt & \multicolumn{1}{c|}{$0.0883\pm0.0283$} & \multicolumn{1}{c|}{$0.9984\pm0.0012$} \\ \hline
FP-FFNN pt & \multicolumn{1}{c|}{$0.0575\pm0.0207$} & \multicolumn{1}{c|}{$0.9977\pm0.0010$} \\ \hline
D-MPNN pt & \multicolumn{1}{c|}{$0.0566\pm0.0096$} & \multicolumn{1}{c|}{$0.9966\pm0.0014$} \\ \hline
OH-FFNN bin & \multicolumn{1}{c|}{$0.0811\pm0.0134$} & \multicolumn{1}{c|}{$0.9985\pm0.0008$} \\ \hline
FP-FFNN bin & \multicolumn{1}{c|}{$0.0535\pm0.0136$} & \multicolumn{1}{c|}{$0.9956\pm0.0032$} \\ \hline
D-MPNN bin & \multicolumn{1}{c|}{$0.0090\pm0.0006$} & \multicolumn{1}{c|}{$0.9629\pm0.0263$} \\ \hline
random guess & \multicolumn{1}{c|}{$8\mathrm{e}{-5}\pm1\mathrm{e}{-5}$} & \multicolumn{1}{c|}{$0.4991\pm0.0127$} \\ \hline
\end{tabular}
\label{tbl:SIRT2_AUCs}
\end{table}

\subsection{Binary classification baseline comparisons (multiple thresholds)}

\begin{figure}[H] 
    \centering
        \includegraphics[scale=1]{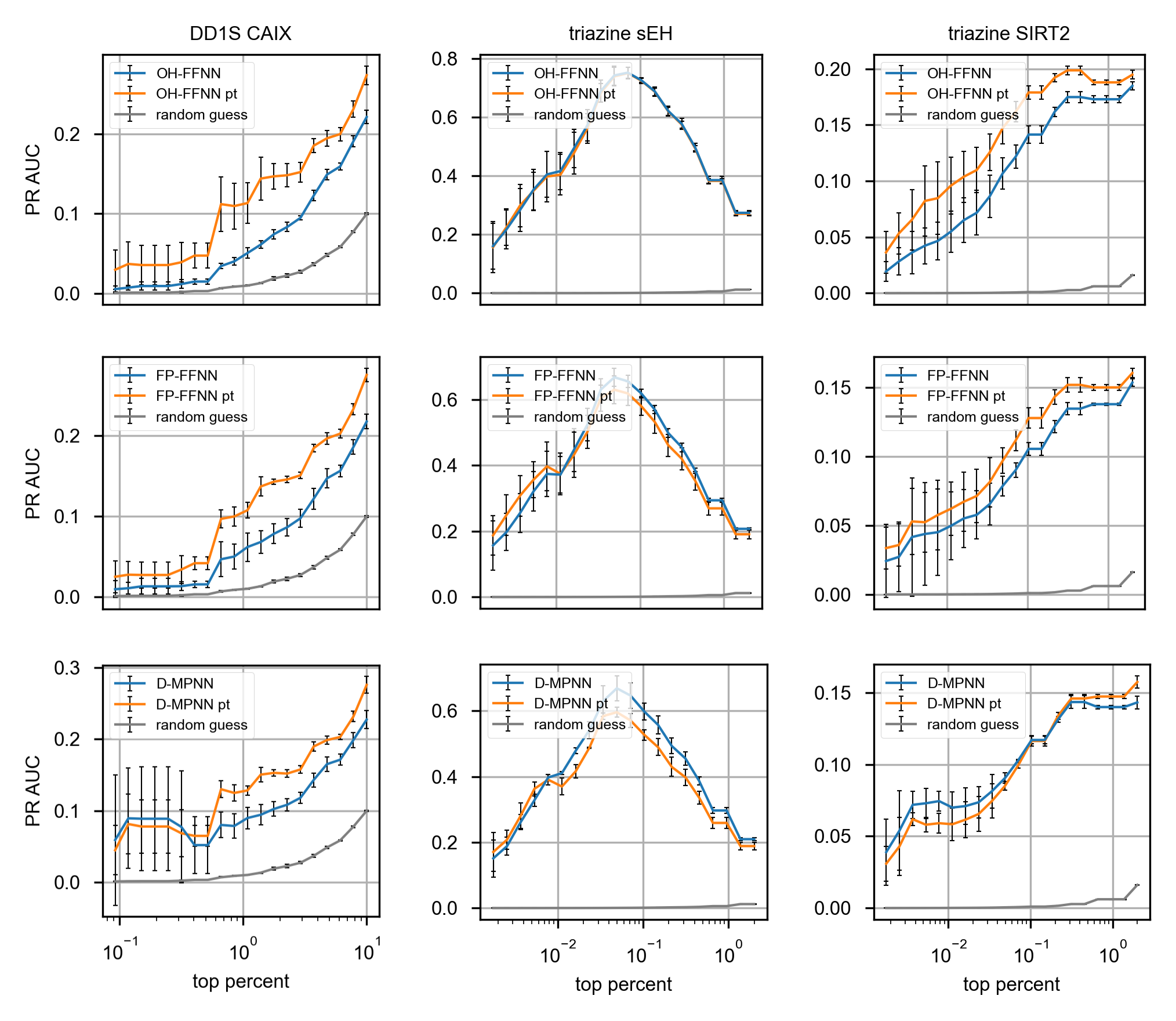}
    \caption{PR AUC scores (mean $\pm$ standard deviation) for the regression models evaluated as classifiers, at various thresholds defined by the percentage of compounds in the training set considered to be enriched. OH-FFNN, FP-FFNN, OH-FFNN pt, FP-FFNN pt, and random guess results are averaged over five trials for each dataset; D-MPNN and D-MPNN pt results are averaged over five trials for the DD1S CAIX dataset and over three trials for the triazine sEH and triazine SIRT2 datasets. For each trial, the random-guess baseline was generated by randomly shuffling the predictions of the FP-FFNN for that trial.}
    \label{fig:bin_plots_multiple_thresholds_PR_AUC}
\end{figure}

\begin{figure}[H] 
    \centering
        \includegraphics[scale=1]{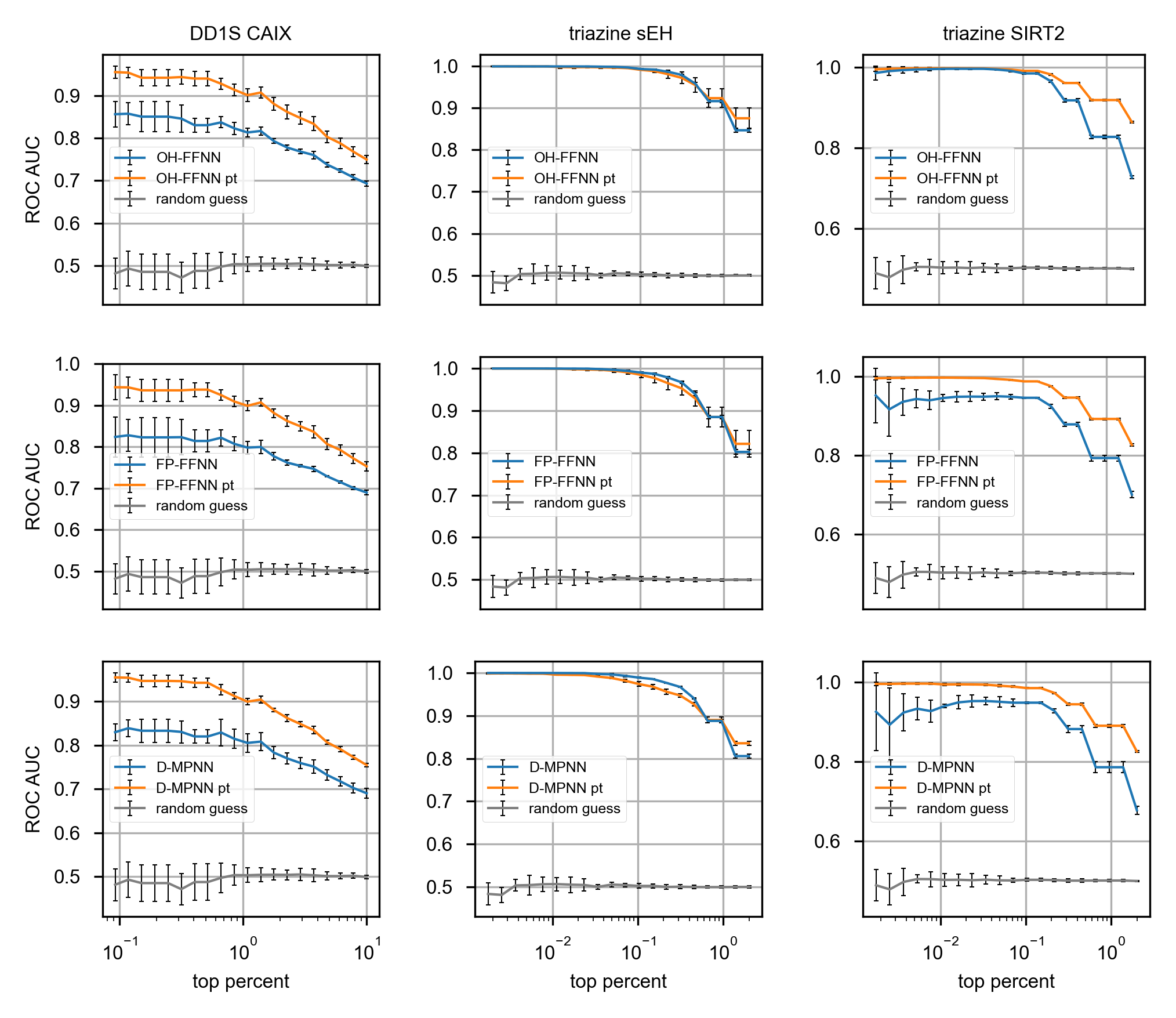}
    \caption{ROC AUC scores (mean $\pm$ standard deviation) for the regression models evaluated as classifiers, at various thresholds defined by the percentage of compounds in the training set considered to be enriched. OH-FFNN, FP-FFNN, OH-FFNN pt, FP-FFNN pt, and random guess results are averaged over five trials for each dataset; D-MPNN and D-MPNN pt results are averaged over five trials for the DD1S CAIX dataset and over three trials for the triazine sEH and triazine SIRT2 datasets. For each trial, the random-guess baseline was generated by randomly shuffling the predictions of the FP-FFNN for that trial.}
    \label{fig:bin_plots_multiple_thresholds_ROC_AUC}
\end{figure}

\subsection{2D histograms of predicted vs. calculated enrichment}

\begin{figure}[H]
    \centering
        \includegraphics[scale=1]{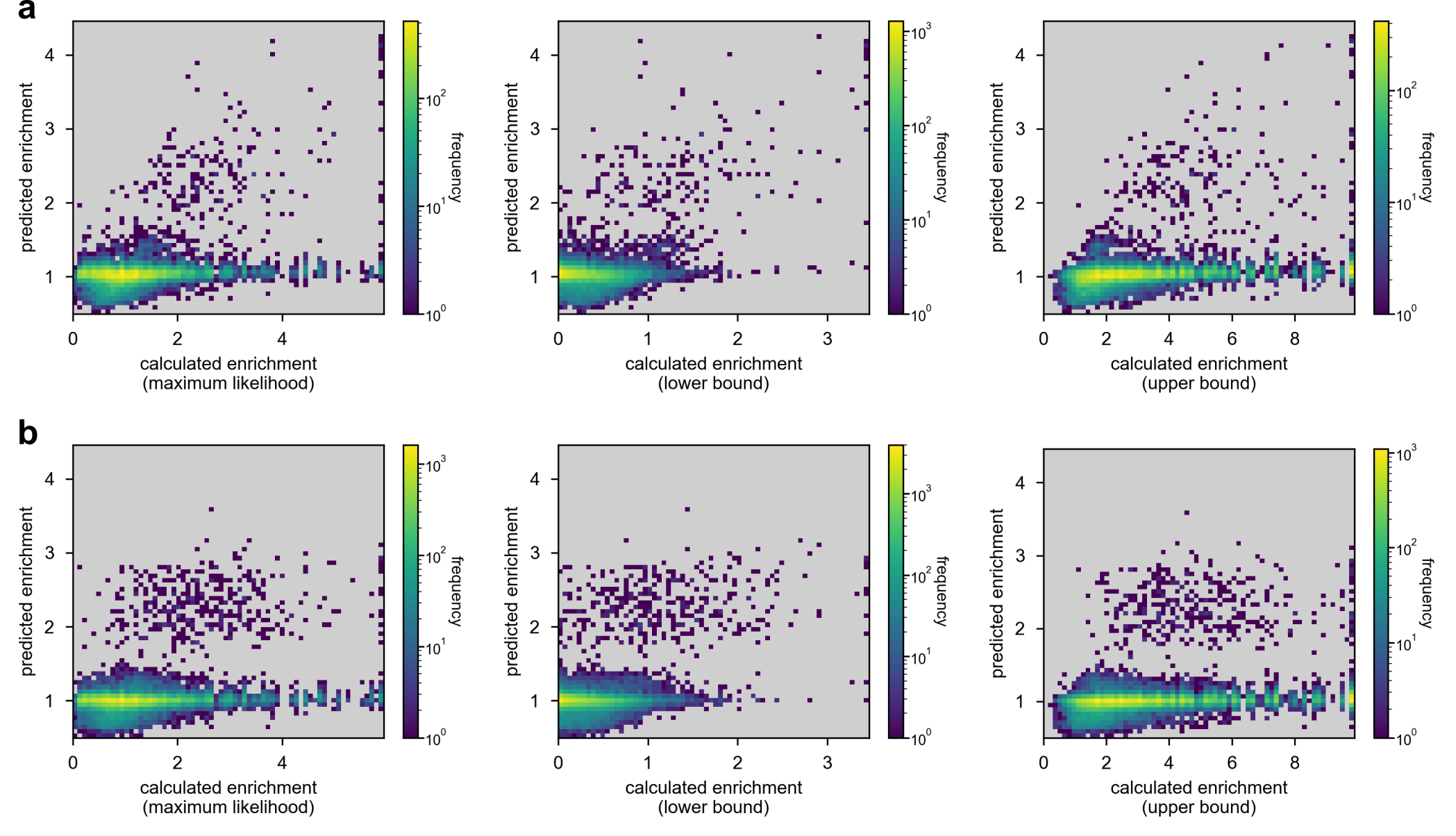}
    \caption{Histograms of calculated and predicted enrichments for the test-set compounds of a FP-FFNN on a \textbf{(a)} random split and \textbf{(b)} cycle-1+2 split (\emph{cf.} Figure~\ref{fig:reps_models_loss_fns_splits}d) of the DD1S CAIX dataset, excluding compounds for which the sum of the experimental and beads counts is less than 3.}
    \label{fig:CAIX_2D_hists}
\end{figure}

\begin{figure}[H]
    \centering
        \includegraphics[scale=1]{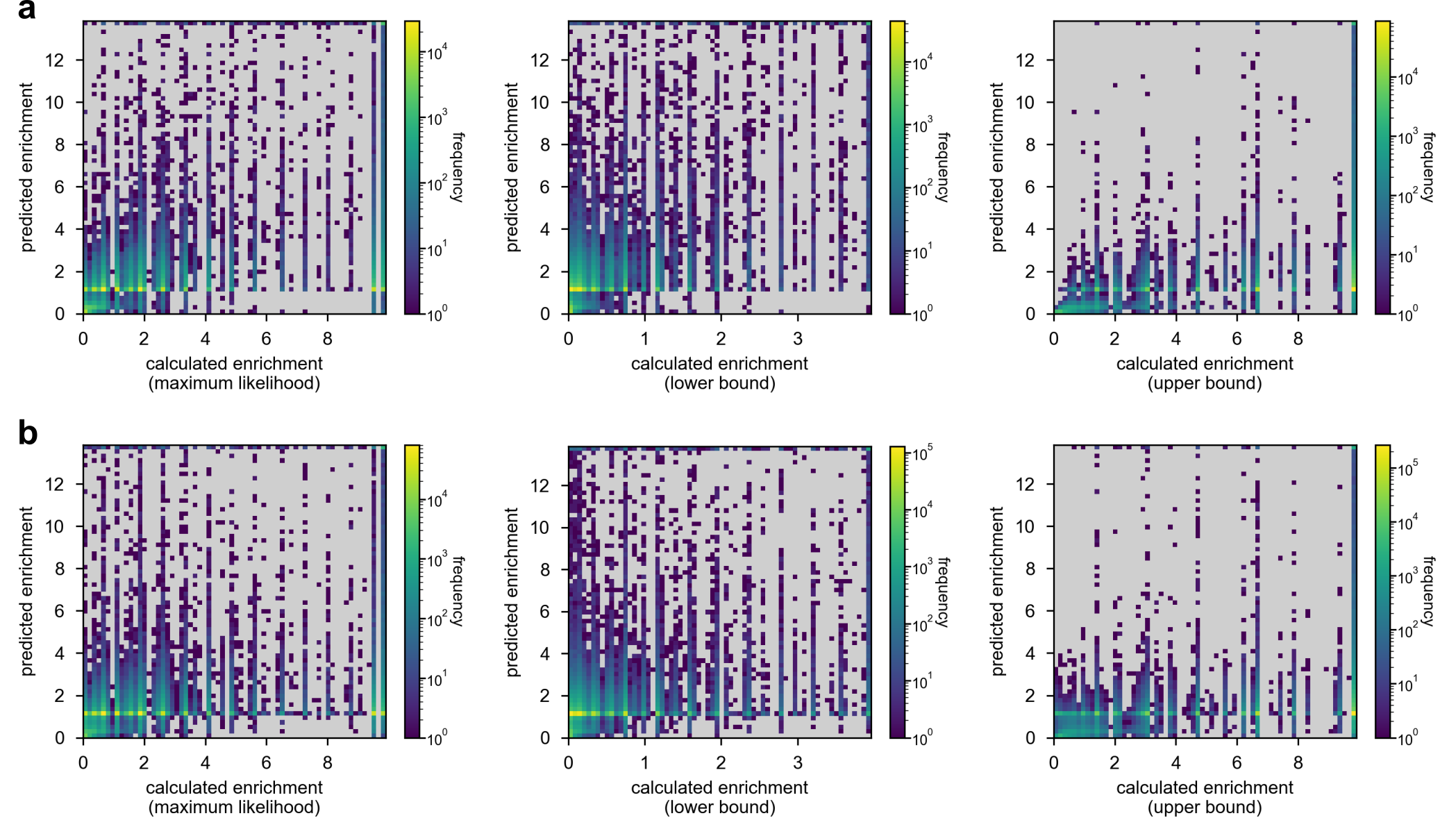}
    \caption{Histograms of calculated and predicted enrichments for the test-set compounds of a FP-FFNN on a \textbf{(a)} random split and \textbf{(b)} cycle-1+2+3 split (\emph{cf.} Figure~\ref{fig:reps_models_loss_fns_splits}d) of the triazine sEH dataset, excluding compounds for which the sum of the experimental and beads counts is less than 3.}
    \label{fig:sEH_2D_hists}
\end{figure}

\begin{figure}[H]
    \centering
        \includegraphics[scale=1]{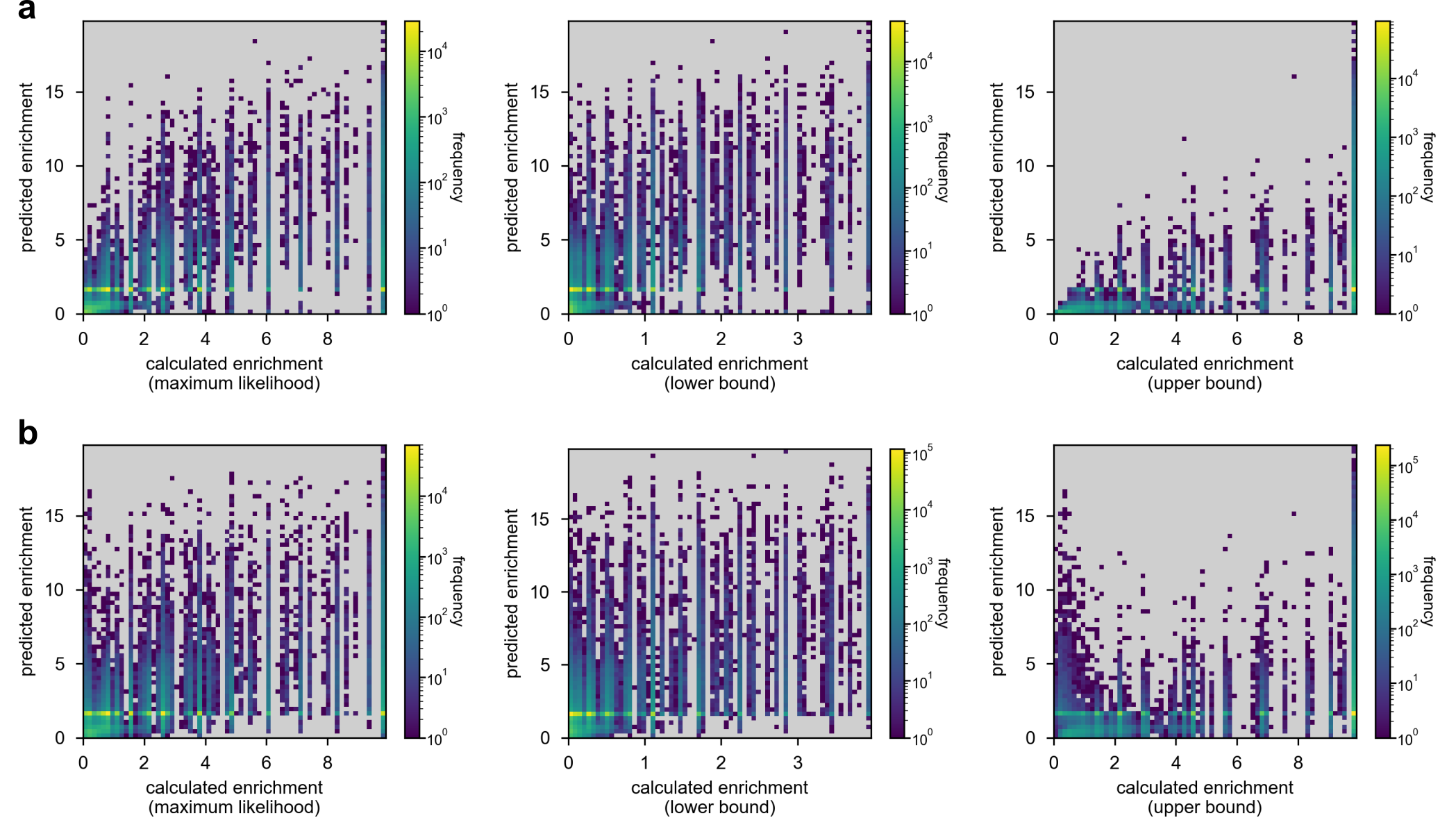}
    \caption{Histograms of calculated and predicted enrichments for the test-set compounds of a FP-FFNN on a \textbf{(a)} random split and \textbf{(b)} cycle-1+2+3 split (\emph{cf.} Figure~\ref{fig:reps_models_loss_fns_splits}d) of the triazine SIRT2 dataset, excluding compounds for which the sum of the experimental and beads counts is less than 3.}
    \label{fig:SIRT2_2D_hists}
\end{figure}

In the above histograms, one would expect linear correlation between the predicted and calculated enrichments, with better correlation for the models trained on a random split (evaluating interpolation) than for models trained on a cycle split (evaluating generalization). However, such a trend is not apparent in the plots. One possibility is that a linear correlation is present but obscured by noise, for instance due to compounds with relatively low counts. For example, in each of the plots for the DD1S CAIX dataset (Figure \ref{fig:CAIX_2D_hists}), the mass of datapoints at a predicted enrichment of about 1 may be attributed to noise. However, it is difficult to validate this hypothesis in the absence of ground truth enrichment values (these would come from experimental validation of the compounds). For the plot of predicted enrichment versus upper bound of calculated enrichment for a random split of the triazine SIRT2 dataset (Figure \ref{fig:SIRT2_2D_hists}a), the fact that most of the datapoints fall below the parity line indicates that the model is not grossly overestimating enrichment. 

\subsection{Bit and substructure importance}

\subsubsection{Single-substructure analysis}

\begin{figure} [H]
    \centering
        \includegraphics[scale=1]{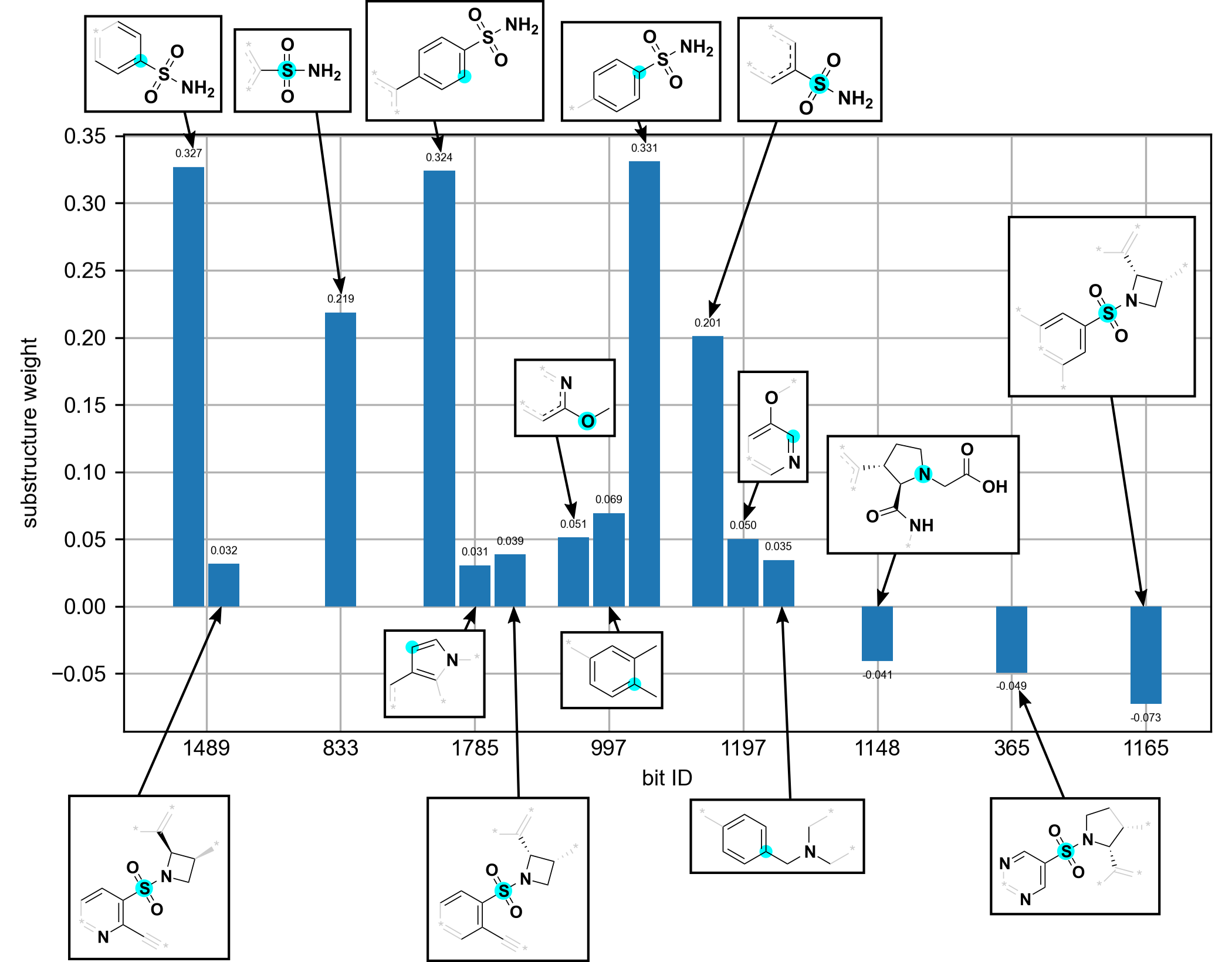}
    \caption{Single-substructure analysis on the DD1S CAIX dataset (random split, seed 0; \emph{cf.} Figure~\ref{fig:reps_models_loss_fns_splits}d), including substructures mapped to the top 5 and bottom 3 bits.}
    \label{fig:CAIX_onebit_seed_0_bars}
\end{figure}

\begin{longtable}{m{1.5cm}|m{4.5cm}|m{8.5cm}}
\caption{SMARTS for single-substructure analysis on the DD1S CAIX dataset (random split; seeds 0, 1, 2). The substructures for the different replicates differ only in the ones mapped to a few of the bottom bits; all substructures across all replicates are included below.} \label{tbl:SMARTS_DD1S}\\

bit ID & substructure example & SMARTS\\ \hline

1489 & \includegraphics[scale=1]{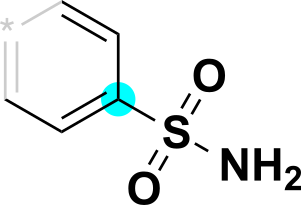} & [\#6;D2;H1;R1;+0]:[\#6;D2;H1;R1;+0]:\newline[\#6;D3;H0;R1;+0](:[\#6;D2;H1;R1;+0]:\newline[\#6;D2;H1;R1;+0])-[\#16;D4;H0;R0;+0]\newline(-[\#7;D1;H2;R0;+0])(=[\#8;D1;H0;R0;+0])\newline=[\#8;D1;H0;R0;+0]\\ \hline

1489 & \includegraphics[scale=1]{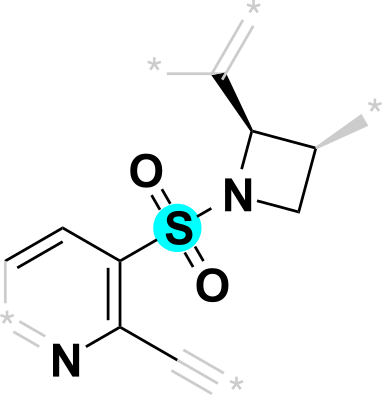} & [\#6;D2;H0;R0;+0]-[\#6;D3;H0;R1;+0]\newline(:[\#7;D2;H0;R1;+0]):[\#6;D3;H0;R1;+0]\newline(:[\#6;D2;H1;R1;+0]:[\#6;D2;H1;R1;+0])-[\#16;D4;H0;R0;+0](=[\#8;D1;H0;R0;+0])\newline(=[\#8;D1;H0;R0;+0])-[\#7;D3;H0;R1;+0]1-[\#6;D2;H2;R1;+0]-[\#6;D3;H1;R1;+0]-[\#6;D3;H1;R1;+0]-1-[\#6;D3;H0;R0;+0] \\\hline

833 & \includegraphics[scale=1]{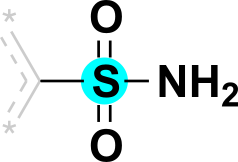} & [\#6;D3;H0;R1;+0]-[\#16;D4;H0;R0;+0](-[\#7;D1;H2;R0;+0])(=[\#8;D1;H0;R0;+0])\newline=[\#8;D1;H0;R0;+0] \\\hline

1785 & \includegraphics[scale=1]{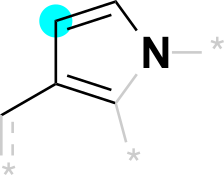} & [\#6;D2;H1;R1;+0]:[\#6;D3;H0;R2;+0]\newline(:[\#6;D3;H0;R2;+0]):[\#6;D2;H1;R1;+0]:\newline[\#6;D2;H1;R1;+0]:[\#7;D3;H0;R1;+0] \\\hline

1785 & \includegraphics[scale=1]{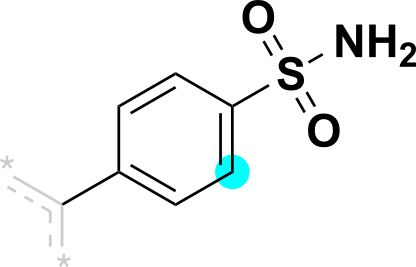} & [\#6;D3;H0;R1;+0]-[\#6;D3;H0;R1;+0]1:\newline[\#6;D2;H1;R1;+0]:[\#6;D2;H1;R1;+0]:\newline[\#6;D3;H0;R1;+0](-[\#16;D4;H0;R0;+0](-[\#7;D1;H2;R0;+0])(=[\#8;D1;H0;R0;+0])\newline=[\#8;D1;H0;R0;+0]):[\#6;D2;H1;R1;+0]:\newline[\#6;D2;H1;R1;+0]:1 \\\hline

1785 & \includegraphics[scale=1]{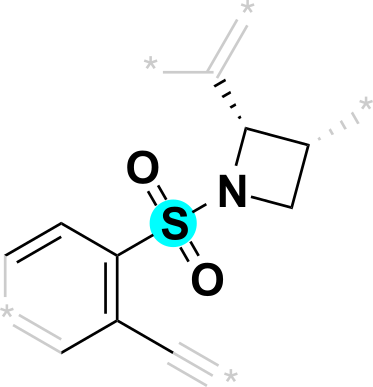} & [\#6;D2;H0;R0;+0]-[\#6;D3;H0;R1;+0]\newline(:[\#6;D2;H1;R1;+0]):[\#6;D3;H0;R1;+0]\newline(:[\#6;D2;H1;R1;+0]:[\#6;D2;H1;R1;+0])-[\#16;D4;H0;R0;+0](=[\#8;D1;H0;R0;+0])\newline(=[\#8;D1;H0;R0;+0])-[\#7;D3;H0;R1;+0]1-[\#6;D2;H2;R1;+0]-[\#6;D3;H1;R1;+0]-[\#6;D3;H1;R1;+0]-1-[\#6;D3;H0;R0;+0] \\\hline

997 & \includegraphics[scale=1]{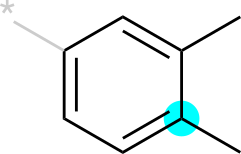} & [\#6;D1;H3;R0;+0]-[\#6;D3;H0;R1;+0]1:\newline[\#6;D2;H1;R1;+0]:[\#6;D2;H1;R1;+0]:\newline[\#6;D3;H0;R1;+0]:[\#6;D2;H1;R1;+0]:\newline[\#6;D3;H0;R1;+0]:1-[\#6;D1;H3;R0;+0] \\\hline

997 & \includegraphics[scale=1]{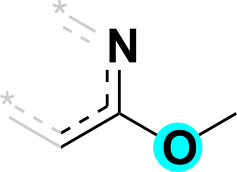} & [\#6;D1;H3;R0;+0]-[\#8;D2;H0;R0;+0]-[\#6;D3;H0;R1;+0](:[\#6;D2;H1;R1;+0]):\newline[\#7;D2;H0;R1;+0] \\\hline

997 & \includegraphics[scale=1]{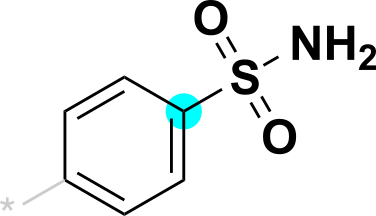} & [\#7;D1;H2;R0;+0]-[\#16;D4;H0;R0;+0]\newline(=[\#8;D1;H0;R0;+0])(=[\#8;D1;H0;R0;+0])-\newline[\#6;D3;H0;R1;+0]1:[\#6;D2;H1;R1;+0]:\newline[\#6;D2;H1;R1;+0]:[\#6;D3;H0;R1;+0]:\newline[\#6;D2;H1;R1;+0]:[\#6;D2;H1;R1;+0]:1 \\\hline

1197 & \includegraphics[scale=1]{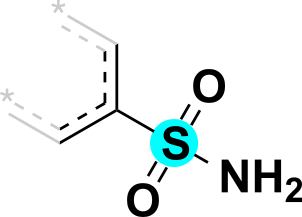} & [\#6;D2;H1;R1;+0]:[\#6;D3;H0;R1;+0]\newline(:[\#6;D2;H1;R1;+0])-[\#16;D4;H0;R0;+0]\newline(-[\#7;D1;H2;R0;+0])(=[\#8;D1;H0;R0;+0])\newline=[\#8;D1;H0;R0;+0] \\\hline

1197 & \includegraphics[scale=1]{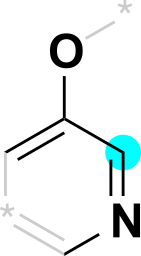} & [\#6;D2;H1;R1;+0]:[\#7;D2;H0;R1;+0]:\newline[\#6;D2;H1;R1;+0]:[\#6;D3;H0;R1;+0]\newline(:[\#6;D2;H1;R1;+0])-[\#8;D2;H0;R0;+0] \\\hline

1197 & \includegraphics[scale=1]{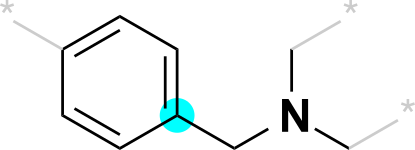} & [\#6;D2;H2;R0;+0]-[\#7;D3;H0;R0;+0]\newline(-[\#6;D2;H2;R0;+0])-[\#6;D2;H2;R0;+0]-[\#6;D3;H0;R1;+0]1:[\#6;D2;H1;R1;+0]:\newline[\#6;D2;H1;R1;+0]:[\#6;D3;H0;R1;+0]:\newline[\#6;D2;H1;R1;+0]:[\#6;D2;H1;R1;+0]:1 \\\hline

1148 & \includegraphics[scale=1]{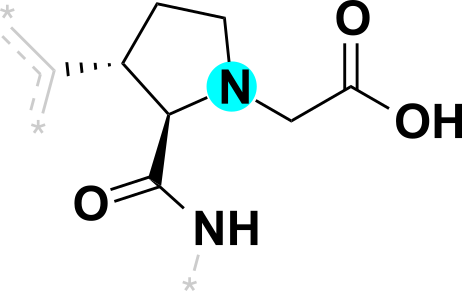} & [\#6;D3;H0;R1;+0]-[\#6;D3;H1;R1;+0]1-\newline[\#6;D2;H2;R1;+0]-[\#6;D2;H2;R1;+0]-\newline[\#7;D3;H0;R1;+0](-[\#6;D2;H2;R0;+0]-\newline[\#6;D3;H0;R0;+0](=[\#8;D1;H0;R0;+0])-\newline[\#8;D1;H1;R0;+0])-[\#6;D3;H1;R1;+0]-1-\newline[\#6;D3;H0;R0;+0](-[\#7;D2;H1;R0;+0])=\newline[\#8;D1;H0;R0;+0] \\\hline

365 & \includegraphics[scale=1]{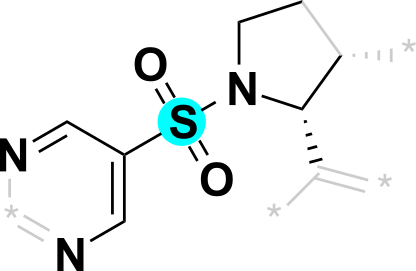} & [\#6;D2;H2;R1;+0]-[\#6;D2;H2;R1;+0]-\newline[\#7;D3;H0;R1;+0](-[\#6;D3;H1;R1;+0](-\newline[\#6;D3;H0;R0;+0])-[\#6;D3;H1;R1;+0])-\newline[\#16;D4;H0;R0;+0](=[\#8;D1;H0;R0;+0])\newline(=[\#8;D1;H0;R0;+0])-[\#6;D3;H0;R1;+0](:\newline[\#6;D2;H1;R1;+0]:[\#7;D2;H0;R1;+0]):\newline[\#6;D2;H1;R1;+0]:[\#7;D2;H0;R1;+0] \\\hline

1165 & \includegraphics[scale=1]{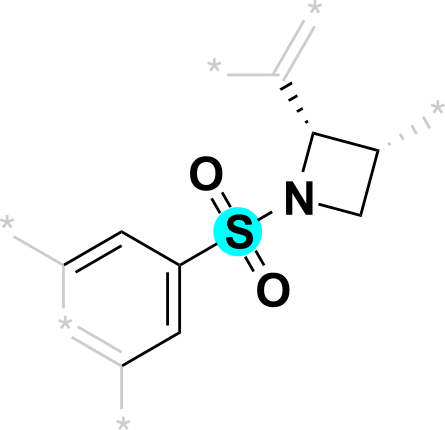} & [\#6;D3;H0;R0;+0]-[\#6;D3;H1;R1;+0]1-\newline[\#6;D3;H1;R1;+0]-[\#6;D2;H2;R1;+0]-\newline[\#7;D3;H0;R1;+0]-1-[\#16;D4;H0;R0;+0](=\newline[\#8;D1;H0;R0;+0])(=[\#8;D1;H0;R0;+0])-\newline[\#6;D3;H0;R1;+0](:[\#6;D2;H1;R1;+0]:\newline[\#6;D3;H0;R1;+0]):[\#6;D2;H1;R1;+0]:\newline[\#6;D3;H0;R1;+0] \\\hline

1736 & \includegraphics[scale=1]{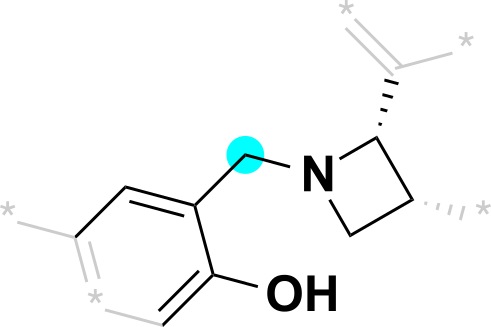} & [\#6;D2;H1;R1;+0]:[\#6;D3;H0;R1;+0](-\newline[\#8;D1;H1;R0;+0]):[\#6;D3;H0;R1;+0](:\newline[\#6;D2;H1;R1;+0]:[\#6;D3;H0;R1;+0])-\newline[\#6;D2;H2;R0;+0]-[\#7;D3;H0;R1;+0]1-\newline[\#6;D2;H2;R1;+0]-[\#6;D3;H1;R1;+0]-\newline[\#6;D3;H1;R1;+0]-1-[\#6;D3;H0;R0;+0]\\\hline

258 & \includegraphics[scale=1]{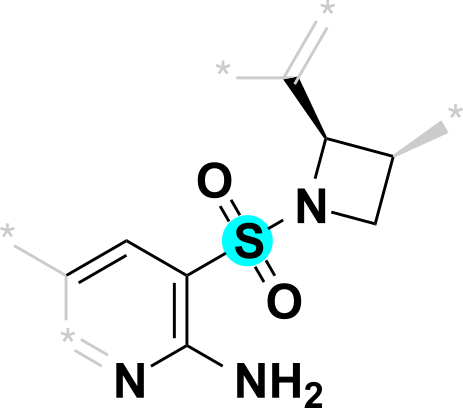} & [\#6;D3;H0;R0;+0]-[\#6;D3;H1;R1;+0]1-\newline[\#6;D3;H1;R1;+0]-[\#6;D2;H2;R1;+0]-\newline[\#7;D3;H0;R1;+0]-1-[\#16;D4;H0;R0;+0](=\newline[\#8;D1;H0;R0;+0])(=[\#8;D1;H0;R0;+0])-\newline[\#6;D3;H0;R1;+0](:[\#6;D2;H1;R1;+0]:\newline[\#6;D3;H0;R1;+0]):[\#6;D3;H0;R1;+0](-\newline[\#7;D1;H2;R0;+0]):[\#7;D2;H0;R1;+0]\\\hline

1844 & \includegraphics[scale=1]{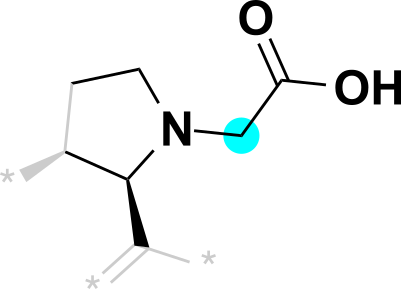} & [\#6;D2;H2;R1;+0]-[\#6;D2;H2;R1;+0]-\newline[\#7;D3;H0;R1;+0](-[\#6;D2;H2;R0;+0]-\newline[\#6;D3;H0;R0;+0](=[\#8;D1;H0;R0;+0])-\newline[\#8;D1;H1;R0;+0])-[\#6;D3;H1;R1;+0](-\newline[\#6;D3;H0;R0;+0])-[\#6;D3;H1;R1;+0]\\\hline

\end{longtable}

\begin{figure} [H]
    \centering
        \includegraphics[scale=1]{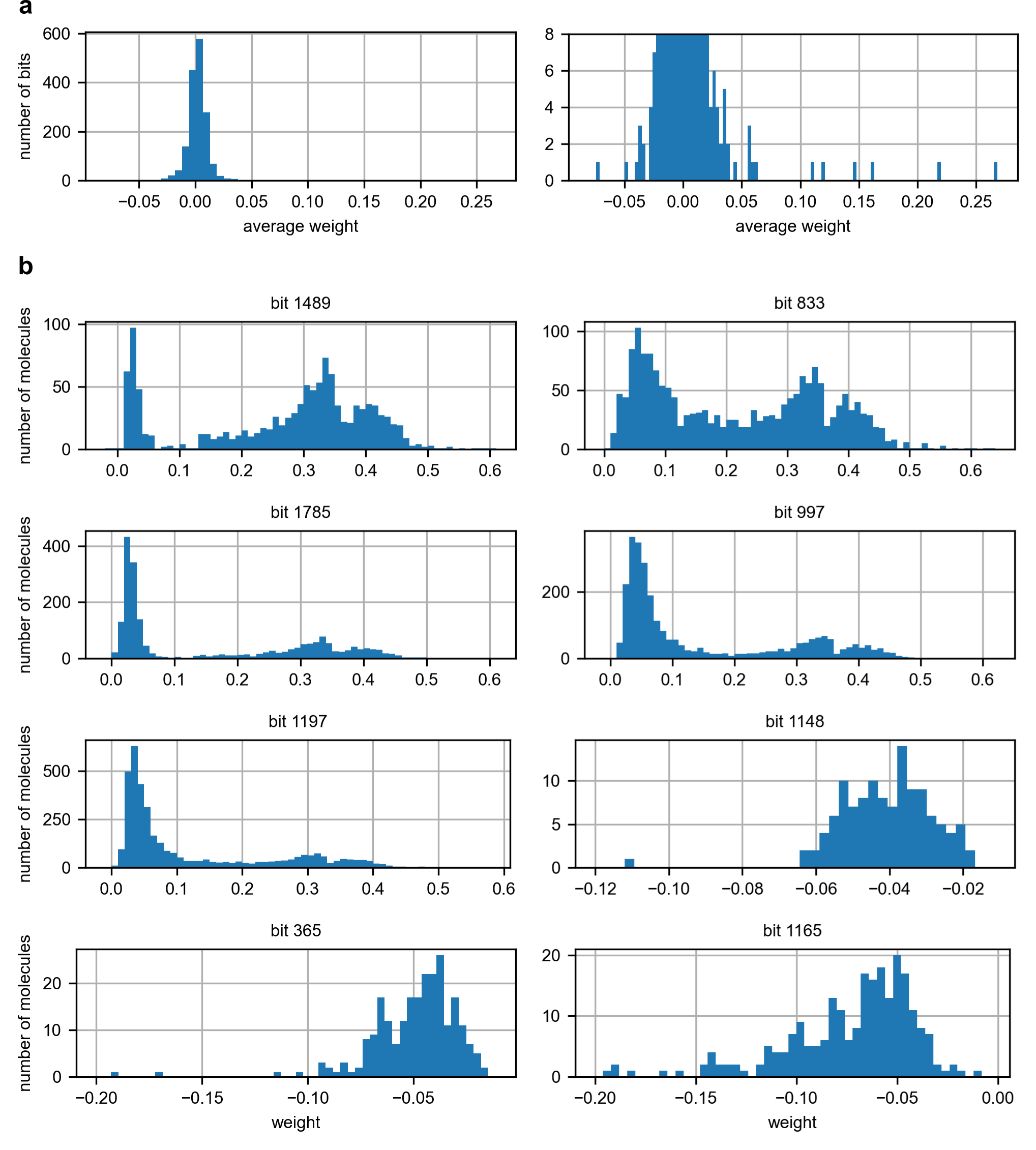}
    \caption{Histograms of \textbf{(a)} bit weights (only including bits set by at least one molecule in the dataset; plot shown in full and zoomed in) and \textbf{(b)} molecule-level bit weights for single-substructure analysis on the DD1S CAIX dataset (random split, seed 0; \emph{cf.} Figure~\ref{fig:reps_models_loss_fns_splits}d), including the top 5 and bottom 3 bits.}
    \label{fig:CAIX_onebit_seed_0_hists}
\end{figure}

\begin{figure} [H]
    \centering
        \includegraphics[scale=1]{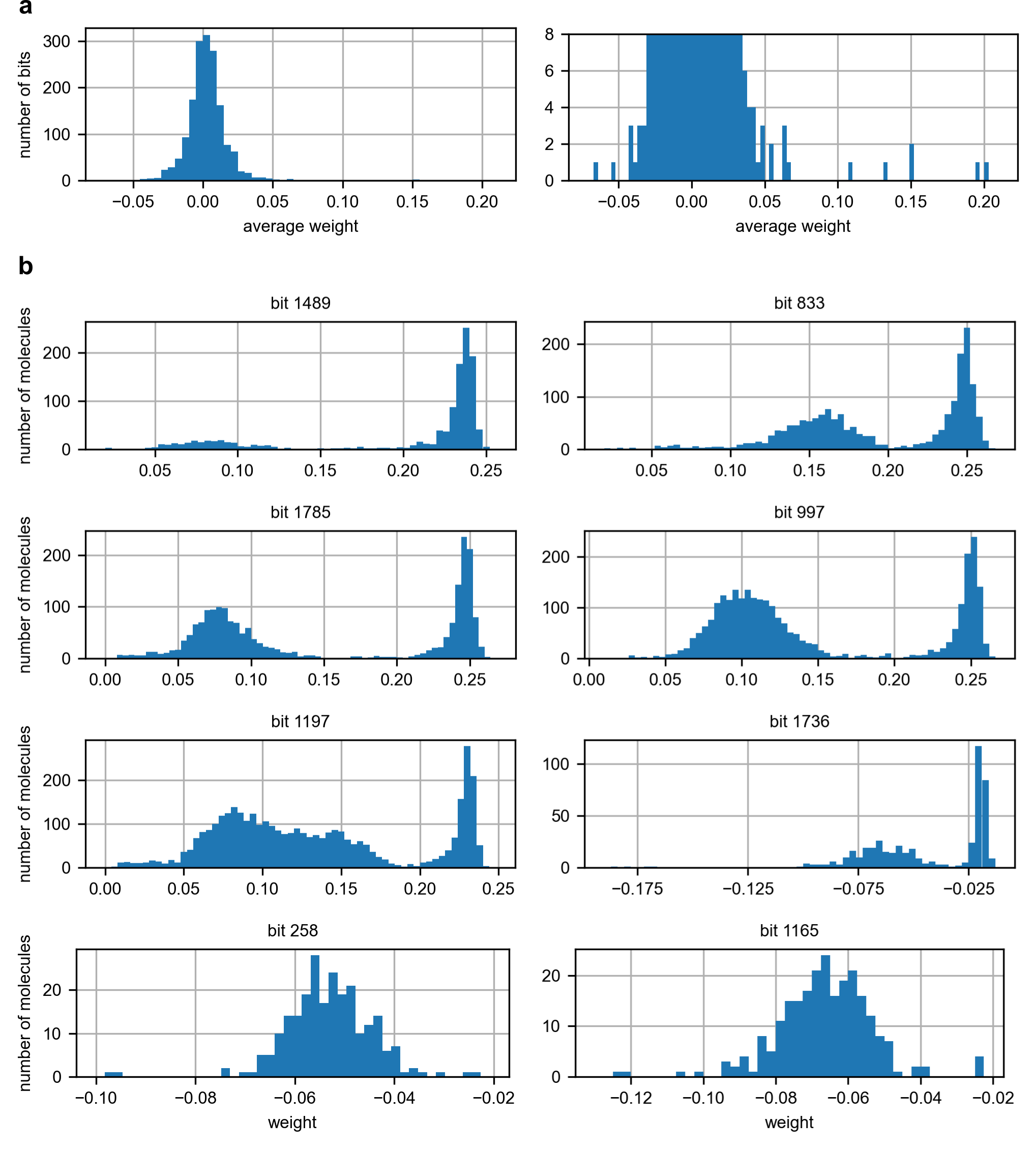}
    \caption{Histograms of \textbf{(a)} bit weights (only including bits set by at least one molecule in the dataset; plot shown in full and zoomed in) and \textbf{(b)} molecule-level bit weights for single-substructure analysis on the DD1S CAIX dataset (random split, seed 1; \emph{cf.} Figure~\ref{fig:reps_models_loss_fns_splits}d), including the top 5 and bottom 3 bits.}
    \label{fig:CAIX_onebit_seed_1_hists}
\end{figure}

\begin{figure} [H]
    \centering
        \includegraphics[scale=1]{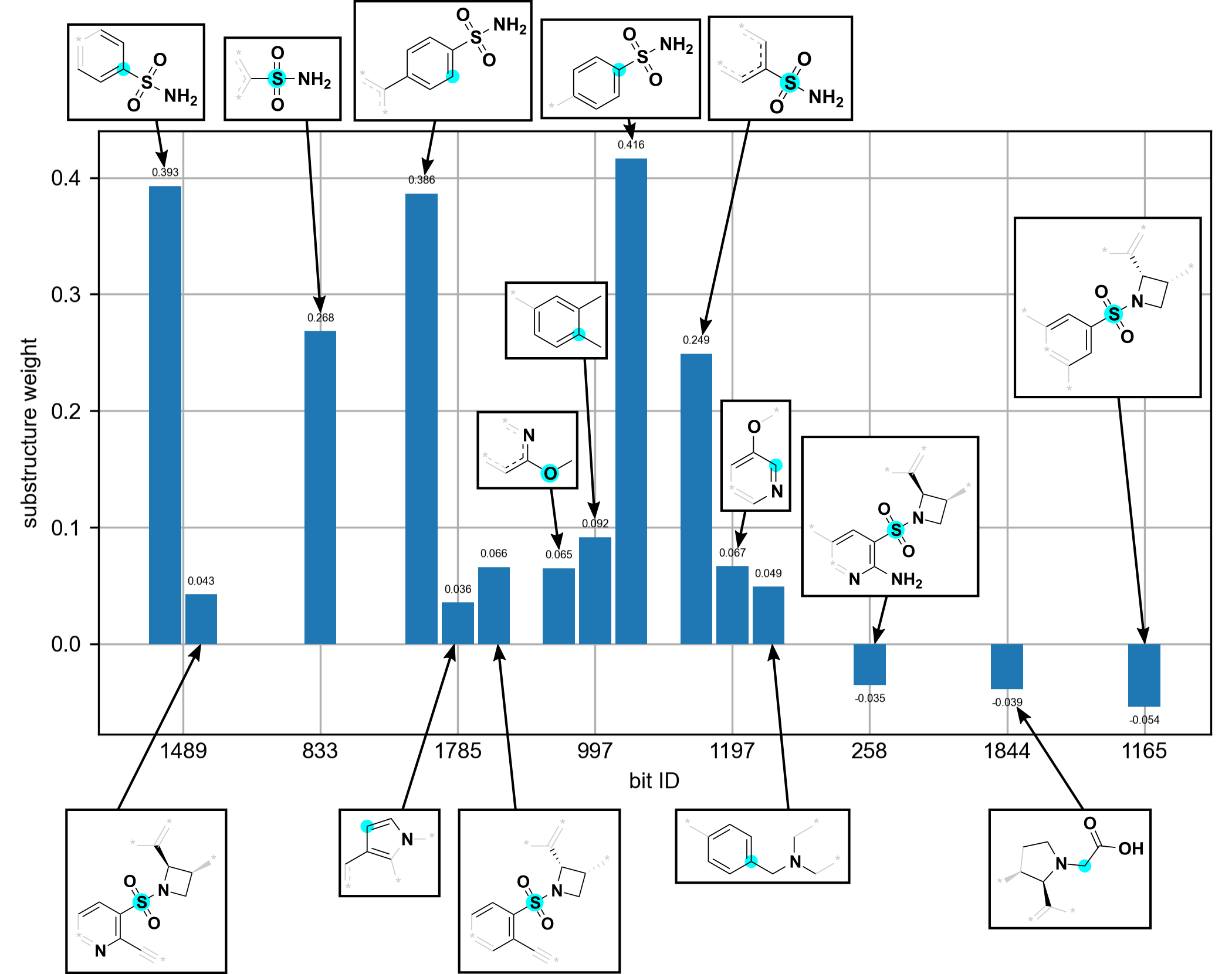}
    \caption{Single-substructure analysis on the DD1S CAIX dataset (random split, seed 2; \emph{cf.} Figure~\ref{fig:reps_models_loss_fns_splits}d), including substructures mapped to the top 5 and bottom 3 bits.}
    \label{fig:CAIX_onebit_seed_2_bars}
\end{figure}

\begin{figure} [H]
    \centering
        \includegraphics[scale=1]{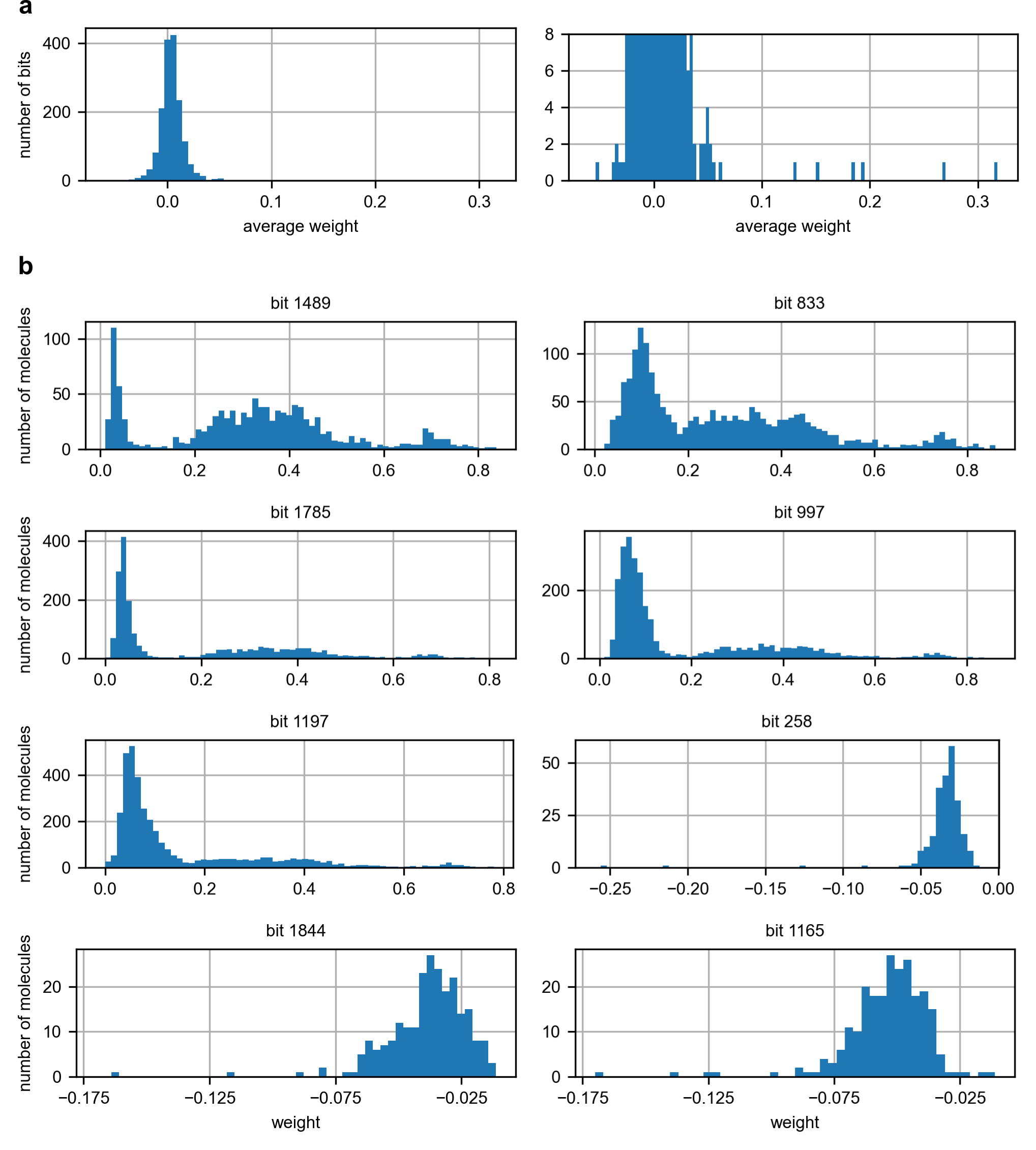}
    \caption{Histograms of \textbf{(a)} bit weights (only including bits set by at least one molecule in the dataset; plot shown in full and zoomed in) and \textbf{(b)} molecule-level bit weights for single-substructure analysis on the DD1S CAIX dataset (random split, seed 2; \emph{cf.} Figure~\ref{fig:reps_models_loss_fns_splits}d), including the top 5 and bottom 3 bits.}
    \label{fig:CAIX_onebit_seed_2_hists}
\end{figure}

\begin{figure} [H]
    \centering
        \includegraphics[scale=1]{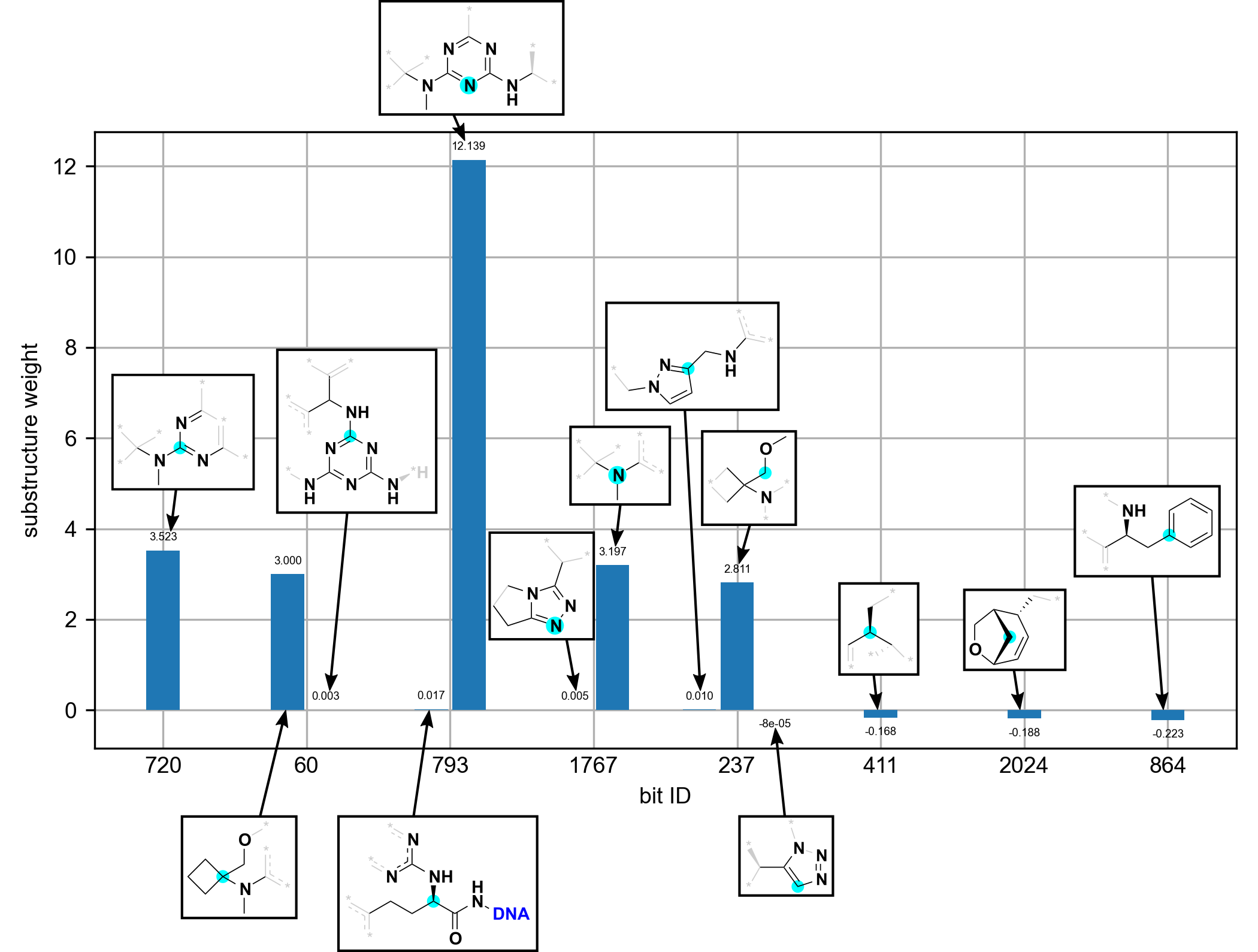}
    \caption{Single-substructure analysis on the triazine sEH dataset (random split, seed 0; \emph{cf.} Figure~\ref{fig:reps_models_loss_fns_splits}d), including substructures mapped to the top 5 and bottom 3 bits.}
    \label{fig:sEH_onebit_bars}
\end{figure}

\begin{longtable}{m{1.5cm}|m{4.5cm}|m{8.5cm}} 
\caption{SMARTS for single-substructure analysis on the triazine sEH dataset (random split, seed 0).}\label{tbl:SMARTS_sEH} \\
bit ID & substructure example & SMARTS\\ \hline
720 & \includegraphics[scale=1]{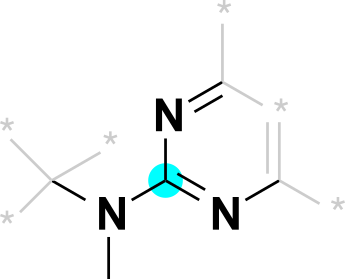} & [\#6;D1;H3;R0;+0]-[\#7;D3;H0;R0;+0](-\newline[\#6;D4;H0;R1;+0])-[\#6;D3;H0;R1;+0](:\newline[\#7;D2;H0;R1;+0]:[\#6;D3;H0;R1;+0]):\newline[\#7;D2;H0;R1;+0]:[\#6;D3;H0;R1;+0] \\\hline

60 & \includegraphics[scale=1]{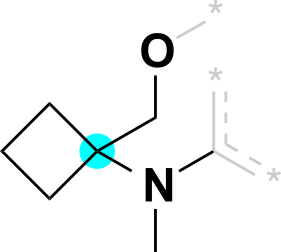} & [\#6;D1;H3;R0;+0]-[\#7;D3;H0;R0;+0](-[\#6;D3;H0;R1;+0])-[\#6;D4;H0;R1;+0]1(-[\#6;D2;H2;R0;+0]-[\#8;D2;H0;R0;+0])-[\#6;D2;H2;R1;+0]-[\#6;D2;H2;R1;+0]-[\#6;D2;H2;R1;+0]-1 \\\hline

60 & \includegraphics[scale=1]{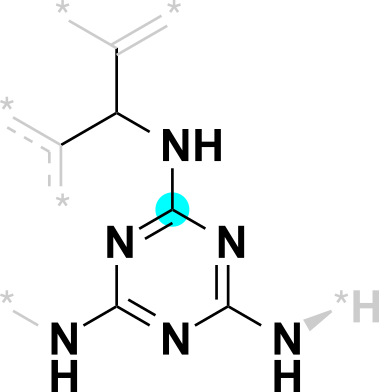} & [\#6;D3;H0;R0;+0]-[\#6;D3;H1;R0;+0](-[\#6;D3;H0;R1;+0])-[\#7;D2;H1;R0;+0]-[\#6;D3;H0;R1;+0]1:[\#7;D2;H0;R1;+0]:\newline[\#6;D3;H0;R1;+0](-[\#7;D2;H1;R0;+0]):\newline[\#7;D2;H0;R1;+0]:[\#6;D3;H0;R1;+0](-\newline[\#7;D2;H1;R0;+0]):[\#7;D2;H0;R1;+0]:1 \\\hline

793 & \includegraphics[scale=1]{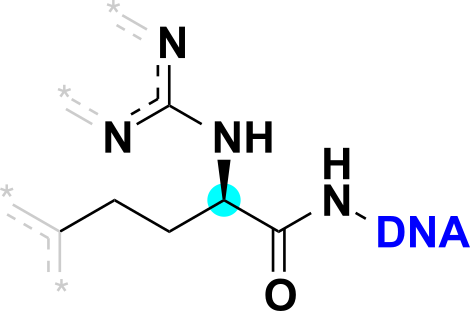} & [\#102;D1;H0;R0;+0]-[\#7;D2;H1;R0;+0]-[\#6;D3;H0;R0;+0](=[\#8;D1;H0;R0;+0])-[\#6;D3;H1;R0;+0](-[\#6;D2;H2;R0;+0]-\newline[\#6;D2;H2;R0;+0]-[\#6;D3;H0;R1;+0])-\newline[\#7;D2;H1;R0;+0]-[\#6;D3;H0;R1;+0](:\newline[\#7;D2;H0;R1;+0]):[\#7;D2;H0;R1;+0] \\\hline

793 & \includegraphics[scale=1]{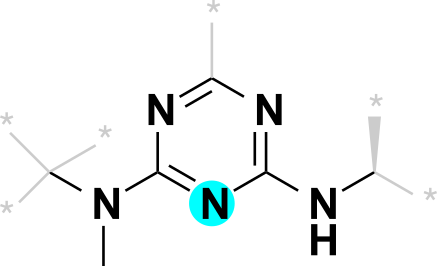} & [\#6;D1;H3;R0;+0]-[\#7;D3;H0;R0;+0](-\newline[\#6;D4;H0;R1;+0])-[\#6;D3;H0;R1;+0]1:\newline[\#7;D2;H0;R1;+0]:[\#6;D3;H0;R1;+0]:\newline[\#7;D2;H0;R1;+0]:[\#6;D3;H0;R1;+0](-\newline[\#7;D2;H1;R0;+0]-[\#6;D3;H1;R0;+0]):\newline[\#7;D2;H0;R1;+0]:1 \\\hline

1767 & \includegraphics[scale=1]{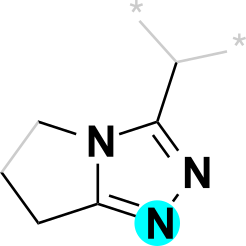} & [\#6;D2;H2;R1;+0]-[\#6;D2;H2;R1;+0]-\newline[\#6;D3;H0;R2;+0]1:[\#7;D2;H0;R1;+0]:\newline[\#7;D2;H0;R1;+0]:[\#6;D3;H0;R1;+0](-\newline[\#6;D3;H1;R0;+0]):[\#7;D3;H0;R2;+0]:1-\newline[\#6;D2;H2;R1;+0] \\\hline

1767 & \includegraphics[scale=1]{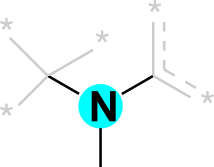} & [\#6;D1;H3;R0;+0]-[\#7;D3;H0;R0;+0](-\newline[\#6;D3;H0;R1;+0])-[\#6;D4;H0;R1;+0] \\\hline

237 & \includegraphics[scale=1]{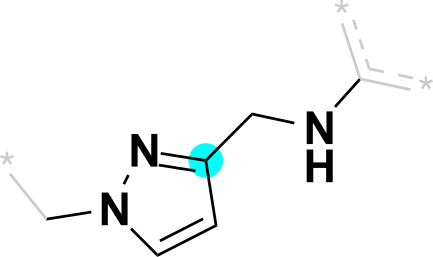} & [\#6;D2;H2;R0;+0]-[\#7;D3;H0;R1;+0]1:\newline[\#6;D2;H1;R1;+0]:[\#6;D2;H1;R1;+0]:\newline[\#6;D3;H0;R1;+0](-[\#6;D2;H2;R0;+0]-[\#7;D2;H1;R0;+0]-[\#6;D3;H0;R1;+0]):\newline[\#7;D2;H0;R1;+0]:1 \\\hline

237 & \includegraphics[scale=1]{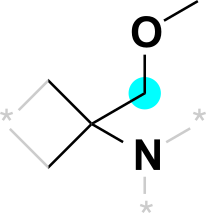} & [\#6;D1;H3;R0;+0]-[\#8;D2;H0;R0;+0]-\newline[\#6;D2;H2;R0;+0]-[\#6;D4;H0;R1;+0](-\newline[\#6;D2;H2;R1;+0])(-[\#6;D2;H2;R1;+0])-\newline[\#7;D3;H0;R0;+0] \\\hline

237 & \includegraphics[scale=1]{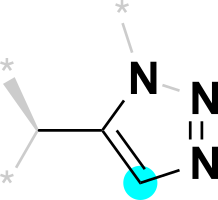} & [\#6;D3;H1;R0;+0]-[\#6;D3;H0;R1;+0](:\newline[\#7;D3;H0;R1;+0]):[\#6;D2;H1;R1;+0]:\newline[\#7;D2;H0;R1;+0]:[\#7;D2;H0;R1;+0] \\\hline

411 & \includegraphics[scale=1]{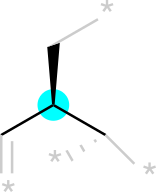} & [\#6;D2;H1;R1;+0]-[\#6;D3;H1;R1;+0](-\newline[\#6;D2;H2;R0;+0])-[\#6;D3;H1;R2;+0] \\\hline

2024 & \includegraphics[scale=1]{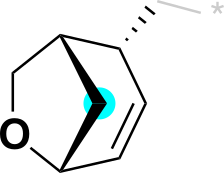} & [\#6;D2;H2;R0;+0]-[\#6;D3;H1;R1;+0]1-\newline[\#6;D2;H1;R1;+0]=[\#6;D2;H1;R1;+0]-\newline[\#6;D3;H1;R2;+0]2-[\#6;D2;H2;R2;+0]-\newline[\#6;D3;H1;R2;+0]-1-[\#6;D2;H2;R1;+0]-\newline[\#8;D2;H0;R1;+0]-2 \\\hline

864 & \includegraphics[scale=1]{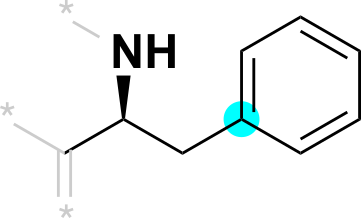} & [\#6;D3;H0;R0;+0]-[\#6;D3;H1;R0;+0](-\newline[\#7;D2;H1;R0;+0])-[\#6;D2;H2;R0;+0]-\newline[\#6;D3;H0;R1;+0]1:[\#6;D2;H1;R1;+0]:\newline[\#6;D2;H1;R1;+0]:[\#6;D2;H1;R1;+0]:\newline[\#6;D2;H1;R1;+0]:[\#6;D2;H1;R1;+0]:1 \\\hline
\end{longtable}

\begin{figure} [H]
    \centering
        \includegraphics[scale=1]{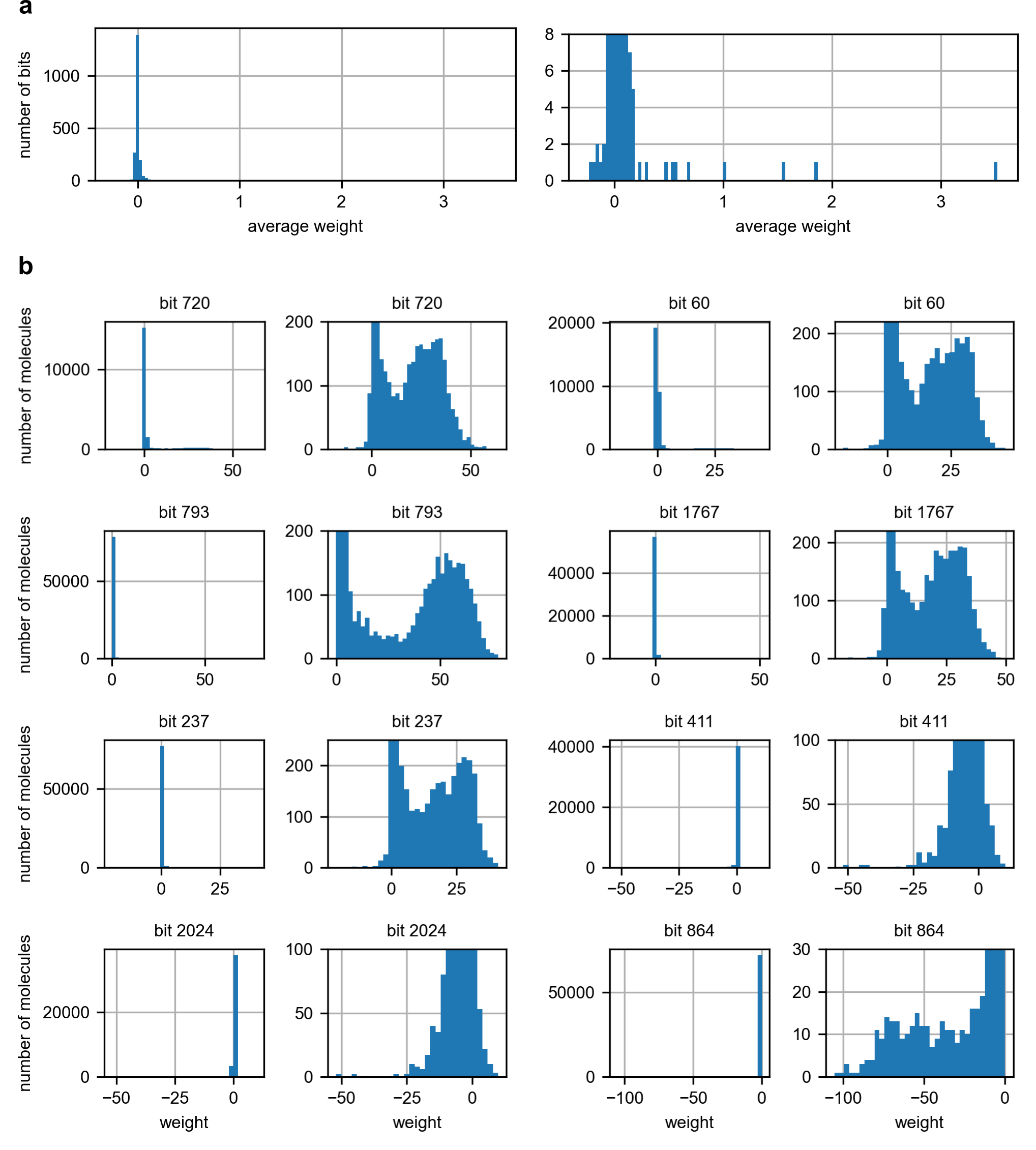}
    \caption{Histograms of \textbf{(a)} bit weights (only including bits set by at least one molecule in the dataset) and \textbf{(b)} molecule-level bit weights for single-substructure analysis on the triazine sEH dataset (random split, seed 0; \emph{cf.} Figure~\ref{fig:reps_models_loss_fns_splits}d), including the top 5 and bottom 3 bits. Plots are shown in full and zoomed in.}
    \label{fig:sEH_onebit_hists}
\end{figure}

\begin{figure} [H]
    \centering
        \includegraphics[scale=1]{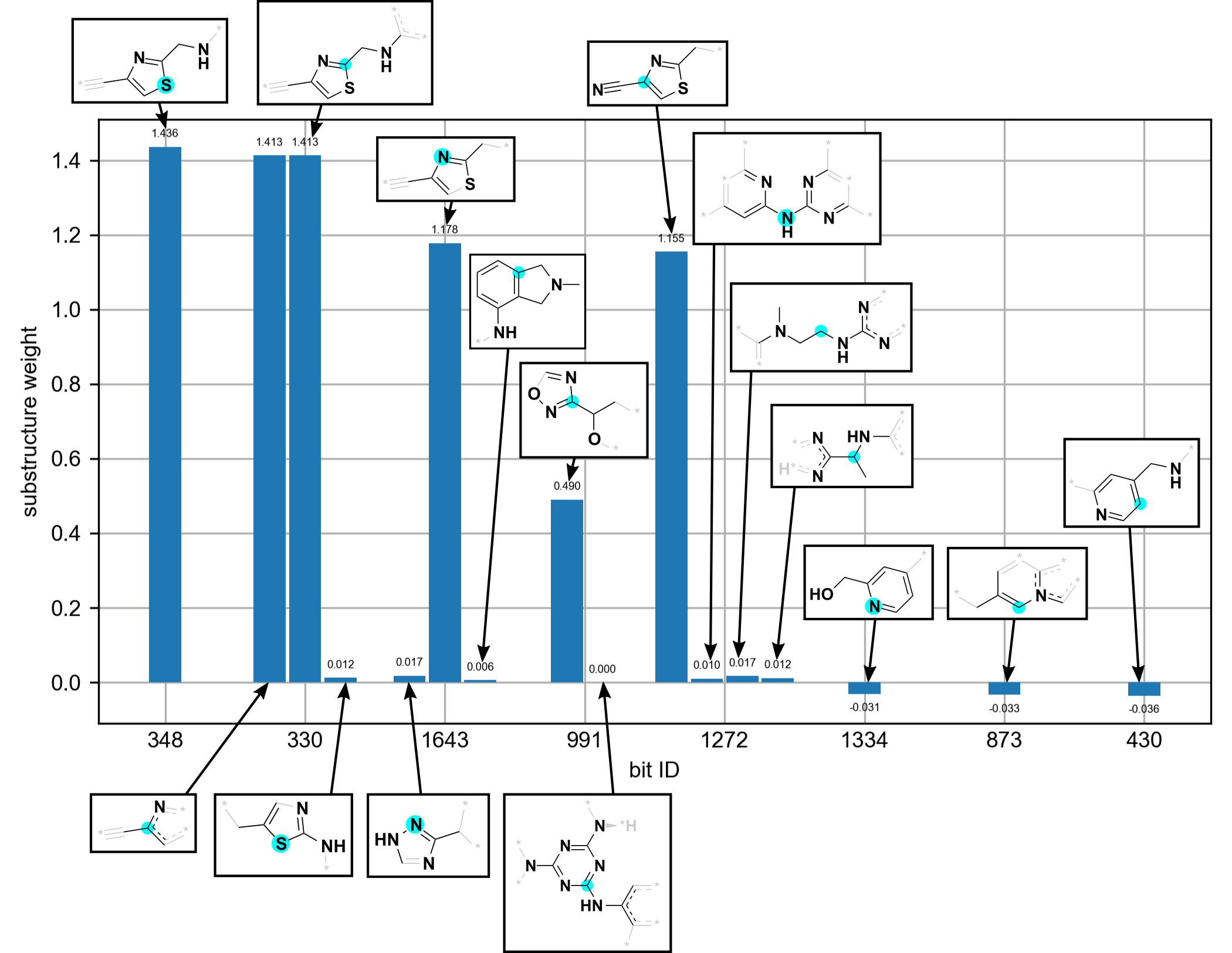}
    \caption{Single-substructure analysis on the triazine SIRT2 dataset (random split, seed 0; \emph{cf.} Figure~\ref{fig:reps_models_loss_fns_splits}d), including substructures mapped to the top 5 and bottom 3 bits.}
    \label{fig:SIRT2_onebit_bars}
\end{figure}

\begin{longtable}{m{1.5cm}|m{4.5cm}|m{8.5cm}} 
\caption{SMARTS for single-substructure analysis on the triazine SIRT2 dataset (random split, seed 0).} \label{tbl:SMARTS_SIRT2} \\
bit ID & substructure example & SMARTS\\ \hline

348 & \includegraphics[scale=1]{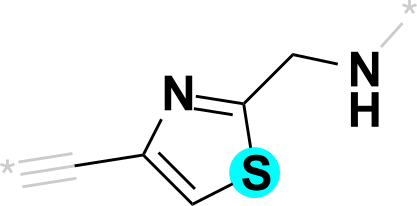} & [\#6;D2;H0;R0;+0]-[\#6;D3;H0;R1;+0]1:\newline[\#6;D2;H1;R1;+0]:[\#16;D2;H0;R1;+0]:\newline[\#6;D3;H0;R1;+0](-[\#6;D2;H2;R0;+0]-\newline[\#7;D2;H1;R0;+0]):[\#7;D2;H0;R1;+0]:1\\\hline

330 & \includegraphics[scale=1]{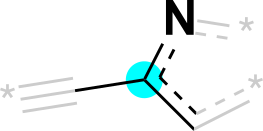} & [\#6;D2;H0;R0;+0]-[\#6;D3;H0;R1;+0](:\newline[\#6;D2;H1;R1;+0]):[\#7;D2;H0;R1;+0] \\\hline

330 & \includegraphics[scale=1]{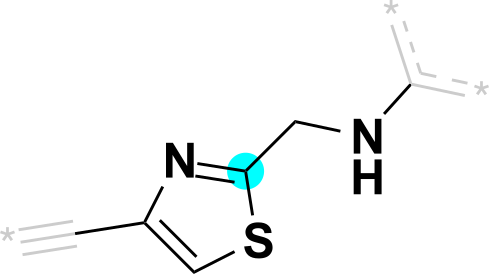} & [\#6;D2;H0;R0;+0]-[\#6;D3;H0;R1;+0]1:\newline[\#6;D2;H1;R1;+0]:[\#16;D2;H0;R1;+0]:\newline[\#6;D3;H0;R1;+0](-[\#6;D2;H2;R0;+0]-\newline[\#7;D2;H1;R0;+0]-[\#6;D3;H0;R1;+0]):\newline[\#7;D2;H0;R1;+0]:1 \\\hline

330 & \includegraphics[scale=1]{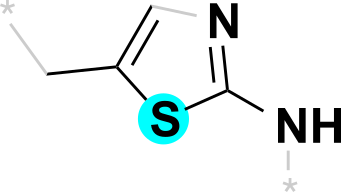} & [\#6;D2;H1;R1;+0]:[\#6;D3;H0;R1;+0](-\newline[\#6;D2;H2;R0;+0]):[\#16;D2;H0;R1;+0]:\newline[\#6;D3;H0;R1;+0](:[\#7;D2;H0;R1;+0])-\newline[\#7;D2;H1;R0;+0] \\\hline

1643 & \includegraphics[scale=1]{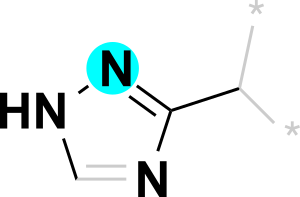} & [\#6;D2;H1;R1;+0]:[\#7;D2;H1;R1;+0]:\newline[\#7;D2;H0;R1;+0]:[\#6;D3;H0;R1;+0](-\newline[\#6;D3;H1;R0;+0]):[\#7;D2;H0;R1;+0] \\\hline

1643 & \includegraphics[scale=1]{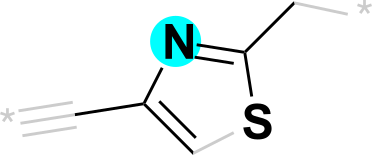} & [\#16;D2;H0;R1;+0]:[\#6;D3;H0;R1;+0](-\newline[\#6;D2;H2;R0;+0]):[\#7;D2;H0;R1;+0]:\newline[\#6;D3;H0;R1;+0](-[\#6;D2;H0;R0;+0]):\newline[\#6;D2;H1;R1;+0] \\\hline

1643 & \includegraphics[scale=1]{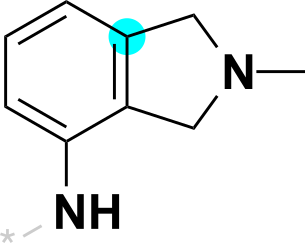} & [\#6;D1;H3;R0;+0]-[\#7;D3;H0;R1;+0]1-\newline[\#6;D2;H2;R1;+0]-[\#6;D3;H0;R2;+0]2:\newline[\#6;D2;H1;R1;+0]:[\#6;D2;H1;R1;+0]:\newline[\#6;D2;H1;R1;+0]:[\#6;D3;H0;R1;+0](-\newline[\#7;D2;H1;R0;+0]):[\#6;D3;H0;R2;+0]:2-\newline[\#6;D2;H2;R1;+0]-1 \\\hline

991 & \includegraphics[scale=1]{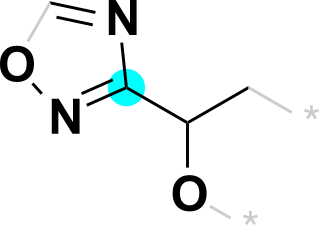} & [\#6;D2;H1;R1;+0]:[\#7;D2;H0;R1;+0]:\newline[\#6;D3;H0;R1;+0](:[\#7;D2;H0;R1;+0]:\newline[\#8;D2;H0;R1;+0])-[\#6;D3;H1;R1;+0](-\newline[\#6;D2;H2;R1;+0])-[\#8;D2;H0;R1;+0] \\\hline

991 & \includegraphics[scale=1]{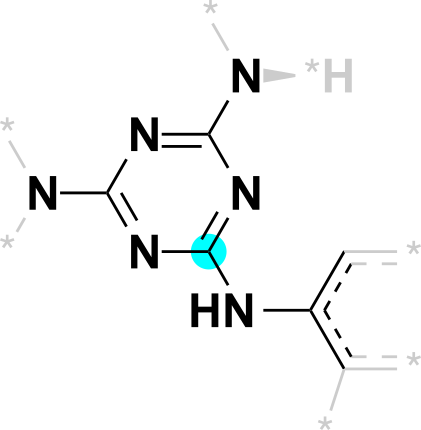} & [\#6;D2;H1;R1;+0]:[\#6;D3;H0;R1;+0](:\newline[\#6;D3;H0;R2;+0])-[\#7;D2;H1;R0;+0]-\newline[\#6;D3;H0;R1;+0]1:[\#7;D2;H0;R1;+0]:\newline[\#6;D3;H0;R1;+0](-[\#7;D3;H0;R0;+0]):\newline[\#7;D2;H0;R1;+0]:[\#6;D3;H0;R1;+0](-\newline[\#7;D3;H0;R0;+0]):[\#7;D2;H0;R1;+0]:1 \\\hline

1272 & \includegraphics[scale=1]{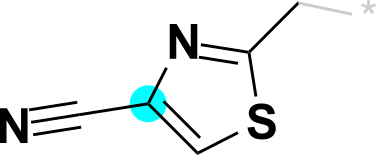} & [\#6;D2;H2;R0;+0]-[\#6;D3;H0;R1;+0]1:\newline[\#16;D2;H0;R1;+0]:[\#6;D2;H1;R1;+0]:\newline[\#6;D3;H0;R1;+0](-[\#6;D2;H0;R0;+0]\#\newline[\#7;D1;H0;R0;+0]):[\#7;D2;H0;R1;+0]:1 \\\hline

1272 & \includegraphics[scale=1]{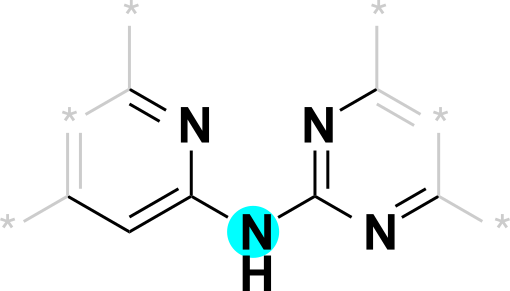} & [\#6;D3;H0;R1;+0]:[\#6;D2;H1;R1;+0]:\newline[\#6;D3;H0;R1;+0](:[\#7;D2;H0;R1;+0]:\newline[\#6;D3;H0;R1;+0])-[\#7;D2;H1;R0;+0]-\newline[\#6;D3;H0;R1;+0](:[\#7;D2;H0;R1;+0]:\newline[\#6;D3;H0;R1;+0]):[\#7;D2;H0;R1;+0]:\newline[\#6;D3;H0;R1;+0] \\\hline

1272 & \includegraphics[scale=1]{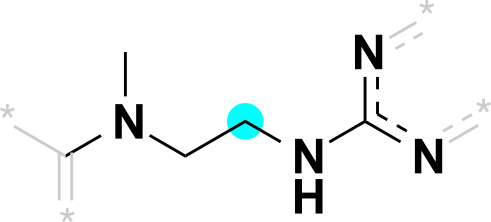} & [\#6;D1;H3;R0;+0]-[\#7;D3;H0;R0;+0](-\newline[\#6;D3;H0;R0;+0])-[\#6;D2;H2;R0;+0]-\newline[\#6;D2;H2;R0;+0]-[\#7;D2;H1;R0;+0]-\newline[\#6;D3;H0;R1;+0](:[\#7;D2;H0;R1;+0]):\newline[\#7;D2;H0;R1;+0] \\\hline

1272 & \includegraphics[scale=1]{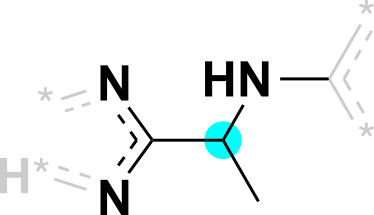} & [\#6;D1;H3;R0;+0]-[\#6;D3;H1;R0;+0](-\newline[\#7;D2;H1;R0;+0]-[\#6;D3;H0;R1;+0])-\newline[\#6;D3;H0;R1;+0](:[\#7;D2;H0;R1;+0]):\newline[\#7;D2;H0;R1;+0] \\\hline

1334 & \includegraphics[scale=1]{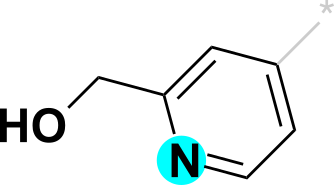} & [\#8;D1;H1;R0;+0]-[\#6;D2;H2;R0;+0]-\newline[\#6;D3;H0;R1;+0]1:[\#6;D2;H1;R1;+0]:\newline[\#6;D3;H0;R1;+0]:[\#6;D2;H1;R1;+0]:\newline[\#6;D2;H1;R1;+0]:[\#7;D2;H0;R1;+0]:1 \\\hline

873 & \includegraphics[scale=1]{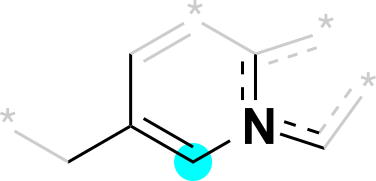} & [\#6;D2;H1;R1;+0]:[\#6;D3;H0;R1;+0](-\newline[\#6;D2;H2;R0;+0]):[\#6;D2;H1;R1;+0]:\newline[\#7;D3;H0;R2;+0](:[\#6;D2;H1;R1;+0]):\newline[\#6;D3;H0;R2;+0] \\\hline

430 & \includegraphics[scale=1]{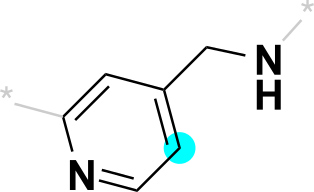} & [\#7;D2;H1;R0;+0]-[\#6;D2;H2;R0;+0]-\newline[\#6;D3;H0;R1;+0]1:[\#6;D2;H1;R1;+0]:\newline[\#6;D2;H1;R1;+0]:[\#7;D2;H0;R1;+0]:\newline[\#6;D3;H0;R1;+0]:[\#6;D2;H1;R1;+0]:1 \\\hline

\end{longtable}

\begin{figure} [H]
    \centering
        \includegraphics[scale=1]{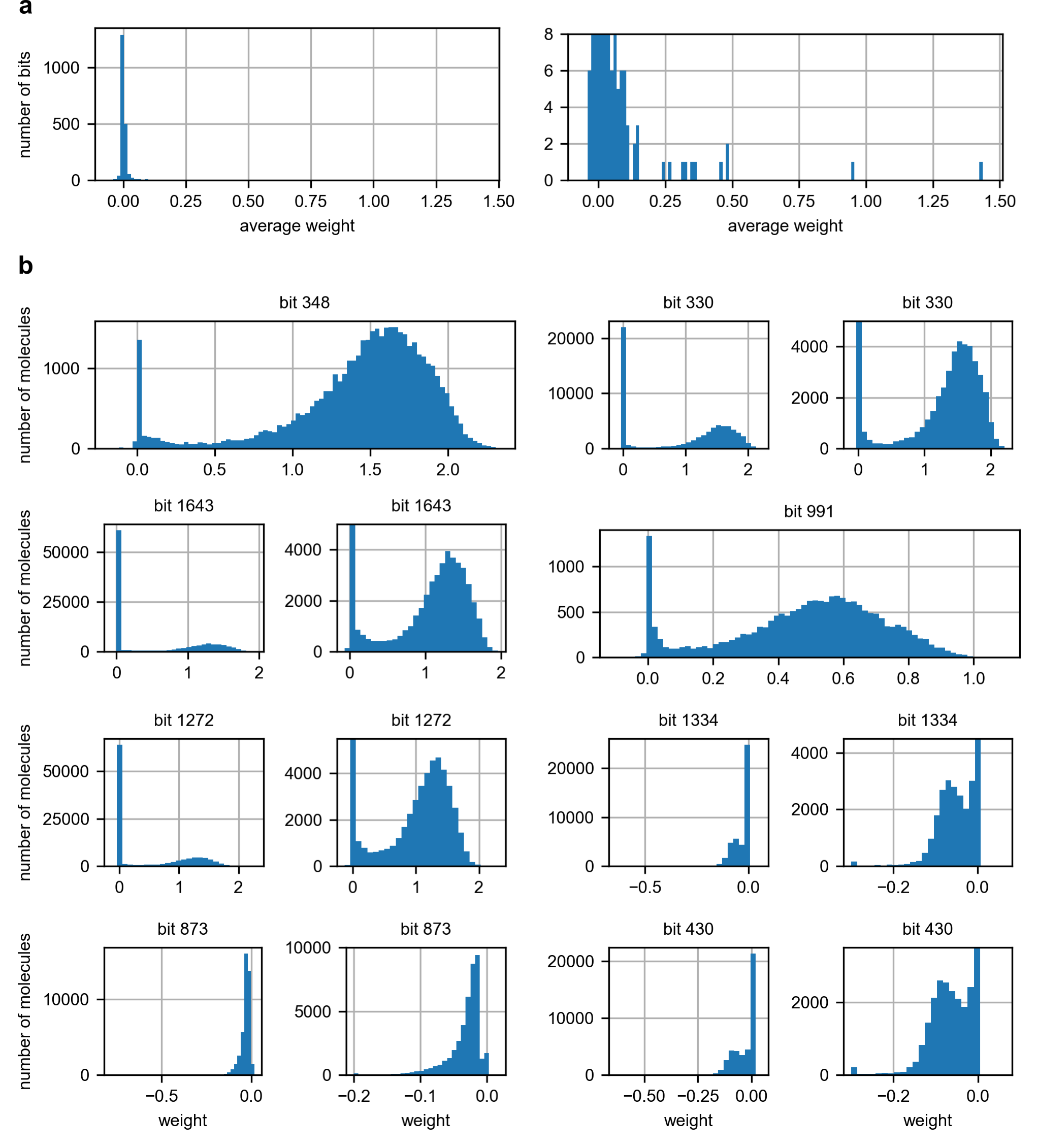}
    \caption{Histograms of \textbf{(a)} bit weights (only including bits set by at least one molecule in the dataset) and \textbf{(b)} molecule-level bit weights for single-substructure analysis on the triazine SIRT2 dataset (random split, seed 0; \emph{cf.} Figure~\ref{fig:reps_models_loss_fns_splits}d), including the top 5 and bottom 3 bits. Plots are shown in full and zoomed in.}
    \label{fig:SIRT2_onebit_hists}
\end{figure}

\subsubsection{Substructure-pair analysis}

\begin{longtable}{m{1.5cm}|m{4.5cm}|m{8.5cm}} 
\caption{SMARTS for substructure-pair analysis on the DD1S CAIX dataset (random split, seed 0).} \label{tbl:SMARTS_twobits_DD1S} \\
bit ID & substructure example & SMARTS\\ \hline

1262 & \includegraphics[scale=1]{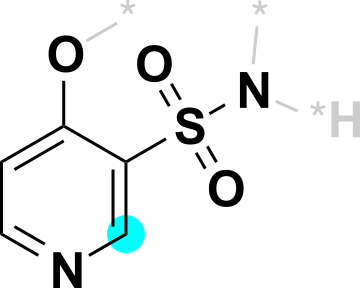} & [\#7;D3;H0;R1;+0]-[\#16;D4;H0;R0;+0](=\newline[\#8;D1;H0;R0;+0])(=[\#8;D1;H0;R0;+0])-\newline[\#6;D3;H0;R1;+0]1:[\#6;D2;H1;R1;+0]:\newline[\#7;D2;H0;R1;+0]:[\#6;D2;H1;R1;+0]:\newline[\#6;D2;H1;R1;+0]:[\#6;D3;H0;R1;+0]:1-\newline[\#8;D2;H0;R0;+0] \\\hline

1262 & \includegraphics[scale=1]{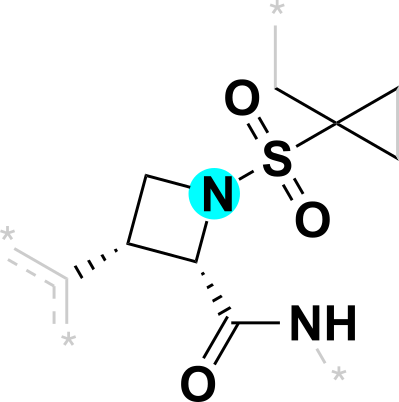} & [\#6;D2;H2;R0;+0]-[\#6;D4;H0;R1;+0](-\newline[\#6;D2;H2;R1;+0])(-[\#6;D2;H2;R1;+0])-\newline[\#16;D4;H0;R0;+0](=[\#8;D1;H0;R0;+0])(=\newline[\#8;D1;H0;R0;+0])-[\#7;D3;H0;R1;+0]1-\newline[\#6;D2;H2;R1;+0]-[\#6;D3;H1;R1;+0](-\newline[\#6;D3;H0;R1;+0])-[\#6;D3;H1;R1;+0]-1-\newline[\#6;D3;H0;R0;+0](-[\#7;D2;H1;R0;+0])=\newline[\#8;D1;H0;R0;+0] \\\hline

1802 & \includegraphics[scale=1]{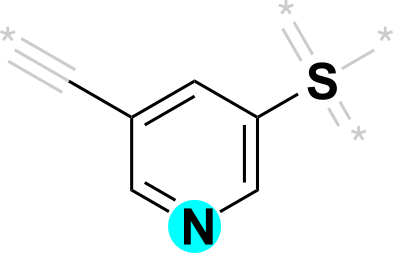} & [\#16;D4;H0;R0;+0]-[\#6;D3;H0;R1;+0]1:\newline[\#6;D2;H1;R1;+0]:[\#7;D2;H0;R1;+0]:\newline[\#6;D2;H1;R1;+0]:[\#6;D3;H0;R1;+0](-\newline[\#6;D2;H0;R0;+0]):[\#6;D2;H1;R1;+0]:1 \\\hline

1953 & \includegraphics[scale=1]{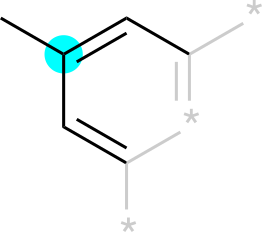} & [\#6;D1;H3;R0;+0]-[\#6;D3;H0;R1;+0](:\newline[\#6;D2;H1;R1;+0]:[\#6;D3;H0;R1;+0]):\newline[\#6;D2;H1;R1;+0]:[\#6;D3;H0;R1;+0] \\\hline

833 & \includegraphics[scale=1]{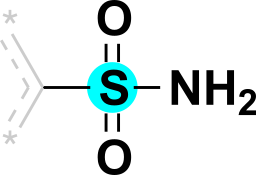} & [\#6;D3;H0;R1;+0]-[\#16;D4;H0;R0;+0](-\newline[\#7;D1;H2;R0;+0])(=[\#8;D1;H0;R0;+0])=\newline[\#8;D1;H0;R0;+0] \\\hline

1489 & \includegraphics[scale=1]{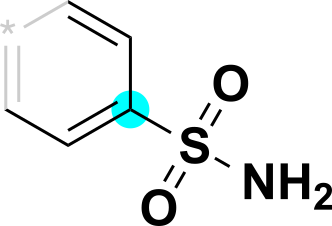} & [\#6;D2;H1;R1;+0]:[\#6;D2;H1;R1;+0]:\newline[\#6;D3;H0;R1;+0](:[\#6;D2;H1;R1;+0]:\newline[\#6;D2;H1;R1;+0])-[\#16;D4;H0;R0;+0](-\newline[\#7;D1;H2;R0;+0])(=[\#8;D1;H0;R0;+0])=\newline[\#8;D1;H0;R0;+0] \\\hline

1489 & \includegraphics[scale=1]{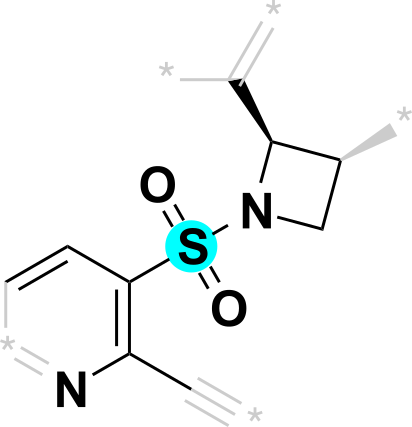} & [\#6;D2;H0;R0;+0]-[\#6;D3;H0;R1;+0](:\newline[\#7;D2;H0;R1;+0]):[\#6;D3;H0;R1;+0](:\newline[\#6;D2;H1;R1;+0]:[\#6;D2;H1;R1;+0])-\newline[\#16;D4;H0;R0;+0](=[\#8;D1;H0;R0;+0])(=\newline[\#8;D1;H0;R0;+0])-[\#7;D3;H0;R1;+0]1-\newline[\#6;D2;H2;R1;+0]-[\#6;D3;H1;R1;+0]-\newline[\#6;D3;H1;R1;+0]-1-[\#6;D3;H0;R0;+0] \\\hline

1736 & \includegraphics[scale=1]{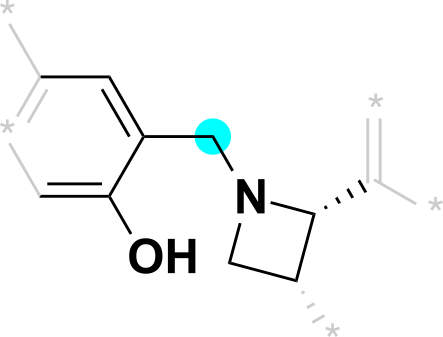} & [\#6;D2;H1;R1;+0]:[\#6;D3;H0;R1;+0](-\newline[\#8;D1;H1;R0;+0]):[\#6;D3;H0;R1;+0](:\newline[\#6;D2;H1;R1;+0]:[\#6;D3;H0;R1;+0])-\newline[\#6;D2;H2;R0;+0]-[\#7;D3;H0;R1;+0]1-\newline[\#6;D2;H2;R1;+0]-[\#6;D3;H1;R1;+0]-\newline[\#6;D3;H1;R1;+0]-1-[\#6;D3;H0;R0;+0] \\\hline

1844 & \includegraphics[scale=1]{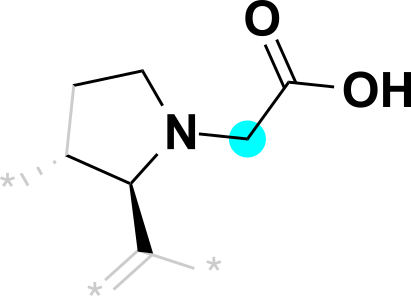} & [\#6;D2;H2;R1;+0]-[\#6;D2;H2;R1;+0]-\newline[\#7;D3;H0;R1;+0](-[\#6;D2;H2;R0;+0]-\newline[\#6;D3;H0;R0;+0](=[\#8;D1;H0;R0;+0])-\newline[\#8;D1;H1;R0;+0])-[\#6;D3;H1;R1;+0](-\newline[\#6;D3;H0;R0;+0])-[\#6;D3;H1;R1;+0] \\\hline

1135 & \includegraphics[scale=1]{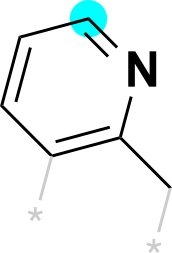} & [\#6;D2;H2;R0;+0]-[\#6;D3;H0;R1;+0]1:\newline[\#6;D3;H0;R1;+0]:[\#6;D2;H1;R1;+0]:\newline[\#6;D2;H1;R1;+0]:[\#6;D2;H1;R1;+0]:\newline[\#7;D2;H0;R1;+0]:1 \\\hline

1135 & \includegraphics[scale=1]{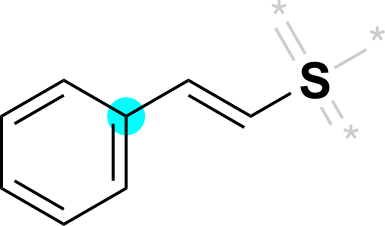} & [\#16;D4;H0;R0;+0]-[\#6;D2;H1;R0;+0]=\newline[\#6;D2;H1;R0;+0]-[\#6;D3;H0;R1;+0]1:\newline[\#6;D2;H1;R1;+0]:[\#6;D2;H1;R1;+0]:\newline[\#6;D2;H1;R1;+0]:[\#6;D2;H1;R1;+0]:\newline[\#6;D2;H1;R1;+0]:1 \\\hline

\end{longtable}

\begin{figure} [H]
    \centering
        \includegraphics[scale=1]{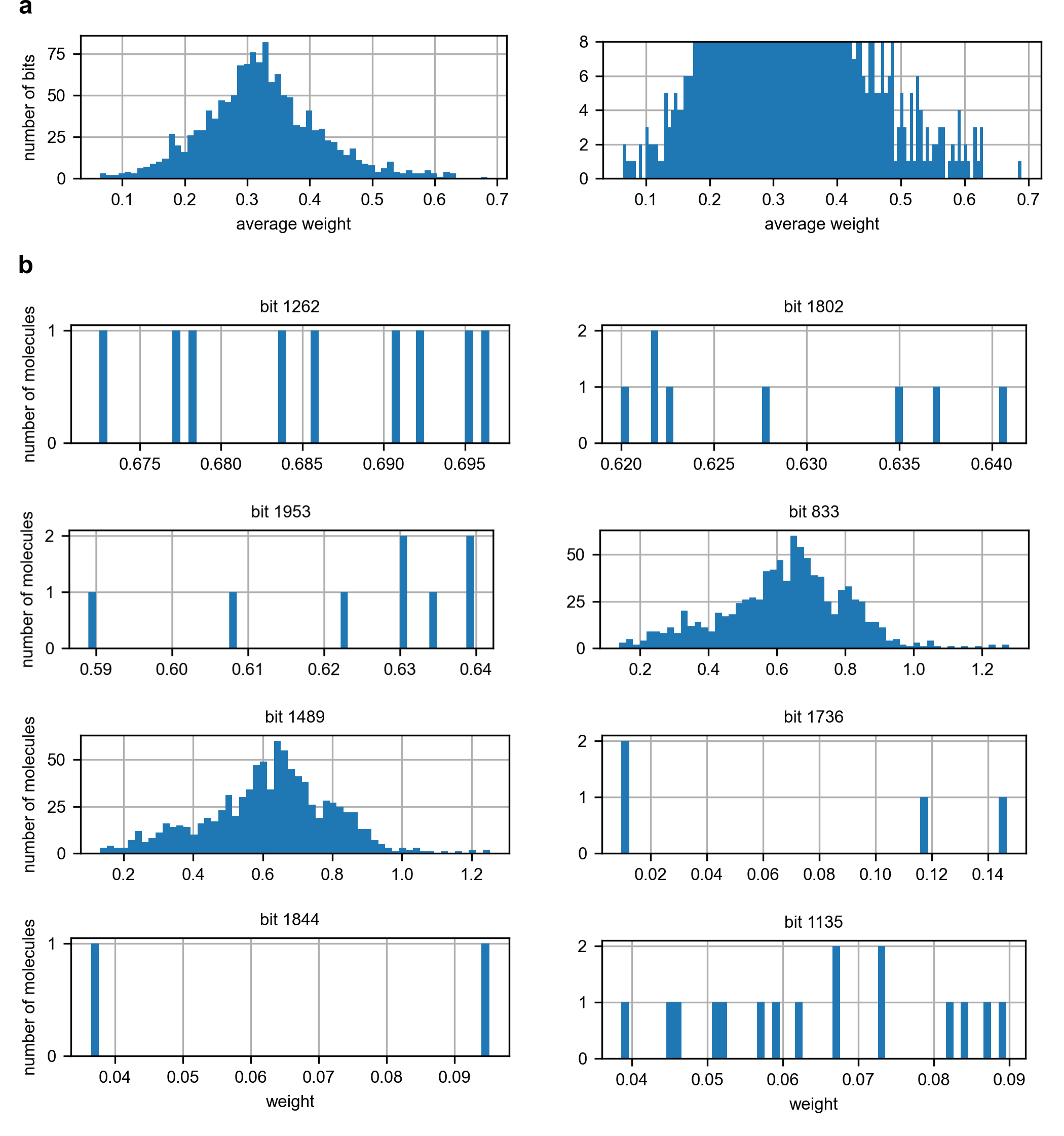}
    \caption{Histograms of \textbf{(a)} bit weights (only including bits set by at least one molecule in the dataset; plot shown in full and zoomed in) and \textbf{(b)} molecule-level bit weights for substructure-pair analysis on the DD1S CAIX dataset (random split, seed 0; \emph{cf.} Figure~\ref{fig:reps_models_loss_fns_splits}d), including the top 5 and bottom 3 bits. }
    \label{fig:CAIX_twobits_seed_0_hists}
\end{figure}

\begin{longtable}{m{1.5cm}|m{4.5cm}|m{8.5cm}} 
\caption{SMARTS for substructure-pair analysis on the DD1S CAIX dataset (random split, seed 1).} \label{tbl:SMARTS_twobits_DD1S_seed_1} \\
bit ID & substructure example & SMARTS\\ \hline

833 & \includegraphics[scale=1]{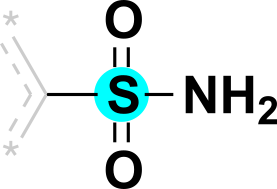} & [\#6;D3;H0;R1;+0]-[\#16;D4;H0;R0;+0](-\newline[\#7;D1;H2;R0;+0])(=[\#8;D1;H0;R0;+0])=\newline[\#8;D1;H0;R0;+0]\\\hline

1785 & \includegraphics[scale=1]{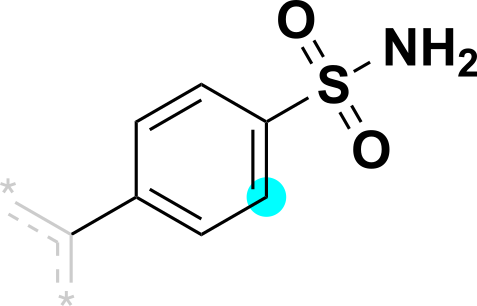} & [\#6;D3;H0;R1;+0]-[\#6;D3;H0;R1;+0]1:\newline[\#6;D2;H1;R1;+0]:[\#6;D2;H1;R1;+0]:\newline[\#6;D3;H0;R1;+0](-[\#16;D4;H0;R0;+0](-\newline[\#7;D1;H2;R0;+0])(=[\#8;D1;H0;R0;+0])=\newline[\#8;D1;H0;R0;+0]):[\#6;D2;H1;R1;+0]:\newline[\#6;D2;H1;R1;+0]:1\\\hline

1785 & \includegraphics[scale=1]{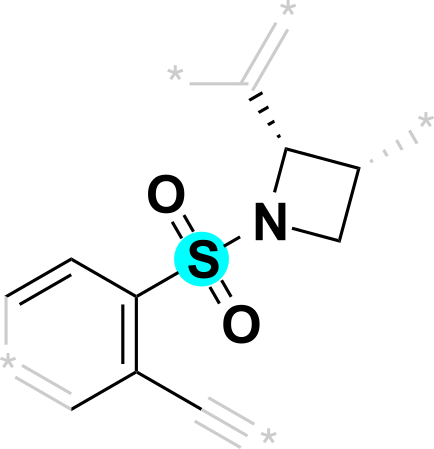} & [\#6;D2;H0;R0;+0]-[\#6;D3;H0;R1;+0](:\newline[\#6;D2;H1;R1;+0]):[\#6;D3;H0;R1;+0](:\newline[\#6;D2;H1;R1;+0]:[\#6;D2;H1;R1;+0])-\newline[\#16;D4;H0;R0;+0](=[\#8;D1;H0;R0;+0])(=\newline[\#8;D1;H0;R0;+0])-[\#7;D3;H0;R1;+0]1-\newline[\#6;D2;H2;R1;+0]-[\#6;D3;H1;R1;+0]-\newline[\#6;D3;H1;R1;+0]-1-[\#6;D3;H0;R0;+0]\\\hline

1489 & \includegraphics[scale=1]{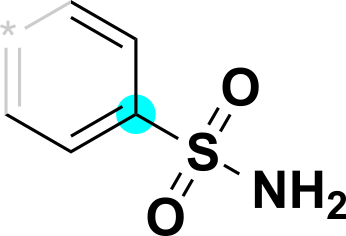} & [\#6;D2;H1;R1;+0]:[\#6;D2;H1;R1;+0]:\newline[\#6;D3;H0;R1;+0](:[\#6;D2;H1;R1;+0]:\newline[\#6;D2;H1;R1;+0])-[\#16;D4;H0;R0;+0](-\newline[\#7;D1;H2;R0;+0])(=[\#8;D1;H0;R0;+0])=\newline[\#8;D1;H0;R0;+0]\\\hline

1489 & \includegraphics[scale=1]{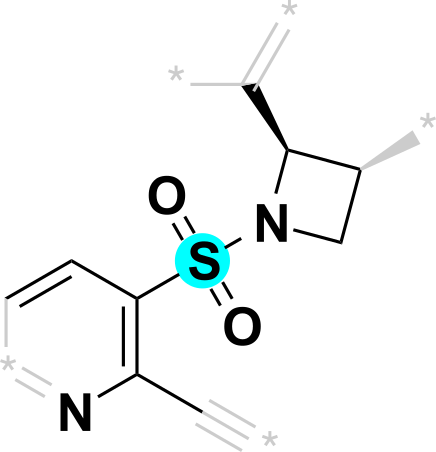} & [\#6;D2;H0;R0;+0]-[\#6;D3;H0;R1;+0](:\newline[\#7;D2;H0;R1;+0]):[\#6;D3;H0;R1;+0](:\newline[\#6;D2;H1;R1;+0]:[\#6;D2;H1;R1;+0])-\newline[\#16;D4;H0;R0;+0](=[\#8;D1;H0;R0;+0])(=\newline[\#8;D1;H0;R0;+0])-[\#7;D3;H0;R1;+0]1-\newline[\#6;D2;H2;R1;+0]-[\#6;D3;H1;R1;+0]-\newline[\#6;D3;H1;R1;+0]-1-[\#6;D3;H0;R0;+0]\\\hline

1197 & \includegraphics[scale=1]{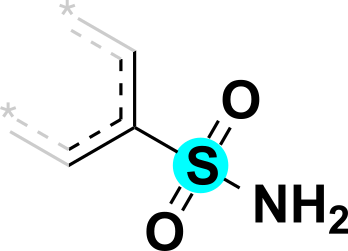} & [\#6;D2;H1;R1;+0]:[\#6;D3;H0;R1;+0](:\newline[\#6;D2;H1;R1;+0])-[\#16;D4;H0;R0;+0](-\newline[\#7;D1;H2;R0;+0])(=[\#8;D1;H0;R0;+0])=\newline[\#8;D1;H0;R0;+0]\\\hline

1197 & \includegraphics[scale=1]{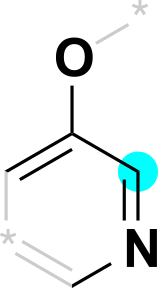} & [\#6;D2;H1;R1;+0]:[\#7;D2;H0;R1;+0]:\newline[\#6;D2;H1;R1;+0]:[\#6;D3;H0;R1;+0](:\newline[\#6;D2;H1;R1;+0])-[\#8;D2;H0;R0;+0]\\\hline

168 & \includegraphics[scale=1]{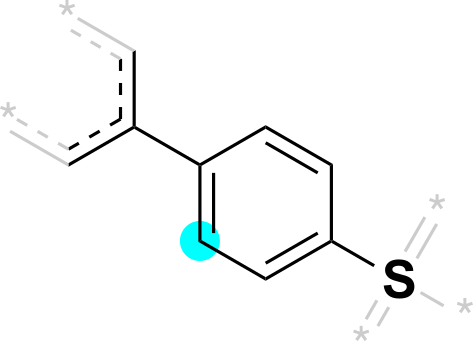} & [\#16;D4;H0;R0;+0]-[\#6;D3;H0;R1;+0]1:\newline[\#6;D2;H1;R1;+0]:[\#6;D2;H1;R1;+0]:\newline[\#6;D3;H0;R1;+0](-[\#6;D3;H0;R1;+0](:\newline[\#6;D2;H1;R1;+0]):[\#6;D2;H1;R1;+0]):\newline[\#6;D2;H1;R1;+0]:[\#6;D2;H1;R1;+0]:1\\\hline

168 & \includegraphics[scale=1]{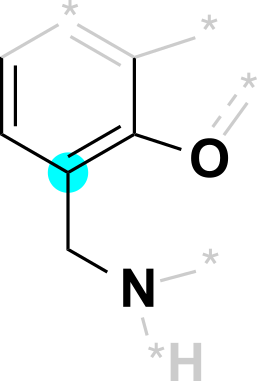} & [\#6;D2;H1;R1;+0]:[\#6;D2;H1;R1;+0]:\newline[\#6;D3;H0;R1;+0](-[\#6;D2;H2;R0;+0]-\newline[\#7;D3;H0;R1;+0]):[\#6;D3;H0;R2;+0](:\newline[\#6;D3;H0;R2;+0]):[\#8;D2;H0;R1;+0]\\\hline

168 & \includegraphics[scale=1]{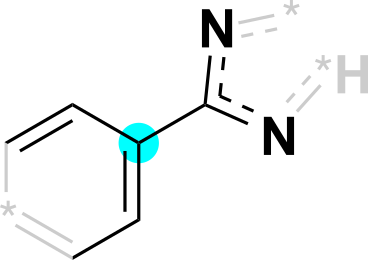} & [\#6;D2;H1;R1;+0]:[\#6;D2;H1;R1;+0]:\newline[\#6;D3;H0;R1;+0](:[\#6;D2;H1;R1;+0]:\newline[\#6;D2;H1;R1;+0])-[\#6;D3;H0;R1;+0](:\newline[\#7;D2;H0;R1;+0]):[\#7;D2;H0;R1;+0]\\\hline

1705 & \includegraphics[scale=1]{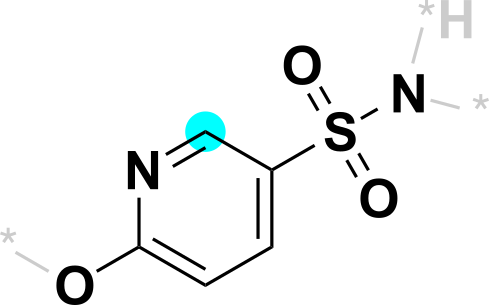} & [\#7;D3;H0;R1;+0]-[\#16;D4;H0;R0;+0](=\newline[\#8;D1;H0;R0;+0])(=[\#8;D1;H0;R0;+0])-\newline[\#6;D3;H0;R1;+0]1:[\#6;D2;H1;R1;+0]:\newline[\#6;D2;H1;R1;+0]:[\#6;D3;H0;R1;+0](-\newline[\#8;D2;H0;R0;+0]):[\#7;D2;H0;R1;+0]:\newline[\#6;D2;H1;R1;+0]:1\\\hline

1705 & \includegraphics[scale=1]{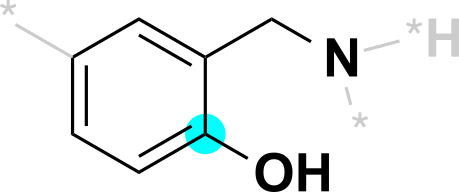} & [\#7;D3;H0;R1;+0]-[\#6;D2;H2;R0;+0]-\newline[\#6;D3;H0;R1;+0]1:[\#6;D2;H1;R1;+0]:\newline[\#6;D3;H0;R1;+0]:[\#6;D2;H1;R1;+0]:\newline[\#6;D2;H1;R1;+0]:[\#6;D3;H0;R1;+0]:1-\newline[\#8;D1;H1;R0;+0]\\\hline

1537 & \includegraphics[scale=1]{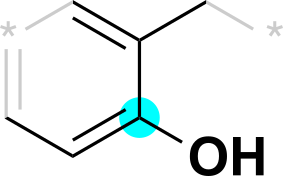} & [\#6;D2;H1;R1;+0]:[\#6;D2;H1;R1;+0]:\newline[\#6;D3;H0;R1;+0](-[\#8;D1;H1;R0;+0]):\newline[\#6;D3;H0;R1;+0](:[\#6;D2;H1;R1;+0])-\newline[\#6;D2;H2;R0;+0]\\\hline

1537 & \includegraphics[scale=1]{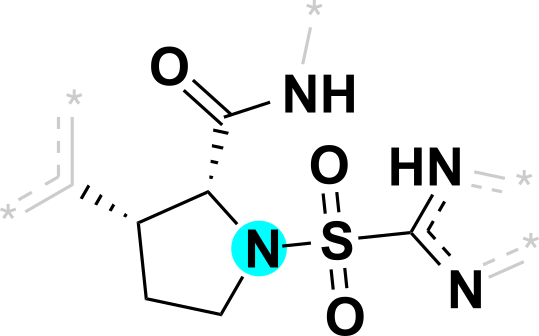} & [\#6;D3;H0;R1;+0]-[\#6;D3;H1;R1;+0]1-\newline[\#6;D2;H2;R1;+0]-[\#6;D2;H2;R1;+0]-\newline[\#7;D3;H0;R1;+0](-[\#16;D4;H0;R0;+0](=\newline[\#8;D1;H0;R0;+0])(=[\#8;D1;H0;R0;+0])-\newline[\#6;D3;H0;R1;+0](:[\#7;D2;H0;R1;+0]):\newline[\#7;D2;H1;R1;+0])-[\#6;D3;H1;R1;+0]-1-\newline[\#6;D3;H0;R0;+0](-[\#7;D2;H1;R0;+0])=\newline[\#8;D1;H0;R0;+0]\\\hline

1537 & \includegraphics[scale=1]{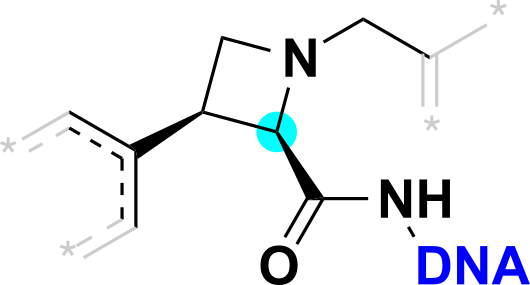} & [\#102;D1;H0;R0;+0]-[\#7;D2;H1;R0;+0]-\newline[\#6;D3;H0;R0;+0](=[\#8;D1;H0;R0;+0])-\newline[\#6;D3;H1;R1;+0]1-[\#6;D3;H1;R1;+0](-\newline[\#6;D3;H0;R1;+0](:[\#6;D2;H1;R1;+0]):\newline[\#6;D2;H1;R1;+0])-[\#6;D2;H2;R1;+0]-\newline[\#7;D3;H0;R1;+0]-1-[\#6;D2;H2;R0;+0]-\newline[\#6;D3;H0;R0;+0]\\\hline

1736 & \includegraphics[scale=1]{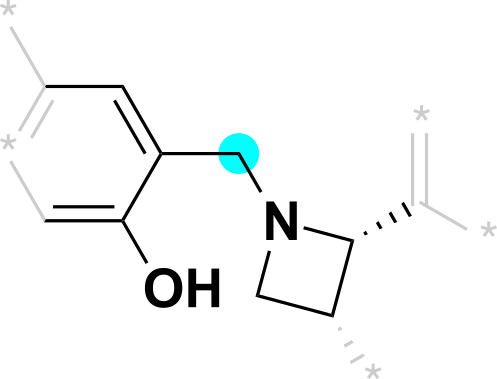} & [\#6;D2;H1;R1;+0]:[\#6;D3;H0;R1;+0](-\newline[\#8;D1;H1;R0;+0]):[\#6;D3;H0;R1;+0](:\newline[\#6;D2;H1;R1;+0]:[\#6;D3;H0;R1;+0])-\newline[\#6;D2;H2;R0;+0]-[\#7;D3;H0;R1;+0]1-\newline[\#6;D2;H2;R1;+0]-[\#6;D3;H1;R1;+0]-\newline[\#6;D3;H1;R1;+0]-1-[\#6;D3;H0;R0;+0]\\\hline

\end{longtable}

\begin{figure} [H]
    \centering
        \includegraphics[scale=1]{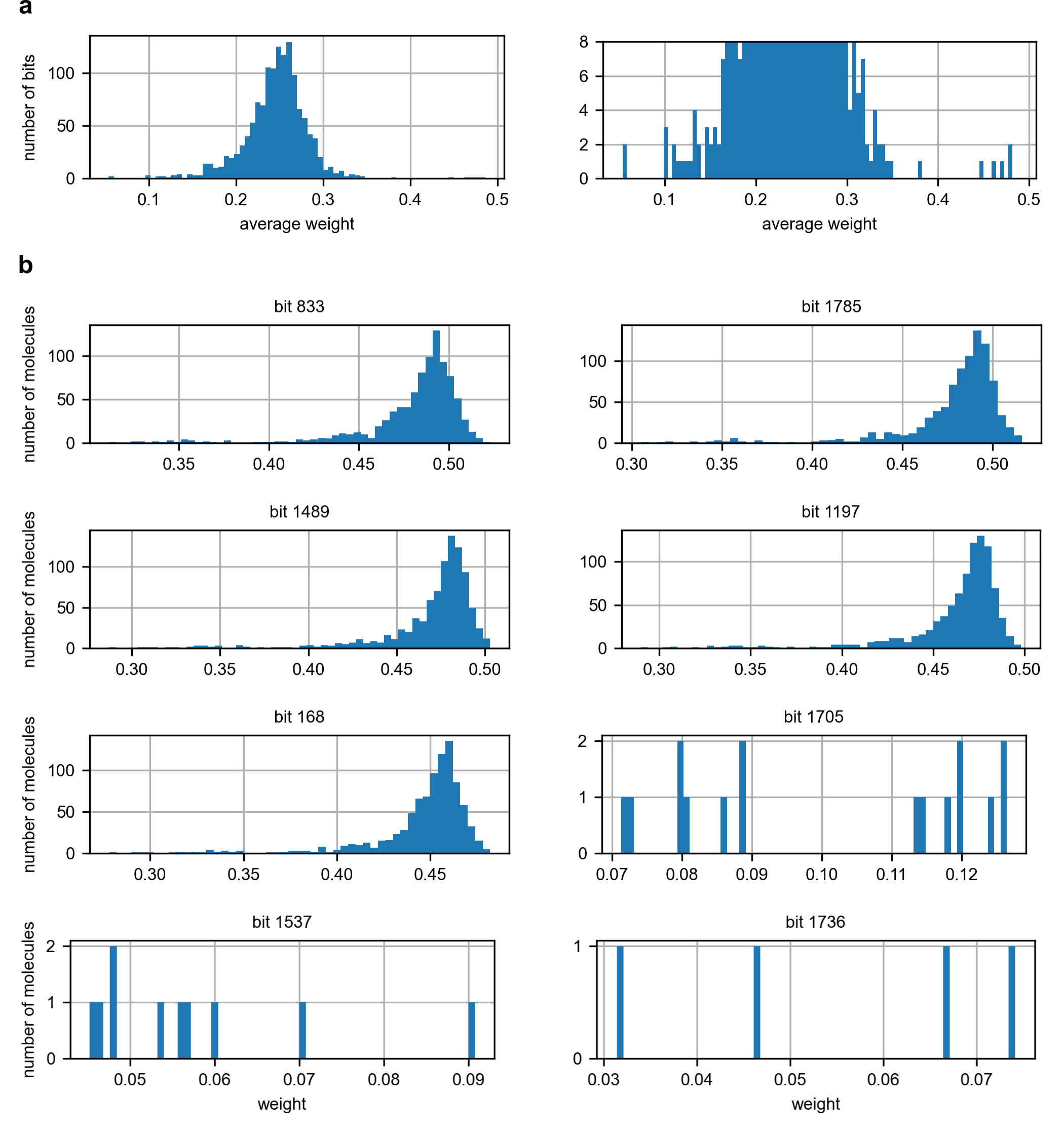}
    \caption{Histograms of \textbf{(a)} bit weights (only including bits set by at least one molecule in the dataset; plot shown in full and zoomed in) and \textbf{(b)} molecule-level bit weights for substructure-pair analysis on the DD1S CAIX dataset (random split, seed 1; \emph{cf.} Figure~\ref{fig:reps_models_loss_fns_splits}d), including the top 5 and bottom 3 bits. }
    \label{fig:CAIX_twobits_seed_1_hists}
\end{figure}

\begin{longtable}{m{1.5cm}|m{5cm}|m{8.5cm}} 
\caption{SMARTS for substructure-pair analysis on the DD1S CAIX dataset (random split, seed 2).} \label{tbl:SMARTS_twobits_DD1S_seed_2} \\
bit ID & substructure example & SMARTS\\ \hline

1127 & \includegraphics[scale=1]{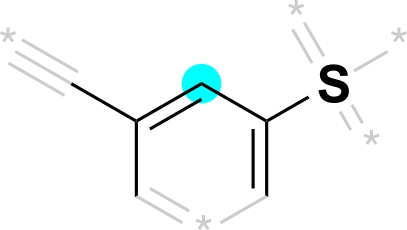} & [\#16;D4;H0;R0;+0]-[\#6;D3;H0;R1;+0](:\newline[\#6;D2;H1;R1;+0]):[\#6;D2;H1;R1;+0]:\newline[\#6;D3;H0;R1;+0](-[\#6;D2;H0;R0;+0]):\newline[\#6;D2;H1;R1;+0]\\\hline

670 & \includegraphics[scale=1]{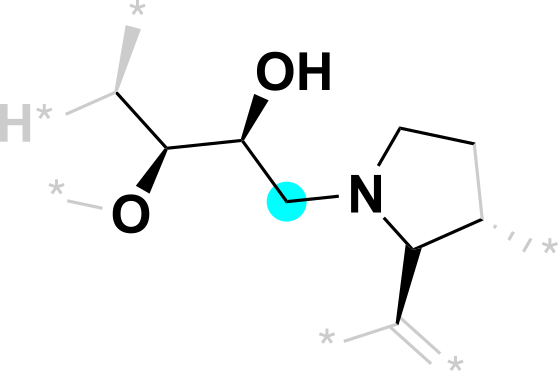} & [\#6;D2;H2;R1;+0]-[\#6;D2;H2;R1;+0]-\newline[\#7;D3;H0;R1;+0](-[\#6;D2;H2;R0;+0]-\newline[\#6;D3;H1;R0;+0](-[\#8;D1;H1;R0;+0])-\newline[\#6;D3;H1;R1;+0](-[\#6;D3;H1;R1;+0])-\newline[\#8;D2;H0;R1;+0])-[\#6;D3;H1;R1;+0](-\newline[\#6;D3;H0;R0;+0])-[\#6;D3;H1;R1;+0]\\\hline

1605 & \includegraphics[scale=1]{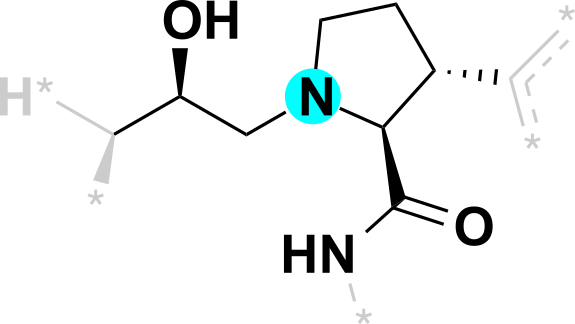} & [\#6;D3;H0;R1;+0]-[\#6;D3;H1;R1;+0]1-\newline[\#6;D2;H2;R1;+0]-[\#6;D2;H2;R1;+0]-\newline[\#7;D3;H0;R1;+0](-[\#6;D2;H2;R0;+0]-\newline[\#6;D3;H1;R0;+0](-[\#6;D3;H1;R1;+0])-\newline[\#8;D1;H1;R0;+0])-[\#6;D3;H1;R1;+0]-1-\newline[\#6;D3;H0;R0;+0](-[\#7;D2;H1;R0;+0])=\newline[\#8;D1;H0;R0;+0]\\\hline

1802 & \includegraphics[scale=1]{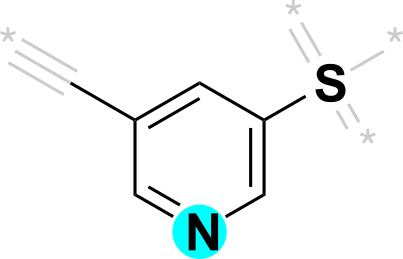} & [\#16;D4;H0;R0;+0]-[\#6;D3;H0;R1;+0]1:\newline[\#6;D2;H1;R1;+0]:[\#7;D2;H0;R1;+0]:\newline[\#6;D2;H1;R1;+0]:[\#6;D3;H0;R1;+0](-\newline[\#6;D2;H0;R0;+0]):[\#6;D2;H1;R1;+0]:1\\\hline

1262 & \includegraphics[scale=1]{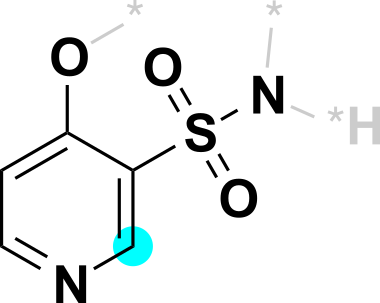} & [\#7;D3;H0;R1;+0]-[\#16;D4;H0;R0;+0](=\newline[\#8;D1;H0;R0;+0])(=[\#8;D1;H0;R0;+0])-\newline[\#6;D3;H0;R1;+0]1:[\#6;D2;H1;R1;+0]:\newline[\#7;D2;H0;R1;+0]:[\#6;D2;H1;R1;+0]:\newline[\#6;D2;H1;R1;+0]:[\#6;D3;H0;R1;+0]:1-\newline[\#8;D2;H0;R0;+0]\\\hline

1262 & \includegraphics[scale=1]{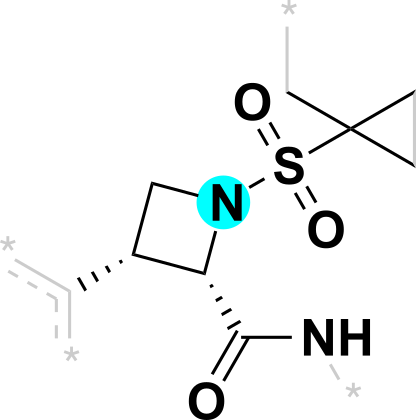} & [\#6;D2;H2;R0;+0]-[\#6;D4;H0;R1;+0](-\newline[\#6;D2;H2;R1;+0])(-[\#6;D2;H2;R1;+0])-\newline[\#16;D4;H0;R0;+0](=[\#8;D1;H0;R0;+0])(=\newline[\#8;D1;H0;R0;+0])-[\#7;D3;H0;R1;+0]1-\newline[\#6;D2;H2;R1;+0]-[\#6;D3;H1;R1;+0](-\newline[\#6;D3;H0;R1;+0])-[\#6;D3;H1;R1;+0]-1-\newline[\#6;D3;H0;R0;+0](-[\#7;D2;H1;R0;+0])=\newline[\#8;D1;H0;R0;+0]\\\hline

1844 & \includegraphics[scale=1]{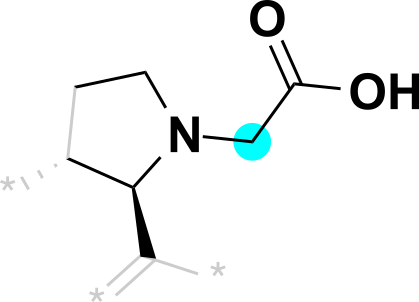} & [\#6;D2;H2;R1;+0]-[\#6;D2;H2;R1;+0]-\newline[\#7;D3;H0;R1;+0](-[\#6;D2;H2;R0;+0]-\newline[\#6;D3;H0;R0;+0](=[\#8;D1;H0;R0;+0])-\newline[\#8;D1;H1;R0;+0])-[\#6;D3;H1;R1;+0](-\newline[\#6;D3;H0;R0;+0])-[\#6;D3;H1;R1;+0]\\\hline

1523 & \includegraphics[scale=1]{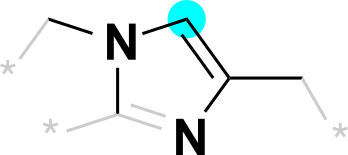} & [\#6;D2;H2;R0;+0]-[\#6;D3;H0;R1;+0](:\newline[\#7;D2;H0;R1;+0]):[\#6;D2;H1;R1;+0]:\newline[\#7;D3;H0;R2;+0](-[\#6;D2;H2;R1;+0]):\newline[\#6;D3;H0;R2;+0]\\\hline

1317 & \includegraphics[scale=1]{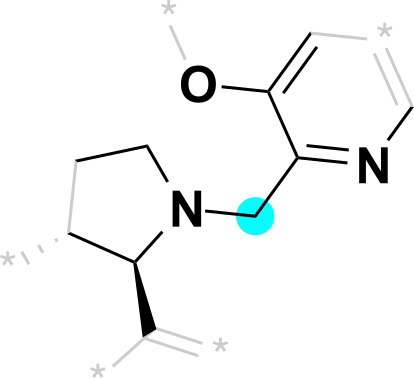} & [\#6;D2;H1;R1;+0]:[\#7;D2;H0;R1;+0]:\newline[\#6;D3;H0;R1;+0](-[\#6;D2;H2;R0;+0]-\newline[\#7;D3;H0;R1;+0](-[\#6;D2;H2;R1;+0]-\newline[\#6;D2;H2;R1;+0])-[\#6;D3;H1;R1;+0](-\newline[\#6;D3;H0;R0;+0])-[\#6;D3;H1;R1;+0]):\newline[\#6;D3;H0;R1;+0](:[\#6;D2;H1;R1;+0])-\newline[\#8;D2;H0;R0;+0]\\\hline

\end{longtable}

\begin{figure} [H]
    \centering
        \includegraphics[scale=1]{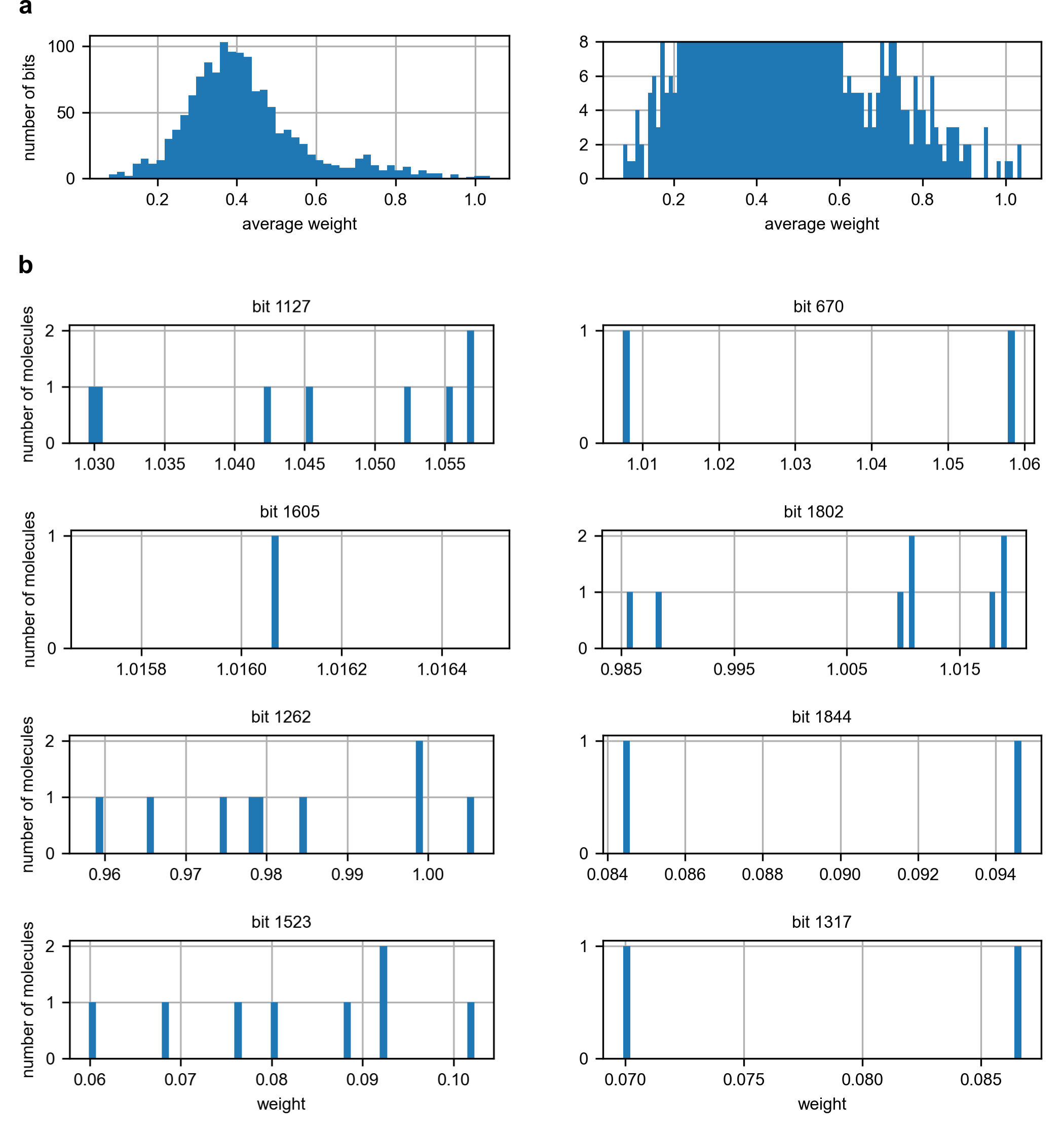}
    \caption{Histograms of \textbf{(a)} bit weights (only including bits set by at least one molecule in the dataset; plot shown in full and zoomed in) and \textbf{(b)} molecule-level bit weights for substructure-pair analysis on the DD1S CAIX dataset (random split, seed 2; \emph{cf.} Figure~\ref{fig:reps_models_loss_fns_splits}d), including the top 5 and bottom 3 bits. }
    \label{fig:CAIX_twobits_seed_2_hists}
\end{figure}

\begin{longtable}{m{1.5cm}|m{4.5cm}|m{8.5cm}} 
\caption{SMARTS for substructure-pair analysis on the triazine sEH dataset (random split, seed 0).} \label{tbl:SMARTS_twobits_sEH} \\
bit ID & substructure example & SMARTS\\ \hline

1402 & \includegraphics[scale=1]{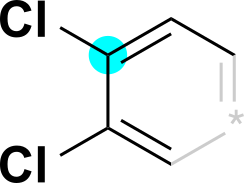} & [\#17;D1;H0;R0;+0]-[\#6;D3;H0;R1;+0](:\newline[\#6;D2;H1;R1;+0]):[\#6;D3;H0;R1;+0](-\newline[\#17;D1;H0;R0;+0]):[\#6;D2;H1;R1;+0]:\newline[\#6;D2;H1;R1;+0] \\\hline

1554 & \includegraphics[scale=1]{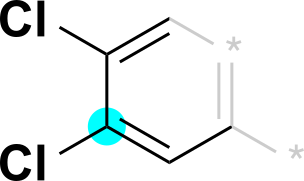} & [\#17;D1;H0;R0;+0]-[\#6;D3;H0;R1;+0](:\newline[\#6;D2;H1;R1;+0]):[\#6;D3;H0;R1;+0](-\newline[\#17;D1;H0;R0;+0]):[\#6;D2;H1;R1;+0]:\newline[\#6;D3;H0;R1;+0] \\\hline

1198 & \includegraphics[scale=1]{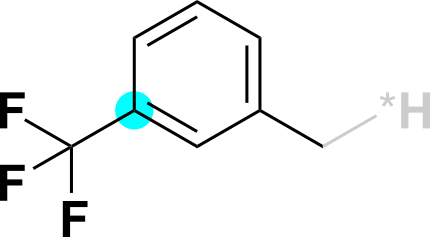} & [\#6;D2;H2;R0;+0]-[\#6;D3;H0;R1;+0]1:\newline[\#6;D2;H1;R1;+0]:[\#6;D2;H1;R1;+0]:\newline[\#6;D2;H1;R1;+0]:[\#6;D3;H0;R1;+0](-\newline[\#6;D4;H0;R0;+0](-[\#9;D1;H0;R0;+0])(-\newline[\#9;D1;H0;R0;+0])-[\#9;D1;H0;R0;+0]):\newline[\#6;D2;H1;R1;+0]:1 \\\hline

1245 & \includegraphics[scale=1]{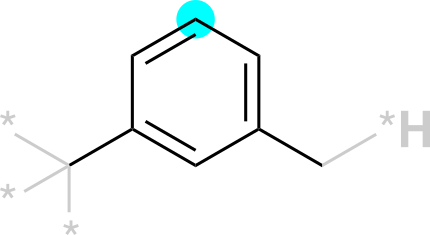} & [\#6;D2;H2;R0;+0]-[\#6;D3;H0;R1;+0]1:\newline[\#6;D2;H1;R1;+0]:[\#6;D2;H1;R1;+0]:\newline[\#6;D2;H1;R1;+0]:[\#6;D3;H0;R1;+0](-\newline[\#6;D4;H0;R0;+0]):[\#6;D2;H1;R1;+0]:1 \\\hline

1245 & \includegraphics[scale=1]{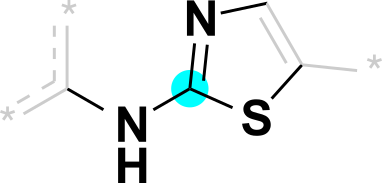} & [\#6;D2;H1;R1;+0]:[\#7;D2;H0;R1;+0]:\newline[\#6;D3;H0;R1;+0](:[\#16;D2;H0;R1;+0]:\newline[\#6;D3;H0;R1;+0])-[\#7;D2;H1;R0;+0]-\newline[\#6;D3;H0;R1;+0] \\\hline

1368 & \includegraphics[scale=1]{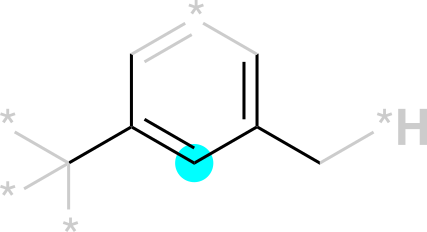} & [\#6;D2;H1;R1;+0]:[\#6;D3;H0;R1;+0](-\newline[\#6;D2;H2;R0;+0]):[\#6;D2;H1;R1;+0]:\newline[\#6;D3;H0;R1;+0](:[\#6;D2;H1;R1;+0])-\newline[\#6;D4;H0;R0;+0] \\\hline

1368 & \includegraphics[scale=1]{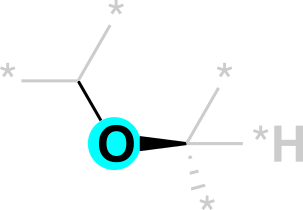} & [\#6;D3;H1;R0;+0]-[\#8;D2;H0;R0;+0]-\newline[\#6;D4;H0;R1;+0] \\\hline

1923 & \includegraphics[scale=1]{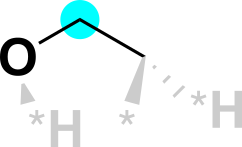} & [\#6;D3;H1;R2;+0]-[\#6;D2;H2;R1;+0]-\newline[\#8;D2;H0;R1;+0] \\\hline

853 & \includegraphics[scale=1]{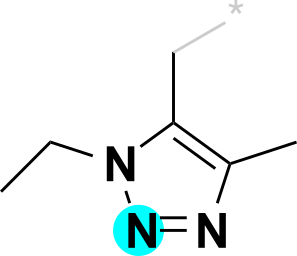} & [\#6;D1;H3;R0;+0]-[\#6;D2;H2;R0;+0]-\newline[\#7;D3;H0;R1;+0]1:[\#7;D2;H0;R1;+0]:\newline[\#7;D2;H0;R1;+0]:[\#6;D3;H0;R1;+0](-\newline[\#6;D1;H3;R0;+0]):[\#6;D3;H0;R1;+0]:1-\newline[\#6;D2;H2;R0;+0] \\\hline

864 & \includegraphics[scale=1]{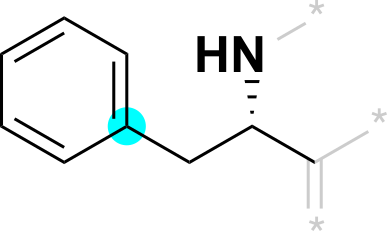} & [\#6;D3;H0;R0;+0]-[\#6;D3;H1;R0;+0](-\newline[\#7;D2;H1;R0;+0])-[\#6;D2;H2;R0;+0]-\newline[\#6;D3;H0;R1;+0]1:[\#6;D2;H1;R1;+0]:\newline[\#6;D2;H1;R1;+0]:[\#6;D2;H1;R1;+0]:\newline[\#6;D2;H1;R1;+0]:[\#6;D2;H1;R1;+0]:1 \\\hline

\end{longtable}

\begin{figure} [H]
    \centering
        \includegraphics[scale=1]{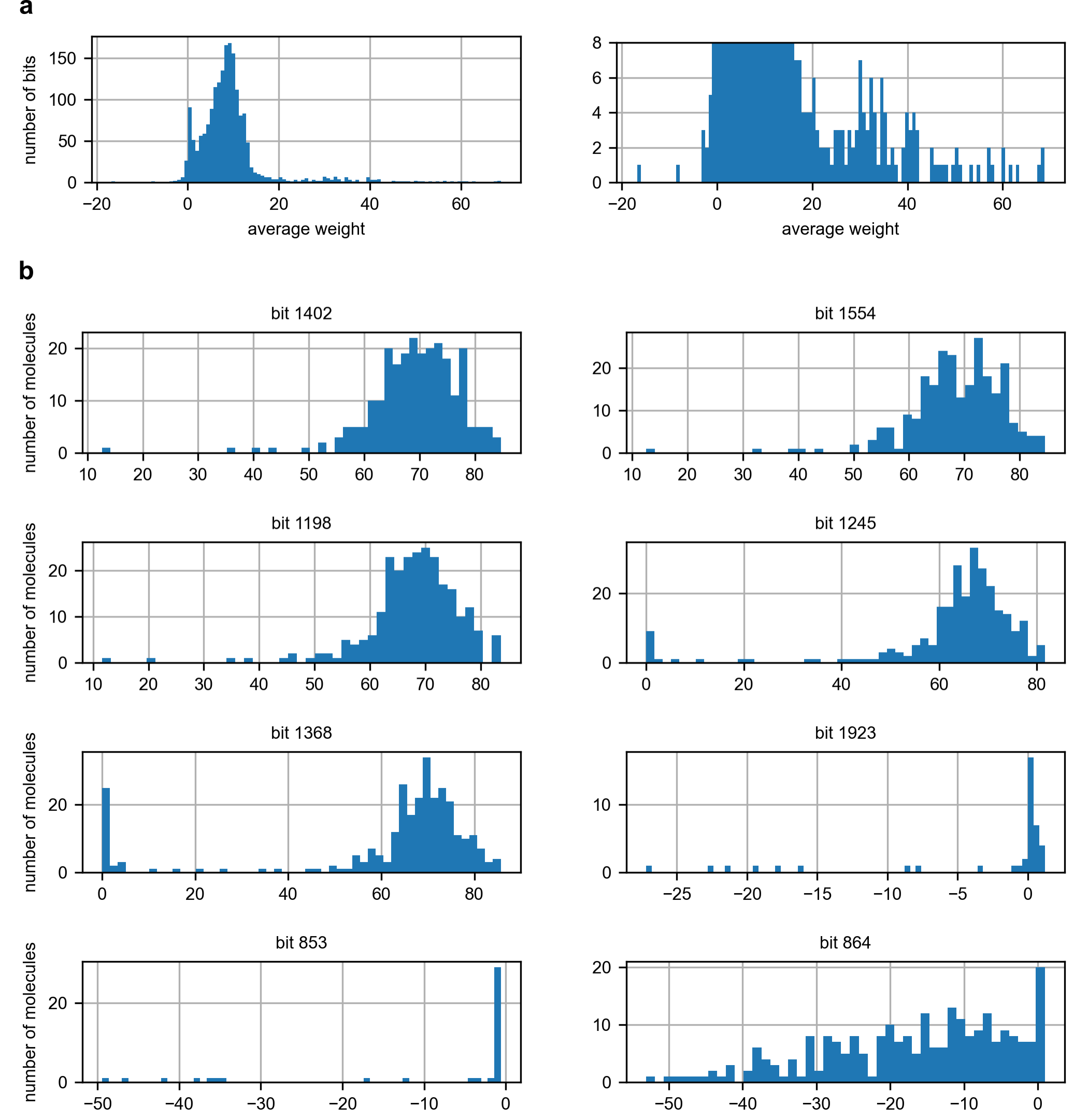}
    \caption{Histograms of \textbf{(a)} bit weights (only including bits set by at least one molecule in the dataset; plot shown in full and zoomed in) and \textbf{(b)} molecule-level bit weights for substructure-pair analysis on the triazine sEH dataset (random split, seed 0; \emph{cf.} Figure~\ref{fig:reps_models_loss_fns_splits}d), including the top 5 and bottom 3 bits. }
    \label{fig:sEH_twobits_hists}
\end{figure}

\begin{longtable}{m{1.5cm}|m{4.5cm}|m{8.5cm}} 
\caption{SMARTS for substructure-pair analysis on the triazine SIRT2 dataset (random split, seed 0).} \label{tbl:SMARTS_twobits_SIRT2} \\
bit ID & substructure example & SMARTS\\ \hline

1024 & \includegraphics[scale=1]{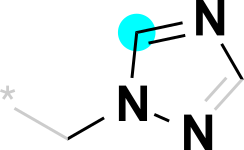} & [\#6;D2;H1;R1;+0]:[\#7;D2;H0;R1;+0]:\newline[\#6;D2;H1;R1;+0]:[\#7;D3;H0;R1;+0](-\newline[\#6;D2;H2;R0;+0]):[\#7;D2;H0;R1;+0] \\\hline

1024 & \includegraphics[scale=1]{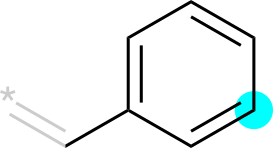} & [\#6;D2;H1;R0;+0]-[\#6;D3;H0;R1;+0]1:\newline[\#6;D2;H1;R1;+0]:[\#6;D2;H1;R1;+0]:\newline[\#6;D2;H1;R1;+0]:[\#6;D2;H1;R1;+0]:\newline[\#6;D2;H1;R1;+0]:1 \\\hline

330 & \includegraphics[scale=1]{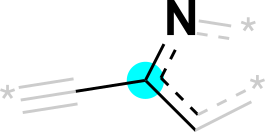} & [\#6;D2;H0;R0;+0]-[\#6;D3;H0;R1;+0](:\newline[\#6;D2;H1;R1;+0]):[\#7;D2;H0;R1;+0] \\\hline

330 & \includegraphics[scale=1]{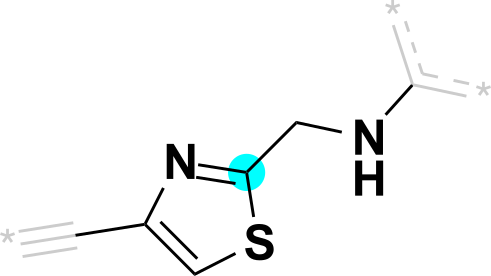} & [\#6;D2;H0;R0;+0]-[\#6;D3;H0;R1;+0]1:\newline[\#6;D2;H1;R1;+0]:[\#16;D2;H0;R1;+0]:\newline[\#6;D3;H0;R1;+0](-[\#6;D2;H2;R0;+0]-\newline[\#7;D2;H1;R0;+0]-[\#6;D3;H0;R1;+0]):\newline[\#7;D2;H0;R1;+0]:1 \\\hline

330 & \includegraphics[scale=1]{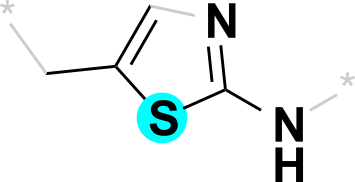} & [\#6;D2;H1;R1;+0]:[\#6;D3;H0;R1;+0](-\newline[\#6;D2;H2;R0;+0]):[\#16;D2;H0;R1;+0]:\newline[\#6;D3;H0;R1;+0](:[\#7;D2;H0;R1;+0])-\newline[\#7;D2;H1;R0;+0] \\\hline

1643 & \includegraphics[scale=1]{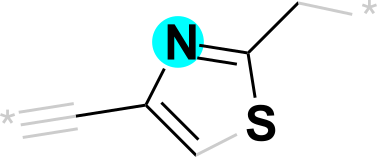} & [\#16;D2;H0;R1;+0]:[\#6;D3;H0;R1;+0](-\newline[\#6;D2;H2;R0;+0]):[\#7;D2;H0;R1;+0]:\newline[\#6;D3;H0;R1;+0](-[\#6;D2;H0;R0;+0]):\newline[\#6;D2;H1;R1;+0] \\\hline

1643 & \includegraphics[scale=1]{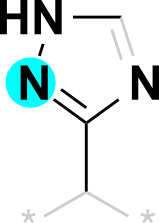} & [\#6;D2;H1;R1;+0]:[\#7;D2;H1;R1;+0]:\newline[\#7;D2;H0;R1;+0]:[\#6;D3;H0;R1;+0](-\newline[\#6;D3;H1;R0;+0]):[\#7;D2;H0;R1;+0] \\\hline

1643 & \includegraphics[scale=1]{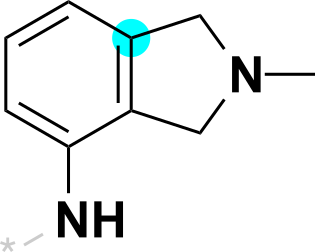} & [\#6;D1;H3;R0;+0]-[\#7;D3;H0;R1;+0]1-\newline[\#6;D2;H2;R1;+0]-[\#6;D3;H0;R2;+0]2:\newline[\#6;D2;H1;R1;+0]:[\#6;D2;H1;R1;+0]:\newline[\#6;D2;H1;R1;+0]:[\#6;D3;H0;R1;+0](-\newline[\#7;D2;H1;R0;+0]):[\#6;D3;H0;R2;+0]:2-\newline[\#6;D2;H2;R1;+0]-1 \\\hline

1272 & \includegraphics[scale=1]{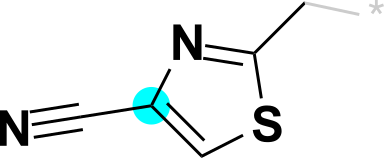} & [\#6;D2;H2;R0;+0]-[\#6;D3;H0;R1;+0]1:\newline[\#16;D2;H0;R1;+0]:[\#6;D2;H1;R1;+0]:\newline[\#6;D3;H0;R1;+0](-[\#6;D2;H0;R0;+0]\#\newline[\#7;D1;H0;R0;+0]):[\#7;D2;H0;R1;+0]:1 \\\hline

1272 & \includegraphics[scale=1]{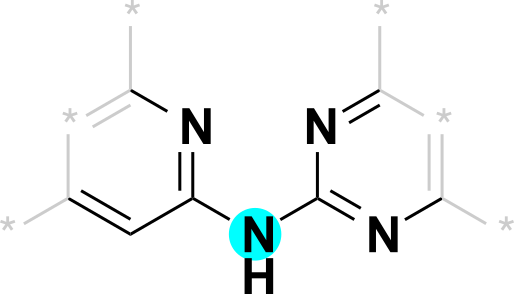} & [\#6;D3;H0;R1;+0]:[\#6;D2;H1;R1;+0]:\newline[\#6;D3;H0;R1;+0](:[\#7;D2;H0;R1;+0]:\newline[\#6;D3;H0;R1;+0])-[\#7;D2;H1;R0;+0]-\newline[\#6;D3;H0;R1;+0](:[\#7;D2;H0;R1;+0]:\newline[\#6;D3;H0;R1;+0]):[\#7;D2;H0;R1;+0]:\newline[\#6;D3;H0;R1;+0] \\\hline

1272 & \includegraphics[scale=1]{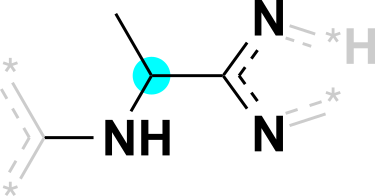} & [\#6;D1;H3;R0;+0]-[\#6;D3;H1;R0;+0](-\newline[\#7;D2;H1;R0;+0]-[\#6;D3;H0;R1;+0])-\newline[\#6;D3;H0;R1;+0](:[\#7;D2;H0;R1;+0]):\newline[\#7;D2;H0;R1;+0] \\\hline

1272 & \includegraphics[scale=1]{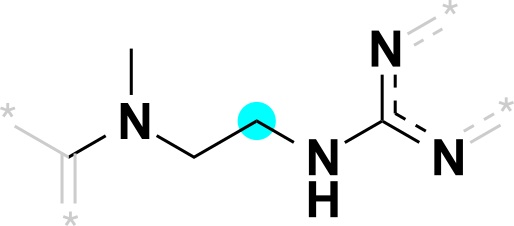} & [\#6;D1;H3;R0;+0]-[\#7;D3;H0;R0;+0](-\newline[\#6;D3;H0;R0;+0])-[\#6;D2;H2;R0;+0]-\newline[\#6;D2;H2;R0;+0]-[\#7;D2;H1;R0;+0]-\newline[\#6;D3;H0;R1;+0](:[\#7;D2;H0;R1;+0]):\newline[\#7;D2;H0;R1;+0] \\\hline

1404 & \includegraphics[scale=1]{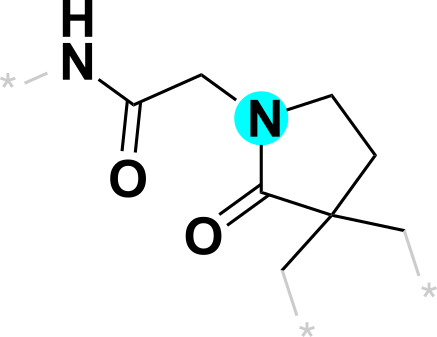} & [\#6;D2;H2;R1;+0]-[\#6;D4;H0;R2;+0]1(-\newline[\#6;D2;H2;R1;+0])-[\#6;D2;H2;R1;+0]-\newline[\#6;D2;H2;R1;+0]-[\#7;D3;H0;R1;+0](-\newline[\#6;D2;H2;R0;+0]-[\#6;D3;H0;R0;+0](-\newline[\#7;D2;H1;R0;+0])=[\#8;D1;H0;R0;+0])-\newline[\#6;D3;H0;R1;+0]-1=[\#8;D1;H0;R0;+0] \\\hline

1404 & \includegraphics[scale=1]{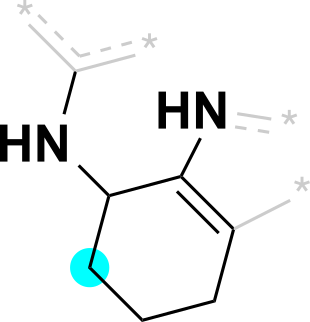} & [\#6;D3;H0;R1;+0]-[\#7;D2;H1;R0;+0]-\newline[\#6;D3;H1;R1;+0]1-[\#6;D2;H2;R1;+0]-\newline[\#6;D2;H2;R1;+0]-[\#6;D2;H2;R1;+0]-\newline[\#6;D3;H0;R2;+0]:[\#6;D3;H0;R2;+0]-1:\newline[\#7;D2;H1;R1;+0] \\\hline

121 & \includegraphics[scale=1]{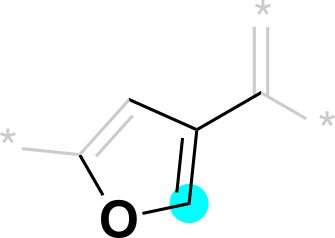} & [\#6;D2;H1;R1;+0]:[\#6;D3;H0;R1;+0](-\newline[\#6;D3;H0;R0;+0]):[\#6;D2;H1;R1;+0]:\newline[\#8;D2;H0;R1;+0]:[\#6;D3;H0;R1;+0] \\\hline

1187 & \includegraphics[scale=1]{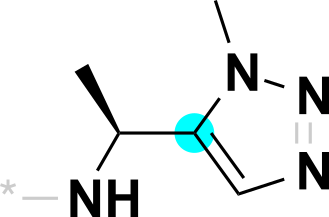} & [\#6;D1;H3;R0;+0]-[\#6;D3;H1;R0;+0](-\newline[\#7;D2;H1;R0;+0])-[\#6;D3;H0;R1;+0](:\newline[\#6;D2;H1;R1;+0]:[\#7;D2;H0;R1;+0]):\newline[\#7;D3;H0;R1;+0](-[\#6;D1;H3;R0;+0]):\newline[\#7;D2;H0;R1;+0] \\\hline

234 & \includegraphics[scale=1]{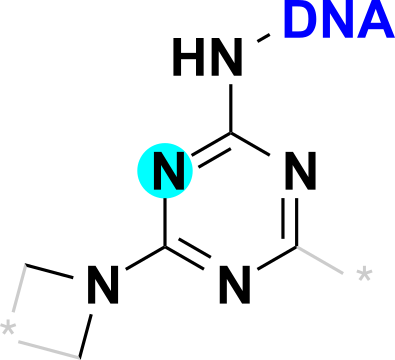} & [\#102;D1;H0;R0;+0]-[\#7;D2;H1;R0;+0]-\newline[\#6;D3;H0;R1;+0]1:[\#7;D2;H0;R1;+0]:\newline[\#6;D3;H0;R1;+0]:[\#7;D2;H0;R1;+0]:\newline[\#6;D3;H0;R1;+0](-[\#7;D3;H0;R1;+0](-\newline[\#6;D2;H2;R1;+0])-[\#6;D2;H2;R1;+0]):\newline[\#7;D2;H0;R1;+0]:1 \\\hline

234 & \includegraphics[scale=1]{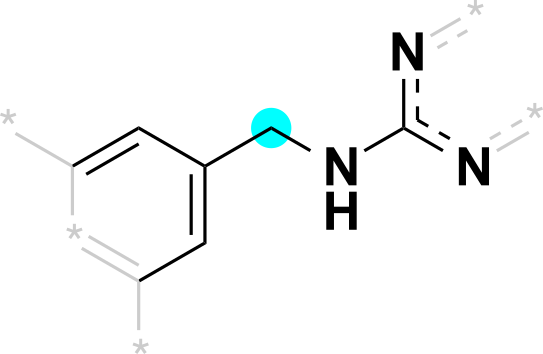} & [\#6;D3;H0;R1;+0]:[\#6;D2;H1;R1;+0]:\newline[\#6;D3;H0;R1;+0](:[\#6;D2;H1;R1;+0]:\newline[\#6;D3;H0;R1;+0])-[\#6;D2;H2;R0;+0]-\newline[\#7;D2;H1;R0;+0]-[\#6;D3;H0;R1;+0](:\newline[\#7;D2;H0;R1;+0]):[\#7;D2;H0;R1;+0] \\\hline

234 & \includegraphics[scale=1]{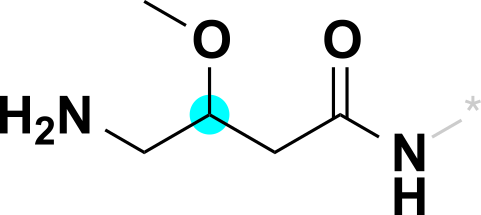} & [\#6;D1;H3;R0;+0]-[\#8;D2;H0;R0;+0]-\newline[\#6;D3;H1;R0;+0](-[\#6;D2;H2;R0;+0]-\newline[\#7;D1;H2;R0;+0])-[\#6;D2;H2;R0;+0]-\newline[\#6;D3;H0;R0;+0](-[\#7;D2;H1;R0;+0])=\newline[\#8;D1;H0;R0;+0] \\\hline

\end{longtable}

\begin{figure} [H]
    \centering
        \includegraphics[scale=1]{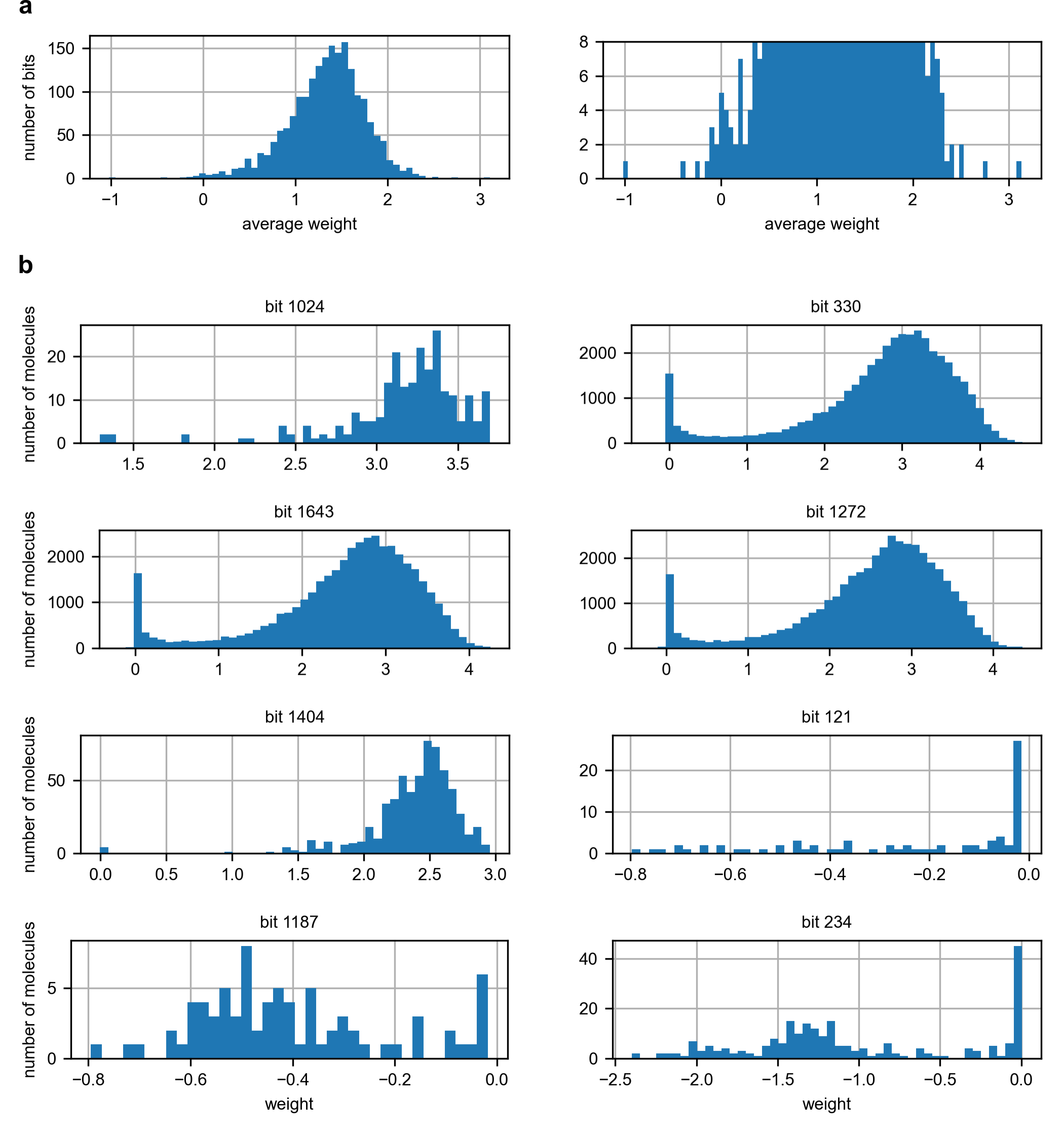}
    \caption{Histograms of \textbf{(a)} bit weights (only including bits set by at least one molecule in the dataset; plot shown in full and zoomed in) and \textbf{(b)} molecule-level bit weights for substructure-pair analysis on the triazine SIRT2 dataset (random split, seed 0; \emph{cf.} Figure~\ref{fig:reps_models_loss_fns_splits}d), including the top 5 and bottom 3 bits. }
    \label{fig:SIRT2_twobits_hists}
\end{figure}

\subsection{Parity plots to evaluate ability to generalize on the triazine sEH and SIRT2 datasets}

\begin{figure}[H]
    \centering
        \includegraphics[scale=1]{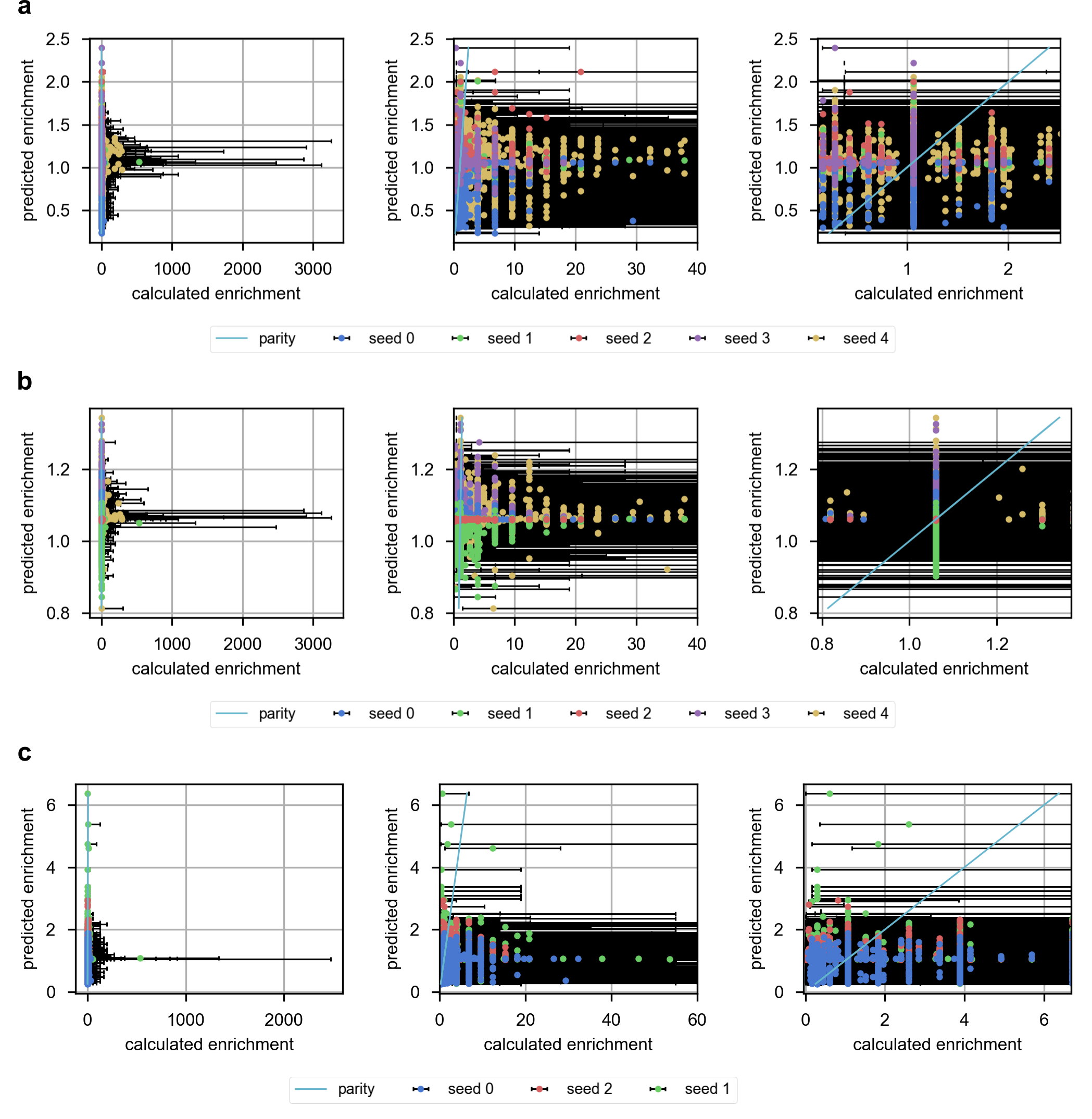}
    \caption{Scatter plots (full and zoomed-in) of predicted versus calculated enrichments for \textbf{(a)} FP-FFNN, \textbf{(b)} OH-FFNN, \textbf{(c)} D-MPNN on five cycle-1+2+3 splits (\emph{cf.} Figure~\ref{fig:reps_models_loss_fns_splits}d) of the triazine sEH dataset. Only the subset of test-set compounds with all building blocks in the test set are included for each split. The light blue parity line is the identity function, for reference. Error bars represent 95\% confidence intervals for calculated enrichments.}
    \label{fig:sEH_c123_parity_plots}
\end{figure}

\begin{figure}[H]
    \centering
        \includegraphics[scale=1]{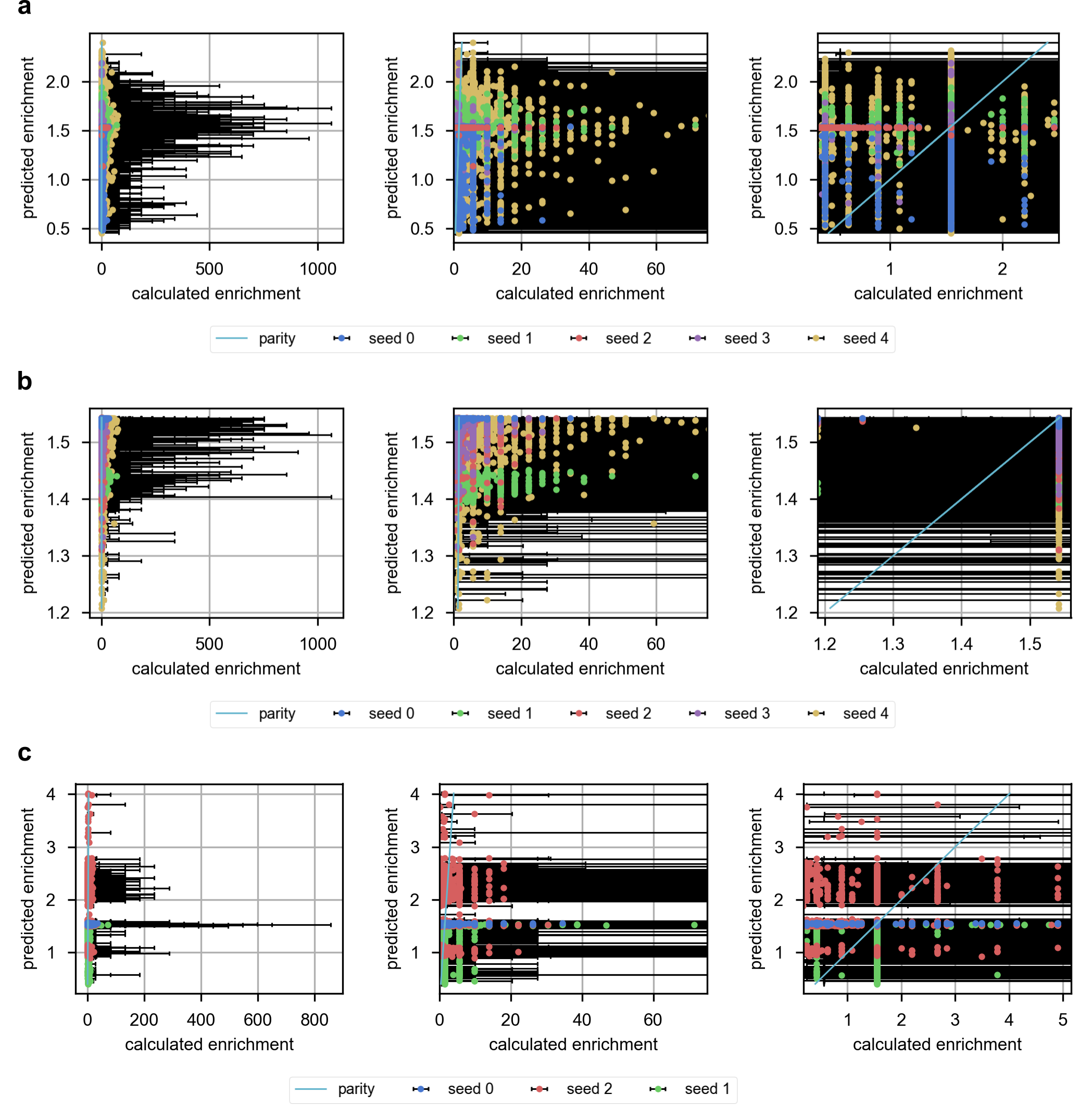}
    \caption{Scatter plots (full and zoomed-in) of predicted versus calculated enrichments for \textbf{(a)} FP-FFNN, \textbf{(b)} OH-FFNN, \textbf{(c)} D-MPNN on five cycle-1+2+3 splits (\emph{cf.} Figure~\ref{fig:reps_models_loss_fns_splits}d) of the triazine SIRT2 dataset. Only the subset of test-set compounds with all building blocks in the test set are included for each split. The light blue parity line is the identity function, for reference. Error bars represent 95\% confidence intervals for calculated enrichments.}
    \label{fig:SIRT2_c123_parity_plots}
\end{figure}

\subsection{Atom-centered Gaussian visualizations for top predicted compounds}

\subsubsection{Random split}

\begin{figure} [H]
    \centering
        \includegraphics[scale=1]{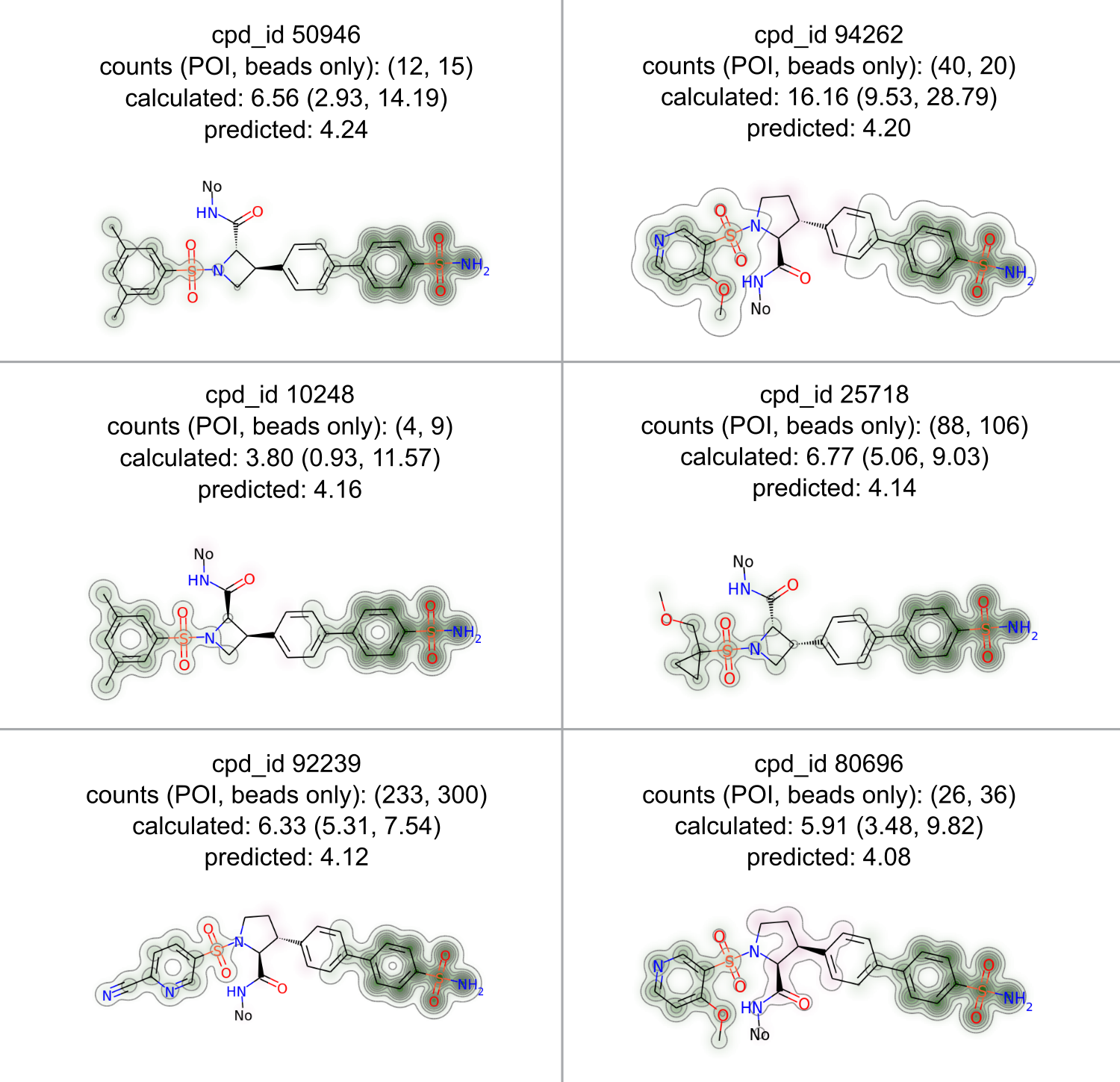}
    \caption{Atom-centered Gaussian visualizations for the top 6 predicted compounds from the test set of a FP-FFNN on a random split (seed 0; \emph{cf.} Figure~\ref{fig:reps_models_loss_fns_splits}d) of the DD1S CAIX dataset. Atoms contributing positively to enrichment are highlighted in green, and atoms contributing negatively to enrichment are highlighted in pink, with color intensity corresponding to the level of contribution to enrichment. ``No'' represents the DNA linker attachment point. Compound IDs (``cpd\_id'') are sequential based on building block cycle numbers.} 
    \label{fig:CAIX_top_predicted_cpds}
\end{figure}

\begin{figure} [H]
    \centering
        \includegraphics[scale=1]{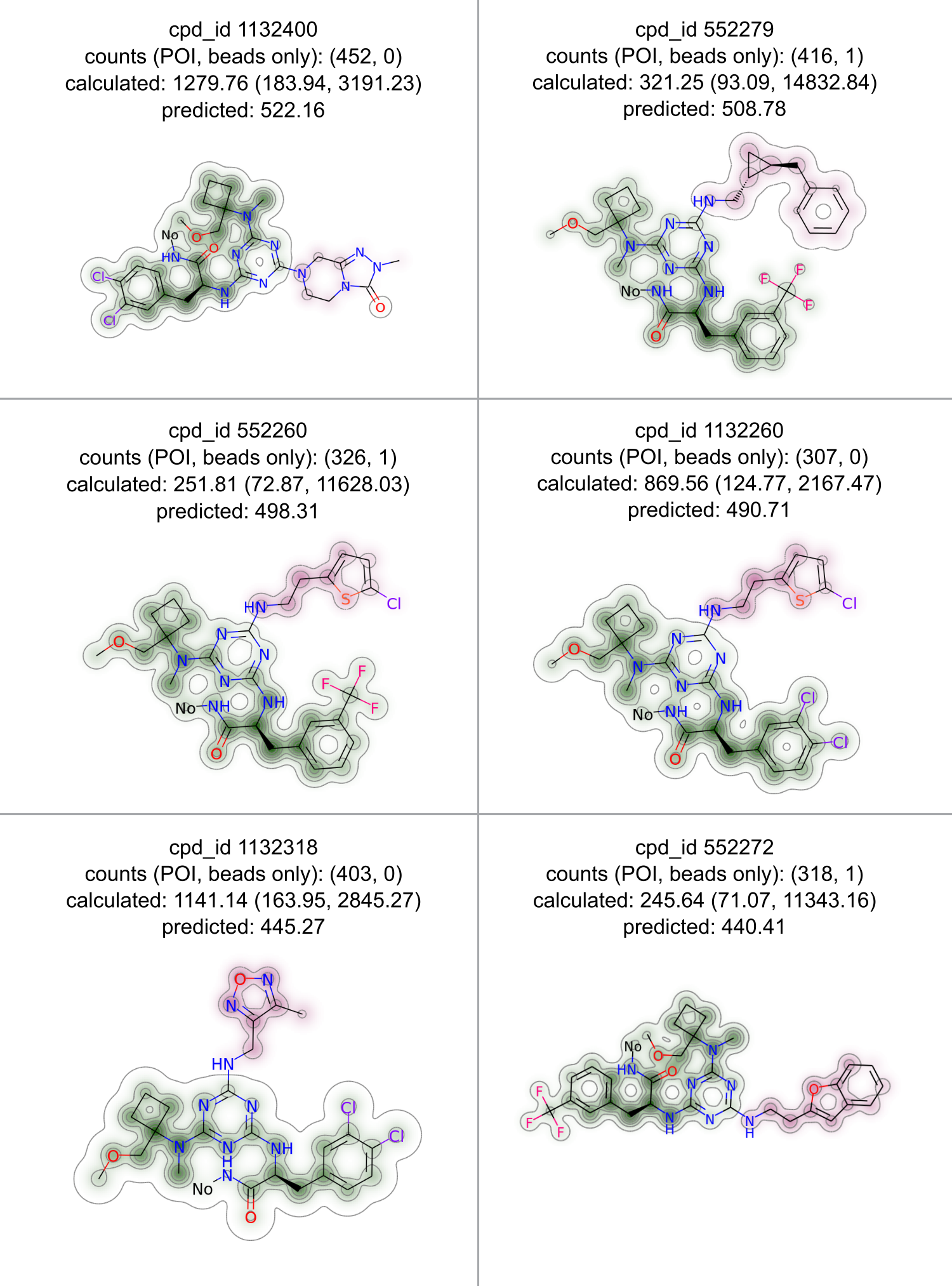}
    \caption{Atom-centered Gaussian visualizations for the top 6 predicted compounds from the test set of a FP-FFNN on a random split (seed 0; \emph{cf.} Figure~\ref{fig:reps_models_loss_fns_splits}d) of the triazine sEH dataset. Atoms contributing positively to enrichment are highlighted in green, and atoms contributing negatively to enrichment are highlighted in pink, with color intensity corresponding to the level of contribution to enrichment. ``No'' represents the DNA linker attachment point. Compound IDs (``cpd\_id'') are sequential based on building block cycle numbers.}
    \label{fig:sEH_top_predicted_cpds}
\end{figure}

\begin{figure} [H]
    \centering
        \includegraphics[scale=1]{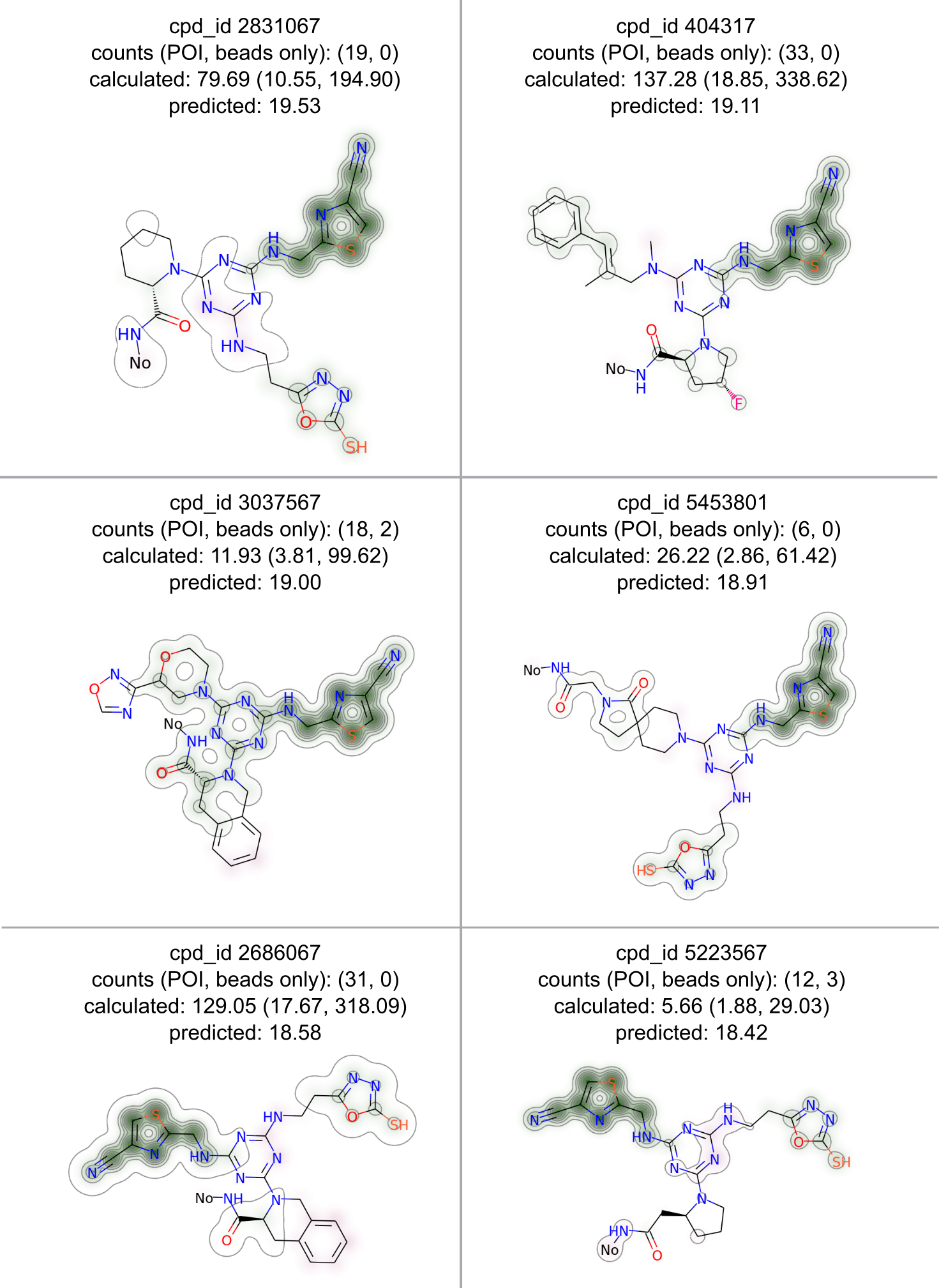}
    \caption{Atom-centered Gaussian visualizations for the top 6 predicted compounds from the test set of a FP-FFNN on a random split (seed 0; \emph{cf.} Figure~\ref{fig:reps_models_loss_fns_splits}d) of the triazine SIRT2 dataset. Atoms contributing positively to enrichment are highlighted in green, and atoms contributing negatively to enrichment are highlighted in pink, with color intensity corresponding to the level of contribution to enrichment. ``No'' represents the DNA linker attachment point. Compound IDs (``cpd\_id'') are sequential based on building block cycle numbers.}
    \label{fig:SIRT2_top_predicted_cpds}
\end{figure}

\subsubsection{Cycle-1+2+3 split}

\begin{figure} [H]
    \centering
        \includegraphics[scale=1]{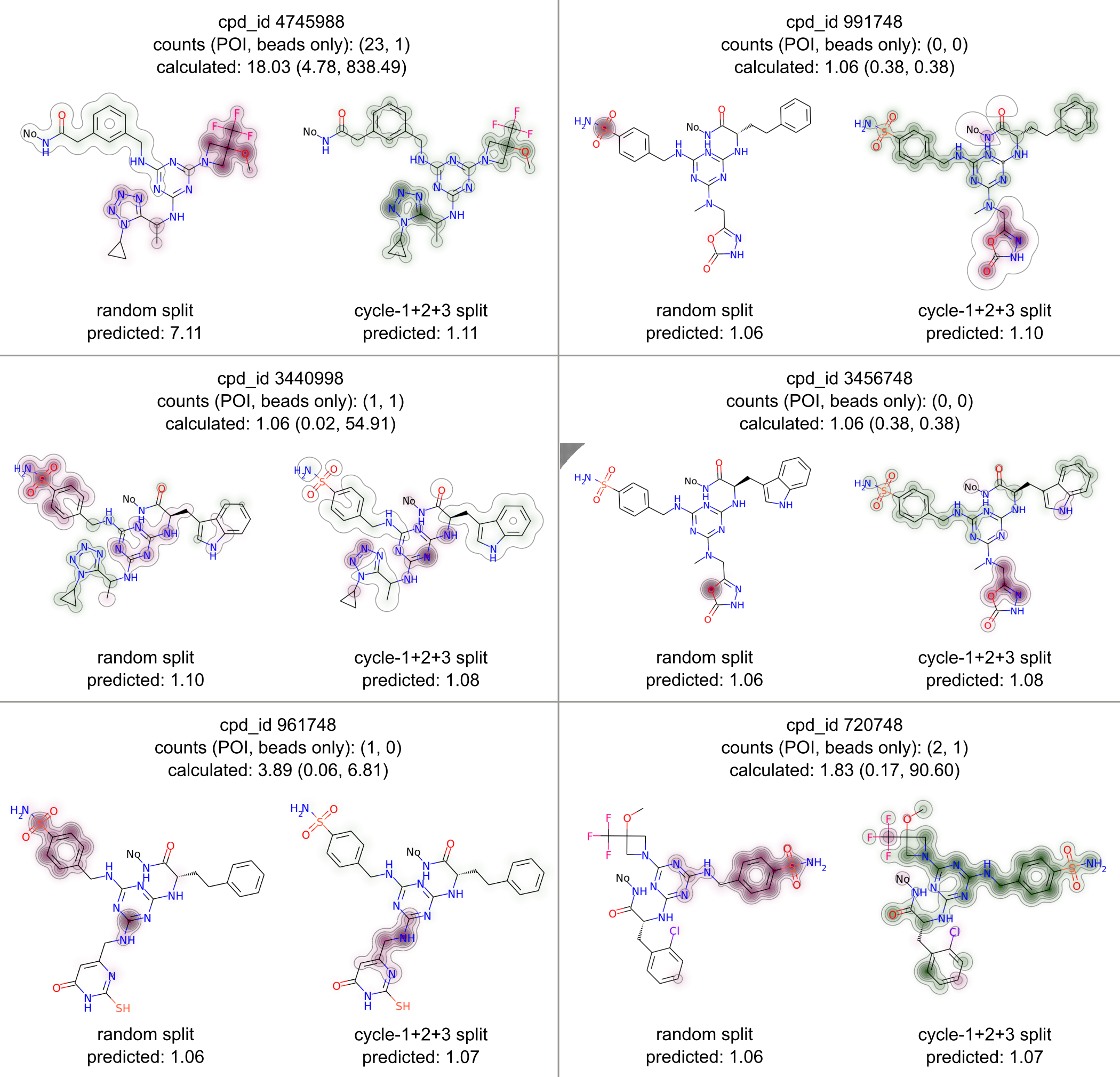}
    \caption{Atom-centered Gaussian visualizations for the top 6 predicted compounds with all building blocks uniquely in the test set for a FP-FFNN on a cycle-1+2+3 split (seed 0; \emph{cf.} Figure~\ref{fig:reps_models_loss_fns_splits}d) of the triazine sEH dataset. These compounds do not contain any of the building blocks the cycle-1+2+3 split model was trained on. For comparison, corresponding visualizations based on the predictions of a FP-FFNN on a random split (seed 0; \emph{cf.} Figure~\ref{fig:reps_models_loss_fns_splits}d) of the dataset are also shown. Atoms contributing positively to enrichment are highlighted in green, and atoms contributing negatively to enrichment are highlighted in pink, with color intensity corresponding to the level of contribution to enrichment. ``No'' represents the DNA linker attachment point. Compound IDs (``cpd\_id'') are sequential based on building block cycle numbers.}
    \label{fig:sEH_new_BB_cpds_full_vis}
\end{figure}

\begin{figure} [H]
    \centering
        \includegraphics[scale=1]{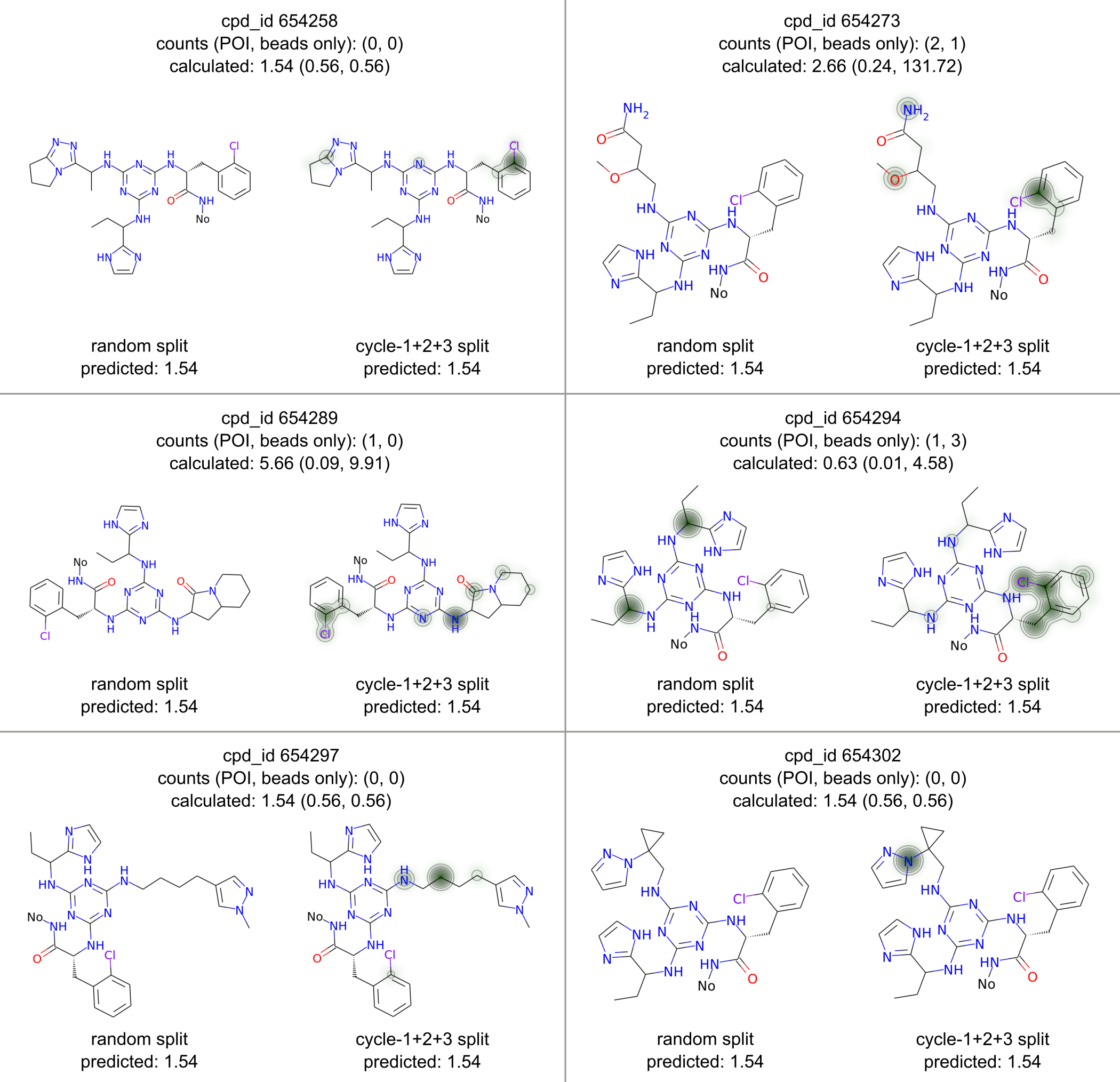}
    \caption{Atom-centered Gaussian visualizations for the top 6 predicted compounds with all building blocks uniquely in the test set for a FP-FFNN on a cycle-1+2+3 split (seed 0; \emph{cf.} Figure~\ref{fig:reps_models_loss_fns_splits}d) of the triazine SIRT2 dataset. These compounds do not contain any of the building blocks the cycle-1+2+3 split model was trained on. For comparison, corresponding visualizations based on the predictions of a FP-FFNN on a random split (seed 0; \emph{cf.} Figure~\ref{fig:reps_models_loss_fns_splits}d) of the dataset are also shown. Atoms contributing positively to enrichment are highlighted in green, and atoms contributing negatively to enrichment are highlighted in pink, with color intensity corresponding to the level of contribution to enrichment. ``No'' represents the DNA linker attachment point. Compound IDs (``cpd\_id'') are sequential based on building block cycle numbers.}
    \label{fig:SIRT2_new_BB_cpds_full_vis}
\end{figure}

\subsection{Chemical similarity between Enamine on-demand libraries and top compounds from DD1S CAIX dataset}

\begin{longtable}{m{6cm}|m{6cm}|m{6cm}}
\caption{Top hits from similarity searches in the Enamine REAL database and Enamine screening collection, for the top 5 compounds (by lower bound of calculated enrichment) in the DD1S CAIX dataset. ``No'' represents the DNA linker attachment point.} \label{tbl:enamine_comparison} \\
DD1S compound & nearest two hits in Enamine REAL database & nearest two hits in Enamine\newline screening collection\\ \hline

\includegraphics[scale=1]{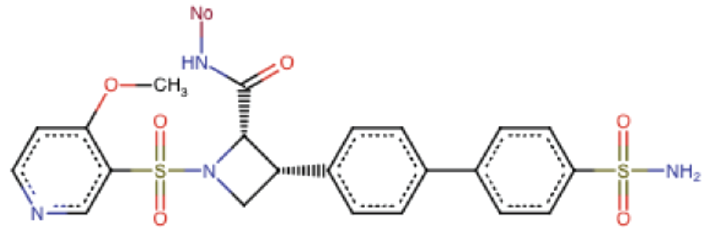} & \includegraphics[scale=1]{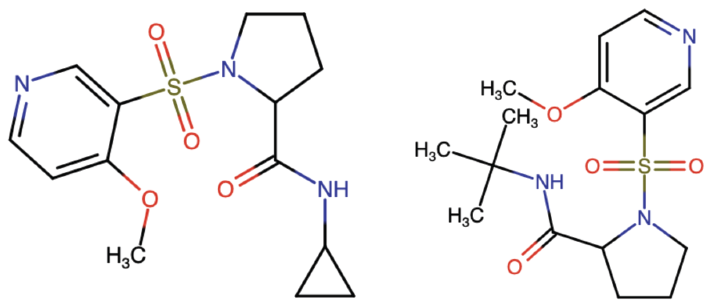} & \includegraphics[scale=1]{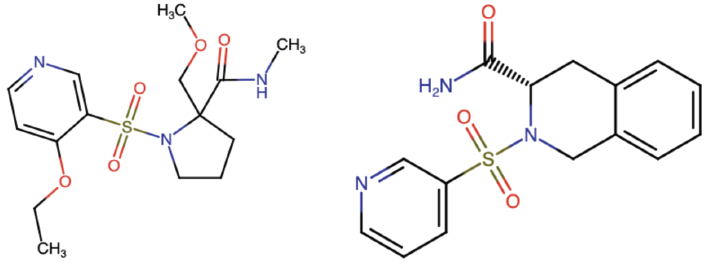}\\ \hline

\includegraphics[scale=1]{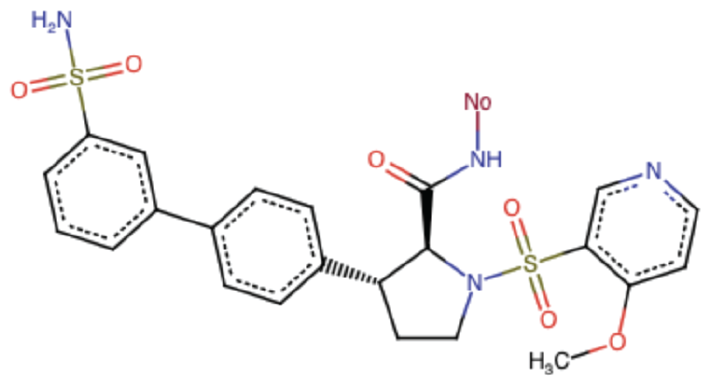} & \includegraphics[scale=1]{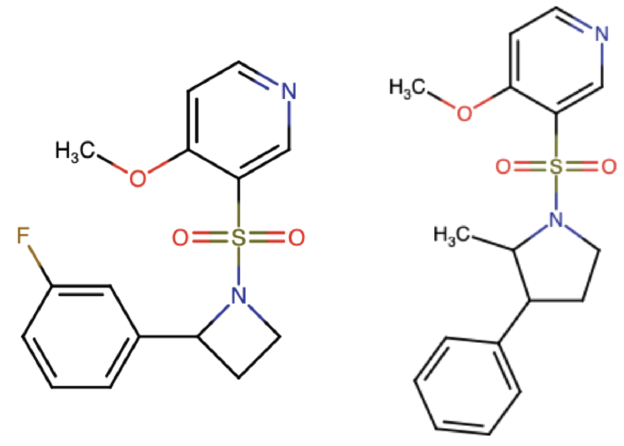} & \includegraphics[scale=1]{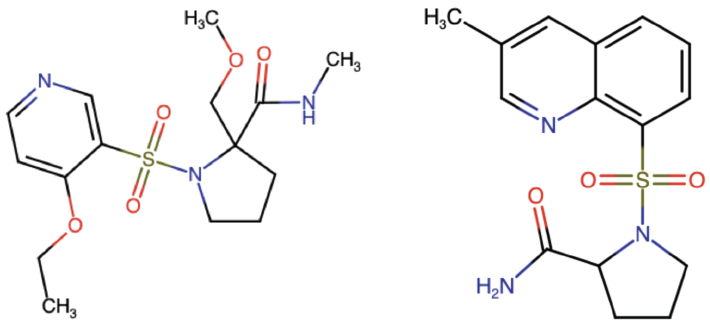}\\ \hline

\includegraphics[scale=1]{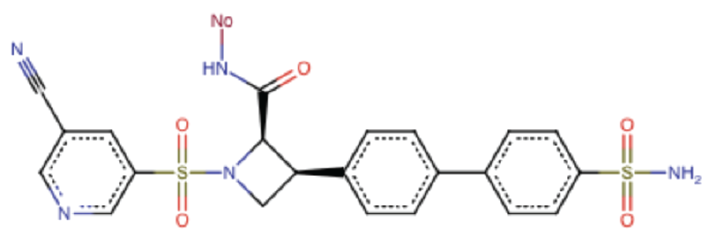} & \includegraphics[scale=1]{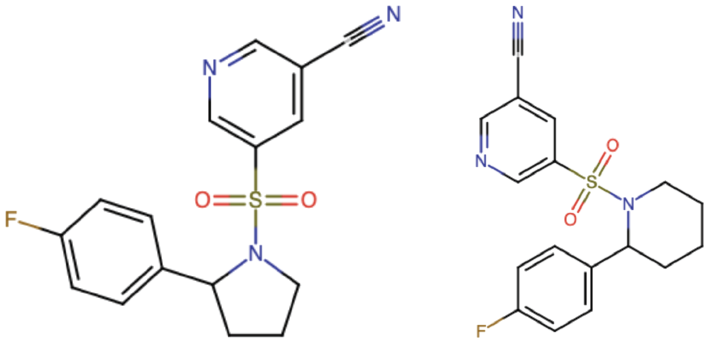} & \includegraphics[scale=1]{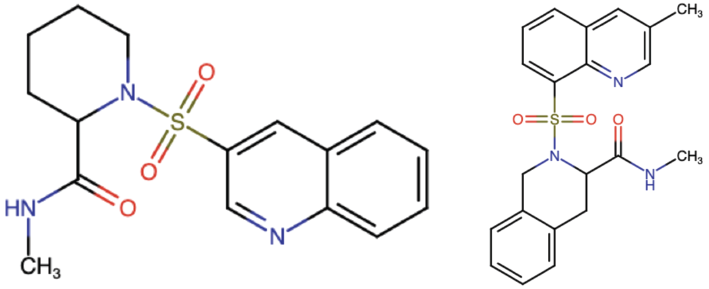}\\ \hline

\includegraphics[scale=1]{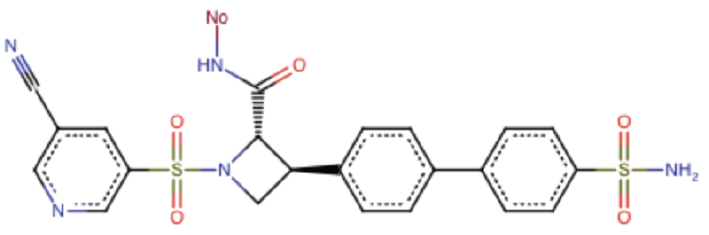} & \includegraphics[scale=1]{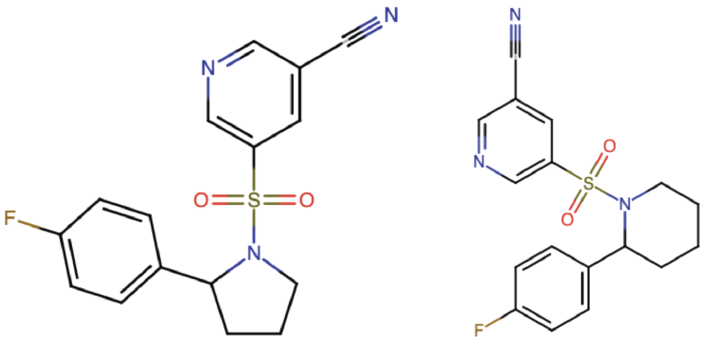} & \includegraphics[scale=1]{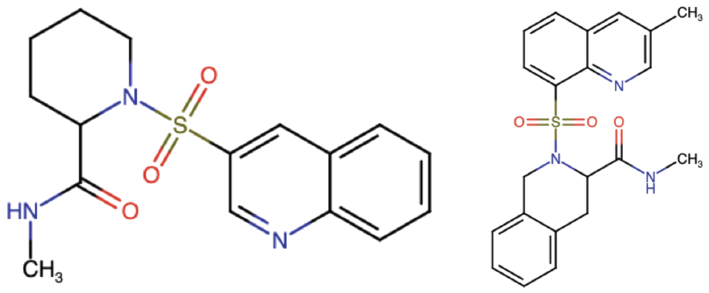}\\ \hline

\includegraphics[scale=1]{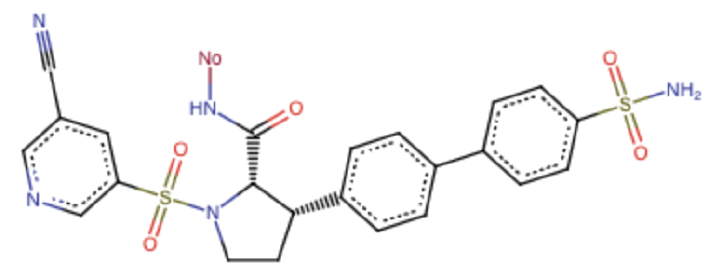} & \includegraphics[scale=1]{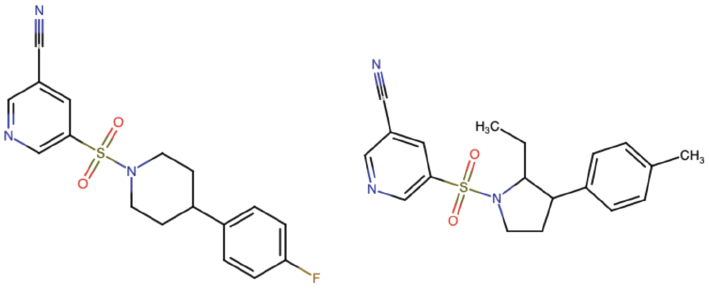} & \includegraphics[scale=1]{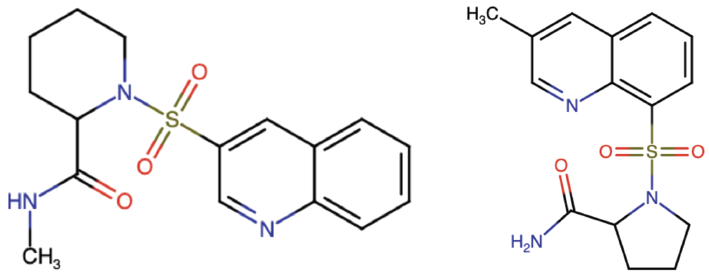}\\ \hline

\end{longtable}

\subsection{DD1S CAIX outliers}

\begin{longtable}{m{6cm}|m{6cm}}
\caption{Example outliers (with high calculated enrichment but low predicted enrichment) and their nearest neighbors (as determined by a FP-KNN model trained on the entire DD1S CAIX dataset) in the test set of a FP-FFNN on a random split (seed 0) of the DD1S CAIX dataset. Compound IDs (``cpd\_id'') are sequential based on building block cycle numbers.} \label{tbl:DD1S_CAIX_outliers} \\
outlier & nearest neighbor\\ \hline

\includegraphics[scale=1]{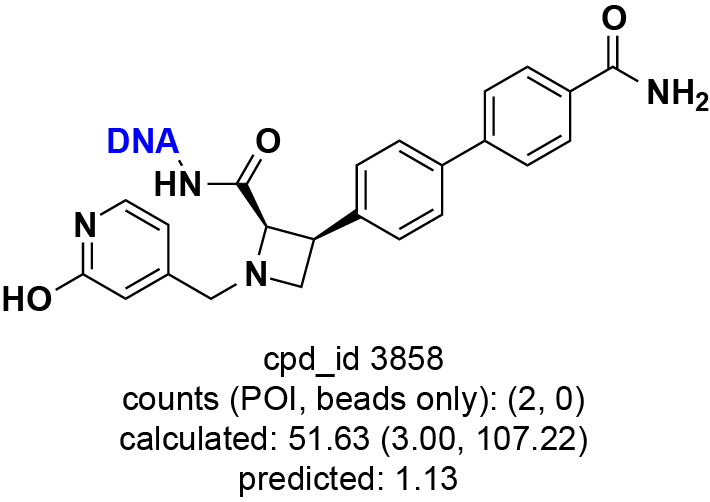} & \includegraphics[scale=1]{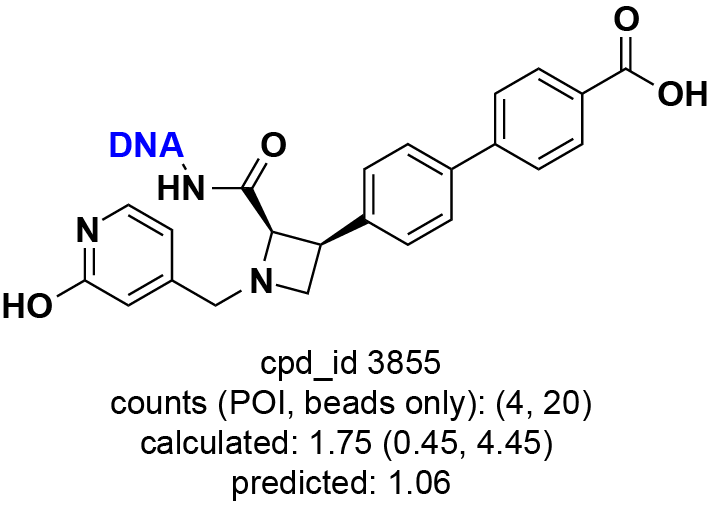}\\ \hline

\includegraphics[scale=1]{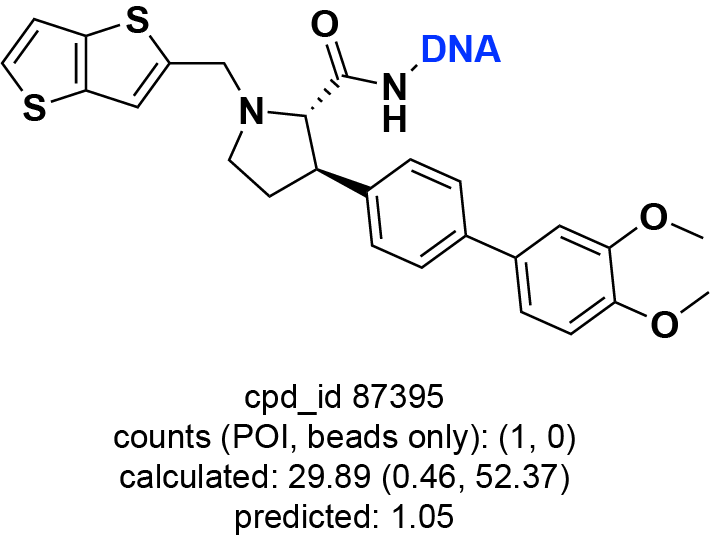} & \includegraphics[scale=1]{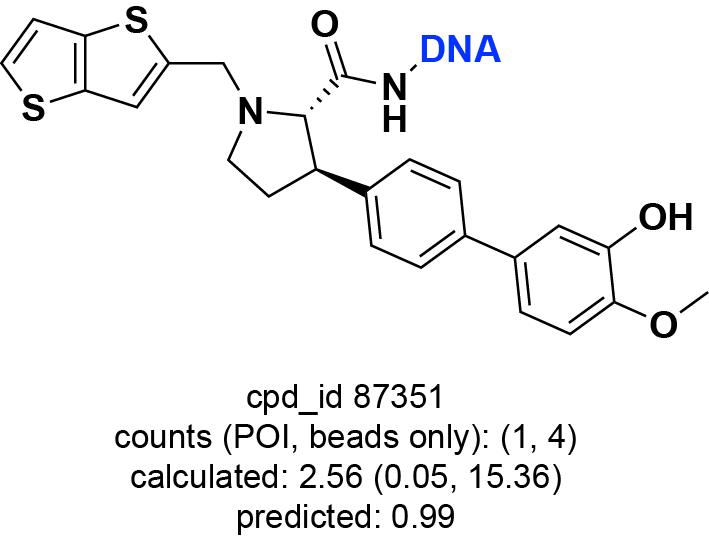}\\ \hline

\includegraphics[scale=1]{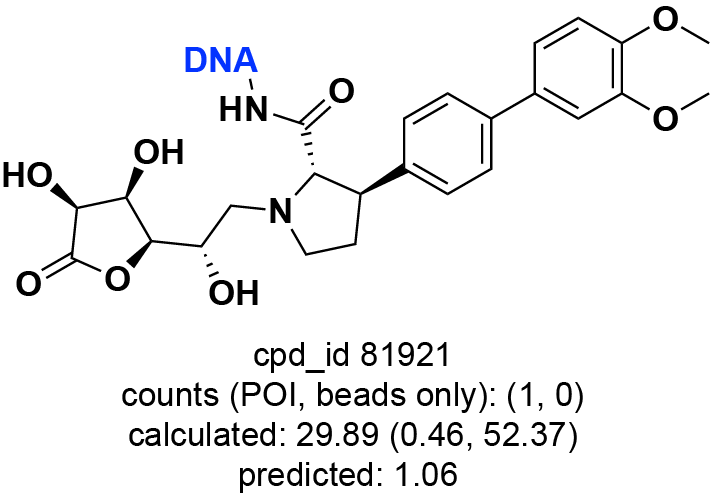} & \includegraphics[scale=1]{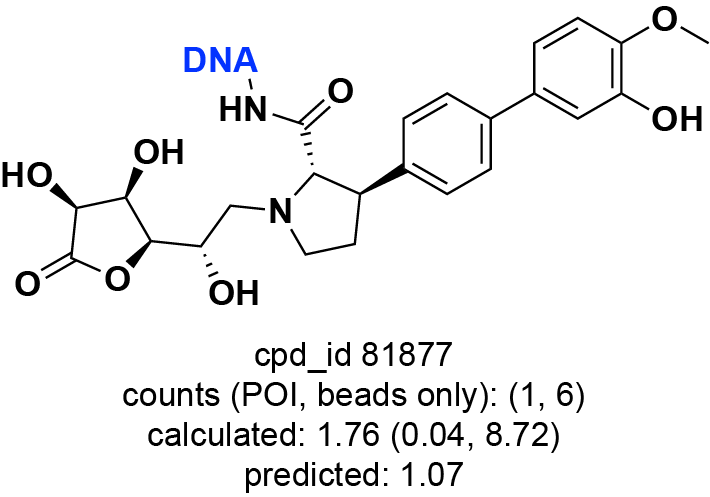}\\ \hline

\includegraphics[scale=1]{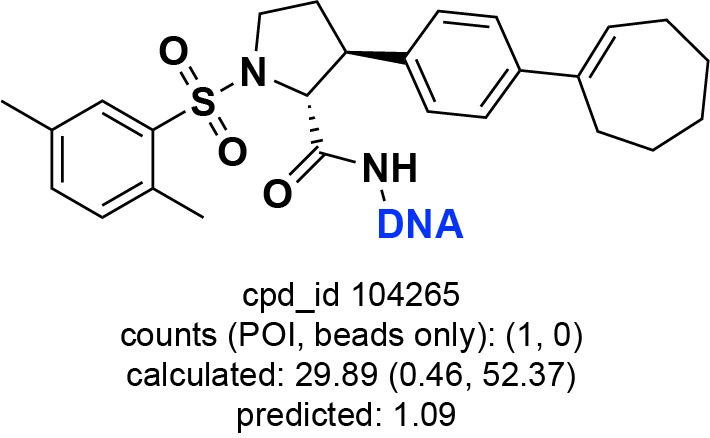} & \includegraphics[scale=1]{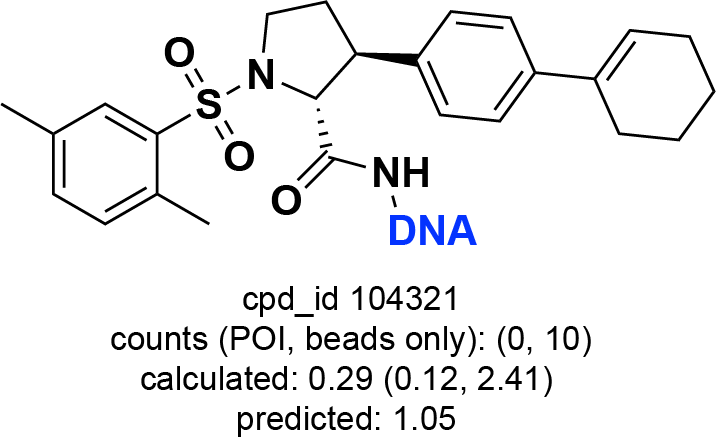}\\ \hline

\includegraphics[scale=1]{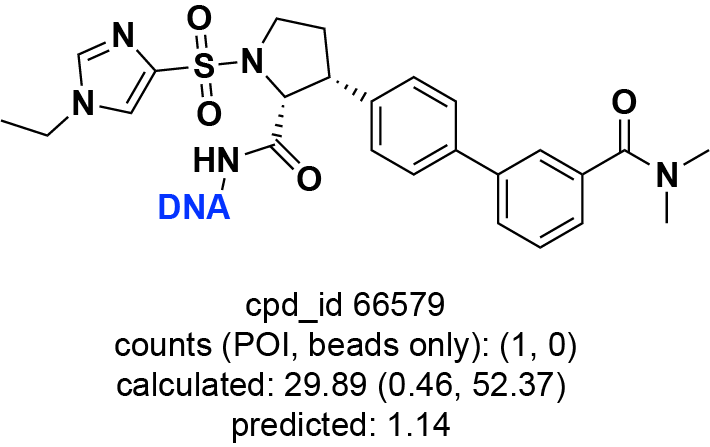} & \includegraphics[scale=1]{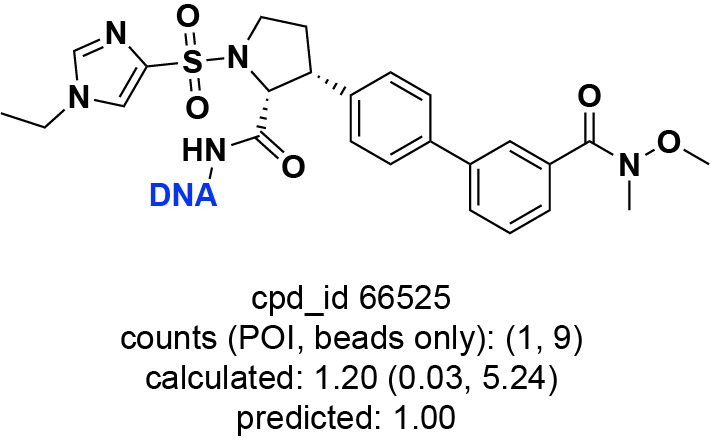}\\ \hline

\end{longtable}

\bibliography{main}